\documentclass[aps,pre,twocolumn,amssymb,amsmath,nofootinbib]{revtex4}
\usepackage{graphicx,latexsym,pifont}
\usepackage{times}
\usepackage{tikz}
\newcommand{\sgn}{{\mathrm{sgn}}}
\newcommand{\rme}{{\mathrm{e}}}
\newcommand{\rmd}{{\mathrm{d}}}

\newcommand{\tr}{{\mathrm{tr}}}
\newcommand{\nn}{\nonumber}
\newcommand {\Eq}[1]{Eq.\hspace{0.55ex}(\ref{#1})}

\newcommand{\fig}[2]{\includegraphics[width=#1]{./figures/#2}}
\newcommand{\Fig}[1]{\includegraphics[width=8.7cm]{./figures/#1}}

\newlength{\bilderlength}
\newcommand{\bilderscale}{0.35}

\newcommand{\bilderskip}{\hspace*{0.8ex}}

\newcommand{\diagram}[1]{\settowidth{\bilderlength}{\bilderskip\includegraphics[scale=\bilderscale]{./figures/#1}\bilderskip}\parbox{\bilderlength}{\bilderskip\includegraphics[scale=\bilderscale]{./figures/#1}\bilderskip}}

\newcommand{\calP}{{\cal P}}

\newcommand{\Tr}{\mathrm{tr}}

\def\be{\begin{equation}}
\def\ee{\end{equation}}

\def\bal{\begin{align}}
\def\eal{\end{align}}

\def\bea{\begin{eqnarray}}
\def\eea{\end{eqnarray}}
\renewcommand{\log}{\ln }
\newcommand{\du}{{\dot u}}

\usepackage{color}

\begin{document}

\bibliographystyle{KAY}

\title{Avalanche dynamics of elastic interfaces{\parbox{0mm}{\raisebox{5mm}[0mm][0mm]{~~~~~~~~~~~~~~~~~~~~~~~~~~~~~\normalsize LPTENS-13/02}}}}

\author{Pierre Le Doussal and Kay J\"org Wiese}
\address{CNRS-Laboratoire de Physique Th\'eorique de l'Ecole
Normale Sup\'erieure, 24 rue Lhomond, 75005 Paris, France
}

\begin{abstract}
Slowly driven elastic interfaces, such as domain walls in dirty magnets, contact lines  wetting a non-homogenous substrate, or cracks  in brittle disordered material  proceed via  intermittent motion, called avalanches. Here we develop a field-theoretic treatment to calculate, from first principles, 
the space-time statistics of instantaneous velocities within an avalanche. For elastic interfaces at (or above) their (internal) upper critical  dimension $d \geq d_{\rm uc}$ 
($d_{\rm uc}=2,4$ respectively for long-ranged and short-ranged elasticity) we show that the field theory for the {\em center of mass} reduces to the motion 
of a {\em point particle} in a random-force landscape, which is itself a random walk (ABBM model). Furthermore, the full {\it spatial} dependence of the
velocity correlations is described by the Brownian-force model (BFM) where each point of the interface sees an independent Brownian-force landscape. Both ABBM and BFM can be solved exactly in any dimension $d$ (for monotonous driving) by summing tree graphs, 
equivalent to solving a (non-linear) {\em instanton} equation. We focus on the limit of slow uniform driving. 
This tree approximation is the mean-field theory (MFT) for
realistic interfaces in short-ranged disorder, up to the renormalization of two parameters at $d=d_{\rm uc}$. We calculate 
a number of observables of direct experimental interest:  Both for the center of mass, and 
for a given Fourier mode $q$, we obtain various correlations and probability distribution functions (PDF's) of the velocity inside an avalanche, as well as the avalanche shape 
and its fluctuations (second shape).
Within MFT we find that velocity correlations at non-zero $q$ are asymmetric under time reversal. Next we calculate, beyond MFT, i.e.\ including loop corrections, 
the 1-time PDF of the center-of-mass velocity $\dot u$ for dimension $d< d_{\rm uc}$. 
The singularity at small velocity \({\cal P}(\dot u)\sim 1/\dot u^{\sf a}\) is substantially reduced from $\sf a=1$ (MFT) 
to \({\sf a} = 1 - \frac{2}{9} (4-d) + ...\)  (short-ranged elasticity) and \({\sf a} = 1 - \frac{4}{9} (2-d) + ...\)  (long-ranged elasticity). We show how
the dynamical theory recovers the avalanche-size distribution, and how the instanton  relates to the response to an infinitesimal step in the force.
\end{abstract}

\maketitle

\section{Introduction}\label{s:Introduction}

Elastic interfaces driven through a disordered medium have been proposed as efficient mesoscopic models for a number of different physical systems and situations, such as the motion of  domain walls in soft magnets  \cite{Barkhausen1919,AlessandroBeatriceBertottiMontorsi1990,AlessandroBeatriceBertottiMontorsi1990b,UrbachMadisonMarkert1995,Colaiori2008,ZapperiCizeauDurinStanley1998,DurinZapperi2000,LemerleFerreChappertMathetGiamarchiLeDoussal1998}, fluid contact lines on a rough surface \cite{LeDoussalWieseMoulinetRolley2009,MoulinetGuthmannRolley2002,RolleyGuthmannGombrowiczRepain1998}, or strike-slip faults in geophysics \cite{FisherDahmenRamanathanBenZion1997,DSFisher1998,BenZionRice1993,BenZionRice1997}. Their response to external driving is not smooth, but exhibits discontinuous and collective \textit{jumps} called \textit{avalanches} which extend over a broad range of space and time scales. Physically, these are detected e.g.~as pulses of Barkhausen noise in magnets \cite{Barkhausen1919,UrbachMadisonMarkert1995,KimChoeShin2003,RepainBauerJametFerreMouginChappertBernas2004,DurinZapperi2006b}, slip instabilities leading to earthquakes on geological faults \cite{Ruina1983,Dieterich1992,Scholz1998,CizeauZapperiDurinStanley1997,Colaiori2008,FisherDahmenRamanathanBenZion1997}, or in fracture experiments \cite{SchmittbuhlMaloy1997,LenglineToussaintSchmittbuhlElkhouryAmpueroTallakstadSantucciMaaloy2011,TallakstadToussaintSantucciSchmittbuhlMaaloy2011,SantucciMaaloyDelaplaceMathiesenHansenHaavigBakkeSchmittbuhlVanelRay2007,MaaloySchmittbuhl2001,BonamyPonsonPradesBouchaudGuillot2006,BonamySantucciPonson2008,Ponson2007,Ponson2008,PonsonBonamyBouchaud2006,PonsonBonamyBouchaud2007}. While the microscopic details of the dynamics are specific to each system,
an important question is whether the large-scale features are universal \cite{SethnaDahmenMyers2001}. The most prominent example are the exponents 
of the power-law distribution of avalanche sizes $P(S) \sim S^{-\tau}$ (for earthquakes, the well-known Gutenberg-Richter distribution 
\cite{GutenbergRichter1944,GutenbergRichter1956,Kagan2002}) and durations, which are believed to be universal. Beyond scaling exponents,
the question of whether the shape of an avalanche is universal is of great current interest \cite{PapanikolaouBohnSommerDurinZapperiSethna2011}.
Understanding whether and how universality arises, and obtaining quantitative predictions for avalanche statistics beyond phenomenological models
are some of the main challenges in the field. 

Historically, the elastic interface model has allowed for analytical progress 
thanks to a powerful method, the Functional Renormalization group (FRG). This method was first developed to calculate either the static (equilibrium) deformations
of an interface pinned by a random potential (e.g.\ the roughness exponent) \cite{DSFisher1986,BalentsDSFisher1993,ChauveLeDoussalWiese2000a,LeDoussalWieseChauve2003}, 
or the critical dynamics at the depinning transition
which occurs when applying an external force $f > f_c$
\cite{NattermannStepanowTangLeschhorn1992,NarayanDSFisher1992b,NarayanDSFisher1993a,ChauveGiamarchiLeDoussal1998,ChauveGiamarchiLeDoussal2000,ChauveLeDoussalWiese2000a,LeDoussalWieseChauve2002,LeDoussalWiese2002a,LeDoussalWieseRaphaelGolestanian2004}. These results are obtained in an expansion in the 
internal spatial dimension $d$ of the interface, around the upper critical dimension $d_{\rm uc}$,
in a loop expansion. Despite these successes the study of {\em avalanches} in elastic systems has remained
centered on toy models \cite{DSFisher1998,AlessandroBeatriceBertottiMontorsi1990,AlessandroBeatriceBertottiMontorsi1990b} 
or on scaling arguments and numerics 
\cite{MiddletonFisher1993,NarayanMiddleton1994,LuebeckUsadel1997,NarayanDSFisher1993a,ZapperiCizeauDurinStanley1998,KoltonRossoGiamarchi2005,KoltonRossoGiamarchiKrauth2006,KoltonRossoAlbanoGiamarchi2006}.
Several other important models have been used to
describe avalanches, such as the random-field Ising model \cite{BanerjeeSantraBose1995,DahmenSethna1996,LiuDahmen2006}
and discrete automata known as sandpile models, for which
analytical results exist \cite{BakTangWiesenfeld1987,IvashkevichPriezzhev1998,Dhar1999,Dhar1999b,BakSneppen1993,MarsiliDeLosRiosMaslov1998,Maslov1996,DorogovtsevMendesPogorelov2000}. 
However, exact results on the avalanche statistics are notably hard to obtain.

One simplifying feature of the interface model in its basic version, i.e.\ with over-damped dynamics,
is that it satisfies the no-crossing rule, or Middleton theorem, which guarantees only forward motion after a finite time, and uniqueness of the sliding state
\cite{Middleton1992,RossoKrauth2001a,Rosso2002}. 
This allows to define unambiguously, at fixed driving velocity $v$, a quasi-static limit $v=0^+$ which we have studied with
high precision both from numerics and using the FRG, testing the agreement up to two-loop accuracy \cite{RossoLeDoussalWiese2006a}.
Recently, we have developed FRG methods \cite{LeDoussalMiddletonWiese2008,LeDoussalWiese2008c,LeDoussalWiese2011b,RossoLeDoussalWiese2009a,LeDoussalRossoWiese2011}
 to calculate the statistics of avalanches for elastic interfaces,
both in a static, and quasi-static framework, obtaining e.g.\ the distribution $P(S)$ of their size, i.e.\ the total area swept during an avalanche.
Initially our calculation focused on static avalanches, i.e.\ switches in the ground state. However, thanks to Middleton's theorem, it can be extended to  quasi-static driving:
Since the system visits a unique sequence of metastable states, we define quasi-static avalanches in a stationary regime (for $v=0^+$)
as jumps from one metastable state to the next. The avalanche size $S$ depends only on the initial and final configuration, 
and is a property of the quasi-static limit. We found \cite{LeDoussalWiese2008c,FedorenkoLeDoussalWiesePREP}
that to 1-loop accuracy $P(S)$ is the same as for depinning as for
the statics, although we expect them to  differ at 2-loop order. 

In this paper we extend our study to the dynamics inside an avalanche; we  calculate the probability distribution of the instantaneous velocity  during an avalanche. 
Although we  focus on the small-driving-velocity limit, it is  a truly dynamical calculation. To properly define the avalanche
statistics, we found it important to separate two very different velocity scales: (i) the small driving velocity $v$, which allows to separate  different avalanches and to define a stationary regime; (ii) the motion inside an avalanche, which is much faster than the driving velocity \(v\), and independent of it for small $v$. It is this fast motion that we study here.

To this aim, we consider the following over-damped equation of motion, which reads, in its simplest form (for short-ranged elasticity of the interface),\begin{equation}\label{1}
\eta_0  \dot u(x,t) = c \nabla^2 u(x,t) +F(u(x,t),x) + m^2 [w(t)-u(x,t)]
\ .
\end{equation}
Here and below, we denote indifferently by $\dot u(x,t)$ or $\partial_t u(x,t)$ the local interface velocity. 
The time-dependent scalar function \(u(x,t)\), \(x \in \mathbb R^d\) describes the displacement of a $d$-dimensional
interface in a $d+1$-dimensional system.
The quenched random force $F(u,x)$ can be taken as a Gaussian random variable, short-ranged in $x$-direction, but with arbitrary correlations in $u$-direction, 
\begin{equation}
\overline{
F(u,x) F(u',x')} = \delta^d(x-x') \Delta_0(u-u')\ .
\label{2}
\end{equation}
In most applications, the disorder $\Delta_0(u)$ is a short-ranged function. The interface is driven and confined by a parabolic well of curvature $m^2$, which advances according to
\be 
w(t)=vt
\ .\ee  This model, and this type of driving, is of experimental relevance for the systems mentioned above. 
In some cases, it requires an extension of the elastic kernel to non-local elasticity, which amounts to
replacing in Eq.\ (\ref{1}), in Fourier space, 
\be 
c q^2 + m^2  \to \epsilon(q) = g(q)^{-1} 
\ .\label{4}\ee  
The combination $\epsilon(q) |u_q|^2$ is the energy associated to the mode $q$, which includes the elastic energy {\it plus
the coupling to the quadratic well}. We have defined
its inverse $g(q),$ i.e.\ the (static) propagator, which we use extensively below. One example 
is $\epsilon(q) = c (q^2 + \mu^2)^{\gamma/2}$, or more complicated kernels,
and we always denote $g(q=0)=1/m^2$ and $\epsilon(q) \sim q^{\gamma}$ at large $q$.
For a contact line, $m$ is related to the inverse capillary length $\mu$ (usually called $\kappa$), 
set by surface tension and gravity \cite{LeDoussalWiese2009a} and $\gamma=1$. For a magnet, 
$m$ is set by the so-called demagnetizing field \cite{UrbachMadisonMarkert1995,ZapperiCizeauDurinStanley1998,DurinZapperi2000} and $\gamma=1$
in some situations dominated by dipolar forces, while $\gamma=2$ in others. 
In fracture experiments, e.g.\ when breaking apart two plates which have been sintered together \cite{SchmittbuhlMaloy1997,LenglineToussaintSchmittbuhlElkhouryAmpueroTallakstadSantucciMaaloy2011,TallakstadToussaintSantucciSchmittbuhlMaaloy2011,SantucciMaaloyDelaplaceMathiesenHansenHaavigBakkeSchmittbuhlVanelRay2007,MaaloySchmittbuhl2001}, $m^2$ is proportional to the inverse thickness of the plates,
and usually $\gamma=1$. 

A toy model to describe the avalanche dynamics which results from Eq.\ (\ref{1}) has been proposed by 
 Alessandro, Beatrice, Bertotti and Montorsi (ABBM) 
\cite{AlessandroBeatriceBertottiMontorsi1990,AlessandroBeatriceBertottiMontorsi1990b}, and further developed in \cite{Colaiori2008,ColaioriZapperiDurin2004,ChenPapanikolaouSethnaZapperiDurin2011,PapanikolaouBohnSommerDurinZapperiSethna2011,LeDoussalWiese2008a}. 
It approximates the motion of the domain wall, i.e.\ a system with many degrees of freedom, by the
motion of a point, at position $u(t)$, which satisfies the equation of motion
\begin{equation}
\label{eq:IntroABBM}
\eta \dot u(t) = F\big(u(t)\big) - m^2\big[u(t)-w(t)\big].
\end{equation}
 In \cite{AlessandroBeatriceBertottiMontorsi1990}, the random pinning force $F(u)$ acting on this point
was postulated to be a Gaussian with the correlations of a random walk,  
\begin{equation}
\label{eq:CorrABBM}
\overline{\left[F(u_1)-F(u_2)\right]^2}=2\sigma |u_1-u_2|\ ,
\end{equation}
where $\sigma>0$ characterizes the disorder strength. One of the motivations for this assumption was that the model becomes solvable.
Although a crude description, it was used extensively to compare with Barkhausen-noise experiments on magnets, with success
in some cases (systems with long-ranged elasticity) and failures in others \cite{ZapperiCizeauDurinStanley1998,CizeauZapperiDurinStanley1997,DurinZapperi2000,Colaiori2008,FisherDahmenRamanathanBenZion1997}. 
The most natural interpretation is that $u(t)$ may represent the average height of the interface, $u(t)=\frac{1}{L^d}\int \rmd^d x\,u(x,t)$, and
that the ABBM model gives a mean-field description of the elastic interface. The random force $F(u)$ is then interpreted
as an {\em effective} random force, sum of the local pinning forces in some correlation volume. 
This is in agreement with the remark \cite{ZapperiCizeauDurinStanley1998,Colaiori2008}
that for infinite-range interactions the effective disorder is indeed correlated as in (\ref{eq:CorrABBM}). 
Thus this view has been taken for granted for a while. However, until now, there was no derivation from first principles
starting from the realistic microscopic model of an elastic interface. 

In this article, we go beyond this simple toy-model description of avalanches, and consider the motion of an elastic interface given by Eq.~(\ref{1}). 
We use the dynamical field theory and methods from the functional renormalization group (FRG). Let us recall that 
the upper critical dimension is $d_{\mathrm uc}=2 \gamma$ in general, hence $d_{\mathrm uc}=4$ for 
short-ranged elasticity, and \(d_{\mathrm uc}=2\) for the most common long-ranged elasticity, i.e.\ magnetic systems with dipolar forces, the contact line 
or fracture. In this article, we will show:

\begin{enumerate} 
\item[(i)] In the small driving-velocity limit, all correlation functions (in time and space) of the instantaneous velocity $\dot u(x,t)$ 
can be computed (to lowest order in $v$) in a dimensional expansion around $d_{\rm uc}$. This is done
by computing averages of exponentials of the velocities (generating functions), whose $O(v)$ contribution
allows to extract the full probability distribution of the velocity field $\dot u(x,t)$ during an avalanche. 

\item[(ii)]  At the upper critical dimension $d=d_{\rm uc}$, and in the small-$m$ limit, the velocity field in an avalanche has the same space-time statistics as 
the Brownian Force Model (BFM) with renormalized parameters $\eta \to \eta_m$ and $\sigma \to \sigma_m$. The
BFM is a model for an interface described by (\ref{1}) where $F(u,x)$ are Brownian motions in $u$, 
of variance $\sigma$,  uncorrelated in $x$. It is a generalization of the ABBM model to a set of elastically coupled ABBM models.
For the BFM the generating functions of the velocity are obtained exactly in any dimension $d$ by summing only tree graphs. 
Furthermore one can consider that the ``tree theory" is the correct mean-field theory 
and describes  the system for $d \geq d_{\rm uc}$, with full universality at $d=d_{\rm uc}$ and small $m$.

\item[(iii)]  The  ABBM model (\ref{eq:IntroABBM}) with the force-force correlator (\ref{eq:CorrABBM}) correctly
describes the avalanche motion of the {\it center of mass of the interface} for $d \geq d_{\rm uc}$ in the limit $v=0^+$. Universality arises
for $d=d_{\rm uc}$ and small $m$, with a dependence of the effective parameters $\eta \to \eta_m$ and $\sigma \to \sigma_m$
that we  computed. 

\item[(iv)] Even for $d = d_{\rm uc}$ the original ABBM model is not sufficient to describe the velocity correlations of different points on the interface, or
the statistics of Fourier modes $q \neq 0$. The latter can however be obtained from the tree theory (i.e.\ the BFM)
which we show to be equivalent to solving a non-linear instanton equation. From this we obtain e.g. the avalanche shape
at finite $q$ at $d=d_{\rm uc}$. 

\item[(v)] Finally, for $d < d_{\rm uc}$ the velocity field in an avalanche has universal statistics not given by the
BFM, nor, for the center of mass, by the ABBM model. It can be obtained within an $\epsilon=d_{\rm uc}-d$ expansion.
We show that the one-time center-of-mass velocity distribution diverges at small velocity not as $P(\dot u)\sim 1/\dot u$, but 
with a modified exponent \begin{equation}
P(\dot u) \sim \frac1{\dot u^{\sf a}}\ .
\end{equation}
For short-ranged elasticity the exponent is (with \(\epsilon = 4-d\)):
\bea
&& {\sf a} = 1 - \frac{2}{9} \epsilon + O (\epsilon^{2})   \quad  \mbox{ non-periodic, RF} ~~~~\\
&& {\sf a} = 1 - \frac{1}{3} \epsilon + O (\epsilon^{2})   \quad  {\rm periodic\ .}
\eea
For long-ranged elasticity (\(\gamma=1\)), the exponent is (with \(\epsilon = 2-d\)): \bea
&& {\sf a} = 1 - \frac{4}{9} \epsilon + O (\epsilon^{2})   \quad  \mbox{
non-periodic, RF} ~~~~\\
&& {\sf a} = 1 - \frac{2}{3} \epsilon + O (\epsilon^{2})   \quad  {\rm periodic\
.}
\eea 
\end{enumerate} 
A  short report of some of our results has already appeared as a Letter  \cite{LeDoussalWiese2011a}.
The present study is the starting point of a calculation of the
avalanche shape and duration to order $O(\epsilon)$ \cite{DobrinevskiLeDoussalWieseprep}. 

Since the methods used here (based on the dynamical MSR path integral) are quite
different from the usual Fokker-Planck approach to solve the ABBM model \cite{AlessandroBeatriceBertottiMontorsi1990,AlessandroBeatriceBertottiMontorsi1990b}, 
our study also provides a new way to solve the ABBM model. In particular, we find that generating functions can be
obtained from the solution of the non-linear instanton equation. This new connection has been exploited and
extended in \cite{DobrinevskiLeDoussalWiese2011b} to derive new results for the ABBM model (and elastically coupled
ABBM models) for finite $v>0$ and for a non-stationary avalanche dynamics. 

One should emphasize that the methods introduced in the present work strongly rely on the Middleton
theorem. Although specific results are obtained  for an over-damped dynamics, the present methods
can be extended to any dynamics which satisfies the Middleton theorem. As an example, we have recently studied
the ABBM model in presence of retardation \cite{DobrinevskiLeDoussalWieseprep}.  A much greater challenge for the future would be to extend these 
methods to models where the no-passing rule does not apply, such as models with inertia or relaxation
which have been proposed, e.g.\ to study earthquake dynamics \cite{JaglaKolton2009}. 
There the very existence of a quasi-static limit is much less clear, and may depend on \
details of the dynamics. Some steps in that directions have been taken in \cite{LeDoussalPetkovicWiese2012}. 
Finally, let us also mention related studies of static avalanches in
spin glasses using Replica Symmetry Breaking \cite{LeDoussalMuellerWiese2010,LeDoussalMuellerWiese2011},
and in the Random-field Ising model \cite{TarjusBaczykTissier2012}.

\medskip

The outline of this article  is as follows:  

In section \ref{s:Model...} we introduce the interface model, define important observables, and explain our strategy for their calculation. 
We also review the expected scaling relations for the avalanche statistics. 

In section \ref{s:tree}, we construct the theory at tree level. We start with calculating the moments of the instantaneous velocity in subsection \ref{s:treeMoments}, before introducing in subsection \ref{simplified}  a non-linear equation, which we call the  instanton equation, to efficiently resum them. In subsection \ref{s:jpdfcomv} we  calculate the joint probability distribution for the center-of-mass velocity at one and several times.
From that we extract various velocity probability distributions, and calculate the average shape of an avalanche, as well as its variance which we call
the {\em\ second shape}. In subsection \ref{sec:interpret} we show that the solution of the instanton equation encodes the response to a small step in the applied force.
In subsection \ref{s:recover-stat} we recover the quasi-static avalanche-size distribution.  
In Section \ref{s:BFM} we discuss the relation between the tree theory and the mean-field theory: We show that the tree theory is equivalent to (i.e.\ is exact for) 
the Brownian force model, and, for the center of mass only, to the ABBM model. We also show that the so-called improved tree theory, 
i.e.\ the tree theory with renormalized values for the disorder and the friction parameters,
is the correct mean-field limit (for $d=d_{\rm uc}$) of the underlying field theory to be discussed in the following section \ref{s:loops}. 
Our approach is based on the Langevin equation
and on the MSR dynamical action; alternatively one can use a Fokker-Planck description, as is explained in subsection \ref{s:ABBM}. It is this latter description which was introduced by ABBM \cite{AlessandroBeatriceBertottiMontorsi1990,AlessandroBeatriceBertottiMontorsi1990b} for a particle, but whose use seems to be restricted to the latter. 
In subsection \ref{sec:spatial} we obtain a number of results beyond the center-of-mass motion, such as the local averaged shape following a local step in the force,
as well as the spatial and time dependence of the second shape.

In section \ref{s:loops}, we study the loop corrections, for $d<d_{\rm uc}$. We explain the general framework in subsection \ref{s:loops-general}, before introducing a simplified theory in section \ref{sec:simplified}, containing all the needed ingredients for the one-loop calculation.
The latter is solved perturbatively in subsection \ref{q1}. 
We then discuss in detail the 1-loop, i.e.\ $O(\epsilon)$, corrections to the velocity distribution in subsection \ref{s:1-point}. 
We derive the necessary counter-terms in subsection \ref{s:counter-terms}. The extension to long-ranged elasticity is detailed in subsection \ref{s:LR}.

The above theory was developed in terms of the velocity $\dot u$ as the dynamical variable. 
In section \ref{s:first-principle} we discuss how to perform the same calculations
using the more standard theory in terms of the position $u$. While this is more involved, 
it avoids certain technical problems which may be present in the velocity theory, and  confirms the validity of 
the latter. 

Several technical issues are presented in appendices \ref{app:laplinv} to \ref{c7}.

\section{Model, observables and program}\label{s:Model...}

\subsection{The bare model}

We consider an elastic interface of internal dimension $d$, with no overhangs, parameterized by a time-dependent 
real valued displacement (or height) field \(u(x,t)\equiv u_{xt} \in 
\mathbb R\), with $x \in \mathbb R^d$. It evolves in presence of a random pinning force $F(u,x)$ according to the simplest possible overdamped equation of motion,
\be \label{eqmogen}
\eta_0 \partial_t u_{xt} = \int_{x'} (g^{-1})_{xx'} ( w_{x' t} - u_{x' t} ) + F(u_{xt},x) 
\ .
\ee 
Here $\eta_0$ is the bare friction coefficient and $(g^{-1})_{xx'}$ is the elastic matrix, with propagator $g_{xx'}=g_{x-x'}$ and $g(q) \equiv g_q=\int_x e^{i q x} g_{x}$ in Fourier space and we define the (squared) mass $m^2=g_{q=0}^{-1}$. Everywhere we denote equivalently 
$\int_x := \int \rmd^d x$ and $\int_t := \int \rmd t$. The interface is driven by an external quadratic potential centered at position $w_{xt}$. The 
total external force acting on the interface is noted\be 
f_{xt} =  \int_{x'}(g^{-1})_{xx'} w_{x' t\ ,}
\ee  
with $f_t=m^2 w_t$ for spatially uniform driving. Equivalently, for inhomogeneous driving, $w_{xt}$ denotes the 
reference interface position in the absence of disorder and in the limit of very slow driving (hence this notation is useful in the statics 
and the quasi-statics). We focus on the
case of {\em local or short range elasticity} $g_q^{-1} := q^2 + m^2$, with an elastic constant set to unity by choice of units.
We will however also give the results for more general non-local elasticity, see the discussion after Eq.~(\ref{4}). We focus on a uniform driving at
fixed velocity $v$, $w_{xt}=w_t=v t$. This leads to Eq.~(\ref{1}) in the introduction.

The pinning force is
chosen as indicated in Eq.~(\ref{2}), where $\Delta_0(u)$ is the microscopic (bare) disorder correlator and  
 $\overline{\cdots}$ denotes disorder averages. 
For realistic disorder the bare disorder correlator is smooth. Note that for the bare model, we always assume
(unless stated otherwise) a small-scale cutoff in $x$, either a lattice spacing $a$, or that
$\Delta_0(u)$ decays on a finite correlation length $r_{\rm f}$. This insures
the existence of a Larkin scale $L_{\rm c}$ \cite{BlatterFeigelmanGeshkenbeinLarkinVinokur1994}, which produces a small-scale cutoff for 
avalanches. We denote $S_0$ the small-scale cutoff on their size. 

The above model exhibits two important properties: Due to 
statistical translational invariance of the disorder and its $\delta$-correlations in
internal space, the model possesses the so-called {\em statistical tilt symmetry} (STS) 
which guarantees that the elasticity $g_q$ is uncorrected by fluctuations (loop corrections), see e.g.\ \cite{LeDoussalWiese2008c} for notations and some definitions in this section. The second important property of the model is the Middleton theorem\footnote{For a model discrete in $x$, this is the case if $(g^{-1})_{xx'} \leq 0$ for $x \neq x'$. Then 
$\dot u_{xt} \geq 0$ if $\dot f_{xt} \geq 0$ and $\dot u_{xt_i} \geq 0$ at some initial time $t_i$.}: If the driving force $m^2 w_t$ is an increasing function of
time, $\dot w_t \geq 0$ (positive driving), and if velocities are all positive at $t=0$, 
$\dot u_{x,t=0} \geq 0 $, then  they
remain so at all times \cite{Middleton1992}. In particular, for a finite interface (of size $L$), submitted to positive driving, 
all velocities become positive after a finite driving distance, and\ the memory of the initial condition
is erased. 

\subsection{Quasi-static observables}

In this paper we focus on the stationary state of the model with fixed driving velocity $w_t= v t$, hence 
$\overline{ \dot u_{xt} } = v$. We focus on the
small-velocity limit $v=0^+$, i.e.\ on  the vicinity of the
quasi-static depinning transition. At a qualitative level, it is expected that because of disorder, at scales larger than
the Larkin length $L_c$, the interface is rough at all times, i.e.\ self-affine $\overline{(u_{xt} - u_{x't})^2} \sim |x-x'|^{2 \zeta}$, with the roughness exponent $\zeta=\zeta_{\rm dep}$ of the depinning transition \cite{RossoKrauth2001b,RossoKrauth2002,RossoHartmannKrauth2002}. Because of the mass term, the interface flattens for scales $|x-x'| > L_m$, $\overline{(u_{xt} - u_{x't})^2} \sim L_m^{2 \zeta}$ with $L_m \sim 1/m$ for local elasticity.  We are interested in the universality
which arises in the small-$m$ limit,  i.e.\ for $L_m \gg L_c$.

It is also expected that on scales larger than the Larkin scale, the motion is not smooth but proceeds
by avalanches, i.e.\ the system jumps from one rough metastable state to the next one.  Thanks to the Middleton theorem there is a
well-defined quasi-static limit, i.e.\ a function $u_x(w)$ such that for $v=0^+$ one has $u_{xt} = u_x(w_t)$ where $w_t= v t$ is the position of the center of the quadratic well. The sequence of visited states is unique. The quasi-static process $u_x(w)$ was defined in \cite{LeDoussalWiese2006a} and studied numerically in
\cite{RossoLeDoussalWiese2006a,RossoLeDoussalWiese2009a}, see also \cite{LeDoussalWieseMoulinetRolley2009} for an experimental
realization. Note that the process $u_x(w)$ is different from $u^{\rm stat}_x(w)$ defined in the statics \cite{LeDoussalWiese2008c} 
which describes shocks, i.e.\ switches in the ground state\footnote{{$u^{\rm stat}_x(w)$ is the minimum-energy configuration for a given $w$.
In contrast, for a particle $u(w)$ is the smallest root of the equation $m^2 w=F(u)-m^2 u$ and, similarly, for an interface
$u_x(w)$ is the metastable state with the smallest $u_x(w)$ for all $x$.}}. However, there are close analogies, hence similarities in
notations in this section and in Ref.\ \cite{LeDoussalWiese2008c}. The quasi-static process  jumps at a set of discrete locations $w_i$, i.e.
\be \label{uxw}
 u_x(w) = \sum_i S_{ix} \theta(w-w_i) 
\ .
\ee
We  also consider the motion of the center of mass of the interface,  denoted \be
u_{t} := L^{-d} \int_x u_{xt}
\ .\ee
For $v=0^+$, it converges to the quasi-static process $u_t = u(w_t)$ for the center of mass,  denoted
\be
  u(w) = L^{-d} \int_x u_x(w) = L^{-d} \sum_i S_{i} \theta(w-w_i)\ . 
\ee 
Here $S_i $ is the size of the $i$-th avalanche.   In the statics, the statistics of these  shocks was studied in Ref.\ \cite{LeDoussalWiese2008c}. 
Here one can also define their size density (per unit $w$) as
\be
\rho(S) = \rho_0 P(S) = \overline{ \sum_i \delta(S-S_i) \delta(w-w_i) }\ .
\ee 
The probability distribution  $P(S)$   of the size is normalized to unity.  Since one
can show that \cite{LeDoussalWiese2006a,LeDoussalWiese2008a}
\be 
m^2 \overline{[w  - u(w)]} = f_c(m)\ , 
\ee  
the critical force at fixed $m$, it implies $\overline{u(w)} \approx w$, hence the process follows the center of
the well, although with a delay. This shows that the total density $\rho_0$ per unit $w$ is related to the average size as
\be \label{rho0}
\rho_0 = \frac{L^d}{\langle S \rangle} \ ,
\ee
where here and below $\langle f(S) \rangle = \int \rmd S\, P(S) f(S)$ denotes the (normalized) 
average of $f(S)$. Note that the existence of a short-scale cutoff (and a Larkin scale)
guarantees that $\rho_0$ is finite, although it may diverge if these cutoff scales go to zero.

 As shown in \cite{LeDoussalWiese2008c} there is an exact relation between the second moment of the avalanche-size distribution 
and the cusp in the renormalized disorder correlator,
\be \label{Sm}
  S_m:=  \frac{  \langle S^2 \rangle }{2 \langle
S \rangle} = \frac{- \Delta'(0^+)}{m^4} \ .
\ee
It defines the avalanche-size scale $S_m$, which behaves as $S_m \sim m^{-(d+\zeta)}$ at small $m$. The definition of the renormalized disorder correlator $\Delta(u)$ is recalled below
and its salient property is that it is non-analytic, even if the bare disorder is smooth. 
This relation holds in any dimension, for statics and quasi-statics, i.e.\ depinning (with, accordingly different values for $\Delta'(0^+)$
and the roughness exponents). The only assumption is that all motion takes place in
shocks or avalanches, as in (\ref{uxw}), which usually holds for small enough $m$ (see \cite{LeDoussalMuellerWiese2010} for a case where the contribution from the smooth part of $u(w)$ is calculated explicitly).

The convergence to the quasi-statics
in the small-$v$ limit occurs on time scales $t_w := \delta w/v$ where $\delta w \sim w_{i+1}-w_i$ is the typical avalanche separation. $t_w$ is 
called the {\em waiting time} (until the next avalanche). On the other hand, the motion {\it inside} an avalanche
occurs on the so-called {\em duration time scale}
\be  \label{timescales}
\tau \sim L_m^z \ll \delta w/v\ ,
\ee  
where $z$ is the dynamical exponent at depinning. In this paper we always assume $v$ small enough so that 
the order of scales is as given by Eq.\ (\ref{timescales}), i.e., the {\em avalanche duration} is much smaller
than the {\em waiting time} between avalanches, so that successive avalanches are {\it well separated}. 
In practice, when $L \gg L_m$ and at least for SR elasticity, 
it may be sufficient to ask that successive avalanches occurring in the same region
of space be well separated, i.e.\ that (\ref{timescales}) holds when $\delta w$ is the typical separation 
of avalanches in the same region of space. The condition (\ref{timescales}) is equivalent to the condition
$L_m \ll \xi_v$, where $\xi_v$ is the  correlation length near the depinning transition
\cite{NattermannStepanowTangLeschhorn1992,NarayanDSFisher1992b,NarayanDSFisher1993a}.

\begin{figure}
\Fig{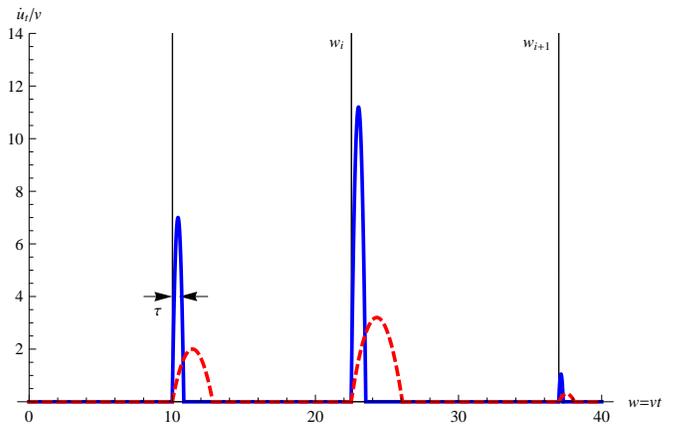}
\caption{Schematic plot of the instantaneous velocity as a function of $v t$ for different $v$. The area under the curve is
the avalanche size hence is constant as $v \to 0^+$. The quasi-static avalanche positions $w_i$ are indicated.}\label{f:velplot}
\end{figure}

\subsection{Dynamical observables}

Our aim is to obtain information about the dynamics in an avalanche. For simplicity we will first consider 
the $n$-times (instantaneous) velocity cumulants $\overline{ \dot u_{t_1} ... \dot u_{t_n} }^{\rm c}$ for the center of mass, 
and discuss space dependence later. The important property about avalanches, and non-smooth motion in general, is that
in the limit $v \to 0^+$
\bea  \label{ordermag}
 \overline{ \dot u_{t_1} ... \dot u_{t_n} }^{\rm c} &=& v f(t_1,...,t_n) + O(v^2) \\
 \overline{ \dot u_{t_1} ... \dot u_{t_n} } &=& v f(t_1,...,t_n) + O(v^2)\ .
\eea 
This means that cumulants {\em and} moments are $O(v) $, and have  the same leading time dependence. This is very different from a smooth motion,
for which they would be $O(v^n)$. Here we are considering times much shorter that the waiting-time scale $\delta w/v$, hence a single avalanche. The result (\ref{ordermag}) can be understood as follows: The above cumulants are non-negligible only when all times are inside the same avalanche. When that occurs, the velocities are $O(v^0)$, with a magnitude studied below. Let us suppose that the separation of the times $t_i$ is of the order of $T$. The above cumulants are thus dominated by the probability that exactly one avalanche occurs in a time interval of duration $T$ (with $T \ll \Delta w/v$). This probability is in terms of the total avalanche density $\rho_0$
\be\label{sum-rule}
\mbox{Prob}(\text{one avalanche in} [-T/2,T/2] ) = \rho_0 v T  \ll 1\ . 
\ee 
More precisely, one can establish the sum rule
\be \label{sumrule}
L^{nd} \int\limits_{[-T/2,T/2]^n} \rmd t_1...\rmd t_n\, \overline{ \dot u_{t_1} ... \dot u_{t_n} } = \rho_0 v T \langle S^n \rangle + O(v^2)\ ,
\ee
which is valid as long as $\rho_0 v T \ll 1$. It comes from the fact that the total displacement $L^d \int \rmd t\, \dot u_{t}$ during the avalanche $i$ is equal to its size, 
\be S_i =  L^d \int \rmd t\, \dot u_{t}
\ .
\ee 
It is clear from the above that the difference between moments and cumulants is at most  of $O(v^2)$. The sum rule (\ref{sum-rule}) thus connects dynamical quantities to quasi-static ones. It  provides a valuable consistency check for our dynamical calculations. \medskip

\subsection{Strategy}

Let us now summarize our strategy. We will calculate the velocity cumulants from perturbation theory in an expansion in the disorder. Naively this expansion is in the bare disorder $\Delta_0(u)$. To lowest order the $n$-times cumulant (\ref{ordermag})
is $O(\Delta_0^{n-1})$ and, as we will see below, is obtained from tree graphs in the graphical representation of perturbation theory. For each $n$, 1-loop graphs only occur at the next order, i.e.\ $O(\Delta_0^n)$, and so on for higher-loop graphs. Hence we start by examining the perturbation theory at tree level in the next section. We compute explicitly the  lowest moments, and then  show that there exists a much more powerful method, based on a simplified field theory, which allows to sum all tree diagrams and compute directly the Laplace transform of the joint probability distribution $P(\dot u_{t_1},...,\dot u_{t_n})$ of the velocities at $n$ times.

In practice it is in fact more accurate to work with the renormalized disorder. 
We recall that the renormalized disorder correlator $\Delta(u)$ is defined in the quasi-static theory from the center-of-mass fluctuations as
\be
m^4 \overline{[u(w) - w][u(w') - w']}^{\rm c} = L^{-d} \Delta(w-w')\ .
\ee
The function $\Delta(u)$ depends on $m$, with $\Delta(u)=\Delta_0(u)$ for $m\to \infty$. 
At small $m$ it takes the universal scaling form  
\bea \label{rescaledDelta}
 \Delta(w) &=& \frac{1}{\epsilon \tilde I_2}  m^{\epsilon - 2 \zeta} \tilde \Delta(m^\zeta w)\ , \\
 \tilde I_2 &=& \int_q \frac{1}{(q^2+1)^2} \ .
\eea
It is given here for SR elasticity. 
The rescaled correlator $\tilde \Delta(w)$ converges to a (non-analytic) fixed-point form $\tilde \Delta(w)\stackrel{m \to 0} {-\!\!\!-\!\!\!\longrightarrow} \tilde \Delta^*(w)=O(\epsilon)$. Here
$\zeta$ is the roughness exponent at depinning. The rescaled correlator $\tilde \Delta(w)$ obeys, as a function of $m$, a 
FRG flow equation which was obtained at the depinning transition, together with its fixed points, to two loops in \cite{ChauveLeDoussalWiese2000a,LeDoussalWieseChauve2002}
and checked  in \cite{RossoLeDoussalWiese2006a} where $\Delta(u)$ was measured from numerics.

Since it is the renormalized disorder, which is small, we then reexpress the perturbative expressions of the velocity cumulants in terms of $\Delta(w)$ directly. Thus we generate an expansion in powers of $\epsilon$ for these quantities. The leading order is  determined solely by tree graphs in the renormalized disorder $\Delta$ (each cumulant being of order $\Delta^{n-1}$) and is valid for $d=d_{\rm uc}$ (to some extent it is also valid for $d>d_{\rm uc}$, see the discussion below). This leads to the tree-level result for the velocity probabilities. Corrections to the tree-level result are obtained in the next section by adding the contribution of one-loop diagrams, i.e.\ the next order in $\tilde \Delta^* = O(\epsilon)$. 

In the remainder of the paper we will switch to the comoving frame, unless explicitly indicated. Hence we define
for $w= v t$\be 
u_{xt} = v t + \breve u_{xt}
\ ,
\ee 
where $\breve u_{xt}$ satisfies the equation of motion:
\begin{equation} \label{eqmo1}
   \eta \partial_t \breve u_{xt} = \nabla_x^2 \breve u_{xt} + F(v t + \breve u_{xt},x) - m^2 \breve u_{xt} - \eta v
\ .
\end{equation}
Below we will  denote $\breve u$ by $u$ for simplicity. 

\subsection{Expected scaling forms for avalanche statistics}

\subsubsection{Size distribution}
The size distribution is by now the best known one. 
Let us first recall our previous results \cite{LeDoussalMiddletonWiese2008,LeDoussalWiese2008c} for the avalanche-size distribution
in the small-$m$ limit, i.e.\ $S_m \gg S_0$, where $S_0$ is the microscopic cutoff,
and $S_m$ the scale of the large avalanches, given by Eq.~(\ref{Sm}). For $S \gg S_0$, the size distribution
 $P(S)$ takes the form 
 \be  \label{PS}
     P_{\rm size} (S) \equiv P (S)= \frac{\langle S \rangle}{S_m^2} p (S/S_m)\ . 
\ee  
Depending on the dimension $d$, $p(s)$ takes different forms: (i) for $d=d_{\rm uc}$
\be  \label{pS}
      p(s) = p^{\rm tree}(s) = \frac{1}{2 \sqrt{\pi}}  s^{-3/2} \rme^{-s/4}\ .
\ee  
 (ii) for $d < d_{\rm uc}$,
\begin{equation}\label{final}
   p(s) = \frac{A}{2 \sqrt{\pi}} s^{-\tau}  \exp\!\left(C \sqrt{s} -
\frac{B}{4} s^\delta\right)\ ,
\end{equation}
to first order in $O(\epsilon = d_{\rm uc}-d)$, where $A-1,B-1,C=O(\epsilon)$ are given in
\cite{LeDoussalWiese2008c}. The exponent $\delta= 1 + \frac{1}{12} (\epsilon - \zeta) + O(\epsilon^2)$ and the avalanche exponent
\be 
\tau = \frac{3}{2} - \frac{1}{8} (\epsilon - \zeta) + O(\epsilon^2)
\ee  
were  agree to first order in $\epsilon$ with the Narayan-Fisher (NF) conjecture \cite{NarayanDSFisher1993a,ZapperiCizeauDurinStanley1998}, which relates the
avalanche-size exponent and the roughness exponent
via\be  \label{NF}
\tau = 2 - \frac{\gamma}{d+ \zeta} 
\ .\ee  
Here $\gamma=2$ for SR elasticity, $\gamma=1$ for LR elasticity\footnote{The exponent $\gamma$ is often called
$\mu$ in the literature, see e.g.\ \cite{DurinZapperi2000}.} and $d_{\rm uc}=2 \gamma$.
This conjecture agrees well with numerics for $d=1,2,3$ 
\cite{LeDoussalMiddletonWiese2008,RossoLeDoussalWiese2009a},
both for the statics and quasi-statics (with the respective values for $\zeta$),
but it is not known if it is exact  (see the discussion in section VIII-A of
\cite{LeDoussalWiese2008c}). It was proposed by NF for depinning
only, but recently we have found a general argument for the 
statics as well, based on droplet considerations \cite{LeDoussalMuellerWiese2010,LeDoussalMuellerWiese2011}.

Here we will recover the above results, within a dynamical calculation, to tree level in $d=d_{\rm uc}$, and to one loop, $O(\epsilon)$, for $d<d_{\rm uc}$.

\subsubsection{Duration distribution}

Assuming one can define unambiguously the duration $T$ of an avalanche  (see the discussion below in Section \ref{sec:cdmabbm}) the duration exponent $\alpha$ is defined through
the small-$T$ behavior of the duration distribution \footnote{{In Section \ref{s:loops} the notation $\alpha$ is  used for a different quantity, see Eq.~(\ref{alpha1}). }}:
\be 
P_{\rm duration}(T) = \frac{1}{T^\alpha} f(T/T_0)\ ,
\ee 
where $T_0$ is a large-time cutoff, and $f(0)$ a constant. This form has been conjectured
in various articles, see e.g.\ \cite{ZapperiCizeauDurinStanley1998}. A simple scaling
argument relates $\alpha$ to $\tau$ and $z$, the dynamical exponent. One writes
$S \sim L^{d+\zeta}$ and $T \sim L^z$, hence $S \sim T^{(d+\zeta)/z}$. Then
\be 
P_{\rm duration}(T) \sim P_{\rm size}\Big(S \sim T^{(d+\zeta)/z}\Big) \frac{\rmd S}{\rmd T} \sim T^{-\alpha}
\ee 
with
\be 
\alpha = 1 + \frac{(\tau-1)(d+\zeta)}{z}\ .
\ee 
If we use in addition the NF conjecture (\ref{NF}) one obtains
\be 
\alpha = 1 + \frac{d+\zeta- \gamma}{z} 
\ ,\ee  
a relation which was conjectured previously, see e.g.\ \cite{ZapperiCizeauDurinStanley1998}.
It is not known at present whether these conjectures are exact. The methods of the
present paper allow to determine $P_{\rm duration}(T)$. Here we obtain  it to tree level,
and in \cite{DobrinevskiLeDoussalWieseprep} to one loop.

\subsubsection{Velocity distribution}

Here we  obtain the distribution of velocities in an avalanche in the form
\be
{\cal P}(\dot u) \sim  \frac1{\dot u^a}
\ .\ee
We will obtain $a=1$ at the mean-field level (tree theory), as in the ABBM model, and $a<1$ for $d< d_{\rm uc}$.
It turns out that our result for the exponent $a$ is not straightforward to derive from scaling arguments.
Hence it may be a new independent exponent.

\section{Tree-level theory}
\label{s:tree}

In this section we implement the program explained above to lowest order, i.e.\ at tree level.
Hence we construct the proper mean-field theory for the interface. We will use systematically
the notation $\Delta$ for the disorder vertices and $\eta$ for the friction. Hence, if
one substitutes $\Delta_0$ and $\eta_0$ one gets the naive perturbation result, i.e.\ genuine tree graphs. If one
considers $\Delta$ and $\eta$ as the renormalized disorder correlator and  friction,
 one obtains the result using the so-called ``improved action", i.e.\ the limit for $d=d_{\rm uc}$ of the
effective action (see Refs.\ \cite{LeDoussalWiese2008c,LeDoussalWiese2011b} for more details on these definitions). This amounts
to summing  tree graphs plus those loop diagrams which renormalize friction or disorder at $d=d_{\rm uc}$. Sometimes we will
denote $\Delta\to \Delta_m$ and $\eta\to \eta_m$ to remind that these quantities
are $m$ dependent. In simple terms, the results expressed in terms of $\Delta_m$ and $\eta_m$ 
are numerically accurate at $d=d_{\rm uc}$, with the correct, and universal, dependence on $m$
for small $m$.

It is useful to recall here the result of \cite{LeDoussalWiese2008c} for the generating function and avalanche-size distribution 
at  tree level, \begin{eqnarray}Z_{S}(\lambda)&:=&\frac{
\left<\rme^{\lambda S}-1\right>}{\left<S \right>}\ ,\\
Z_{S}^{\mathrm{\rm tree}}(\lambda) &=&\frac 1{2S_m} \left( 1-\sqrt{1-4\lambda S_m } \right)\ . 
\label{Z-tree}\end{eqnarray}  
We have added the subscript $S$ to distinguish
from the notation for the  dynamical generating functions introduced below;  let us also note that we use indistinguishably the three suffixes ``tree'', ``MF'', and  ``$0$'', to indicate tree, i.e.\ mean-field quantities. 
Eq.~(\ref{Z-tree}) holds both for the statics and quasi-statics, and will be recovered below in
the dynamical approach.

\subsection{Calculation of moments} 
\label{s:treeMoments}
The equation of motion (\ref{eqmo1}) in the comoving frame can also be written as\be \label{eqmo2}
u_{xt} = \int_{x't'} R_{xt,x't'} \Big[ F(v t + u_{x't'},x') - \eta v\Big]\ , 
\ee 
where
\be \label{defresponse}
R_{xt,x't'}=R_{x-x',t-t'} = \big(\eta \partial_t \delta_{tt'} + g^{-1}_{xx'} \delta_{tt'}\big)_{xt,xt'}^{-1} 
\ee
is the bare response function with $R_{xt}= \int_q e^{i q x} R_{qt}$.  In Fourier space it reads
\be \label{bareR}
R_{qt}=\frac{1}{\eta} \theta(t) e^{-(q^2+m^2) t/\eta}\ .
\ee

\subsubsection{First moment}
We start with the first moment, which defines the critical force
$f_c=f_c(m,v)$.
Taking the disorder average of (\ref{eqmo2}) we have
\begin{eqnarray}
&& m^2 \overline{u_{xt}} = f_c - \eta v \\
&& f_c:= \overline{F(v t + u_{xt},x)\ } 
\ .\end{eqnarray}
This yields the exact equation
\begin{equation} \label{exact1}
 u_{xt} - \overline{u_{xt}} = \int_{x't'} R_{x-x',t-t'} \big[ F(v t + u_{x't'},x') - f_c\big]\ ,   
\end{equation}
from which we now compute the cumulants to leading order in perturbation theory.

\subsubsection{Second moment}

To lowest order in $\Delta$ one  finds from Eq.~(\ref{exact1}) that
\begin{equation} \label{pt1}
\overline{u_{x_1t_1} u_{x_2 t_2}}^{\rm c} = \int_{x't't''} R_{x_1-x',t_1-t'}  R_{x_2-x',t_2-t''} \Delta\big(v (t'-t'')\big). 
\end{equation}
From this we obtain the cumulant of the center-of-mass velocity,\begin{eqnarray}
&& \overline{\dot u_{t_1} \dot u_{t_2}}^{\rm c} = L^{-d} \partial_{t_1}\partial_{t_2} \frac{1}{\eta^2} \nn\\
&& \times \int_{s_1<t_1,s_2<t_2} \rme^{- \frac{m^2}{\eta} (t_1-s_1) - \frac{m^2}{\eta} (t_2-s_2)}  \Delta\big(v (s_1-s_2)\big) \nn\\
&& = - L^{-d}  \frac{v^2}{\eta^2} \int\limits_{\tau_1>0, \tau_2>0}\!\!\! \!\!\!\!\rme^{- \frac{m^2}{\eta} (\tau_1+\tau_2) }  \Delta''\big(v (t_1-t_2-\tau_1+\tau_2)\big) \nn
.\\
\end{eqnarray}
Let us now consider the limit of $v \to 0^+$, and assume that $\Delta(u)$ has a cusp, i.e.
\begin{equation}
   \Delta''(u) = 2 \Delta'(0^+) \delta(u) + \Delta''(0) + O(|u|)\ .
\ee 
Then we find that\begin{eqnarray} \label{uu3}
 \overline{\dot u_{t_1} \dot u_{t_2}}^{\rm c} &=& 
 - 2 L^{-d}  \Delta'(0^+)  \frac{v}{\eta^2} \int_{\tau_1>0} \rme^{- \frac{m^2}{\eta} (2 \tau_1-t_1+t_2) }  \nn \\
&&  - L^{-d}  \frac{v^2}{\eta^2} \Delta''(0) \int\limits_{\tau_1>0, \tau_2>0} \!\!\!\!\rme^{-
\frac{m^2}{\eta} (\tau_1+\tau_2) }\nn\\ &&+ O(v^3)
\ .
\eea
Hence we obtain
\bea \label{second}
 \overline{\dot u_{t_1} \dot u_{t_2}}^{\rm c}   &=& -  L^{-d}  \Delta'(0^+)  \frac{v}{m^2 \eta} \rme^{- \frac{m^2}{\eta} |t_2-t_1| } \nn\\&&-  L^{-d}   \Delta''(0) \frac{v^2}{m^4}+ O(v^3)\ .  
\end{eqnarray}
Note that the cusp is crucial to get non-smooth, avalanche motion: Since $\overline{\dot u}=v$, the term of order $v$ in the above equation is possible only since the manifold moves with velocity $\dot u$  of order one (i.e.\ independent of $v$) for a time of order $1/v$.  In the absence of a cusp, $ \dot u\sim v$, and  the second cumulant of the velocity is $O(v^2)$ indicating a smooth motion. 
To this order, the typical time scale \(\tau_m\) of an avalanche is read off from the exponential in the first line of Eq.~(\ref{second}), as 
\be
\tau_m = \frac{\eta}{m^2}\ .
\ee
In the improved action, $\eta$ will be  renormalized to  $\eta \equiv \eta_m$, as is discussed below.

Using that the size of an avalanche is $S=L^d\int_t \dot u_t$, we can now integrate over the time difference
to obtain \be
 \rho_0 v \langle S^2 \rangle \equiv L^{2d} \int_{-\infty}^{\infty} \rmd t ~ \overline{\dot u_{0} \dot u_{t}}^{\rm c} = -  2 v L^{d}  \frac{\Delta'(0^+)}{m^4}
\ .\ee
 Using Eq.~(\ref{rho0}), i.e.\ $\rho_0=L^d/\left<S\right>$, this exact relation agrees with the general sum rule for $n=2$, provided Eq.~(\ref{Sm}) holds. 
This  is indeed an exact relation obtained both in the statics and in 
the quasi-static limit in \cite{LeDoussalWiese2011a,LeDoussalWiese2008c}; it relates the cusp to the second moment of the
avalanche-size distribution.

In order to simplify the notations for the calculation of higher cumulants,
we now switch to dimensionless units. They amount to replacing
\be 
x \to x/m, \quad L \to L/m, \quad t \to t \tau_m , \quad v \to v/\tau_m
\ee  
and $\Delta'(0^+) \to m^{4-d} \Delta'(0^+)$. 
In effect this is equivalent  to setting $\eta=m^2=1$. 

We now reproduce the above result, introducing a graphical representation
which will be useful for the calculation of the higher cumulants. Let us consider Eq.~(\ref{pt1}) integrated over space  
 and rewrite it graphically as
\be \label{uu}
 L^d \overline{u_{t_{1}} u_{t_{2}}}^{\rm c} =  \rule[-2mm]{0mm}{7mm}^{t_{1}}_{s_{1}}\!\!\diagram{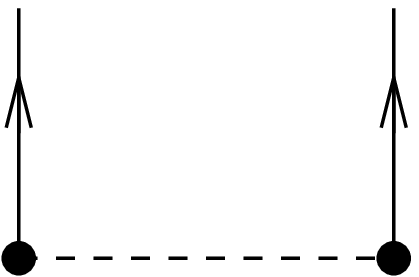} \!\rule[-2mm]{0mm}{7mm}^{t_{2}}_{s_{2}}\ . 
\ee
Here the dashed line represents the disorder vertex $\Delta$ which is bilocal in time  and the full lines are response functions (\ref{bareR}), here taken at zero momentum $q=0$.  (For details on
 this standard graphical representation see  e.g.\ \cite{LeDoussalWieseChauve2002}.) The second velocity cumulant thus reads
\be  \label{uu2}
 \!\!\! L^d \overline{\dot u_{t_{1}} \dot u_{t_{2}}}^{\rm c} =  \partial_{t_1} \partial_{t_2} \rule[-2mm]{0mm}{7mm}^{t_{1}}_{s_{1}}\!\!\diagram{cumul2d} \!\rule[-2mm]{0mm}{7mm}^{t_{2}}_{s_{2}} \ .
\ee 
Hence the time derivatives act on the external legs.
We now use the fact that the response function depends only on the time difference, i.e.,
\be
\partial_{t_1} R_{q,t_1-s_1} = - \partial_{s_1} R_{q,t_1-s_1} 
\ ,
\ee
where here and below we denote $R_t := R_{q=0,t}=\theta(t) e^{-t}$ in our dimensionless units.
Hence, by partial integration,  we can move both time derivatives to act on the disorder vertex as $\partial_{s_1} \partial_{s_2}$
which produces the term $- v^2 \Delta''(v(s_1-s_2))$ as in Eq.~(\ref{uu3}). To lowest order in $v$ this can be replaced by
$- 2 v \Delta'(0^+) \delta(s_1-s_2)$, hence the two internal times are identified. This can be represented as
\bea \label{uu4}  \nn
&&\!\!\! L^d \overline{\dot u_{t_{1}} \dot u_{t_{2}}}^{\rm c} =  - 2 v \Delta'(0^+) \diagram{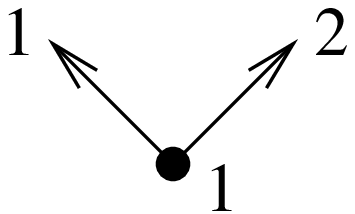} \\ \nn
&& =  - 2 v \Delta'(0^+) \int_{s_1<\min(t_1,t_2)} e^{-(2 s_1 - t_1-t_2)} \\
&& = - v \Delta'(0^+)\,\rme^{-|t_{1}-t_{2}|} \ ,
\eea
recovering the above result (\ref{second}) to lowest order in $v$.

\subsubsection{Third moment}\label{a28}
We are now ready to compute the third cumulant. Here and below we label external times by
$t_{i}$ and internal times by $s_{i}$ (black dots). To lowest order in the disorder, one finds from
 Eq.~(\ref{exact1}):
\begin{equation}\label{a29}
L^{2 d} \overline{\dot {u}_{t_1} \dot {u}_{t_2} \dot {u}_{t_3}   }^{\rm c} =  \partial_{t_1}\partial_{t_2}\partial_{t_3} \left[ 6\,\mbox{Sym}\diagram{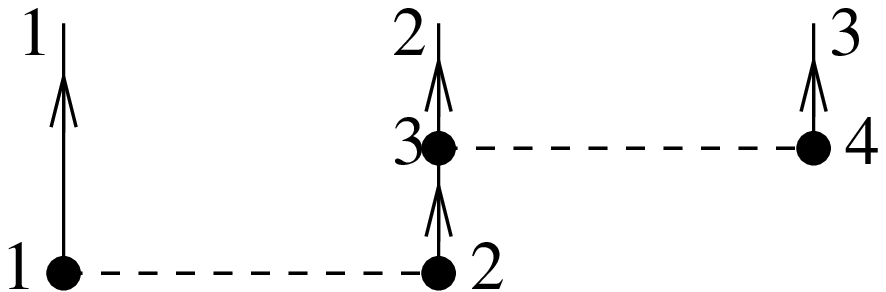} \right]
\end{equation}
where ${\rm Sym}$ denotes symmetrization w.r.t.\ the external times $t_i$. Hence
one has
\begin{equation}\label{a30}
L^{2 d} \overline{\dot{ {u}}_{t_1} \dot{ {u}}_{t_2}\dot{ {u}}_{t_3}   }=
6\,\mbox{Sym}\,  \partial_{t_{1}} \partial_{t_{2}}  \partial_{ t_{3}}\diagram{cumul3d}  \ ,
\ .
\end{equation} 
The first thing one could do is to perform the $\partial_{t_3}$ derivative, using partial integrations
\begin{align}\label{a31}
&\int_{s_{4}}\partial_{t_{3}} R_{t_{3}-s_{4}} \Delta' (s_{3}-s_{4})\nonumber \\
&\qquad = -\int_{s_{4}}\partial_{s_{4}} R _{t_{3}-s_{4}} \Delta' (s_{3}-s_{4})\nonumber \\
&\qquad = \int_{s_{4}} R_{t_{3}-s_{4}} \partial_{s_{4}} \Delta' (s_{3}-s_{4})\nonumber \\
&\qquad = - 2\Delta' (0^{+})\int_{s_{4}} R_{t_{3}-s_{4}} \delta (s_{3}-s_{4})\ .~~~
\end{align}
Note that we have safely replaced $\Delta'(v(s_3-s_4))$ by $\Delta'(s_3-s_4)$ since we anticipate that
to lowest order we will need only $\Delta'(u) = \Delta'(0^+) {\rm sgn}(u) + O(u)$. 
Note that there is no boundary term if time integrals are performed from
$[-\infty,\infty]$ and the theta function is included in $R$. By this procedure, the term   $\Delta
(s_{3}-s_{4})$ will have exactly two derivatives. However, to be able to proceed further,
it is better to consider $\partial_{t_2} \partial_{t_3}$ simultaneously, while symmetrizing at
the same time leading instead to (passing always one external derivative onto each disorder
vertex-end): 
\begin{eqnarray}\label{a35}
\lefteqn{\!\!\! \frac{1}{2} \partial_{t_{2}} \partial_{t_{3}} \left[
\diagram{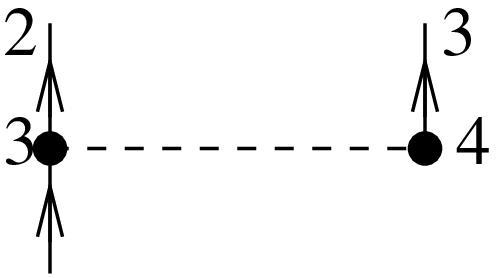}+
\diagram{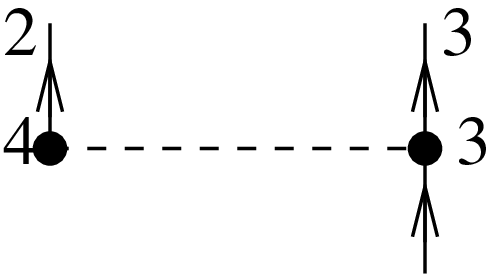} \right] } \nonumber \\
&=& -  \Delta' (0^{+})  \int_{s_{4}} \Big[\partial_{t_2} R
(t_{2}-s_{3}) R (t_{3}-s_{4})\delta (s_{3}-s_{4})\nn \\
&& ~~~~~~~~~~~~~~~~~~~~~~~+ R
(t_{2}-s_{4})  \partial_{t_{3}} R (t_{3}-s_{3})\delta (s_{3}-s_{4})\Big] \nonumber \\
&=&  -  \Delta' (0^{+}) \left(\partial_{t_2}+\partial_{t_{3}} \right)
\Big[R(t_{2}-s_{3}) R (t_{3}-s_{3})\Big]  \nonumber \\
&=&   \Delta' (0^{+}) \partial_{s_{3}} \Big[
R(t_{2}-s_{3}) R (t_{3}-s_{3})   \Big]
\ .
\end{eqnarray}
Integration by part w.r.t.\ $s_3$ is then possible, and together with
taking $\partial_{t_1}$ on the left branch and using time translational
invariance of $R_{s_3-s_2}$ and $R_{t_1-s_1}$ respectively
leads to two derivatives on the lower vertex $\Delta(s_1-s_2)$.

In summary, we find that the surplus external derivatives can always be passed down
in the tree, so that at the end each vertex receives exactly two
derivatives. This means that we can rewrite (\ref{a30}) as 
\begin{equation}\label{a36}
L^{2 d} \overline{\dot{{u}}_{t_1} \dot{ {u}}_{t_2}\dot{ {u}}_{t_3}
}= 6 v \Delta' (0^{+})^{2}\, \mbox{Sym}\,\int_{s_{1}}\int_{s_{2}}  \diagram{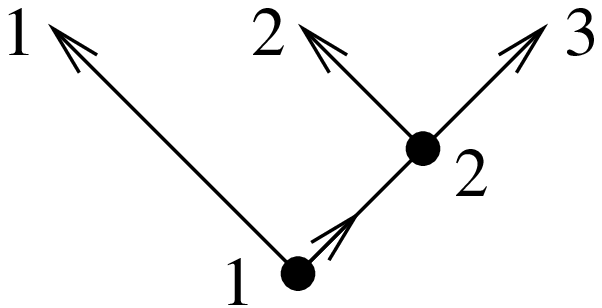}\ ,
\end{equation}
where the points are intermediate times, and the arrows standard
response functions. We now have to compute this new diagram, with the huge simplification
that vertices are now {\it local} in time and which apart from the vertices contains only response functions.

We also note that the single $v$ factor comes from the lower vertex: This can be interpreted as the point in space and time, where an avalanche is triggered with rate $v$. 

Let us now complete the integration over internal times. To this aim, let us fix the smallest
internal time $s_{1}$, and integrate over $s_{2}$: 
\begin{eqnarray}\label{a37}
&& \int_{s_{2}}  \diagram{cumul3-sum} = R_{t_{1}-s_{1}} \int_{s_{2}} R_{t_{2}-s_{2}} R_{t_{3}-s_{2}} R_{s_{2}-s_{1}}  \nn \nonumber \\
&=&R_{t_{1}-s_{1}} \left[\rme^{ -[\max ( t_{2},t_{3})-s_{1}]} - \rme^{- ( t_{2}-s_{1})- ( t_{3}-s_{1})}  \right] \nn
\\
&& \times \Theta (s_{1}<\min (t_{2},t_{3})) 
\ .
\end{eqnarray}
Integrating once more gives 
\begin{eqnarray}\label{a38}
 \int_{s_{1},s_{2}}  \diagram{cumul3-sum} 
&=& \frac{1}{2} \rme^{2\min (t_{1},t_{2},t_{3}) -t_{1}-\max (t_{2},
t_{3})}  \nn \\
&-& \frac{1}{3}\rme^{3 \min (t_{1},t_{2},t_{3}) -t_{1}-t_{2}-t_{3}}
\ .
\end{eqnarray}
Finally, after symmetrization it simplifies into \begin{eqnarray}\label{a39}
&&\!\!\!6\, \mbox{Sym}\int_{s_{1},s_{2}}  \diagram{cumul3-sum} \nonumber \\
&&=   \rme^{\min (t_{1},t_{2},t_{3})-\max (t_1,t_{2},t_{3})}
\ .
\end{eqnarray}
Hence, assuming that the external times are ordered as
$t_{1}<t_{2}<t_{3}$ we obtain our final result for the third velocity cumulant
as
\begin{eqnarray}\label{a33}
 L^{2 d} \overline{\dot{ {u}}_{t_1} \dot{ {u}}_{t_2}\dot{ {u}}_{t_3}  }^{\rm c} &=&  2 v \Delta' (0^{+})^{2} \rme^{t_{1}-t_{3}}   \\ \nonumber
&=& 2  v \Delta' (0^{+})^{2}
\rme^{- (|t_{1}-t_{2}| + |t_{1}-t_{3}|+|t_{2}-t_{3}|)/2}\ .\quad 
\end{eqnarray}
Note that the final expression is  simple, while the starting one was quite non-trivial. 

We can now check that the sum rule (\ref{sumrule}) is satisfied. Indeed 
\begin{eqnarray}
 v 
\frac{\langle S^{3} \rangle}{\langle S \rangle}&=&L^{2 d} \int_{t_{2}}\int_{t_{3}} 
\overline{\dot{{u}}_{t_1} \dot{{u}}_{t_2}\dot{ {u}}_{t_3}   }^{\rm c} \nn\\ &=&  12\, v \Delta' (0^{+})^{2} \int_{0=t_{1}<t_{2}<t_{3}}\rme^{t_{1}-t_{3}}~~~~~~~~ \nn \\
& =&  12 v \Delta' (0^{+})^{2}  
\end{eqnarray}
recovering the result of \cite{LeDoussalWiese2008c}, and which can be obtained by expanding (\ref{Z-tree}) for the third moment of the avalanche-size distribution. 

\subsubsection{Fourth moment} 

The higher moments can be computed using the same method, as the same 
simplifying features can be generalized. The result
for the fourth cumulant is, supposing the times are ordered as
$t_{1}<t_{2}<t_{3}<t_{4}$: 
\begin{eqnarray}\label{a41}
\lefteqn{L^{3d} \overline{\dot{ {u}}_{t_1} \dot{{u}}_{t_2}\dot{ {u}}_{t_3} \dot{ {u}}_{t_4}
}}\\ &=& - 24 v \Delta'(0^+)^3  \,\mbox{Sym}\left[
\diagram{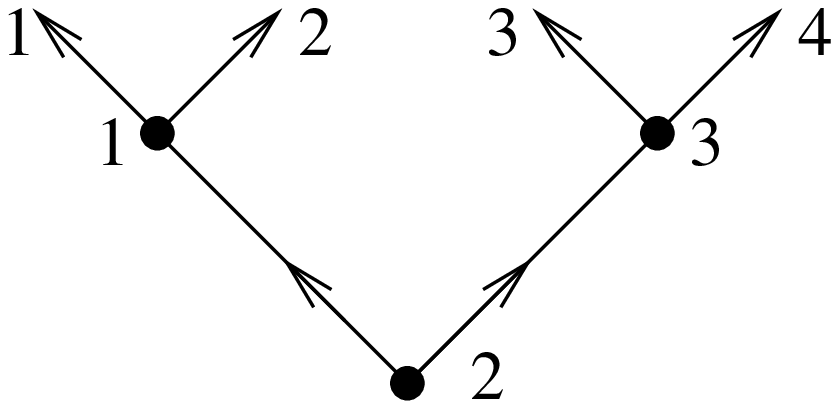} \right.\nn\\ &&
\hphantom{= - 24 v \Delta'(0^+)^3  \,\mbox{Sym}}
+4\left.\diagram{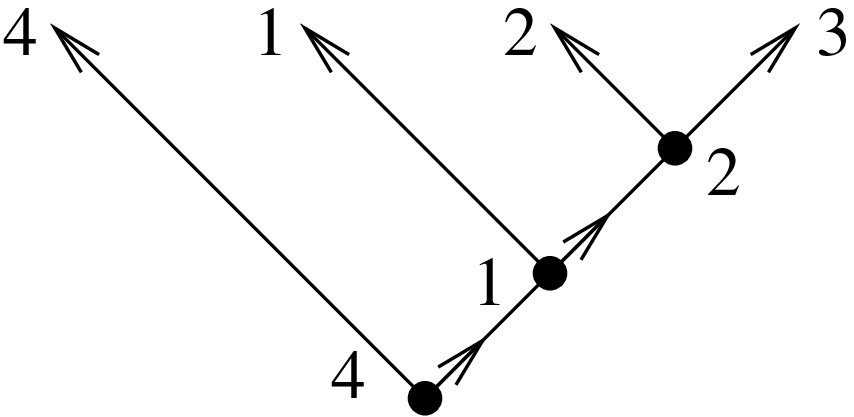}  \right] \nn \\
&& = v |\Delta'(0^+)|^3 [ 4 \rme^{t_{1}-t_{4}} +2 \rme^{t_{1}+t_{2}-t_{3}-t_{4}} ]
\end{eqnarray}
The sum rule gives 
\begin{equation}\label{a42}
v \frac{\langle S^{4} \rangle}{\langle S \rangle} = L^{3d}\int_{t_{2},t_{3},t_{4}}\overline{\dot{ {u}}_{t_1} \dot{{u}}_{t_2}\dot{{u}}_{t_3} \dot{{u}}_{t_4}
}  =  120 v |\Delta'(0^+)|^3  
\ ,
\end{equation}
which coincides with the result for the fourth moment of Eq.~(\ref{Z-tree}). 
\subsubsection{Fifth moment}
Finally, we give the fifth moment
\begin{widetext}
\begin{eqnarray}\label{a44}
&& L^{4 d}  \overline{\dot{{u}}_{t_1}
\dot{{u}}_{t_2}\dot{{u}}_{t_3} \dot{{u}}_{t_4} \dot{
{u}}_{t_5}} \nn\\
&& =  v \Delta'(0^+)^4  5! \,\mbox{Sym} \left[8 \diagram{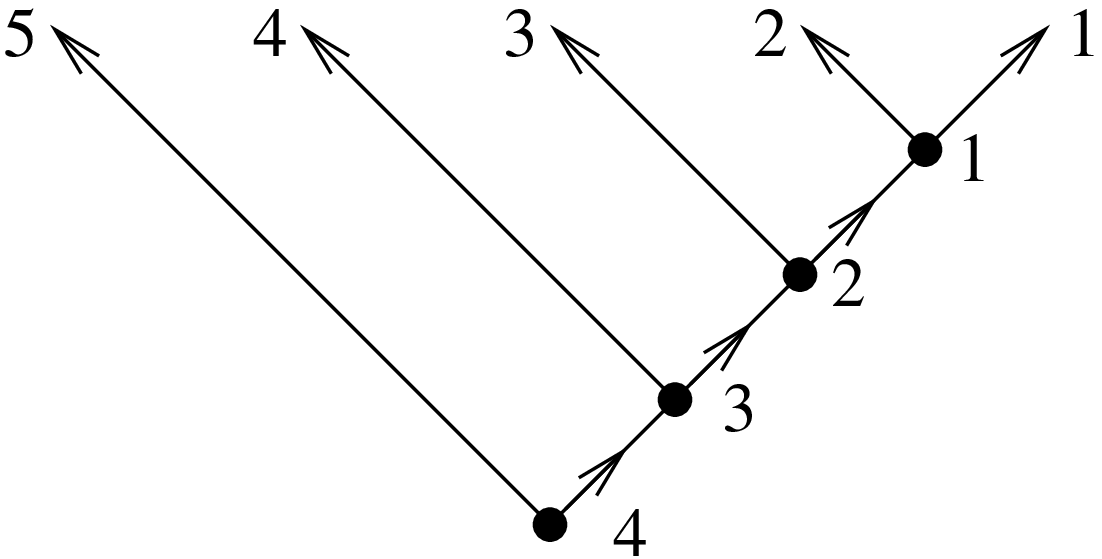}+2
\diagram{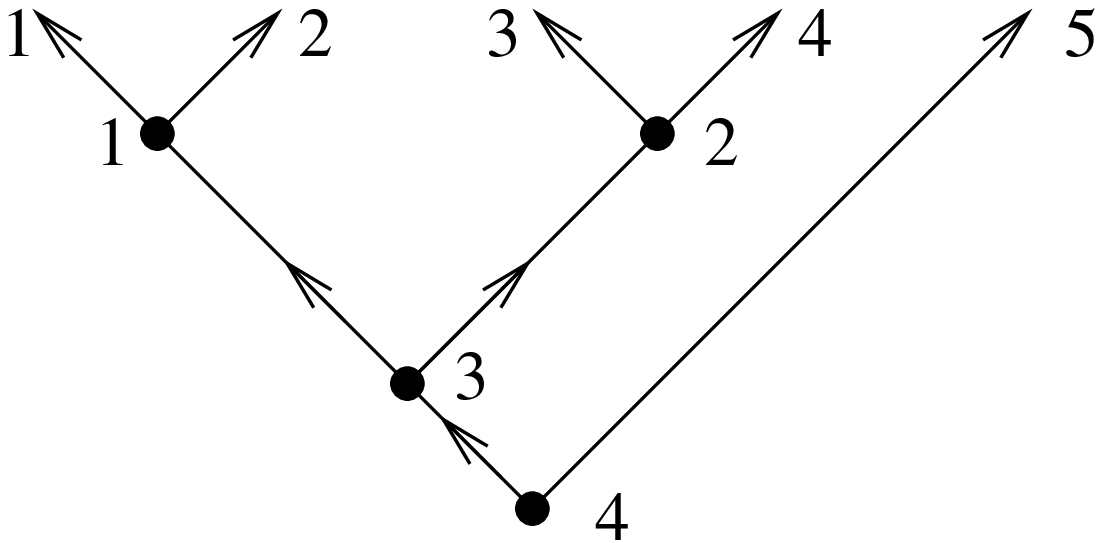}+4\diagram{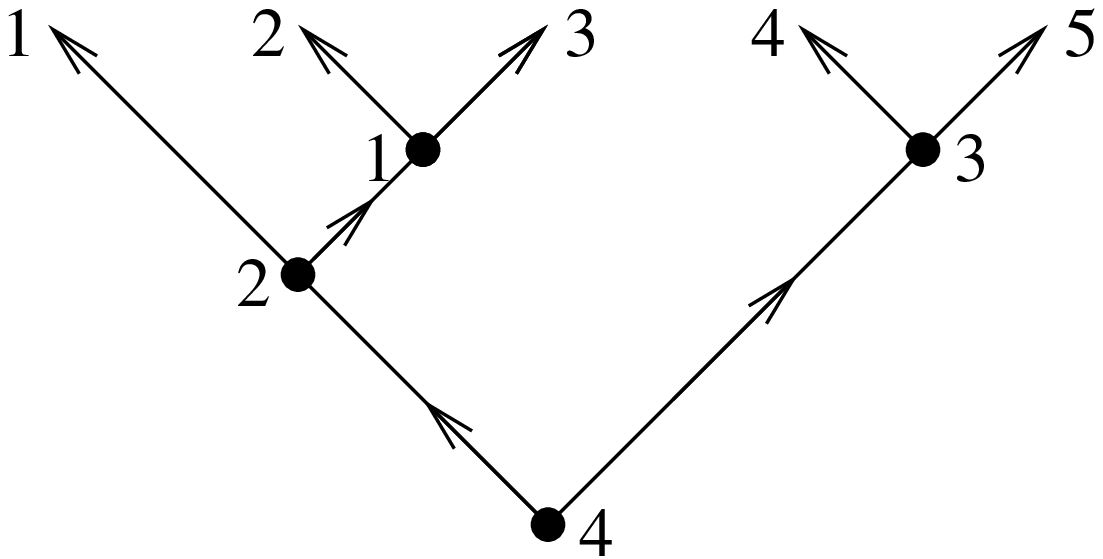}\right] \nonumber \\
&& = v \Delta'(0^+)^4 [ 8 \rme^{t_1-t_{4}}+4 \rme^{t_{1}+t_{2}-t_{3}-t_{5}} +8
\rme^{t_{1}+t_{2}-t_{4}-t_{5}} +4   \rme^{t_{1}+t_{3}-t_{4}-t_{5}} ]
\end{eqnarray}
\end{widetext}
We check that 
\begin{equation}\label{a45} v \frac{\langle S^{4} \rangle}{\langle S \rangle}
=L^{4d}\int_{t_{2},t_{3},t_{4},t_{5}}   \overline{\dot{{u}}_{t_1}
\dot{{u}}_{t_2}\dot{ {u}}_{t_3} \dot{ {u}}_{t_4} \dot{{u}}_{t_5}} = 5! \times 14 v \Delta'(0^+)^4  
\end{equation}
coincides with the result for the fifth moment of (\ref{Z-tree}). 

The above results suggest that there is an underlying simplification at the level of tree diagrams of the
original field theory, which is non-local in time, into a field theory which is local in time. We now show how
the latter arises.

\subsection{Generating function and instanton equation: Simplified (tree) field theory}
\label{simplified}

Since here we want to study the temporal and spatial statistics of the instantaneous
velocity field, we define the following generating functional of a (possibly space- and time-dependent) source field $\lambda_{xt}$,\be  \label{genfunctionG}
G[\lambda] := \overline{\rme^{\int_t \lambda_{xt} (v + \dot u_{xt})}}\ .
\ee 
We remind that we are working in the comoving frame, i.e.\ \(v + \dot u_{xt}\) is the velocity of the manifold in the laboratory frame. The functional
\(G[\lambda]\) encodes all possible information. In particular, all moments can be recovered by differentiation w.r.t.\ the source.
In this article we focus on the small driving-velocity limit. In view of the results of the previous sections, it will be sufficient to
compute the  generating function,
\begin{equation} \label{genfunctionZ}
   Z[\lambda] := L^{-d} \partial_v G[\lambda] \Big|_{v=0^+} \ ,
\ee 
which contains the leading $O(v)$  dependence of all moments
in the limit of small velocity $v=0^+$.

It turns out that, within the tree level theory, it is possible to compute these generating functions and
obtain all cumulants at once, as well as the velocity distribution. 
We now show how this simplification occurs.

We start not from the equation of motion (\ref{1}), but from its time derivative in the comoving frame\footnote{Below, when indicated, we
will alternatively use this equation in the laboratory frame, which amounts to setting $v=0$ in Eq.~(\ref{eqmo}).}
\be \label{eqmo}
(\eta \partial_t - \nabla^2_x + m^2) \dot u_{xt} = \partial_t F(v t + u_{xt} , x) + \dot f_{xt} - m^2 v
\ .
\ee
For completeness we wrote it for arbitrary driving $f_{xt}=(m^2-\nabla^2_x) w_{xt}$, however we will mostly specialize 
to uniform driving, i.e.\ $\dot w_{xt}=v$, $\dot f_{xt} = m^2 v$, in which case the last term is zero. We denote indifferently time derivatives by $\dot u$ or $\partial_t u$, and
for now we use the original (microscopic) units. Again,  one has to set
$\eta \to \eta_0$, $\Delta \to \Delta_0$ for a derivation starting from the
bare model, or the renormalized parameters if one deals with the
improved action. 

We now average over disorder (and initial conditions) using the MSR dynamical action $S$ associated to the equation of motion (\ref{eqmo}):\bea
 {\cal S} &=& {\cal S}_0 + {\cal S}_{\mathrm{dis}} \\
  {\cal S}_0 &=& \int_{xt} \tilde u_{xt} (\eta \partial_t - \nabla_x^2 + m^2) \dot u_{xt}  \label{msr} \\
  {\cal S}_{\rm dis} &=&  - \frac{1}{2} \int_{xtt'}  \tilde u_{xt}  \tilde u_{xt'}  \partial_t \partial_{t'} \Delta\big(v (t{-}t') + u_{xt} {-}u_{xt'} \big) .~~~~~~~~~ \label{msr2}
\eea
 Note that this is the dynamical action associated to
the velocity theory, i.e.\ in terms of $\tilde u_{xt}$ and $\dot u_{xt}$ to be distinguished from the one usually considered,
associated to the position theory, in terms of $\hat u_{xt}$ and $u_{xt}$, to be discussed below.

The generating function (\ref{genfunctionG}) can then be written as
\bea
 G[\lambda] &=& \int{ \cal D} [\dot u] {\cal D} [ \tilde u] e^{-  {\cal S}_\lambda } \\
  {\cal S}_\lambda &=&  {\cal S} - \int_{xt} \lambda_{xt} (v + \dot u_{xt})\ ,
\eea 
with $G[0]=1$ and $Z[0]=0,$ since the dynamical partition function is normalized to unity. We can rewrite for the time-derivatives appearing in Eq.~(\ref{msr2})
\bea
 \lefteqn{\partial_t \partial_{t'} \Delta(v (t-t') + u_{xt}-u_{xt'} )} \nn\\
&& =
(v + \dot u_{xt}) \partial_{t'} \Delta'(v (t-t') + u_{xt }-u_{xt'}) \nn\\
&& = (v + \dot u_{xt})  \Delta'(0^+) \partial_{t'}  {\rm sgn}(t-t') + ...\label{70} 
\ .
\eea 
Here we have used that ${\rm sgn}(v(t-t') +u_{xt}-u_{xt'})={\rm sgn}(t-t')$, i.e.\ the motion for $v>0$ is monotonously forward,
as guaranteed by the Middleton theorem \cite{Middleton1992}. The neglected terms in Eq.~(\ref{70}) are  higher derivatives of $\Delta(u)|_{u=0^+}$. As we discuss below at length, they
contribute only to $O(\epsilon=4-d)$ to $Z[\lambda],$ hence they can  be
 neglected at tree level. This is consistent with our findings in the previous
section that  only $\Delta'(0^+)$ appears at tree level. Hence we can  rewrite the disorder 
part $ {\cal S}_{\mathrm{dis}}$ of the dynamical action,  which is a priori non-local in time, as
$ {\cal S}_{\mathrm{dis}}= {\cal S}^{\mathrm{tree}}_{\mathrm{dis}}+...$, where
\be  \label{cubic}
    {\cal S}^{\mathrm{tree}}_{\mathrm{dis}} =   \Delta'(0^+)  \int_{xt}  \tilde u_{xt}  \tilde u_{xt}  (v + \dot u_{xt}) 
\ee 
is an action {\it local in time}. Furthermore we recognize   the cubic vertex which
generates  the simple graphs obtained in the previous sections by a
  systematic
perturbation expansion. The action\be  \label{Stree}
    {\cal S}^{\mathrm{tree}}:= {\cal S}_0 +  {\cal S}_{\mathrm{dis}}^{\mathrm{tree}}
\ee  
is the so-called tree-level, or mean-field, action. Note that if we use the improved action, it then includes the loop corrections to $\eta$ and $\Delta$, and yields the correct result for $d=d_{\rm uc}=4$, making the dependence in $m$ explicit as $\eta \to \eta_m$ and $\Delta \to \Delta_m$, see the discussion below and in Ref.\ \cite{LeDoussalWiese2008c}. 
Note that due to the STS symmetry mentioned above, $m^2$, the elastic coefficient in front of $\nabla^2u_{xt}$, and $v$ are 
not corrected. 

We can now study algebraically the tree approximation
\bea
 Z^{\mathrm{tree}}[\lambda] &=& L^{-d} \partial_v G^{\mathrm{tree}}[\lambda]\Big|_{v=0^+} \label{Ztree} \\
 G^{\mathrm{tree}}[\lambda] &=& \int {\cal D} [\dot u]{\cal  D} [\tilde u] e^{-  {\cal S}^{\mathrm{tree}}_\lambda } \label{Gtree} \\
  {\cal S}^{\mathrm{tree}}_\lambda &=&  {\cal S}^{\mathrm{tree}} - \int_t \lambda_{xt} (v + \dot u_{xt})\ .   \label{Streelambda}
\eea
Note that the highly non-linear action (\ref{msr})  (\ref{msr2}) has been reduced to a much simpler cubic theory.
 Cubic theories among others describe branching processes, such as the Reggeon field theory \cite{CardySugar1980} for directed percolation. The present theory however is simpler, and
can be reduced to a non-linear equation as we now explain.

Remarkably, considering (\ref{Ztree}), one notices that $\dot u_{xt}$ appears in $ {\cal S}^{\mathrm{tree}}_\lambda$ only linearly, i.e.\
in the form $\int_{xt} \dot u_{xt} O_{xt}[\tilde u,\lambda]$. It can thus be  integrated out, leading to a $\delta$-function constraint\footnote
{Equivalently one can view $\dot u_{xt}$ as a response field associated to the equation $O_{xt}[\tilde u,\lambda]=0.$} $\prod_{xt} \delta(O_{xt}[\tilde u,\lambda])$.  Hence in the tree-level theory the field $\tilde u_{xt}$ is not fluctuating, but given by the  non-linear equation
\begin{equation} \label{mfnonlinearxt}
 ( \eta \partial_t + \nabla_x^2 - m^2) \tilde u_{xt} - \Delta'(0^+) \tilde u_{xt}^2  + \lambda_{xt} = 0\ .
\end{equation}
This equation is the saddle-point equation w.r.t.\ $\dot u$ of $ {\cal S}^{\mathrm{tree}}_\lambda$ in presence of a source, and
 is satisfied exactly. We also call it {\it the instanton equation}. 
We denote $\tilde u^\lambda_{xt}$ the solution of this equation for a given source field $\lambda_{xt}$
with $\tilde u^{\lambda=0}=0$.
After integration over $\dot u_{xt}$, we thus obtain from Eqs.\ (\ref{Stree}) to (\ref{Streelambda}):
\bea
 G[\lambda] &=& e^{v L^d Z[\lambda]} \label{Gv} \\
 Z[\lambda] &=& L^{-d} \int_{xt} [ \lambda_{xt} - \Delta'(0^+)   (\tilde u_{xt}^\lambda)^2 ] \nn\\
& =& -  L^{-d}  \int_{xt} (\eta \partial_t + \nabla_x^2 - m^2) \tilde u^\lambda_{xt} \nn\\ &=& m^2 L^{-d} \int_{xt}  \tilde u^\lambda_{xt} 
\ . \label{78}
\eea
Here we have used the saddle-point equation (\ref{mfnonlinearxt}) and, in the last equality, assumed
that $\tilde u_{xt}^\lambda$ (resp. $\nabla_x \tilde u^\lambda$) vanishes at large $t$ (resp.\ $x$). 
This is insured if the source vanishes at infinity which we assume in the following. Note that since
$Z[\lambda]$ is independent of the velocity, Eq.~(\ref{Gv}) gives the full dependence at finite $v$,
a fact which is exploited and studied in detail in Ref.\ \cite{DobrinevskiLeDoussalWiese2011b}. 

In summary we find that the calculation of $Z[\lambda]$, i.e.\ of all cumulants of the velocity field, is equivalent to
solving the non-linear equation (\ref{mfnonlinearxt}). The solution $\tilde u^{\lambda}_{xt}$ can be 
constructed perturbatively in an expansion in powers of the source $\lambda_{xt}$. To lowest order
\be  \label{fot}
   \tilde u^\lambda_{x't'} = \int_{x,t} \lambda_{xt} R_{xt,x't'}  + O(\lambda^2) 
\ ,
\ee 
where $R_{xt,x't'}$ is the usual  bare response function (\ref{defresponse}). 
Integrating Eq.~(\ref{mfnonlinear}) or (\ref{fot}), one finds
\be 
   Z(\lambda) = L^{-d} \int_{xt} \lambda_{xt} + O(\lambda^2)
\ ,
\ee 
which is consistent with $\overline{\dot u_{xt}}=0$ ($v$ is uncorrected). 
Pursuing to $O(\lambda^2)$ and higher orders, one recovers the velocity cumulants obtained in the previous sections, and in addition obtains their full spatial dependence. Instead of working perturbatively, we obtain and analyze in the next subsection the (joint) probability distributions of the velocity at one (and several) times,
focusing on the simplest observable,  the center-of-mass velocity $\dot u_t$. 

Let us note that the simplified (tree) theory defined above does not contain {\it all} tree graphs. 
There are other tree graphs involving $\Delta''(0)$ and higher derivatives, as e.g.\ the following configurations of order $v^2$,  \be
\diagram{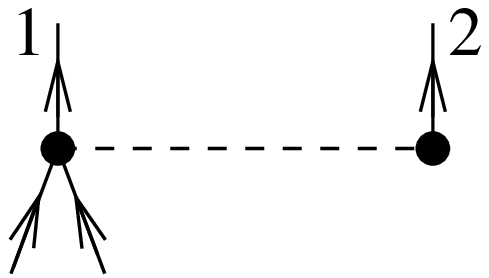}+\diagram{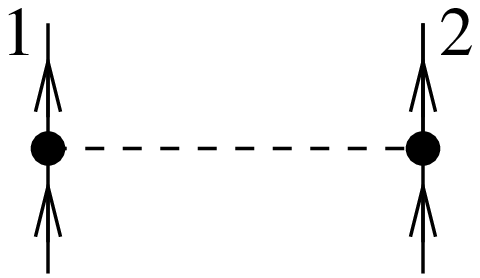}+\diagram{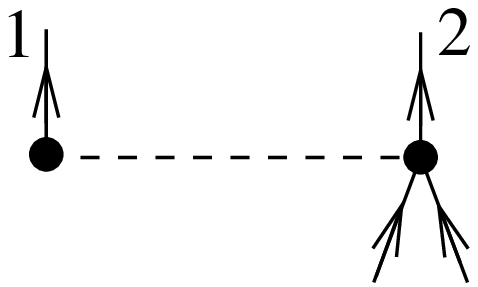}.
\ee While they are  similar to those in Eq.\ (\ref{uu}), different classes of trees appear starting at the fourth moment, as e.g.
\be
\diagram{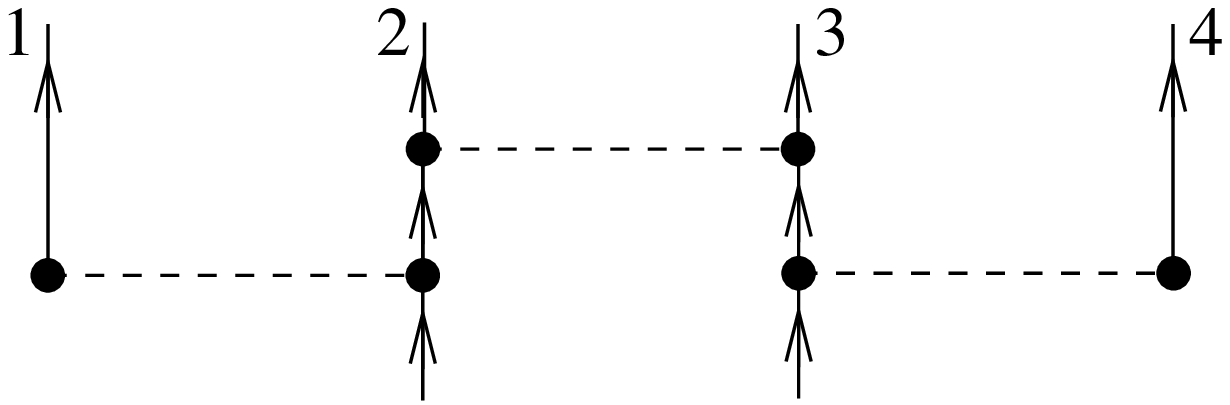}\ .
\ee  
These diagrams are characterized by the fact that they have {\em two} (or more) roots (lowest vertices), and are of order $v^2$ (or higher). 
The full tree theory is studied in section \ref{s:first-principle} and can  be reduced to two non-linear saddle-point equations. However since these additional tree graphs lead to contributions 
which are of higher order in $v$, to study a single avalanche  in the small-$v$ limit,  they are not needed. 

Finally, it is important to stress that the above simplified tree theory corresponds to
the problem of an elastic manifold in a random-force landscape made out of uncorrelated Brownian
motions, for which it is exact for monotonous driving. This is the BFM, 
discussed  in Section \ref{s:BFM}.

\subsection{Joint probability distributions for the center-of-mass velocity}
\label{s:jpdfcomv}
To analyze the results, it is convenient to use dimensionless equations, 
hence replacing $x \to x/m$, and 
$t \to \tau_m t$. In mean field
$\tau_m = \eta/m^2$, $\lambda \to \lambda\tau_m/S_m$,
$\tilde u_{xt} \to \tilde u_{xt}/(m^2 S_m)$, and $v \to v v_m$, where 
$v_m=S_m m^{d}/\tau_m$, $L \to L/m$. We start by using these units and,
whenever indicated, switch back to dimension-full units in discussing the final
results. We also keep the factor of $L^d$ in the beginning, but later on we find
it convenient to suppress it. That amounts to a further change of 
units as $v \to v \tilde v_m$ with $\tilde v_m=(m L)^{-d} v_m$ whenever
indicated below.

\subsubsection{1-time center-of-mass velocity distribution}
\label{s:MF-1pt}

The center-of-mass velocity distribution is obtained by choosing a uniform $\lambda_{xt} = \lambda_t$. The 1-time probability is obtained   from the inverse Laplace transform of $\tilde Z(\lambda)$,  choosing $\lambda_t:=\lambda \delta(t)$, 
\begin{equation}
   \tilde Z(\lambda) = L^{-d} \partial_v \overline{e^{L^d \lambda (v + \dot u)}}|_{v=0^+} 
\ .\ee 
Here $\dot u=\dot u_{t=0}$, and the tilde on $\tilde Z(\lambda)$ reminds us that we use dimensionless units.
The saddle-point equation (\ref{mfnonlinearxt}) admits a spatially uniform solution  $\tilde u_{xt}=\tilde u_t$, thus we need to solve
\begin{equation} \label{mfnonlinear}
   (\partial_t - 1) \tilde u_{t} + \tilde u_{t}^2  = - \lambda \delta(t)\ . 
\end{equation}
The boundary condition is  $\tilde u_{t} \to 0$ at $t=\pm \infty$, leading to
\be \label{soluinstanton}
 \tilde u_t  = \frac{\lambda}{ \lambda + (1-\lambda) e^{-t}} \theta(-t)\ . 
\ee
This gives the generating function  \begin{equation} \label{ztilde}
   \tilde Z(\lambda)  =  \int_t \tilde u_t =  - \ln(1 -  \lambda)\ . 
\end{equation}
We now want to infer from this the 1-time velocity distribution in an avalanche. Before doing so, let us restore dimension-full units. We assume
that in the limit $v=0^+$ there are times when the velocity is exactly zero, i.e.\ $v + \dot u=0$ (since we use  the co-moving frame) and times (when an avalanche is proceeding) when the velocity is non-zero. This picture is  confirmed
by results below \footnote{ This gives the universal regime for $\dot u \gg v_0$. For velocities smaller than the cutoff 
$v_0$ one expects a dependence on the details of the dynamics.}. 
Hence the 
1-time velocity probability (at say time $t=0$) must take the form
\be \label{prob}
P(\dot u) = (1-p_a) \delta(v+\dot u) + p_a {\calP}(\dot u)\ .
\ee
Here $p_a$ is the probability that $t=0$ belongs to an avalanche, and ${\calP}(\dot u)$ is the  conditional probability of velocity, given that $t=0$ belongs to an avalanche. Both ${\calP}$ and $P$ are normalized to unity. One notes the two (always) exact relations $\langle \dot u \rangle_P=0$ and $p_a  \langle v + \dot u \rangle_{{\calP}}=v$. It is easy to see that
\be \label{pa}
p_a=\rho_0 v \langle \tau \rangle\ .
\ee
The mean duration of an avalanche is $\langle \tau \rangle=\frac{1}{N_a} \sum_i \tau_i$ where $N_a$ is the total number of avalanches and $\tau_i$ the duration of the $i$-th avalanche\footnote{Note that we are implicitly working  to lowest order in $v$, at
small $v$. Hence the fact that  \(p_a\) increases linearly with  $v$, while \(\left<\tau\right>\) remains constant, does not conflict with the requirement that \(p_a<1\), since we study here the regime of small $p_a$. At larger $v$, 
avalanches will merge, and formula (\ref{pa}) ceases to be valid.}. Now from Eq.\ (\ref{prob}) one has\be
\overline{ e^{\lambda L^d (v + \dot u)} } = 1 + p_a \int \rmd \dot u\, {\calP}(\dot u) 
\left(e^{L^d \lambda (v+\dot u)}-1\right)\ .
\ee
Taking a derivative w.r.t.\ $v$, one obtains to leading order in $v =0^+$
\bea \label{Zt}
 Z(\lambda) &=& \frac{1}{m^{-d} v_m} \tilde Z( m^{-d} v_m \lambda) \nn\\
& =& L^{-d} \rho_0 \langle \tau \rangle \int \rmd\dot  u \,{\calP}(\dot u)\left 
(e^{L^d \lambda \dot u}-1\right)\ . 
\eea
The identity 
\be 
\tilde Z(\lambda) = - \ln(1-\lambda) = 
\int_0^\infty \frac{\rmd {\sf x}}{{\sf x}} e^{-{\sf x}} \left(e^{\lambda {\sf x}} -1\right)\ee 
allows to perform the inverse Laplace transform\footnote{In practice one
performs the Laplace inversion on $\tilde Z'(\lambda)$ which yields $\dot u {\calP}(\dot u) $, thus has no singularity at $\dot u=0$.} of Eq.\ (\ref{ztilde}).
We thus obtain, in the slow-driving
limit, the distribution of the instantaneous velocity of the center of mass for $v_0 \ll \dot u $
(where $v_0$ is a small-velocity cutoff) as 
\begin{equation} \label{1time}
   {\calP}(\dot u) = \frac{1}{\rho_0 \langle \tau \rangle \tilde v_m^2} \,p\!\left(\frac{\dot u}{\tilde v_m}\right)\ , \qquad p({\sf x}) = \frac{1}{{\sf x}} e^{-{\sf x}\ }\ .
\end{equation}
We have defined $\tilde v_m = (m L)^{-d} v_m = L^{-d} S_m/\tau_m$. This agrees with the
above exact relation which becomes $\rho_0 \langle \tau \rangle \langle \dot u \rangle_{{\calP}} =1$ in the
limit of $v=0^+$. One notes that the distribution of small velocities diverges with a non-integrable $1/\dot u$
weight. Since ${\calP}(\dot u)$ should be normalized to unity, the ensuing logarithmic divergence requires a small-velocity
cutoff $v_0$. This leads to the additional relation
\be \label{tautau}
\rho_0  \langle \tau \rangle \tilde v_m \approx \ln\left(\frac{\tilde v_m}{v_0}\right)
\ .\ee
Hence we already anticipate that the average avalanche duration will exhibit a logarithmic 
dependence on the small-scale cutoff, as  confirmed below. Let us note that
the rescaled function  $p({\sf x})$ is not a bona-fide probability, rather it is normalized such that 
$\int \rmd {\sf x}\, {\sf x}\, p({\sf x}) =1$. Finally let us comment on the typical scale of the center-of-mass
velocity, $\tilde v_m$. Since $\dot u_t = L^{-d} \int_x \dot u_{xt}$ we find that the
scaling variable ${\sf x}$ entering $p({\sf x})$ is the ratio of the instantaneous increase in the total
area swept by the interface, $\int_x \dot u_{xt}$, divided by  its typical value $S_m/\tau_m$ (hence it
does not contain the factor of $L^{-d}$).

\begin{figure}[t]
\Fig{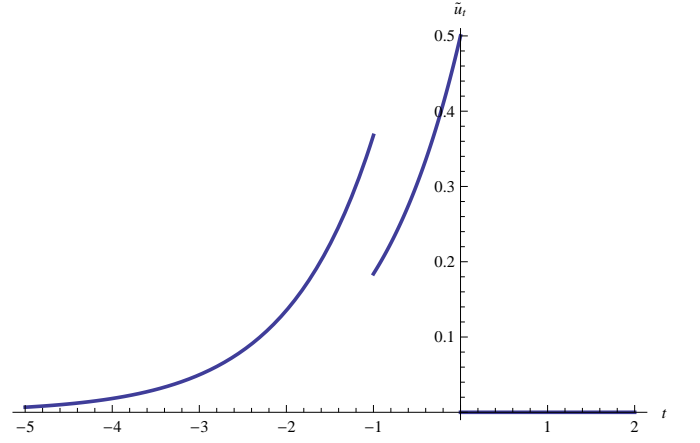}
\caption{Solution $\tilde u_{t}$ of the instanton equation (\ref{mfnonlinear}) as a function of $t$ 
for a source $\lambda(t)=\lambda_1 \delta(t-t_1) + \lambda_2 \delta(t-t_2)$, with $t_{1}=0$, and
$t_{2}=-1<t_1$. The function $\tilde u_{t}$ has the following properties: (i) It has the form $\tilde u (t) =
\frac{1}{a\rme^{-t}-1}$ on any interval where the source $\lambda(t)$ vanishes. (ii) It is zero for 
for $t>t_{1}$ by causality. (ii) It jumps by $\lambda_{1}$ (here $-0.5$) at $t=t_{1}$  and by $\lambda_{2}$ (here
$-0.155$) at $t=t_{2}$.}
\end{figure}

Let us indicate here for completeness the 1-time instanton solution in dimension-full 
units, as well as the generating function:
\bea 
 \tilde u_t &=& \frac{1}{m^2 S_m} \tilde u_t^{\rm dimless}(t/\tau_m,\lambda S_m/\tau_m) \\
 G (\lambda) &=& e^{v m^2 L^d \int \rmd t\, \tilde u_t} = e^{v \frac{\tau_m}{S_m} L^d \tilde Z(\lambda S_m/\tau_m)} 
\ .
\eea
We recall that  (\ref{1time}), and all formulae concerning the center-of-mass velocity distribution,
assume that the driving velocity $v$ is small enough at fixed $L$ so that only a {\it single} avalanche occurs,
$p_a \ll 1$; hence $v$ scales as $\sim L^{-d}$. If  $L$ goes to infinity first, at fixed small $v$, multiple avalanches 
occur along the interface. For small enough $v$ they occur at far away locations (distances $\gg 1/m$)
and are statistically independent. In that case the center-of-mass velocity distribution can be computed from convolutions 
of the distribution (\ref{1time}). It tends to a Gaussian distribution for large $L$ and fixed $v$. The present results thus 
describes mesoscopic fluctuations.

\subsubsection{Exact result for the $p$-time generating function}

We now obtain the generating function for the $p$-time distribution of the center-of-mass velocity,
\be  \label{defzp}
   \tilde Z_p(\lambda_1,...,\lambda_p) = L^{-d} \partial_v \overline{e^{L^d \sum_{i=1}^p \lambda_i (v + \dot u_{t_i})}}\Big|_{v=0^+} 
\ ,\ee 
by solving Eq.\ (\ref{mfnonlinear}) in presence of the source
$\lambda_t = \sum_{j=1}^p \lambda_j \delta(t-t_j)$. In this subsection we order the times as $t_{p+1}=-\infty<t_p<\dots< t_1$,
although in the following subsections we will choose the opposite order.

The solution reads 
\be  \label{solup}
\tilde u_t = \sum_{j={ 1}}^{p} \frac{\theta(t_{j+1} < t < t_j)\tilde u_{t_j^-}  }{(1-\tilde u_{t_j^-}) e^{t_j-t} +\tilde u_{t_j^-}}
\ee 
with $t_{p+1}=-\infty<t_p<\dots< t_1$, $\tilde u_{t_{j}^{-}}=\lambda_j + \tilde u_{t_{j}^+}$ and 
$ \tilde u_{t_1^+}=0$. Integration of (\ref{solup}) leads to $\tilde Z_p := \tilde Z_p(\lambda_1,...,\lambda_p)$ with\be 
\tilde Z_p = - \sum_{j=1}^{p} \ln(1- z_{j+1,j} \tilde u_{t_j^-} ) 
\ .
\ee 
We used  the  definition\be  \label{defz}
z_{i,j} \equiv z_{ij}:= 1 - \rme^{- |t_{i}-t_{j}|}\ ,
\ee 
hence in this section $z_{ij}=1-\rme^{t_{i}-t_{j}}$ with $i>j$. To generate $\tilde Z_p$ one can 
construct a recursion relation for the argument of the logarithm. From the above,
one finds \bea
 \Pi_{j+1} &=& A_j \Pi_j + B_j \Pi_{j-1} \\
 A_j &=& \frac{z_{j+2,j}}{z_{j+1,j}} - z_{j+2,j+1} \lambda_{j+1} \\
 B_j &=& 1- \frac{z_{j+2,j}}{z_{j+1,j}}
\eea 
with $\Pi_0=1$ and $\Pi_1=1-z_{21} \lambda_1$, so that 
\be 
\tilde Z_p = - \ln \Pi_p \big|_{z_{p+1,j} \to 1\ ;}
\ee 
here $t_{p+1}$ is set to $-\infty$. This leads to
\begin{eqnarray}\label{a96}
 \tilde Z_{2} &=& -\ln (1-\lambda_{1}-\lambda_{2} +\lambda_{1}\lambda_{2} z_{21})\\
 \tilde Z_{3} &=& -\ln\Big (1- \lambda_{1}-\lambda_{2}-\lambda_{3}\nn\\&&\qquad ~+\sum_{i>j}
\lambda_{i}\lambda_{j} z_{ij} -\lambda_{1}\lambda_{2}\lambda_{3}
z_{32}z_{21}\Big)   
\end{eqnarray}
\begin{figure*}
\begin{center}
\fig{4.5cm}{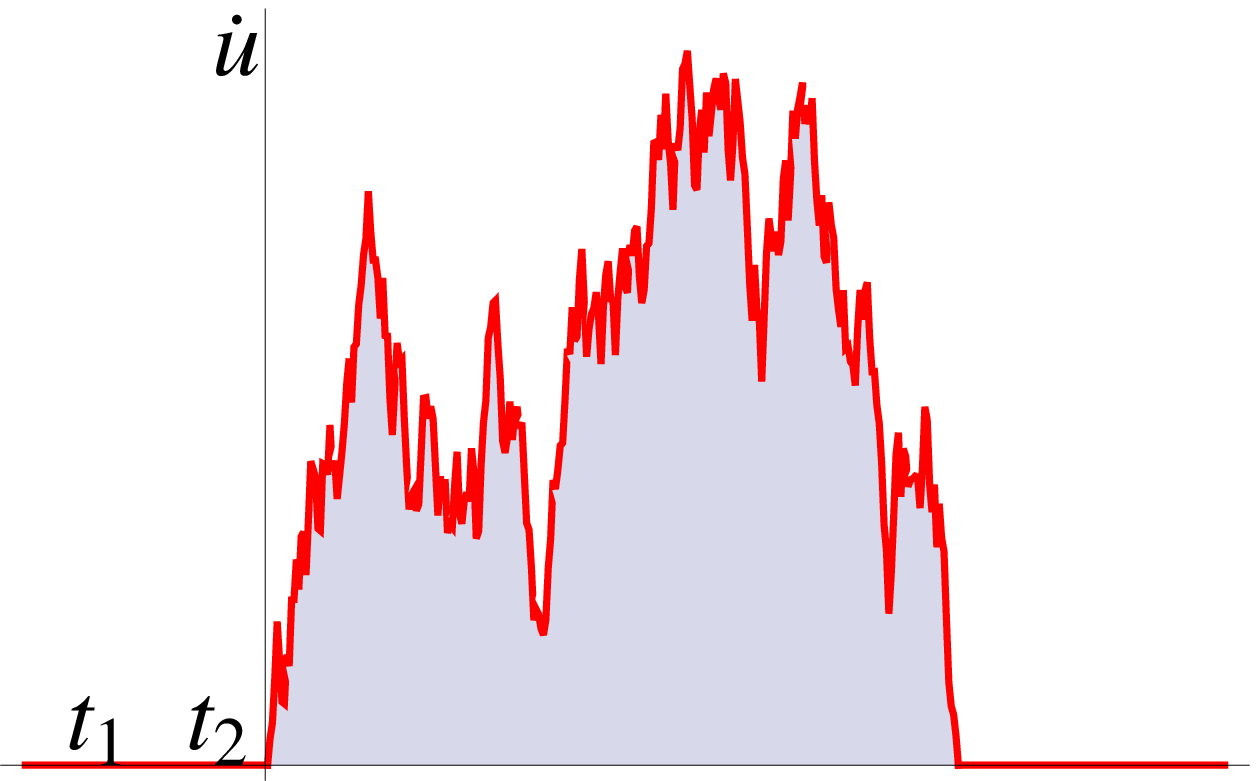}\fig{4.5cm}{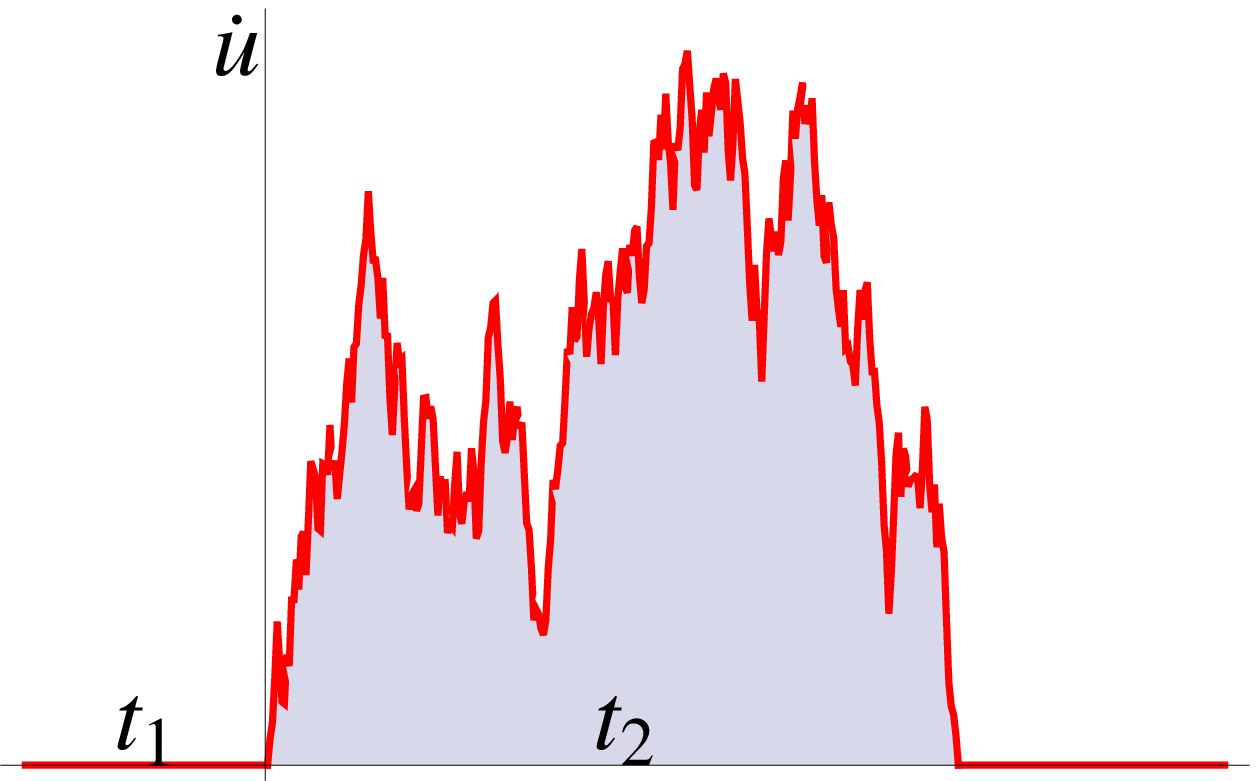}\fig{4.5cm}{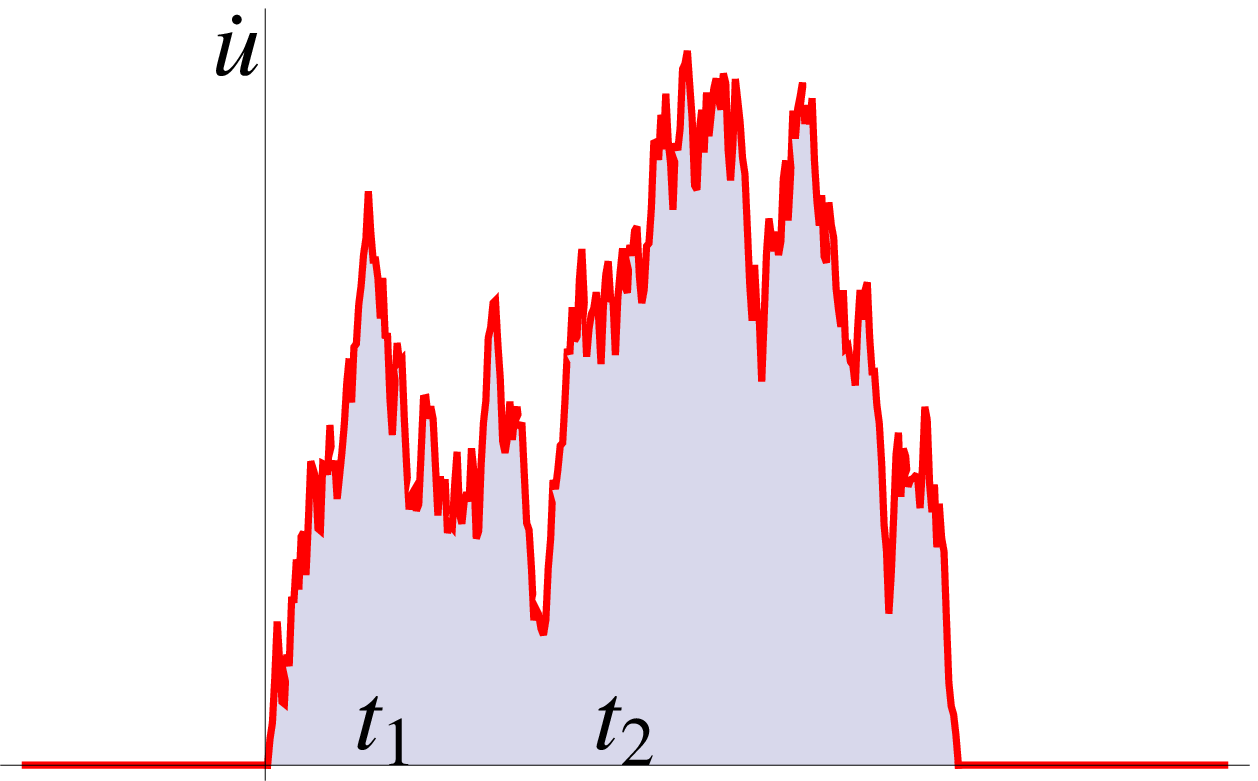}\fig{4.5cm}{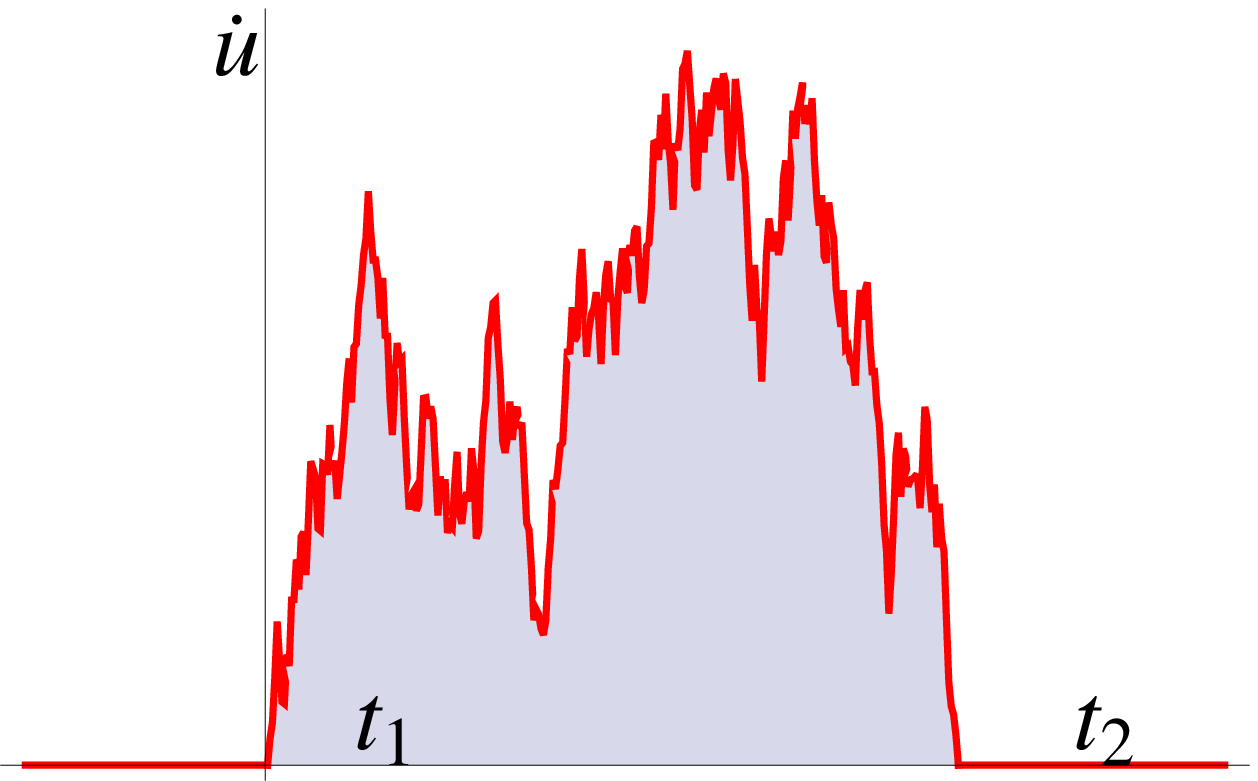}\\
(i)\hspace{4.5cm}(ii)\hspace{4.5cm}(iii)\hspace{4.5cm}(iv)
\end{center}
\caption{The four cases in Eq.~(\ref{decomp}): both times outside the avalanche
(i),  only $t_2$ inside  the avalanche (ii), both times inside the avalanhe
(iii), only $t_1$ inside  the avalanche (iv).}\label{f:4cases}
\end{figure*}By inspection of the higher-order results, we arrive at the 
following conjecture for $t_p<...<t_1$\bea  \label{zpresult}
\tilde Z_p = &-& \ln \left( 1 - \sum_{i=1}^p \lambda_i\right. \\
&+&\left. \sum_{q=2}^p \sum_{1 \leq i_1 < i_2 <...<i_q \leq p} \prod_{j=1}^q (-\lambda_{i_j})
z_{i_2 i_1} z_{i_3i_2} ... z_{i_qi_{q-1}}   \right) \nn
\eea 
 Note that this expression corrects a misprint in an earlier version of the result, Eq.~(17) in \cite{LeDoussalWiese2011a}.
This can also be written as 
\bea
  \tilde  Z_p &=& - \ln \Big( 1 - \sum_{k=0}^{p-1} (-1)^k \Tr(N M^k) \Big) \nn\\
 &=& - \ln \Big( 1 - \Tr\big(N (1+M)^{-1}\big) \Big) \\
 M_{ij} &=& \lambda_j z_{ij} \theta(i>j) \\
 N_{ij}&=&\lambda_j\ . 
\eea
The functions $\tilde Z_p$ possess an interesting factorization property, which we demonstrate on the simplest example \(\tilde Z_3\): Suppose 
that we choose $\lambda_2 =  -\tilde u_{t_2^+}=-\frac{\lambda_1}{{(1-\lambda_1 ) e^{t_1-t_2}}+\lambda_1}$, then one finds that
$\tilde u_t=0$ in the interval $t_3<t<t_2$. This leads to  \begin{equation}
Z_3(\lambda_1,\lambda_2,\lambda_3)\Big|_{\textstyle\lambda_2 =-\frac{\lambda_1}{{(1-\lambda_1 ) e^{t_1-t_2}}+\lambda_1}} = Z(\lambda_3) Z(\lambda_1,\lambda_2)
\ ,
\end{equation}
which we have checked explicitly. It implies that the observable $e^{\lambda_2 \dot u_{t_2} + \lambda_1 \dot u_{t_1}}$ for this particular
relation between $\lambda_2$ and $\lambda_1$ is strictly statistically independent from the velocity at any time in its past. It would be
interesting to investigate further the consequences of this property. 

\subsubsection{2-time probability}

Here we consider the joint velocity distributions at two times, and  choose $t_1<t_2$ {(from now one we
choose the notations of times in the more natural order $t_{i} < t_{i+1}$)}.
We expect that in the limit $v\to 0^+$ the 2-time probability takes the form (with $\dot u_j:=\dot u_{t_j}$):
\bea \label{decomp}
 P(\dot u_1,\dot u_2) &=& (1-q_1-q_2-q_{12}) \delta(v+\dot u_1) \delta(v+\dot u_2) \nn \\
&& + q_2 \delta(v+\dot u_1)
{\calP}_2(\dot u_2)+ q_{12} {\calP}(\dot u_1,\dot u_2) \nn\\&& \ + q_1 \delta(v+\dot u_2)
{\calP}_1(\dot u_1)\ . \
\eea
The four terms, in the order of their appearance, are plotted on Fig.~\ref{f:4cases}. The expression $q_{12}=v q'_{12}$ is the probability that both $t_1$ and $t_2$ belong to an avalanche (case (iii) of Fig.~\ref{f:4cases}). In the  small-$v$ limit we are studying here,
it must then be the same avalanche, and \(q_{12}\) must be proportional to $v$. 
The quantity  ${\calP}(\dot u_1,\dot u_2)$
is the normalized velocity distribution, conditioned to that event. $q_1=v q'_1$ (resp. $q_2=v q'_2$) are the probabilities 
that $t_1$ (resp.\ $t_2$) belongs to an avalanche but not $t_2$ (resp.\ $t_1$), and ${\calP}_1(\dot u_1)$ (resp. 
${\calP}_2(\dot u_1)$) the distribution conditioned to that event, (cases (ii) and (iv) of Fig.~\ref{f:4cases}). The first term in the decomposition  (\ref{decomp}) ensures that the probability is correctly normalized.\\ Integrating over $\dot u_2$, one 
 recovers the single-time distribution; hence comparing with Eq.\ (\ref{prob}) we have
\bea
&& p_a=q_1+q_{12}=q_2+q_{12}\ , \label{sumrule1} \\
&& p_a {\calP}(\dot u_1) = q_1 
{\calP}_1(\dot u_1)  + q_{12} \int d\dot u_2 {\calP}(\dot u_1,\dot u_2)\ ,~~~~~~~~~~~~~  \label{sumrule2}
\eea 
and similarly for $\dot u_1$. Hence, $q_1=q_2$. 
From the definition (\ref{defzp}) of $Z_2=Z_2(\lambda_1,\lambda_2)$ and Eq.\ (\ref{decomp}) we now 
have
\bea \label{big}
 Z_2 &=&  \partial_v \overline{e^{\lambda_1 (v + \dot u_1)+\lambda_2 (v + \dot u_2) }-1}\Big|_{v=0^+}  \nn\\
&=& q'_1 \int \rmd \dot u_1 {\calP}_1(\dot u_1) \big(e^{\lambda_1 \dot u_1 }-1\big) \nn \\&&+
 q'_2 \int \rmd \dot u_2 {\calP}_2(\dot u_2)\big (e^{\lambda_2 \dot u_2 }-1\big) \nn\\&&+
q'_{12} \int \rmd \dot u_1 \rmd\dot  u_2 {\calP}(\dot u_1,\dot u_2) \big( e^{\lambda_1 \dot u_1 +\lambda_2  \dot u_2 }-1\big)\ . \qquad 
\eea
We remind that here and below (until stated otherwise) we have suppressed all
factors of $L^d$. The latter are restored below, when going to the result in dimension-full units\footnote{ Units of the center-of-mass velocity are then $\tilde v_m$ which does contain the factor $L^{-d}$, see the remark at
the beginning of section \ref{s:jpdfcomv}.}. Note that the symmetry of $Z_2(\lambda_1,\lambda_2)$ in its arguments further implies that 
${\calP}_1(\dot u)={\calP}_2(\dot u)$ and that $ {\calP}(\dot u_1,\dot u_2)$ is also a symmetric function of its arguments. 
 Hence there is no way to tell the arrow of time from the velocity distribution {\em of the center of mass at the
mean-field level}. Below  we will however show that an asymmetry in time arises for finite Fourier modes, or
local velocities, already at the mean-field level. As a consequence, it will also arise for the center of mass
at one-loop order \cite{DobrinevskiLeDoussalWieseprep}, i.e.\ for $d< d_{\rm uc}$.

Taking now one derivative w.r.t.\ $\lambda_1$ of (\ref{big}), one  obtains from  the
formula (\ref{a96}) for $\tilde Z_2$ via  Laplace inversion the combination 
\bea \label{combi}
\lefteqn{ \dot u_1 \left[ q'_1 {\calP}_1(\dot u_1) \delta(\dot u_2) + q'_{12} {\calP}(\dot u_1,\dot u_2) \right]}\nn
 \\
 && ={\mathrm{LT}}^{-1}_{s_i \to u_i} \partial_{\lambda_1} Z_2(\lambda_1,\lambda_2)\Big|_{\lambda_i \to - s_i\nn} \\
 && ={\mathrm{LT}}^{-1}_{s_i \to u_i} \frac{1+s_2 z}{1+s_1+s_2+s_1 s_2 z\nn} \\
 && =  {\mathrm{LT}}^{-1}_{s_2 \to u_2} e^{- \frac{\dot u_1 (1+s_2)}{1+z s_2}} 
\ .\eea 
We denote $z:=z_{12}=1 - e^{- |t_2-t_1|}$. We now use the general result
\be  \label{laplaceformula}
\mathrm {LT}^{-1}_{s \to u} e^{d + \frac{a}{b+s}} = e^d \delta(u) + \sqrt{\frac{a}{u}} \,I_1(2 \sqrt{a u}) e^{d-b u} 
\ee  
with $d=-\dot u_1/z$, $a=\dot u_1 (1-z)/z^2$, $b=1/z$, and \(I_1\) the Bessel-\(I\) function. This yields 
the smooth part, in dimensionless units, as $q'_{12} {\calP}(\dot u_1,\dot u_2) = p(\dot u_1, \dot u_2)$
with
\be \label{pdef}
  p_2(\dot u_1, \dot u_2) =  {\rme^{-\frac{\dot u_{1}+\dot u_{2}}z}}  
\frac{\sqrt{1-z}}{z{\sqrt{\dot u_{1}\dot u_{2}}}} 
\,I_{1}\!\left(2  {\sqrt{\dot u_{1}\dot u_{2}} } \frac{\sqrt{1-z}}{z}\right)\ . 
\ee
In dimensionfull units\bea \label{2times}
&& q'_{12} {\calP}(\dot u_1,\dot u_2) = \frac{1}{\tilde v_m^3} p_2\left(\frac{\dot u_1}{\tilde v_m},\frac{\dot u_2}{\tilde v_m}\right) \ ,\\
&& z = 1 - e^{- |t_2-t_1|/\tau_m} \ .\label{defz2}
\eea
Since ${\calP}(\dot u_1,\dot u_2)$ is normalized to
unity, integrating Eq.\ (\ref{2times}) over both variables, one obtains the probability that both $t_1$ and $t_2$ belong to an avalanche,\be  \label{q12b}
q_{12} = v q'_{12} = \frac{v}{\tilde v_m} \ln(1/z)\ .
\ee 
For consistency we can  check that the combination which involves only $q'_{12} {\calP}(\dot u_1,\dot u_2)$ leads to a relation (in dimensionless units)
\bea
&& \partial_v \overline{ (e^{\lambda_1 (v+\dot u_1)} -1) (e^{\lambda_2 (v+\dot u_2)} -1) }\Big|_{v=0^+} \nn \\
&& =\tilde Z_2(\lambda_1,\lambda_2) - \tilde Z_1(\lambda_1) - \tilde Z_1(\lambda_2) \nn \\
&& = - \ln\left(\frac{1 - \lambda_1 - \lambda_2 + z \lambda_1 \lambda_2}{(1-\lambda_1)(1-\lambda_2)}\right) \nn\\
&& = \int \rmd \dot u_1 \rmd  \dot u_2 \,p(\dot u_1,\dot u_2)  (e^{\lambda_1 \dot u_1}-1) (e^{\lambda_2 \dot u_2}-1) 
~~~~~~~~\eea 
which is indeed satisfied by the function (\ref{pdef}). 

The $\delta$-function piece in (\ref{laplaceformula}) allows to obtain $q'_2 {\calP}_2(\dot u_2)$ in (\ref{combi}) 
 (in dimensionfull units) as
\be  \label{tildeP1}
q'_1 {\calP}_1(\dot u_1) = \frac{1}{\tilde v_m^2} p_1'\!\left(\frac{\dot u_1}{\tilde v_m}\right) \quad , \quad p'_1({\sf x}) = \frac{1}{{\sf x}} e^{-{\sf x}/z}
\ .\ee  
Normalization leads to $q_1 = (v/\tilde v_m) \ln(z \tilde v_m/v_0)$, in agreement with the
results (\ref{q12b}), (\ref{pa}), (\ref{tautau})  and the sum rule (\ref{sumrule1}). Note that 
(\ref{tildeP1}) can be obtained directly from Laplace inversion (in dimensionless units) of $\lim_{\lambda_2 \to - \infty} \partial_{\lambda_1} \tilde Z_2=z/(1-z \lambda_1)$
since that limit selects\footnote{\label{foontote-6} Recall that the Laplace transform $\hat f(\lambda)={\mbox{LT}}_{\dot u \to-\lambda= s }
f(\dot u):=\int \rmd\dot u\, e^{\lambda \dot u} f(\dot u)$ satisfies: (i) $\hat f(\lambda)=1$ for $f(\dot u)=\delta(\dot u)$, (ii) $\hat f(\lambda)-\hat f(0)= -\ln (1-\lambda)$
for $f(\dot u) = \frac{ \rme^{-\dot u}}{\dot u}$ and (iii) $\hat f(\lambda)=\frac{\Gamma (\alpha +1)}{
(1-\lambda )^{\alpha +1}}$ for $f(\dot u) = \dot u^\alpha \rme^{-\dot u}$, $\alpha>-1$. 
Second, the behavior of $f(\dot u)$ at $\dot u$ near zero is related to the behavior at
$\lambda \to - \infty$ of $\hat f(\lambda)$: if the limit of $\lambda\to -\infty $ in $\hat f(\lambda)$ exists, 
and is non-vanishing, it picks out the term $\sim \delta(\dot u)$. The term $f(0^+)$ is extracted, 
in the same limit, from the term $\sim 1/(-\lambda)$ in a large-$\lambda$ expansion.} 
the $\delta(\dot u_2)$ piece in (\ref{combi}); equivalently, the first terms in (\ref{big})
are 
\be 
q'_1 \int\rmd \dot u_1 {\calP}_1(\dot u_1) (e^{\lambda_1 \dot u_1 }-1) = - \ln(1- z \lambda_1)\ .
\ee 
Finally (\ref{sumrule2}) follows from the trivial identity $Z_2(\lambda_1,0)=Z_1(\lambda_1)$.

\subsubsection{Avalanche duration}
The distribution of avalanche durations can be obtained by several methods. 
Let us recall that  avalanche durations are well-defined as time intervals where the velocity
is strictly positive. Consider then the
probability that  there exists an avalanche starting in $[t_1,t_1+\rmd t_1]$ and ending in $[t_2,t_2+\rmd t_2]$.
On the one hand, this is equal to
\bea
&& P(t_1,t_2)\rmd t_1\, \rmd t_2\nn\\
&&=
\rho_0 v  P_{\rm duration}(\tau=t_2-t_1)  \rmd\tau  \, \rmd\!\left(\frac{t_1+t_2}{2}\right)\ ,
\eea
where $P_{\rm duration}(\tau)$ is the probability distribution of  avalanche durations. On the other hand
it also equals
\be \label{ttq}
-\rmd t_1\, \rmd t_2 \,\partial_{t_1} \partial_{t_2} q_{12} \ ,
\ee  
where $q_{12}$
computed above is the probability that $t_1$ and $t_2$ belong to the same avalanche. From Eqs.\ (\ref{q12b}) and (\ref{defz}) we obtain the distribution of durations as \be p_{\rm duration}(\tau)=\frac{(1-z)}{z^2}\ ,\ee 
where we recall $z=z_{21}=1-e^{- |t_2-t_1|}$, and in dimensionfull units
\bea \label{duration}
 P_{\rm duration}(\tau) &=& \frac{1}{\rho_0 \tilde v_m \tau_m^2} \frac{e^{-\tau/\tau_m}}{(1-e^{-\tau/\tau_m})^2} \nn\\
&=& \frac{1}{\rho_0 \tilde v_m \tau_m^2} \frac{1}{4 \sinh^2(\frac{\tau}{2\tau_m})}
\ .\eea 
This probability distribution has a power-law divergence for small durations $\tau \ll \tau_m$,
\be 
 P_{\rm duration}(\tau) \simeq \frac{1}{\rho_0 \tilde v_m \tau^2} \ ,
\ee  
i.e.\ there are many short avalanches. We assume a microscopic cutoff time $\tau_0$. 
The mean duration exhibits a divergence, i.e.\ 
\be  \label{tauav}
\langle \tau \rangle \approx \frac{1}{\rho_0 \tilde v_m} \ln\left(\frac{\tau_m}{\tau_0}\right)
\ ,
\ee 
as a function of $\tau_0$. However, higher moments are well-defined (i.e.\ independent of short scales).
The expression (\ref{tauav}) is in good agreement with our previous result (\ref{tautau})
if one assumes $\ln(\frac{v_m}{v_0}) \approx \ln(\frac{\tau_m}{\tau_0})$. 

There are several other ways to obtain the duration distribution. First one
notes that performing the limit $\lim_{\lambda_2 \to- \infty} \partial_{t_2}$ constrains the
avalanche to end at $t_2$, and similarly $- \lim_{\lambda_1 \to- \infty} \partial_{t_1}$ 
constrains it to start at $t_1$. Hence, in dimensionless units one recovers
\be  \label{dur3}
 \lim_{\lambda_1,\lambda_2 \to- \infty} - \partial_{t_1} \partial_{t_2} \tilde Z_2 = p_{\rm duration}(t_2-t_1)
\ .\ee  
It  also yields another method to obtain \(q'_{12}\) from (\ref{ttq}), writing 
\bea \nn
 q'_{12} &=& \int_{-\infty}^{t_1}\rmd s_1  \int^{\infty}_{t_2}\rmd s_2 ~ p_{\rm duration}(s_2-s_1) \\
& =& \lim_{\lambda_1,\lambda_2\to -\infty} \Big[Z_{2}(\lambda_1,\lambda_2,t_1-t_2)- Z_{2}(\lambda_1,\lambda_2,\infty)\Big]\ \nn \\
& =& -\ln (z)\ ,  
\eea 
inserting Eq.~(\ref{dur3}) (second line) and $\tilde Z_2$ from Eq.\ (\ref{a96}), in agreement with Eq.\ (\ref{q12b})
in dimensionless units.

Another way to obtain the duration is as follows: We note that when the avalanche starts and ends, the velocity must vanish. Hence the duration distribution  can be recovered from ${\calP}(0^+,0^+)$
which should be proportional to the probability that an avalanche starts at $t_1$ and ends at $t_2$. We can indeed
check on our result (\ref{pdef}), (\ref{2times}) that
\be 
q'_{12} {\calP}(0^+,0^+) = \frac{\rho_0 \tau_m^2}{\tilde v_m^2} P_{\rm duration}(\tau=t_2-t_1)\ ; 
\ee  
hence this is true, up to a normalization. We note that this term can also be obtained as 
the coefficient of $1/(\lambda_1 \lambda_2)$ in an  expansion of $\tilde Z_2$ at
large (negative) $\lambda_i$.

To study the  temporal avalanche statistics, it turns out to be more efficient to use two properties simultaneously:
(i) $\dot u_i=0$ outside the avalanche, an event whose probability can be selected by taking the
limit $\lambda_{i} \to - \infty$; (ii) taking a $\partial_{\lambda_i}$ on the generating function
multiplies by $\dot u_{t_i}$, hence is non-zero only if $t_i$ belongs to the avalanche. 
Using these properties we will now show how to generate the $p$-times distribution of velocities inside an avalanche 
conditioned to start and end at some given times. In particular, we recover the
duration distribution, from the normalization, and we compute shape functions, which
are of high interest in view of experiments.

\subsubsection{1-time velocity distribution at fixed duration and mean avalanche-shape}

We start with the information contained in the joint $3$-time  distribution, which can
be obtained from $\tilde Z_3$ in (\ref{a96}). Choosing again $t_1<t_2<t_3$, and generalizing 
the form (\ref{decomp}), we expect that the joint distribution contains a piece\  
\be 
v q'_{13,2} {\calP}_{13,2}(\dot u_2) \delta(\dot u_1) \delta(\dot u_3)
\ ,
\ee  
where $v q'_{13,2}$ is the probability that $t_1$ and $t_3$ do not belong to
an avalanche while $t_2$ does, and ${\calP}_{13,2}(\dot u_2)$ is the velocity distribution
conditioned to this event. From the above remarks, to obtain this piece we  need to inverse-Laplace
transform
\bea
&& \lim_{\lambda_1,\lambda_3 \to - \infty} \partial_{\lambda_2} \tilde Z_3 = \frac{1}{b  - \lambda_2} \\
&& b = \frac{z_{31}}{z_{21} z_{32}} = \frac{1}{z_{21}} + \frac{1}{z_{32}} -1\ . \label{b}
\eea
Hence we  find  in dimensionless units
\be 
q'_{13,2}  {\calP}_{13,2}(\dot u_2) = \frac{1}{\dot u_2} e^{- b \dot u_2} 
\ .\ee  
Integration over $\dot u_2$, in presence of a small-velocity cutoff $v_0$, leads to
\be 
q'_{13,2} = - \ln b v_0
\ .
\ee 
Taking two time derivatives we recover  the duration distribution
\be  \label{duration2}
- \partial_{t_1} \partial_{t_3} q'_{13,2} = - \frac{\partial_{t_1} b \partial_{t_3} b}{b^2} = 
P_{\rm duration}(\tau=t_{3}-t_1)\ , 
\ee 
using that $\partial_{t_1} \partial_{t_3} b = 0$. We also find the
distribution of the velocity at $t_2$ {\it conditioned} s.t.\ the avalanche
starts at $t_1$ and ends at $t_3$,
\bea
 P(\dot u_2 |13)  &=& \frac{ - \partial_{t_1} \partial_{t_3} [ q'_{13,2}  {\calP_{13,2}}(\dot u_2) ] }{P_{\rm duration}(\tau=t_3-t_1)}
\nn\\ &=& \frac{ - \partial_{t_1} \partial_{t_3} [ q'_{13,2}  {\calP}_{13,2}(\dot u_2) ] }{- \partial_{t_1} \partial_{t_3} q'_{13,2} } 
\ .
\eea
This leads to
\be  \label{meanP}
   P(\dot u_2 |13) =  \dot u_2 
b^2 e^{- b \dot u_2} 
\ .
\ee  
From this one obtains the shape function
\bea \label{shapeshape}
 \langle \dot u_2 \rangle_{13} &:=& \int \rmd  \dot u_2 \dot u_2 P(\dot u_2 |13) = \frac{2}{b} \nn\\
& =& \tilde v_m \frac{4 \sinh\big(\frac{t}{2 \tau_m}\big) \sinh\big(\frac{\tau}{2 \tau_m}(1 - \frac{t}{\tau})\big)}{\sinh\big(\frac{\tau}{2 \tau_m}\big)}~~~~~
\eea 
for a fixed avalanche duration $\tau=t_3-t_1$, denoting $t=t_2-t_1$.  We have restored all units 
in the last line. This form interpolates from a parabola for small $\tau \ll \tau_m$ to a flat shape for the longest avalanches (see Fig.\ \ref{f:pulse-shape}). The result holds for
an interface at or above its upper critical dimension, which previously
was used \cite{PapanikolaouBohnSommerDurinZapperiSethna2011} on the basis of the ABBM model. 
\begin{figure}{
\setlength{\unitlength}{0.87mm}
\fboxsep0mm 
\mbox{\begin{picture} (100,52)
\put(14,0){\fig{7.1cm}{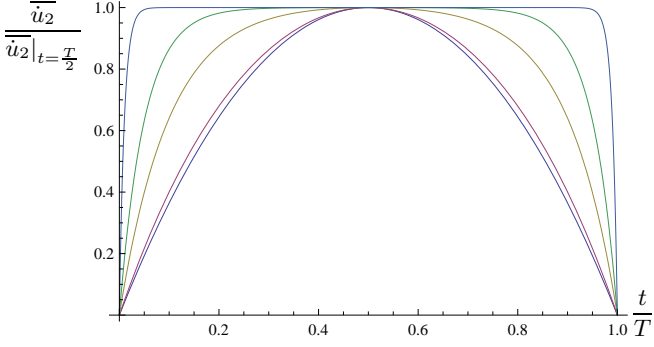}}
\put(96,2.8){$\displaystyle \frac{t}{T}$}
\put(0,46){$\displaystyle \frac{\overline{\dot{u}_{2} }}{\overline{\dot{u}_{2} }|_{t=\frac{T}2}}$} 
\end{picture}}}
\caption{``Pulse-shape'': The normalized velocity at time $t$ in an avalanche of
duration $T$ for $T \ll \tau_{m}$ (lower curve) to $T \gg \tau_{m}$ (upper
curve). }
\label{f:pulse-shape}
\end{figure}

An alternative approach is to obtain \(p_{3}(0^+,\dot u_2,0^+)\) from $\tilde Z_3(\lambda_1,\lambda_2,\lambda_3)$. 
As discussed above, one needs to extract the coefficient of $1/(\lambda_1 \lambda_3)$ in the large
$\lambda_1,\lambda_3$ expansion of $\tilde Z_3$. Hence we first need to calculate  
\bea
\tilde Z_{2|13}(\lambda_2)&:=&
\lim_{\lambda_3\to -\infty}  \frac{\lambda_3^2\, \rmd} {\rmd {\lambda_3}}
\lim_{\lambda_1\to -\infty}  \frac{\lambda_1^2\, \rmd} {\rmd {\lambda_1}} \tilde Z_3(\lambda_1,\lambda_2,\lambda_3)\nn\\
& =& \frac{\lambda _2 \left(z_{31}+z_{21} \left(z_{32}-1\right)-z_{32}\right)-z_{31}+1}{\left(z_{31}-\lambda _2
   z_{21} z_{32}\right){}^2} \nn\\
&=&\left[\frac{1}{2 \sinh\left(\frac{t_1-t_3}2\right) } \frac{ b}{b-\lambda_2} \right]^2
\ .
\eea 
$b$ is defined in Eq.\ (\ref{b}).
The inverse Laplace transform (in dimensionless units) gives 
\bea \label{alt}
\mathrm {LT}^{-1}_{-\lambda_2 \to u} \tilde Z_{2|13}(\lambda_2) &=& \frac{1}{4 \sinh^2((\frac{t_1-t_3}2)} \times P(\dot u_2 |13)  \nn\\
&=&  P(\dot u_2 |13) P_{\rm duration}(t_1-t_3) \nn
\ ,
\eea 
where \(P(\dot u_2 |13)\) is given in Eq.\ (\ref{meanP}), and \( P_{\rm duration}\) in Eq.~(\ref{duration}). 

\subsubsection{2-time velocity distribution at fixed duration and fluctuations of the shape of an 
avalanche: The ``second shape"}
\label{s:lambda-to-inf}
We  now derive the 2-time velocity distribution at fixed avalanche duration. For that we consider the term $\delta(\dot u_1) \delta(\dot u_4) q_{14,23} {\calP}_{14,23}(\dot u_2,\dot u_3)$ in the joint $ 4$-time  distribution (with $t_1< t_2< t_3< t_4$) which can be obtained from $\tilde Z_4$. We recall that 
\be \label{p2314}
P(\dot u_2,\dot u_3|14) = \frac{ - \partial_{t_1} \partial_{t_4} [q'_{14,23} {\calP}_{14,23}(\dot u_2,\dot u_3)]}{- \partial_{t_1} 
\partial_{t_4} q'_{14,23} } 
\ee
is the 2-time velocity distribution at fixed avalanche duration $\tau=t_4-t_1$. We expect, and will check
below, that $- \partial_{t_1} \partial_{t_4} q'_{14,23} = P_{\rm duration}(\tau=t_4-t_1)$, i.e.\ comparing with (\ref{duration2}), the number of intermediate points does not matter. 

The simplest quantity to obtain is the 2-time shape function. Indeed multiplying (\ref{p2314}) by $\dot u_2 \dot u_3$ and integrating, one finds
\be
 \langle \dot u_2 \dot u_3  \rangle_{14} =  \frac{ - \partial_{t_1} \partial_{t_4} 
[ \lim_{\lambda_1,\lambda_4 \to - \infty} \partial_{\lambda_2} \partial_{\lambda_3} \tilde Z_4|_{\lambda_2=0,\lambda_3=0} ] }{P(\tau=t_4-t_1)}
\ . \label{143}
\ee
It is easy to calculate from (\ref{zpresult})
\bea
\lefteqn{ \lim_{\lambda_1,\lambda_4 \to - \infty} \partial_{\lambda_2} \partial_{\lambda_3} \tilde Z_4\big|_{\lambda_2=0,\lambda_3=0}}  \nn\\
&& = -   \frac{z_{21} z_{43}}{z_{41}^2} (z_{32} z_{41} - z_{31} z_{42}) \nn\\
&& = 4 \frac{\sinh^2\!\big(\frac{1}{2} (t_2-t_1)\big) \sinh^2\!\big(\frac{1}{2} (t_4-t_3)\big) }{\sinh^2\!\big(\frac{1}{2} (t_4-t_1)\big)} \label{sh}
\ .
\eea
Taking two derivatives in (\ref{143}) one finds a complicated expression for
$ \langle \dot u_2 \dot u_3  \rangle_{14}$ which however simplifies
greatly if one forms the cumulant combination and uses the
above result  for the shape. Then both results can be 
summarized, introducing the function $h(t) := 4 \sinh(t/2)$, as (in dimensionless units):
\bea
\langle \dot u_2 \rangle_{14} &=& \frac{h(t_4-t_2) h(t_2-t_1)}{h(t_4-t_1)}\ ,  \\
 \langle \dot u_2 \dot u_3 \rangle^{\rm c}_{14} &=& \langle \dot u_2 \dot u_3 \rangle_{14} 
- \langle \dot u_2 \rangle_{14} \langle \dot u_3 \rangle_{14} \nn\\
&=&  \frac{1}{2} 
\bigg(\frac{h(t_4-t_3) h(t_2-t_1)}{h(t_4-t_1)} \bigg)^2\ . \label{cumshape}
~~~~\eea 
Hence the {\it fluctuation of the shape} has a simple expression, and it
would be nice to measure it in experiments. We  call this the {\it ``second shape"} since it
gives more information about the avalanche statistics than the usual shape, the average of the velocity. The {\em second shape} tells about the
variability, i.e.\ fluctuations of the avalanche shape. For $t_2=t_3$ one recovers the relation 
$\langle u^2 \rangle^{\rm c} = \frac{1}{2} \langle u \rangle^2$ 
between second cumulant and mean of the single time velocity distribution (\ref{meanP}).  Note that the
second cumulant always starts quadratically in time near the edges. It is 
quite remarkable that the dimensionless ratio
\be 
\frac{\langle \dot u(t_2)^2  \rangle_{14}}{\langle \dot u(t_2)  \rangle_{14}^2} = \frac{3}{2} 
\ee  
is {\em\ independent} of $t_1,t_2,t_4$. This is an important signature of the mean-field
theory which should be studied in experiments.
On figure \ref{pipipu}, we have plotted \be\label{ppop}
 C(t,T):=\frac{\left< \dot u(t) \dot u(-t)\right>^c}{\left<
\dot u(t) \right> \left<  \dot u(-t)\right>}\bigg|_{t_1=-T/2,t_4=T/2}\ .
\ee
It measures the correlations between the left and right part of the avalanche. 

 One can go further and obtain the full 2-time distribution. For this one notes that 
the function $q'_{14,23} {\calP}_{14,23}$ is obtained (in dimensionless units)
by Laplace inversion as
($i=2,3$)\bea
&& q'_{14,23} \dot u_2 \dot u_3 {\calP}_{14,23}(\dot u_2,\dot u_3) \nn\\
&& ={\mathrm{LT}}^{-1}_{s_i \to \dot u_i} \Big( \lim_{\lambda_1,\lambda_4 \to - \infty} \partial_{\lambda_2} \partial_{\lambda_3} \tilde Z_4 \Big)\Big|_{\lambda_i \to - s_i} \nn
\\
&& ={\mathrm{LT}}^{-1}_{s_i \to \dot u_i} \frac{z_{21} z_{43} (z_{31} z_{42}-z_{32} z_{41}) }{
[z_{41} + s_3 z_{31} z_{43} +  s_2 z_{21} (z_{42} + s_3 z_{32} z_{43})]^2 }\ . \nn \\
&& \label{tobeinv}
\eea
We have used the result (\ref{zpresult}). 
The normalization is obtained by integrating\footnote{There seems to be a non-commutation of limits, hence
we need to take first the large-$\lambda$ limit.} 
the above $\int_0^\infty \rmd s_2 \int_0^\infty \rmd s_3$ leading to
\be 
   q_{14,23}= v q'_{14,23} = \frac{v}{\tilde v_m} \ln \frac{z_{42} z_{31}}{z_{41} z_{32}} \ .
\ee  
This is the probability that there is an avalanche starting in the interval $[t_1,t_2]$ and ending 
in the interval $[t_3,t_4]$. Indeed one can check for consistency that integrating the duration 
distribution (\ref{duration}) we obtain
\be 
v \rho_0 \int_{t_1}^{t_2} \rmd t' \int_{t_3}^{t_4} \rmd t  P_{\rm duration}(t-t') = \frac{v}{\tilde v_m} \ln \frac{z_{42} z_{31}}{z_{41} z_{32}} 
\ .
\ee  
Laplace inversion of (\ref{tobeinv}) w.r.t $s_2$ yields an expression equal to minus the derivative 
$- \partial_b$ of (\ref{laplaceformula}), with other values for $a=\dot u_2 a',b,d=-\dot u_2 d'$. Finally we
find
\be 
q'_{14,23} {\calP}_{14,23}(\dot u_2,\dot u_3) = \sqrt{\frac{a'}{\dot u_2 \dot u_3}} I_1\left(2 \sqrt{a' \dot u_2 \dot u_3}\right) 
e^{- d' \dot u_2 - b \dot u_3} ,
\ee  
with 
\bea
&& d' = \frac{z_{31}}{z_{21} z_{32}}   \quad , \quad  b = \frac{z_{42}}{z_{32} z_{43}} \\
&& a' = \frac{ z_{31} z_{42} -z_{32} z_{41}  }{z_{21} z_{32}^2 z_{43}} = \frac{1}{4 \sinh^2(\frac{t_3-t_2}{2})} 
\ .
~~\eea
This leads to the final expression
\bea
\lefteqn{ q'_{14,23} {\calP}_{14,23}(\dot u_2,\dot u_3) }\\
 &=& \frac{1}{2 \sinh(\frac{t_3-t_2}{2}) \sqrt{\dot u_2 \dot u_3}} 
\,I_1\!\left(  \frac{\sqrt{\dot u_2 \dot u_3}}{\sinh(\frac{t_3-t_2}{2})} \right) \nn\\
&& \times\rme^{-\left(\frac{1}{1-e^{t_1-t_2}} + \frac{1}{e^{t_3-t_2}-1} \right) \dot u_2 - \left( \frac{1}{1-e^{t_2-t_3}} + \frac{1}{e^{t_4-t_3}-1} \right)
 \dot u_3 } \nn
\ .
\eea 
The 2-time velocity distribution at fixed avalanche duration $\tau=t_4-t_1$ is then obtained as
\be 
P(\dot u_2,\dot u_3|14) = \frac{ - \partial_{t_1} \partial_{t_4}[q'_{14,23} {\calP}_{14,23}(\dot u_2,\dot u_3)]}{- \partial_{14} q'_{14,23} } 
\ .
\ee 
This leads to the result\begin{figure}
\Fig{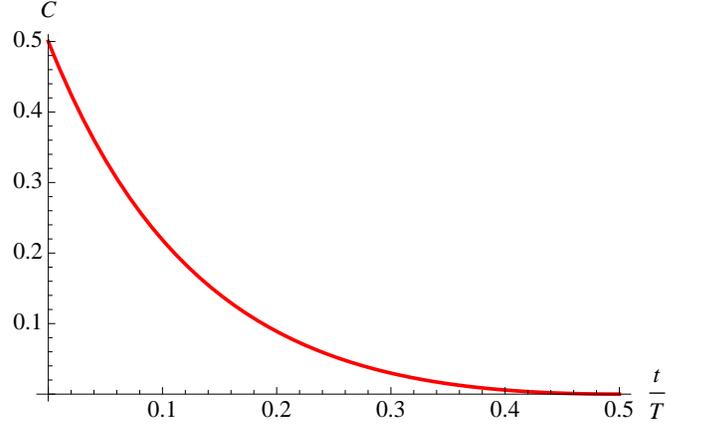}
\caption{The velocity correlation  $C(t,T)$ of Eq.\ (\ref{ppop}) for $T=1$. }\label{pipipu}
\end{figure}
\bea
\lefteqn{ P(\dot u_2,\dot u_3|14) }  \\
& =& \frac{\sqrt{\dot u_2 \dot u_3}}{2 \sinh(\frac{t_3-t_2}{2}) } 
\,I_1\!\!\left(  \frac{\sqrt{\dot u_2 \dot u_3}}{\sinh(\frac{t_3-t_2}{2})} \right)  \nn\\
&&\times \frac{  \sinh^2(\frac{t_4-t_1}{2}) }{4 \sinh^2(\frac{t_1-t_2}{2}) \sinh^2(\frac{t_3-t_4}{2})} \nn \\
&& \times 
e^{- \left (\frac{1}{1-e^{t_1-t_2}} + \frac{1}{e^{t_3-t_2}-1} \right) \dot u_2 - \left( \frac{1}{1-e^{t_2-t_3}} + \frac{1}{e^{t_4-t_3}-1} \right)
 \dot u_3 } 
\ .\nn
\eea 
One can check its normalization using the useful formula
\be 
\int_0^\infty \rmd x \int_0^\infty \rmd y \sqrt{x y}  \,I_1(2 a \sqrt{xy}) e^{-b x - c y} = \frac{a}{(b c - a^2)^2}\ , \nn
\ee  
while derivatives w.r.t.\ $b$ and $c$ allow to recover shape cumulants such as (\ref{cumshape}). 
For instance one finds the third cumulant of the shape as
\bea
&&\!\!\! \langle \dot u_2^2 \dot u_3 \rangle^{\rm c} = \langle \dot u_2^2 \dot u_3 \rangle -  \langle \dot u_2^2 \rangle \langle \dot u_3 \rangle 
- 2 \langle \dot u_2 \rangle \langle \dot u_2 \dot u_3 \rangle  + 2 \langle \dot u_2 \rangle^2 \langle \dot u_3 \rangle \nonumber \\
&& = \frac{1}{2} \frac{h(t_2-t_1)^3}{h(t_4-t_1)^3} h(t_4-t_2) h(t_3-t_4)^2
\eea 
This procedure can be pursued to obtain higher $p$-time distributions at fixed avalanche duration. We will stop here, and just point out that one can check explicitly that (\ref{zpresult}) satisfies\be 
\lim_{\lambda_1,\lambda_3 \to \infty} \partial_{\lambda_2} \partial_{\lambda_4} \tilde Z_4 = 0\ , 
\ee
consistent with the fact that there is only a single avalanche to this order, since a non-zero value would require 
that $t_2$ and $t_4$ are in two separate avalanches, since the limit $\lambda_3 \to - \infty$ selects $\dot u_3=0$.

\subsection{Interpretation of the instanton solution: response to a small step in the force}
\label{sec:interpret}
 
Here we examine the question of what  is the physical meaning of the instanton solution $\tilde u^\lambda_{xt}$? We show that it encodes
the (linear) response to a small (infinitesimal) step in the applied force at $x,t$, equivalently a small kick in the driving velocity. The inverse Laplace transform
of $\tilde u^\lambda_{xt}$ is then related to the change in the probability distribution of $\dot u$ due to this kick.

First note that the action in presence of the source $\lambda$, noted $ {\cal S}^{\mathrm{tree}}_\lambda$ in (\ref{Streelambda}), is such that
$\tilde u_{xt}$ does not fluctuate. This means that all cumulants of $\dot u$ and
$\tilde u$  involving at least 2 response fields vanish. In other words, in any
expectation value the field $\tilde u_{xt}$ can be replaced by $\tilde u^\lambda_{xt}$. Hence
from Eq.~(\ref{Gtree}) \bea 
 \tilde u^\lambda_{x't'} &=& \langle \tilde u_{x't'} \rangle_{ {\cal S}^{\mathrm{tree}}_\lambda} = 
 \frac{1}{G[\lambda]} \left<  \tilde u_{x't'} e^{\int_{xt} \lambda_{xt} (v+\dot u_{xt})} \right>_{ {\cal S}^{\mathrm{tree}}} \nn \\
& =& \frac{1}{G[\lambda]} \overline{ \frac{\delta e^{\int_{xt} \lambda_{xt} (v+\dot u_{xt})} }{\delta \dot f_{x't'}} } \nn\\ &=& \frac{1}{G[\lambda]} g_{x'x''} \overline{ \frac{\delta e^{\int_{xt} \lambda_{xt} (v+\dot u_{xt})} }{\delta \dot w_{x''t'}} } \label{instint}
\ .\eea 
By definition of the response field, since $\tilde u_{x't}$ couples to $\dot f_{x't'} = \int_{x''}g^{-1}_{x'x''} \dot w_{x''t'}$, see
Eqs.~(\ref{eqmogen}) and (\ref{eqmo}), it is the response to a change in the driving from $w_{xt}= vt \to v t + \delta w_{xt}$, and more precisely to an infinitesimal kick $\delta \dot w_{x t} = \delta w  \,\delta(x-x')\, \delta(t-t')$ in the velocity at position $x'$ and time $t'$. Note that (\ref{instint}) is independent of $v$, a fact 
which comes from the form (\ref{Gv}) and is a peculiarity of the tree theory (at fixed $\eta$ and $\sigma$). 

Taking a derivative w.r.t.\ $\lambda$ at $\lambda=0$, and comparing with (\ref{fot}) 
yields the property that for the tree theory the exact response 
function ${\cal R}_{xt,x't'}$ (in the velocity theory) is uncorrected by disorder,\be \label{respu}
{\cal R}_{xt,x't'} := \langle  \tilde u_{x't'} \dot u_{xt} \rangle_{ {\cal S}^{\mathrm{tree}}} = R_{xt,x't'} := \langle  \tilde u_{x't'} \dot u_{xt} \rangle_{ {\cal S}_0}
\ ,
\ee  
as clearly the cubic vertex (\ref{cubic}) cannot lead to corrections of the response. This is in agreement with the fact that the effective action
$\Gamma= {\cal S}$ for this theory as discussed in detail in \cite{DobrinevskiLeDoussalWiese2011b}. 
Note that Eq.~(\ref{respu}) is a non-trivial property for $v=0^+$, since then, 
in most realizations of the disorder, the particle is not moving and under a kick it will experience only
a small avalanche (of the order of the cutoff).

Let us now use Eq.~(\ref{instint}) in the limit of $v\to 0^+$, i.e.\ order $0$ in $v$, but to lowest order in the perturbation $\delta \dot f$,
\be
\overline{ e^{\int_{xt} \lambda_{xt} \dot u_{xt} }-1}  = \int_{x't'} \delta \dot f_{x't'} \tilde u^\lambda_{x't'} + O\big((\delta \dot f)^2\big) \label{instint00}
\ .\ee 
We used that $G[\lambda]=1$ for $v=0^+$. The instanton solution thus gives the statistics of the motion induced by the kick. For instance,
let us apply Eq.~(\ref{instint00}) to calculate the center-of-mass velocity  $\dot u_1 \equiv \dot u_{t_1}$ at time $t_1$,
choosing $\lambda_{xt} = \lambda \delta(t-t_1)$, given that there was an infinitesimal uniform kick 
$\delta \dot w_{xt} = \delta w \delta(t-t_0)$ at some time $t_0<t_1$,
on top of the $v=0^+$ stationary state. The instanton solution is uniform $u^\lambda_{xt_0} = u^\lambda_{t_0}$ and precisely encodes
that information
\be  
 \overline{e^{L^d \lambda \dot u_{t_1} }}-1 =
m^2 L^d \delta w\,u^\lambda_{t_0} + O(\delta w^2)\ . \label{instint2}
\ee  
Note  that Eq.~(\ref{instint2}) can be generalized to any source $\lambda(t),$ 
hence the instanton solution $\tilde u^\lambda_{t_0}$ gives the first order in
$\delta w$ of the generating function of velocities at any later times;  $\tilde u^\lambda_{t_0}$ does not depend on the sources at times
smaller than $t_0$. 

Performing the inverse 
Laplace transform of the instanton solution w.r.t. $s:=-\lambda L^d$ gives  
\be 
\mbox{LT}^{-1}_{s \to \dot u_1} L^d m^2 \tilde u^\lambda_{t_0} = \frac{\delta}{\delta w} P(\dot u_1)\ . 
\label{162}\ee  
This is the linear change of the velocity distribution at time $t_1$ as response to an infinitesimal kick at time $t_0<t_1$. Using the explicit
form for the instanton solution (\ref{soluinstanton}) and performing its Laplace inversion
we find from (\ref{instint2}) (restoring all units): 
\bea \nn
 P(\dot u_1)  &=& \delta(\dot u_1)\left ( 1 - \frac{\delta w}{\tilde v_m \tau_m} \frac{1}{e^{(t_1-t_0)/\tau_m}-1} \right) 
\\
&& + \frac{\delta w}{\tilde v_m^2 \tau_m} \frac{e^{- \frac{\dot u_1}{\tilde v_m} \frac{1}{1-e^{(t_0-t_1)/\tau_m}}}}{4 \sinh^2(\frac{t_1-t_0}{2 \tau_m})}
+ O(\delta w^2) \label{u1kick}
\ ,
~~~~~~~\eea
which is interpreted as follows: For $v=0^+$, at a given time $t_0^{-}$, almost surely the particle has zero velocity.
The infinitesimal kick at time \(t_0\) produces an avalanche (it gives a velocity $\dot u_{t_0^+}=\delta w/\tau_m$) 
which most of the times dies out well before time $t_1$ (in a time $\sim \tau_0$, the microscopic cutoff time).
Exceptionally rarely, however, and with probability $O(\delta w)$, this kick produces a larger avalanche, i.e.\ lasting a time of order $\tau_m$.
Hence the result that the response function is unchanged by disorder is not  trivial
at all: For most realizations $\tau_m \delta \dot u_{t_1}/\delta w$ is very small; however for 
some realizations $\delta \dot u_{t_1} = O(1)$ hence $\delta \dot u_{t_1}/\delta w \sim 1/\delta w$.
After averaging over disorder these rare events lead to the bare response function, which is $O(1)$.

Let us now comment on stationary versus non-stationary avalanches. In previous sections, and most
of the paper, we study avalanches in the steady state, obtained by time-uniform driving $w_{xt} =v t$ (with small $v$). 
These can thus be called stationary avalanches. Adding a kick at time $t_0$ leads to non-stationary driving.
Indeed the avalanche generated by the kick appears non-stationary, i.e.\ 
$P(\dot u_1)$ in (\ref{u1kick}) is quite different from the 1-time distribution found in Eqs
(\ref{prob}), (\ref{1time}). It is time (i.e.\ $t_1$) dependent, and for instance the average velocity decays
exponentially, $\overline{\dot u_1} = \frac{\delta w}{\tau_m} e^{-(t_1-t_0)}$. One can ask 
whether such non-stationary avalanches are qualitatively different from the stationary ones.

For an infinitesimal kick, this is not the case. Indeed, if one considers as in Section \ref{s:jpdfcomv} to lowest order in 
$v$ the steady state,   i.e.\ the distribution of probability of $\dot u_1=\dot u(t_1),$
 {\it conditioned} to an avalanche having started at $t_0$, one 
obtains exactly $P(\dot u_1)  $, as given  in Eq.\ (\ref{u1kick}): As usual, this conditional probability is obtained as $- \partial_{t_0} q_1' {\cal P}_1(\dot u_1)$ using
formula (\ref{tildeP1}) ($t_1,t_2$ there are $t_0,t_1$ here, respectively). This, in fact, is more generally true:  
Namely {\it an infinitesimal uniform kick at time $t_0$ produces the same velocity statistics for $t>t_0$ as conditioning} an avalanche in
the steady state to start at time $t_0$. It can be shown at the  mean-field level from the identity
\bea \label{iden}
&& \lim_{\lambda \to - \infty} (- \partial_{t_0}) \int_{-\infty}^{+\infty} dt~ \tilde u_t^{\lambda} = \tilde u^{\mu}_{t_0^+} \\
&& \lambda(t) = \lambda \delta(t-t_0) + \mu(t)\ . 
\eea
Here $\mu(t)=0$ for $t \leq t_0$, but $\mu(t)$ is arbitrary for  $t>t_0$. The r.h.s of (\ref{iden}) 
is related (via Laplace inversion) to the effect of the infinitesimal kick at time $t_0$ on the joint distribution
of the velocities at all later times, while the l.h.s. is related to the velocity distribution conditioned
to the avalanche starting at $t_0$ (the conditioning results from the operation  $- \lim_{\lambda \to - \infty} \partial_{t_0}$
as we learned in Section \ref{s:jpdfcomv}). The proof of this result, which is easy to obtain from the
instanton equation, and more details on these properties will be given  in \cite{DobrinevskiLeDoussalWieseprep}.

\subsection{Finite step in the force and arbitrary monotonous driving}
\label{s:finitek}

For completeness, let us discuss the case of a finite kick, studied in 
\cite{DobrinevskiLeDoussalWiese2011b}. 
First one notes that one can generalize our method to  arbitrary monotonous driving.
Starting from Eq.~(\ref{eqmo}) in the laboratory frame (i.e.\ setting $v=0$),
but with arbitrary driving $\dot f_{xt} \geq 0,$ we follow the same steps 
as in Section \ref{simplified} to obtain for the generating function of velocities 
\bea
 \overline{\rme^{\int_{xt} \lambda_{xt} \dot u_{xt}}} &=& \int {\cal D}[\dot u] {\cal D}[\tilde u] \rme^{\int_{xt} \lambda_{xt} \dot u_{xt} -\tilde
u_{xt} (\eta \partial_t - \nabla^2_x + m^2) \dot u_{xt} } \nn \\
&& ~~~~~~~~~~~~~~~~\times \rme^{ \int_{xt} \tilde u_{xt} \dot f_{xt} + \sigma \tilde
u_{xt}^2 \dot u_{xt}} 
\ .\eea
Here $\sigma =-\Delta'(0^+)$. The Middleton theorem allows to restrict the path integral
to positive velocities $\dot u_{xt} \geq 0$. Again, integrating over $\dot u_{xt}$ enforces the instanton equation to be satisfied.
Inserting its solution thus eliminates all terms proportional to $\dot
u$, such that we are left with \cite{DobrinevskiLeDoussalWiese2011b}
\be\label{magic}
\overline{\rme^{\int_{xt}\lambda_{xt} \dot u_{xt}}} = \rme^{\int_{xt}
\tilde u_{xt}^{\lambda} \dot f_{xt}}\ .
\ee
As written, on an unbounded time domain, this formula holds if and only if all
trajectories are forward for all times. It can thus be applied for $v=0^+$ and an infinitesimal kick 
$\dot f_{xt}=\delta f_{xt} \geq 0$, recovering (\ref{instint00}) and (\ref{instint2}) by expanding
to lowest order in $\delta f$ (and to order 0 in $v$). It also holds for any finite kick, and
allows to study arbitrary non-stationary monotonous driving as done in detail in 
\cite{DobrinevskiLeDoussalWiese2011b}. For instance, one can prepare the system
at $t=t_0$ in the quasi-static Middleton state $u_x(w)$: In the distant past one first
drives monotonously with  $\dot f_{xt} > 0$ to erase the memory of the initial condition, then stops driving.
The above formula implies
\be\label{magic2}
\overline{\rme^{\int_{xt>t_0}\lambda_{xt} \dot u_{xt}}} = \rme^{\int_{xt>t_0}
\tilde u_{xt}^{\lambda} \dot f_{xt}}\ 
\ee
with  initial condition
\be  \label{init} 
\dot u_{x t_0}=0 \quad , \quad u_{xt_0} = u_x(w_{t_0})\ .
\ee 
This can be used to study non-stationary avalanches obtained from the
Middleton state at $t=t_0$, generated by applying a finite kick $\delta f= m^2 \delta w$ at time $t_0$.
Interestingly, these avalanches can also be shown, within mean field, to be equivalent to those of the  steady state,   
under conditioning of the velocity at $t_0$ to be equal to $\dot u_{t_0+}=\delta w$ as will be discussed in
\cite{DobrinevskiLeDoussalWieseprep}. Note however that these formulae  do not say anything about non-monotonous
driving as in the hysteresis loop, which remains to be investigated. They only pertain to avalanches in the
Middleton state. 

Consider now an application to a spatially non-uniform kick at time $t_0$, of arbitrary finite strength $\dot f_{xt} = \delta f_x \delta(t-t_0)$. 
It is interesting to note that any observable involving the centor of mass at later times depends only
on $\int_x \delta f_x$, since the associated source $\lambda_{xt}=\lambda_t$ is spatially uniform; hence the instanton solution is also
spatially uniform, $\tilde u^{\lambda}_{x t}=\tilde u^\lambda_t$. One consequence is that the probability that the avalanche which started at $t_0$ 
has terminated before $t_1$, 
\bea
P(T<t_1) &=& \lim_{\lambda \to - \infty} e^{\int_x \delta f_x \tilde u^{\lambda_t=\lambda \delta(t-t_1)}_{t_0} } =  e^{- \frac{\int_x \delta f_x}{e^{t_1-t_0}-1}  }\nn\\
& =&  1 -  \frac{\int_x \delta f_x}{e^{t_1-t_0}-1} + O(\delta f_x^2)  \nn
\ .\eea
(in dimensionless units)
also depends only on $\int_x \delta f_x$. This is because, although an 
avalanche has ended if and only if all $\dot u_{xt}=0$, thanks to Middleton's theorem
this is equivalent to the center-of-mass velocity being zero. Hence we can
use the uniform source $\lambda_t = \lambda \delta(t-t_1)$, leading to
the above explicit expression, which we use below. 

As a last application, to be discussed again below, consider an arbitrary driving $\dot f_{xt} \geq 0$ 
for $t>t_0$ with the initial condition (\ref{init}). Let us define a kick of finite duration $t_{\rm f}-t_0$ as
a driving such that $\dot f_{xt} >0$ for $t_0<t<t_{\rm f}$ and $\dot f_{xt}=0$ for $t>t_{\rm f}$. 
Consider a source $\lambda_{xt} = \sum_{j=1}^n \lambda_j \delta(t-t_j)$
with $t_0<t_1< \dots <t_n$. The solution of the instanton equation with such a source was studied in
Section \ref{s:jpdfcomv} \footnote{The time ordering there was opposite. }. 
One can check that in the limit of all $\lambda_i \to - \infty$ the instanton solution takes a
very simple form (in dimensionless units), namely
\be 
\tilde u_t = \sum_{j=1}^n \frac{\theta(t_{j-1}<t<t_j)}{1-e^{t_j-t}}
\ .\ee 
Hence we obtain the joint probability
\bea
&& {\rm Prob}( \dot u_{t_1}=0, \dot u_{t_2}=0, \dots  ,\dot u_{t_n}=0 ) \label{jointav} \\
&& ~~~~~= \exp\bigg(- \int_{t_0}^{t_1} \frac{\rmd t \int_x \dot f_{xt}}{e^{t_1-t}-1} - ... - \int_{t_{n-1}}^{t_n} \frac{\rmd t \int_x \dot f_{xt}}{e^{t_n-t}-1} \bigg) \nn
\eea
We can learn a lot from this formula: First, for $n=1$, we see that $\dot u_{t_1}$ can vanish (strictly) only
either when the driving has stopped strictly before $t_1$, e.g.  $\dot f_{xt}=0$ for all $t_0< t_f< t< t_1$, or if it stops
at $t_1$, e.g. $\dot f_{xt} \sim (t-t_1)^a$ with $a>0$ 
such that the integral remains finite. Hence  a kick of finite duration
produces only a single avalanche which lasts longer than $t_{\rm f}-t_0$, more precisely,
taking a derivative w.r.t. $t_1$,
\be \label{durkick}
P(T=t_1) = \int_{t_0}^{t_{\rm f}} \frac{\rmd t \int_x \dot f_{xt}}{4 \sinh^2(\frac{t_1-t}{2})}  
 \,\rme^{- \int_{t_0}^{t_{\rm f}} \frac{\rmd t' \int_y \dot f_{yt}}{e^{t_1-t'}-1} }
\ .\ee
Then, for $n>1,$ formula (\ref{jointav}) allows to analyze the case of a succession of several kicks of finite duration.
Because the joint probability takes the form of a product on each interval $[t_i,t_{i+1}],$ it shows
that for a given $\dot f_{xt}$ the events $\dot u(t_i)$ are statistically independent \footnote{There is no
contradiction with  the fact that for a single kick $\dot u_{t_1}=0$ implies $\dot u_{t_2>t_1}=0$: Indeed,
the probability of the second event is one if the driving vanishes on the interval $t_1,t_2.$}.

To conclude, let us note that the formula (\ref{magic}) being more general, it also allows to study
the properties of stationary avalanches in the steady state with constant driving $\dot w_{xt}=v$
(see e.g.\ \cite{DobrinevskiLeDoussalWiese2011b}). However formulae such as (\ref{jointav}) and (\ref{durkick}) 
do not readily apply (they would lead to divergent integrals). This is because one must perform the
limit of infinite Laplace parameters $\lambda_i$ {\it after} integration over time,
the physics of the single avalanche being restored for $v=0^+$ as 
explained in details in Section \ref{s:jpdfcomv}.

\subsection{Recovering the quasi-static avalanche-size distribution}\label{s:recover-stat}

Here we show how to recover the quasi-static avalanche-size distribution, first within the stationary state at a constant
small driving velocity $v$, by measuring for a finite time, and second in a non-stationary setting, 
by driving the system over a finite distance. The results for the avalanche-size distribution in a finite time window are new and
of experimental interest. Some results at the end about a finite driving are also new. 

\subsubsection{Steady state: Limit of infinite time window}

Consider  the center of mass, i.e.\ the total size $S$ of an avalanche. 
In the limit of small $v$, in the comoving frame, the latter  is 
$S= L^d \int_{-T/2}^{T/2} \rmd t (v+\dot u_t)$, where $T$ is a time much larger 
than the typical single-avalanche duration, but much shorter than the waiting 
time between two consecutive avalanches. We want to compute
\be 
\overline{e^{L^d \lambda \int_{-T/2}^{T/2} (v+\dot u)}} = \overline{e^{\lambda S}} 
\ .\ee 
One would like to take $T\to\infty$, and consider a static source $\lambda_t := \lambda$. 
The instanton equation then admits static solutions,
\be 
\tilde u_t = \tilde u \quad , \quad - m^2 \tilde u + \sigma \tilde u^2 = -\lambda
\ .
\ee 
The one of interest is \be
\tilde u_t = \tilde u(\lambda) = \frac{m^2 - \sqrt{m^4 - 4 \lambda \sigma}}{2 \sigma} 
\ .\label{166}\ee
The other root is not continuously related to $\tilde u=0$ at $\lambda=0$, 
and  for this reason we reject it. The solution (\ref{166})  has to be injected into  Eq.\ (\ref{78}).  Due  to the time integral in the latter, this leads to
an infinite $Z(\lambda)$. Hence to recover the avalanche-size distribution from the dynamics in the
setting of a constant driving, \(w(t)=vt\), one must be more careful
and consider $T$ large, but not infinite. For instance, we may
consider a square source\be 
\lambda_t = \lambda \theta(t_2-t) \theta(t-t_1)
\ee  with $t_1=-T/2$ and 
$t_2=T/2$. If $T$ is large enough, the solution  is expected to look like\bea \label{sfp}
&& \tilde u_t = 0 \quad , \quad t > t_2 \\
&& \tilde u_t = \tilde u(\lambda)     \quad , \quad t_1 < t \ll t_2 \\
&& \tilde u_t = 0  \quad , \quad t \ll t_1 
\ .
\eea
One then finds, expanding {(\ref{Gv})} in small $v$,
\be 
\overline{e^{L^d \lambda \int_{-T/2}^{T/2} (v+\dot u)}-1} = v L^d \left[ T m^2 \tilde
u(\lambda) + O(T^0) \right] + O(v^2) 
\ .
\ee  
We work here in the limit $T \gg \tau_m$, but $\rho_0 v T \ll 1$.
On the other hand, we know that quasi-static avalanches obey \cite{LeDoussalWiese2008c}
\bea
  \overline{e^{\lambda L^d [u(w) - u(0)]}-1} &=& \int \rmd S \rho(S) (e^{\lambda S} -1) w + O(w^2) \nn  \\
 & =& 
 L^d Z_{S}(\lambda) w + O(w^2) 
\ .
\label{Za}\eea
Here  we denoted
(instead of $Z(\lambda)$
as in
Ref.\ \cite{LeDoussalWiese2008c}) \be 
Z_{S}(\lambda) = L^{-d} \left< e^{ \lambda S} -1 \right>_\rho =
\frac{1}{\langle S \rangle} \Big(\left< e^{ \lambda S} \right> - 1\Big)\ \\ \\  
\ee  
the generating function for quasi-static avalanche sizes.  $\left<\dots\right>_\rho$ denotes
the un-normalized average\footnote{Note however that the expression with $\rho$ also holds
for a continuum avalanche process with no cutoff. From (\ref{rho0}) it is normalized to
the volume \(\left<S\right>_\rho =L^d\), see \cite{LeDoussalWiese2008c}.} w.r.t. $\rho$ and we have used
(\ref{rho0}) to transform it into a normalized average over $P(S)$. Identifying 
$w=v T$ and the total displacement $u(w)-u(0)=u_{T/2} - u_{-T/2}$, we obtain \be 
Z_{S}(\lambda) = m^2 \tilde u(\lambda)\ .
\label{173}\ee  
Hence we recover the tree result for the size distribution
 \cite{LeDoussalWiese2008c}\be 
 Z^{\mathrm{tree}}_{S}(\lambda)  =  \frac{1 - \sqrt{1 - 4 \lambda S_m}}{2 S_m} 
 \ .\ee 
 It leads, upon inverse Laplace transformation, to $P(S)$ given by
 Eqs.\ (\ref{PS}) and 
(\ref{pS}).
Note that the same procedure can be performed to recover
the {\it local} avalanche-size distribution, by considering a time independent
but space dependent solution of the instanton equation. One then recovers, for instance,
the results obtained in section IX of \cite{LeDoussalWiese2008c}.

\subsubsection{Steady state: Distribution of
avalanche sizes during a finite time window}

To be complete, we  now show that the solution of the instanton equation indeed has
the form (\ref{sfp}) at large $T$, i.e.\ that the static fixed point is attractive. This also provides a novel physical observable for measurements 
restricted to a finite time window. The effect of finite {\it space} windows has been studied before in the avalanche context in 
\cite{ChenPapanikolaouSethnaZapperiDurin2011}, while a general study of windows in scale invariant Gaussian signals can be 
found in \cite{SantachiaraRossoKRauth2007}. 

We solve the instanton equation in dimensionless units  for a square source $\lambda_t := \lambda \theta(t_2-t) \theta(t-t_1)$,\be 
(\partial_t -1) \tilde u_t + \tilde u_t^2 = - \lambda \theta(t_2-t) \theta(t-t_1) 
\ .
\ee  
Its solution is \bea \label{solusquare}
&& \tilde u_t = 0 \quad , \quad t > t_2 \\
&& \tilde u_t = \frac{ 1}{2} \left[1+ \sqrt{1-4 \lambda} ~ \phi_\lambda\Big( \frac{t-t_2}{2} 
\sqrt{1-4 \lambda} -C_\lambda \Big)\right] , \nn \\
&&  t_1 < t < t_2 \nn \\
&& \phi_\lambda(z) = \tanh (z) , \quad C_\lambda = {\rm arctanh\!}\left(\frac{1}{\sqrt{1-4 \lambda}}\right)  , \quad \lambda<0  \nn \\
&&  \phi_\lambda(z) = \coth (z)  , \quad  C_\lambda = {\rm arcoth\!}\left(\frac{1}{\sqrt{1-4 \lambda}}\right) , \quad \lambda > 0 \nn \\
&& \tilde u_t = \frac1{1+ \big( \frac{1}{u_{t_1^+}}-1\big) e^{t_1-t}} \quad , \quad t< t_1 \nn
\ .
\eea 
The two branches depending on the sign of $\lambda$ are actually identical (by analytic continuation)
since $\tanh(z + i \pi/2) = \coth z$. We see on these solutions that the above fixed-point form (\ref{sfp}) indeed
holds at large $T$. 

We  now study the probability distribution of {\it the total displacement during a time-window size  $T$}, i.e.\ of the 
observable
\be 
U=\int_{-T/2}^{T/2} \rmd t (v + \dot u_t)\ .
\ee 
This  quantity is clearly of experimental interest. (For simplicity  we have suppressed all factors of $L^d$, which can
be restored at the end). It should interpolate
between the distribution of the instantaneous velocity at short times, and the distribution of
sizes of quasi-static avalanches at large times. To check this, we
compute $\tilde Z(\lambda) = \int_t \tilde u_t$ using Eq.\ (\ref{solusquare}),
which leads to\bea
 \tilde Z(\lambda) &=& \frac{T+\log (1-4 \lambda ) }{2} \nn\\
&&-\log \Bigg((1-2 \lambda ) \sinh  
  \bigg(\frac{T}{2} \sqrt{1-4 \lambda }\bigg) \nn\\
\nn && \qquad ~~~~+\sqrt{1-4 \lambda }
   \cosh\bigg(\frac{T}{2} \sqrt{1-4 \lambda }\bigg)\Bigg) \\
   & =& \lambda  T+\frac{\lambda ^2 T^2}{2}+\frac{1}{6} \lambda ^2 (2 \lambda -1)
   T^3+O(T^4) 
\ . ~~~~~~~~~~~
\eea
In the last line we have indicated the behaviour at small $T$. The series expansion in $\lambda$, which gives the moments, is also instructive,   \bea \nn
  \tilde Z(\lambda) &=& \lambda  T+\lambda ^2 (T+e^{-T}-1)\\
   && +2 \big[ T-2  + e^{-T} (2 + T)\big] \lambda^3 + O(\lambda ^4) 
\ .~~~~~~~~
\eea 
It shows that $\overline{U}=v T$ exactly, as expected, and that at large $T$
all moments grow linearly as 
\be  \label{momentsU}
\overline{U^p} = v\left[ c_p T + d_p + O(T^{a_p} e^{-T})\right]\ ,
\ee  
 i.e.\ up to
exponentially decaying terms, and with possible power-law prefactors. 

As in the preceding section, in the small-$v$ limit the probability distribution of $U$ is expected to take the form
\be 
P(U) = (1- \rho_0 v T) \delta(U) +  \rho_0 v T {\calP}(U)
\ .
\ee  
Here $\rho_0 v T$ is the probability that an avalanche has started inside the time window $T$.
Note that if $U$ is non-zero, the avalanche can have started anytime during the time window
and may, or may not, have finished during that time. $U$ thus contains information about
the signal measured in a time window without the necessity to determine when the
avalanche starts or ends. 

Since  $\overline{U} = \rho_0 v T \langle U \rangle = v T$ from the above,  (where and 
below $\langle ... \rangle$ denotes moments w.r.t.\ the distribution ${\calP}(U)$), using 
Eq.~(\ref{rho0}) we obtain the
remarkable property that the first moment of the distribution ${\cal P}(U) $,
\be  \label{equa}
\langle U \rangle = \langle S \rangle = \lim_{T \to \infty} \langle U \rangle
\ee  
is independent of $T$. The distribution ${\calP}(U)$ can then be obtained 
via Laplace inversion, 
\bea
\lefteqn{ \frac{1}{\langle S \rangle} U {\calP}(U) ={\mathrm{LT}}^{-1}_{s \to U} \partial_\lambda Z(\lambda)\Big|_{\lambda = -s} } \nn\\ &&=
\mbox{LT}^{-1}_{s \to U} \frac{1}{T(2 s+1)} \nn \\
&& ~~\times \bigg[ \frac{4 s \left(s T-1\right)}{\sqrt{4 s+1} \left((2 s+1) \tanh
   \left(\frac{1}{2} \sqrt{4 s+1} T\right)+\sqrt{4 s+1}\right)} \nn \\
&& ~~~~~~~+\frac{4 s
   (T+1)+T}{(4 s+1)} \bigg] \label{laplaceU}
\ .
\eea 
For $s=0$, this yields Eq.~(\ref{equa}). The Laplace inversion is
performed in Appendix \ref{app:laplinv}. Here we give some general features and 
limiting behaviors. First note that for any finite $T$ the apparent
singularity at $s=-1/4$ is fictitious, since the LT is analytic there. The closest singularity is
at $s_1(T)<-1/4,$ and the leading exponential decay at large $U$ is proportional to $ e^{s_1(T) U}$
where $s_1(T) = - 1/T$ at small $T$, and $s_1 = -1/4$ at large $T$. 

Examining Eq.~(\ref{laplaceU}) at large $s \gg \max(1, 1/T^2)$ shows that
the small-$U$ behaviour at fixed $T$ is independent of $T$, and given for $U \ll \min(1,T^2)$ by
\be 
{\calP}(U) \simeq  \frac{\langle S \rangle}{2 \sqrt{\pi } U^{3/2}}
\ .\ee  
The persistence of this strong divergence at small $U$, which requires a
short-scale cutoff $U_0 \sim S_0$, is consistent with the property (\ref{equa}),
since demanding normalization to unity of ${\calP}(U)$  leads to $\langle U \rangle \sim \sqrt{U_0}$,
 i.e.\ $\langle U \rangle \sim \sqrt{U_0 S_m}$ in dimensionfull units.

At large $T$ one can set the $\tanh$ in Eq.\ (\ref{laplaceU}) to unity and obtain
\bea
 {\calP}(U) &=& \langle S \rangle \Bigg\{  \frac{e^{-U/4}}{2 \sqrt{\pi } U^{3/2}} \\
&&~~~~~~~~~ + \frac{1}{T U} \left[ 1-\text{erf}\left(\frac{\sqrt{U}}{2}\right)-\frac{e^{-U/4}}{2} \right] + ... \bigg\}\nn 
\ .
\eea 
The neglected terms $...$ give the subdominant exponentially decaying 
part in Eq.~(\ref{momentsU}), while the linear and constant parts (i.e.\  $c_p$ and $d_p$) are reproduced by this
formula. It thus gives the leading correction to a measurement of the avalanche-size
distribution {\it if the time window is not large enough}. Restoring units we  find that these
corrections are decaying quite slowly as $O(\tau_m/T)$. They do exhibit a divergence 
$\sim 1/(2 T U)$ at small $U$, but which is too weak to correct the tail 
$U^{-\tau}$ with $\tau=3/2$ which agrees with the distribution (\ref{PS}), (\ref{pS}).

We note that the above formulae (in Laplace) are reminiscent, but different, from the ones leading 
to the joint distribution of avalanche durations and sizes given in \cite{DobrinevskiLeDoussalWiese2011b}.

\subsubsection{Avalanches size distribution in non-stationary driving}
\label{sec:levy}
In the first part of this section, we have considered what happens when measuring the avalanche-size distribution in the steady state obtained by
constant driving $w_t = v t$, during {\em a finite time}. 
On the other hand, one may also consider what happens when the system is {\em driven only over a finite distance $\delta w$}, i.e.\ in a non-stationary
setting. For this we  recall the discussion of arbitrary monotonous driving in Section \ref{s:finitek} and use  formula (\ref{magic2}). We work  in the laboratory frame 
and focus on the case
where the system is prepared at rest in the Middleton state, as described there and in Ref.\ \cite{DobrinevskiLeDoussalWiese2011b}, i.e.\ 
$w_{t}=w_{t_0}$ for $t_i < t  \leq t_0$ and $t_i \to - \infty$.
The driving is turned back on at $t_0$. Hence at $t=t_0$ one has $u_{xt_0} = u_x(w_{t_0})$, zero velocity $\dot u_{xt_0}=0,$ and  formula (\ref{magic2})  holds
for $t \geq t_0$. Since the particle has been at rest for a while for $t<t_0$ we define the total avalanche size as
\be 
S= L^d \int_{t_0 }^{\infty} dt \dot u_t = L^d ( u_{+\infty} - u_{t_0} )\ .
\ee 
To compute its
distribution we can choose a source $\lambda_{xt}=\lambda$, for $t>t_0$, independent of space and time.
The advantage of this setting is that  the instanton solution is then simply
the constant solution, $\tilde u_{x t} = \tilde u(\lambda)$ for $t>t_0$, given by Eq.\ (\ref{166}).
Hence one has, denoting $w_0=w_{t_0}$:
\bea
 \overline{ e^{\lambda S }} &=& e^{m^2 \int_{xt>t_0} \dot w \tilde u_{xt} } = e^{m^2 L^d \tilde u(\lambda) \int_{t>t_0} \dot w } \nn\\
& =& e^{m^2 L^d \tilde u(\lambda) \delta w} \\
 \delta w &:=& \int_{t_0}^{\infty} \dot w_t  \,\rmd t= w_{\infty}-w_0 
\eea 
Note that at this stage we consider an {\it arbitrary driving} $ \dot w_t \geq 0$ for $t>t_0$, i.e.\ we only assume that $\delta w <  \infty$.
We have not assumed it to be slow or small. To fix ideas, two extreme examples are:
\begin{itemize}
\item A kick $\dot w_{xt} = \delta w \delta(t-t_0)$
\item A constant driving during a finite window, $\dot w_t = v$ for $t_0<t<t_1$ and $\dot w_t=0$ for $t>t_1$, such that $\delta w=v(t_1-t_0)$. 
\end{itemize}
Now we know, from Middleton's theorem, that
\be 
u_{+\infty}  := \lim_{t \to \infty} u_t = u(w_0 + \delta w)\ . 
\ee  
Hence we have found that
\bea \label{res1}
 \overline{ e^{\lambda L^d [u(w_0 + \delta w) - u(w_0)] }} &=& e^{m^2 L^d \tilde u(\lambda) \delta w} \nn\\
& =& e^{L^d  \frac{1 - \sqrt{1 - 4 \lambda S_m}}{2 S_m}  \delta w}\ , 
~~~~~~~~~~~ \eea
with $S_m=\sigma/m^4$, 
for arbitrary $\delta w$. In the limit of small $\delta w$, from the definition (\ref{Za}) of
$Z_a(\lambda)$ we recover again $Z_a(\lambda)=m^2 \tilde u(\lambda)$. But we find
more. By Laplace inversion one obtains the distribution of $S$,
\be  \label{res2}
P_{\delta w}(S) = \frac{L^d \delta w}{2 \sqrt{\pi} S_m^{1/2} S^{3/2}} e^{- \frac{S}{4 S_m} + \frac{L^d \delta w}{2 S_m} - \frac{(L^d \delta w)^2}{4 S S_m}}
\ .\ee  
This is Eq.\ (33) of Ref.\ \cite{DobrinevskiLeDoussalWiese2011b} where it was obtained for the kick and for a particle ($d=0$), but as we see here,
it is {\it independent of the precise form of the driving},  depending only on $\delta w$. What is remarkable is that
the probability (\ref{res2}) is two things in one:

(i) It is the distribution of  size $S=\int_{t_0}^{\infty} \rmd t\, \dot u_t$ of the avalanche, produced by an arbitrary driving resulting
in a total shift of the quadratic well of $\delta w =   \int_{t_0}^{\infty} \dot w_t$. Since the driving velocity can be
arbitrarily large this is a priori a non-trivial dynamical observable. Note that for the kick
one is guaranteed that there is a single avalanche, but if $\dot w_t$ has a more complicated form then 
$S$ may encompass several avalanches, separated by time regions where 
$\dot u =0$, e.g. for a succession of several finite duration kicks, as discussed in Section \ref{s:finitek}.

(ii) It is also the distribution of 
\bea
S&=& L^d \left[ u(w_0+\delta w) - u(w_0)\right]\nn\\
&=&\int_x u_x(w_0+\delta w) - u_x(w_0)\ , 
\eea a quasi-static observable, 
which for finite $\delta w$ may also encompass {\it several quasi-static avalanches},
since e.g. $\langle S \rangle = L^d \delta w$. In the limit of small $L^d \delta w \ll S_m$
one recovers the form (\ref{PS}) of $P(S)$ for a {\it a single avalanche} 
for $S \gg S_{\delta w}$ where $S_{\delta w}=(L^d \delta w)^2/S_m$ acts as a small-scale cutoff. The true {\em\ single-avalanche} limit however is reached only\footnote{In the limit where
the microscopic cutoff
$S_0 \to 0$ there are infinitely many small avalanches.} when
$S_{\delta w} \approx S_0$.

The fact that (i) and (ii) are the same is a simple, but remarkable, consequence of  Middleton's theorem. The fact
that the form for $P(S)$ is given by (\ref{res2}), and the property (\ref{res1}),  are a consequence of the  simplified
tree theory\footnote{The full tree theory with an arbitrary $\Delta(u)$ does {\em not} satisfy  property (\ref{res1}).}. 
As discussed below, its use  is justified for $d \geq d_{\rm uc}$, and  a priori only in the limit of slow driving $\dot w \ll v_m$. 
The property (\ref{res1}) is consistent with $u(w)$ being a Levy process, 
i.e.\ a jump process made of statistically independent avalanches, each 
distributed with the single avalanche distribution $P(S)$ from (\ref{PS}). The 
property recovered here is also present in the statics,  i.e.\ for the process $u^{\rm stat}(w), $ in mean field,  in the BFM and in the Burgers equation. It has  the same $P(S)$, 
as is discussed in  detail in \cite{LeDoussalWiese2011b}. 

Finally, a similar analysis can be performed 
for the probability distribution of the local observable
\be 
S^\phi := \int_x \phi_x [u_x(w_0+\delta w)-u_x(w_0)]\ .
\ee 
One  must then solve the space-dependent instanton equation 
with a source $\lambda_{xt} = \phi(x)$, which is a hard problem. 
In the case $\phi(x) = \delta(x_1)$, i.e.\ a hyperplane in a $d$-dimensional 
space, the time-independent instanton solution is known, see section IX of \cite{LeDoussalWiese2008c}:\bea
&& \overline{ e^{\lambda L^d [u(w_0 + \delta w) - u(w_0)] }} = e^{L^d  \frac{1}{S_m} \tilde Z(S_m \lambda) \delta w} \ ,\\
&&  \lambda(\tilde Z) = \frac{1}{72} \tilde Z (\tilde Z-6) (\tilde Z-12)
\ .\eea 
The Laplace inversion is involved, but a simple generalization of Eq.\ (220) in \cite{LeDoussalWiese2008c}.
The same trick yields the (normalized) probability distribution of $S^\phi=S$,
\bea
&& P^\phi_{\delta w}(S) = \frac{1}{S_m} \,p_{\frac{L^d \delta w}{S_m}}\left(\frac{S}{S_m}\right) \\
&& p_{w}(s) = \frac{2\times 3^{1/3}}{s^{4/3}} e^{6 w} w\, {\rm Ai}\left(\left[\frac{3}{s}\right]^{1/3} [s + 2 w]\right)\ .~~~ ~~~~
\eea 
Here $\langle S^\phi \rangle = w L^d$ and for $L^d \delta w \ll S_m$, i.e.\ $w\ll1$, one recovers 
$p_w(s) \approx w p(s)$ with $p(s) = 2 K_{\frac{1}{3}}(\frac{2 s}{\sqrt{3}})/(\pi s),$ the (rescaled)\ single-avalanche  size distribution obtained in \cite{LeDoussalWiese2008c}. These results are exact for the BFM
(discussed below), an application being a single-site avalanche for a string ($d=1$). 

\subsection{Mean-field theory for avalanches: The Brownian-force model and its ABMM limit}
\label{s:BFM}

We are now ready to discuss the correct mean-field theory for the avalanche motion
of elastic interfaces in the limit $v\to 0^+$, and to identify its universal properties in the
limit of small $m$. 

In a nutshell, the mean-field theory is the tree theory, with however a renormalization
of  two parameters of the model. Hence we first discuss these parameters and their universality.
In a second stage, the tree theory is identified with the BFM and the ABBM model is recovered. 

We recall that the upper critical dimension is $d_{\rm uc}=2 \gamma$ for an arbitrary elastic
kernel behaving as $\epsilon(q) \simeq q^{\gamma}$, i.e.\ $d_{\rm uc}=4$ for usual SR elasticity ($\gamma=2$) and
$d_{\rm uc}=2$ for the most common LR elasticity ($\gamma=1$).

\subsubsection{Improved tree theory and the parameters of the model}
\label{s:ittapotm}
We have shown above that to lowest order in perturbation theory in the bare disorder, all
 generating functions of the velocity, to first order in $v$, are given by the sum of
tree graphs. Equivalently, they can be computed from the simplified tree action $S^{\rm tree}$ defined
in Eq.~(\ref{Stree}). At the bare level, this action only 
contains three parameters $\eta_0$, $m$ and $\sigma_0=-\Delta_0'(0^+)$. 
These bare parameters  are corrected by disorder, and
acquire a dependence on $m$, as we now discuss.

Let us now use well-established results from the FRG approach
to the statics and dynamics of elastic interfaces. First, $m$ is uncorrected
to any order in perturbation theory thanks to the STS symmetry, hence we
can use everywhere the bare mass $m$. Second, perturbation theory converges 
for $d>d_{\rm uc}$ (in a sense recalled in Appendix \ref{app:frg}). Third, at $d=d_{\rm uc}$ there are only two operators which become
marginally relevant. The first one is the local part of the renormalized disorder, $\Delta(u)$, which actually is  a  function of $u$; so in principle
there is an infinity of marginally relevant directions. However, as far as single avalanches are concerned, 
we only need  $\Delta'(0^+)$: It is shown in Section \ref{s:loops} that the higher derivatives lead to
loop corrections, i.e.\ are important only for $d<d_{\rm uc}$. The second parameter is the renormalized friction $\eta$.
Both parameters, $\Delta'(0^+)$ and $\eta,$ receive logarithmically divergent corrections
in $d=d_{\rm uc}$ from 1-loop diagrams. These are cut off by the mass $m$ and can be resummed 
using the FRG flow equation to 1-loop order. 

Let us now determine the renormalized parameters at the upper critical dimension $d=d_{\rm uc}$. 
Define $\ell := \ln(\Lambda/m)$, where $\Lambda$ is a small-scale UV cutoff;
at $d=d_{\rm uc}$, for SR elasticity, set
\be  \label{scal1} 
\Delta(u) = 8 \pi^2 \tilde \Delta(u) = 8 \pi^2 \hat \Delta(u \ell^{-\zeta_1}) \ell^{-1+ 2 \zeta_1} 
\ .\ee  
Then the FRG flow equation for $\hat \Delta(u)$ is (B.14) in \cite{LeDoussalWiese2008c}. As $m \to 0,$ the rescaled
correlator tends to a fixed point $\hat \Delta(u) \to \hat \Delta^*(u),$ which is  the  same one
obtained to  first order in a $\epsilon$-expansion i.e.\ $\hat \Delta^*(u) = \lim_{\epsilon \to 0} \tilde \Delta^*(u)/\epsilon$. 
Similarly, see e.g. \cite{LeDoussalWieseChauve2002}, one obtains\be 
\partial_l \ln \eta = - \tilde \Delta''(0^+) = - \hat \Delta''(0^+) \ell^{-1} 
\ .\ee  
Hence, the two parameters of the model acquire a {\it universal dependence} on $m, $ in the
limit of $m \to 0$ \footnote{ Since $S_m=\sigma_m/m^4$, this corrects a misprint in Eq.\ (108) of \cite{LeDoussalWiese2008c}.}:
\bea
&& \sigma \to \sigma_m = - \Delta_m'(0^+) \simeq 8 \pi^2 |\hat \Delta^{*\prime}(0^+)| \big[\ln(\Lambda/m)\big]^{-1+\zeta_1}  \nn \\
&& \eta \to \eta_m \simeq \eta_0 \big[\ln(\Lambda/m)\big]^{z_1} 
\eea 
Both $z_1$ and $\zeta_1$ are defined by the 1-loop result  for the dynamic and roughness exponents,
\bea
 z&=& 2 - \tilde \Delta^{* \prime \prime}(0^+)  = 2 + z_1 \epsilon + O(\epsilon^2) \\
 \zeta &=& \zeta_1 \epsilon + O(\epsilon^2)
\eea 
with $\zeta_1=1/3$ and $z_1=(\zeta_1-1)/3=-2/9$ for non-periodic SR disorder.

The above formulae extend to LR elasticity by changing everywhere above $m \to \mu$,
defined below in (\ref{defkernelLR}), and the factor $8 \pi^2 \to C_{d=d_{\rm  uc},\gamma}$ (see 
its definition and detailed discussion in section X of \cite{LeDoussalWiese2008c})
with $C_{2,1}=2\pi$, the fixed point $\hat \Delta^*(u)$ being unchanged. 

We can now make a precise statement, based on the effective action  $\Gamma$ of the theory. 
For its definition see \cite{Zinn}, and in the context of FRG e.g.\ \cite{LeDoussal2008,LeDoussalWiese2008c,LeDoussalWiese2011b} (statics)
and \cite{LeDoussalWiese2006a,WieseLeDoussal2007} (dynamics),  summarized in \cite{LeDoussalWiese2008a}, Appendix A.   It is a 
general property of $\Gamma$ that all connected correlations of the theory (here of the velocity field) are 
{\em  tree} diagrams in $\Gamma$: The vertices of the trees are vertices not of the original action
$S$, but vertices of $\Gamma$, i.e.\ renormalized vertices, which contain all loop diagrams.

When $d \to d_{\rm uc}$ and in the limit  of $m \to 0,$ the effective action $\Gamma$ becomes simpler and its limit 
is the so-called improved action. This is discussed in Appendix \ref{app:frg}, where we show how the irrelevant
operators become negligible for $d \approx d_{\rm uc}$, when properly scaled. For instance,
the higher time derivatives in the equation of motion, or higher disorder cumulants, become
negligible, and one can focus  on $\eta$ and $\Delta(u) $ only.

If in addition one considers positive driving only, $\dot f_{xt} \geq 0$, then for $d=d_{\rm uc}$ the effective action of the velocity theory is
$\Gamma = S_{\rm tree}|_{\eta,\Delta'(0^+)}$, i.e.\ the tree action with the renormalized parameters $\sigma \to \sigma_m$ and $\eta \to \eta_m$.
It sums tree graphs except for the renormalization of $\sigma$ and $\eta,$
which contain loop corrections. This remains true for $d > d_{\rm uc}$, where $\sigma$ and $\eta$ 
flow to non-universal limits as $m \to 0$, as discussed in Appendix \ref{app:frg}. 
Note that the statement we make here is only for $v=0^+$: Since we have not analyzed the FRG flow at non-zero $v$, 
we focus   on the limit of small driving. This also means a small step in the force, i.e.\ a small kick, in the 
non-stationary setting discussed in Section \ref{s:finitek}. 

For $d<d_{\rm uc}$ the behavior is universal but different from mean-field, and is analyzed in 
Section \ref{s:loops}.

\subsubsection{Brownian force model (BFM) or elastically coupled ABBM models and universality}

The mean-field tree-level theory has a very simple interpretation. It is clear from
Section \ref{simplified} that what has been done is to replace the original equation of motion
(\ref{eqmo}) in a disorder described by the gaussian force correlator $\Delta_0(u)$ by
a disorder described by a (renormalized) correlator $\Delta(u) =\Delta(0) + \Delta'(0^+) |u|$, since we have
neglected all higher-order derivatives $\Delta^{(n)}(0^+)$;  the latter become important
only upon considering loop corrections to the velocity distributions. This means that this (simplified) tree theory
describes exactly {\it an elastic manifold in a Brownian force landscape} $F(x,u)$ with Gaussian correlations,
\be
\overline{F(x,u) F(x',u')}^{\rm c} = \delta^d(x-x') \Big[\Delta(0) - \sigma |u-u'|\Big] 
\ ,
\label{194}\ee
where $\sigma = - \Delta'(0^+)$. Such a landscape is constructed in a spatially discretized version, by considering that for each $x$, $F(x,u)$ performs a Brownian motion (BM) as a function of
$u$, and that
these BMs are mutually independent for different $x$. Furthermore, they are stationary Brownian motions, hence
they are constructed by considering e.g.\ a much larger periodic system in the $u$ direction. 
An elastic manifold of internal dimension $d$ in such a landscape is called the 
Brownian force model (BFM) 
\cite{LeDoussalWiese2011b}. The {\it statics} of this model was studied in \cite{LeDoussalWiese2011b}.
As we discuss below, a non-stationary BM version can also be considered. 

Hence, from the previous paragraph we conclude that the full statistics of the velocity field in an avalanche for an interface at $d \geq d_{\rm uc}$ 
identifies in the small-$m$, small-$v$ limit with that of the BFM, with parameters $\sigma \to \sigma_m$,
$\eta \to \eta_m$. This BFM  can also be described as a set of ABBM models 
for each $u_{xt}$ with an elastic coupling $g^{-1}_{xx'}$ between them. 

A crucial property of the BFM is that {\it the dynamics of the center of mass of the elastic manifold is described
by the ABBM model} \cite{AlessandroBeatriceBertottiMontorsi1990,AlessandroBeatriceBertottiMontorsi1990b}, i.e.\ by equations (\ref{eq:IntroABBM}) and (\ref{eq:CorrABBM}). Intuitively it is easy to understand why: To compute center-of-mass
observables in perturbation theory we need to consider all graphs with external momenta set to zero, $q=0$. 
However, since we have summed only tree graphs, it implies that all propagators are evaluated at $q=0$.
Hence, apart from the (non-trivial) renormalization of the parameters of the model, in effect,
the avalanche dynamics of the center of mass $\dot u_t$ for $v=0^+$ is described by the ABBM model, i.e.\ a {\it single point} driven
in a long-range correlated random-force landscape, $F(u)$, with {\it Brownian} statistics. It
amounts to suppressing the space dependence in Eq.~(\ref{eqmo}), hence
corresponds in our  general model to the special case $d=0$ and
$\Delta_0(0)-\Delta_0(u)=\sigma |u|$.

Let us now   connect our previous results, obtained
directly for the center of mass of the interface, to the standard analysis of the ABBM model.
Then we will revisit the BFM, and finally  calculate observables beyond the center of mass,  requiring
the full power of the BFM.  

\subsubsection{Center-of-mass observables and ABBM model}
\label{sec:cdmabbm}

Let us recall the original solution \cite{AlessandroBeatriceBertottiMontorsi1990,AlessandroBeatriceBertottiMontorsi1990b} 
of the ABBM model, based on a Fokker-Planck approach (see more details in \cite{LeDoussalWiese2008a}). The equation of
motion (\ref{eqmo}) for the instantaneous velocity  in the laboratory frame ${\sf
v}=\dot u_t$ of a particle in a Brownian landscape (suppressing internal degrees of freedom $x$) can  be 
written as a stochastic equation
\be \label{abbmlangevin}
\eta d {\sf v} = m^2 (v - {\sf v} ) \rmd t\,  + \rmd F 
\ ,
\ee
where $\overline{\rmd F^2}=2 \sigma {\sf v} \rmd t\, $. The associated Fokker-Planck equation for the probability distribution $Q\equiv Q({\sf v},t|{\sf v}_1,0)$  of the
velocity at time $t$,  given velocity $v_1$ at time $t=0 $ is
\begin{equation} \label{fp}
\eta \partial_t Q = \partial_{{\sf v}} \left[\frac{\sigma}{\eta} \partial_{{\sf v}} ({\sf v} Q) + m^2 ({\sf v} - v) Q\right]\ .
\end{equation}
It satisfies $Q({\sf v}_2,0^+|{\sf v}_1,0)=\delta({\sf v}_2 - {\sf v}_1)$. It is normalized to unity at all times upon integration over the final velocity ${\sf v}$, thus it is  the propagator of the system. For $v>0$, it evolves at large times to the
stationary (zero current) distribution $ Q_0:=\lim_{t \to \infty} Q $ with
\be \label{ste}
Q_0({\sf v}) =\frac{1}{{\sf v}}\left(\frac{{\sf v}}{v_m}\right)^{\!{v}/{v_m}} \frac{ e^{- {\sf v}/v_m}}{\Gamma( \frac{v}{v_{m}})}\ .
\ee
Here $v_m=S_m/\tau_m$,   $S_m=\sigma/m^4$ and $\tau_m=\eta/m^2$. Note that here
we study a point particle, hence the velocity scale is $v_m$; if we study the center of mass of an interface,
it is to be replaced by $\tilde v_m$ as discussed in Section \ref{s:ittapotm}. 

One notes that taking $v \to 0^+$ and forgetting the normalization, 
$Q_0$ converges to the single-time velocity distribution obtained above in Eq.~(\ref{1time}) 
by a completely different method. There, the normalization was fixed from considerations
of a small-scale cutoff. Similarly, in the limit $v\to 0^+$, one finds that the 
propagator takes the form
\be
Q({\sf v},t|{\sf v}_1,0)=\frac 1{v_m}
\tilde Q\left(\frac{{\sf v}}{v_m},\frac{t}{\tau_m}\bigg|\frac{{\sf v}_1}{v_m},0\right) 
\ ,\ee with 
\begin{equation}\label{solu2abbm}
\tilde Q(v_2,t|v_1,0) = v_1 e^{v_1} \bigg[ p_2(v_1,v_2) +
\frac{1}{v_1} e^{-\frac{v_1}{1-e^{-t}}} \delta(v_2) \bigg]\ .
\end{equation} The term $p_2 (v_{1},v_{2}),$   given by Eq.~(\ref{pdef}), is indeed a
solution of (\ref{fp}) with $Q({\sf v}_2,0^+|{\sf v}_1,0)=\delta({\sf v}_2 - {\sf v}_1)$. 
We note that the piece $\sim \delta(v_2)$, which corresponds to avalanches which have already terminated at time $t$,
is necessary for $Q$ to conserve  probability, i.e.\ such that $\int_{0^-}^{\infty} \rmd v_2 \tilde Q(v_2,t|v_1,0)=1$ for all $t$.
Since $Q$ is a conditional probability, we can also consider the {\it joint distribution} of velocities, \be   
\tilde Q(v_2,t|v_1,0)p_1(v_1)
 = \tilde Q(v_2,t|v_1,0) \frac{1}{v_1}e^{-v_1}\ .
\ee  
We find that it reproduces the 2-time probabilities given in Eqs.~(\ref{2times}) and  (\ref{tildeP1}).
 More details about the ABBM propagator and how it behaves in the $v\to 0^+$ limit can be found in Appendix \ref{a:AABBM}.

By using the dynamical field theory of interfaces, we have in this paper obtained  a novel, and
completely {\em independent} way to solve the ABBM model. Indeed, our method is even more powerful, since it allows to
 treat  interfaces and spatial degrees of freedom, and it can be extended beyond the tree level, as will be discussed in the following sections. Already its consequences for the ABBM model itself are   quite interesting: By allowing to compute
directly Laplace transforms through the instanton equation (\ref{mfnonlinear}), 
it provides a useful complementary method to the Fokker-Planck approach. For avalanche observables
it is quite efficient, as was shown in the previous sections and Ref.~\cite{DobrinevskiLeDoussalWiese2011b}. For other observables (such as $U=\int_{-T/2}^{T/2} \rmd t\,\dot u_t$), non-locality in time makes it very hard to obtain the result  via the Fokker-Planck method. On the other hand, one advantage of the Fokker-Planck approach  is
that since $\sf{v} (t)$ is a Markov process, the $n$-time velocity probability can   be written 
in a factorized form as 
\be
q'_{1...n} {\calP}(\dot u_1,\ldots ,\dot u_n)=\frac{1}{\dot u_1} e^{-\dot u_1}\prod_{j=1}^{n-1} Q(\dot u_{j+1} t_{j+1}| \dot u_j t_j) 
\ ,
\ee
where $q'_{1...n}$ is the probability that all $n$ times  belong to an avalanche. Curiously, it is not
 easy to recover that property immediately from our general expression for $\tilde Z_n$. In Appendix
\ref{a:3-time-formula} we check it explicitly for $n=3$. 

Let us note that since the ABBM model 
is the zero-dimensional limit of the equation of motion  (\ref{eqmo}) of an interface, 
the dynamical-action method can  be applied. Hence we just found
that, for the ABBM model at $v=0^+$, {\it the tree approximation is
exact}. In the field theory for the velocity it means that the effective action $\Gamma$
equals the bare action $S$, and there are no loop corrections. Hence
$\Delta'(u)=\Delta_0'(u)=- \sigma \, \mbox{sgn}(u)$ is  an exact FRG fixed point with scaling exponent  $\zeta=4-d$, as  
already noted in the statics in \cite{LeDoussalWiese2008c}. Crucial for this remarkable property is
that the force landscape is a Brownian, and even in $d=0$, this is
not valid for any other, e.g.\ shorter-ranged, force landscape. These properties and
a direct solution of the ABBM model at any $v$ are discussed in \cite{DobrinevskiLeDoussalWiese2011b}. 

A word of caution should be said about the notion of the duration of an avalanche. In the 
present tree-level mean-field theory (and similarly in the ABBM model) 
avalanche durations can be defined unambiguously for a continuum version where
the small scale cutoff $S_0 \to 0$, and accordingly the avalanche density $\rho_0 \to \infty$, as the velocity $\dot u$ exactly vanishes at some time for $v=0^+$.
In that version there is an infinite number of infinitely small avalanches and the quasi-static process
is infinitely divisible (a Levy process) as discussed at the end of Section \ref{sec:levy}. On the
other hand, if one studies the original interface model (\ref{1}) with smooth and short-ranged disorder, in the limit $v=0^+$ or in the
limit of a small step in the force $\delta w$, an avalanche has, strictly, an infinite duration (diverging with some power of $1/v$ or $1/\delta w$). 
Indeed the starting point is a metastable state (zero force state) with one marginally unstable direction and the final state is generically a stable zero force state. 
Near both points the motion is very slow, so the duration is very large, but the associated displacement is negligible. 
One must thus focus on the part of the avalanche motion such that $\dot u \gg v_0$, or such that the interface has significantly moved by more than $S_0$. This part
of the motion is universal and described by the ABBM model. It would be interesting to make this statement mathematically
 precise.

\subsubsection{ABBM model: Connection between the instanton equation 
and the Fokker-Planck equation}
\label{s:ABBM}

The Fokker Planck equation can be Laplace-transformed
in $\lambda$, or equivalently one can write the evolution equation  (in the laboratory frame)
for 
\be  G(\lambda,t) :=
\overline{e^{\lambda \dot u_t}} = \int_0^\infty \rmd{\sf v} \,e^{\lambda {\sf v}} P({\sf v},t)  
\ .
\label{202}\ee  
Without specifying  the initial conditions, the evolution equation is 
\be  \label{eqZ}
\frac{\partial G}{\partial t}+\frac{\partial G}{\partial \lambda}(\lambda-\lambda^2)=\lambda G
\ .
\ee 
The solution can be found in the form
\be 
G(\lambda,t) =e^{v \tilde  Z(\lambda,t)}\ ,
\ee 
with $Z(0,t)=0$ since $G(\lambda=0,t) =1$. Then $\tilde Z$ satisfies the equation
\be  \label{eqZ2}
\frac{\partial  \tilde Z}{\partial t}+\frac{\partial \tilde Z}{\partial \lambda}(\lambda-\lambda^2)=\lambda
\ .
\ee 
This equation admits a time-independent solution $\tilde Z(\lambda)\equiv \tilde Z (\lambda,t)  $
\be  \label{resold}
\tilde Z(\lambda)=-\log(1-\lambda)\ .
\ee  
Hence we recover the result (\ref{ztilde}) obtained via the MSR dynamical-action method. 

The connection to the instanton equation can be made as follows.
The equation (\ref{eqZ}) can be solved by the method of characteristics: Define a 
function $\lambda(t)$ which obeys the following differential equation,
\begin{align}\label{eq:instanton}
\frac{\rmd \lambda(t)}{\rmd t}&=\lambda(t) -\lambda^2(t)\ .
\end{align}
Further define $\tilde Z(t) := \tilde Z(\lambda(t),t)$. Then, using Eq.\ (\ref{eqZ2}), the total derivative is
\begin{align}\label{eq:instanton2}
\frac{\rmd\tilde   Z(t)}{\rm d t}&=\lambda(t).
\end{align}
Equation (\ref{eq:instanton}) is exactly the instanton equation (\ref{mfnonlinear}), if one identifies $\lambda(t)=\tilde u(t)$. 
For $t<0$,
it admits the solution \be  \label{soluabbm}
 \lambda(t) = \frac
{\lambda_0 }{\lambda_0+(1-\lambda_0)e^{-t}}
\ee 
with  boundary conditions $\lambda(-\infty)=0$, and $\lambda(0)=\lambda_0$. In addition
\begin{align}\label{eq:Z}
\tilde Z(t):=\int_{-\infty}^t \lambda(t')\, \rmd  t'.
\end{align}
Hence
if we express $Z(\lambda_0) := Z(t=0)$ as a function of $\lambda_0=\lambda(0)$
we obtain precisely (\ref{resold}). 

Eq.\ (\ref{eqZ2}) is  solved for any initial condition $Z(\lambda,t=0)=Z_0(\lambda)$ as
\be
\tilde Z(\lambda,t) =  - \ln(1-\lambda+\lambda e^{-t}) + \tilde  Z_0\!\left( \frac
{\lambda }{\lambda+(1-\lambda)e^{t}} \right)\ .
\ee Note that the argument of \(\tilde Z_0\) is $\lambda(-t)|_{\lambda_0 \to \lambda}$.
Hence from Eq.~(\ref{202}) we get
\bea \label{decay} 
G(\lambda,t) &=& (1-\lambda+\lambda e^{-t})^{-v} G_0\left(
\frac
{\lambda }{\lambda+(1-\lambda)e^{t}} \right)~~~~~~~~~\\
G_0(\lambda)&=& \rme^{v \tilde Z_0(\lambda)}
\ .
\eea
This gives the decay to the steady state as
\bea
 \overline{\dot u(t)} &=& v (1-e^{-t}) + e^{-t} \overline{\dot u(0)} \\
 \overline{\dot u(t)^2}^{\rm c} &=& v (1-e^{-t})^2 + 2 \overline{\dot u(0)} e^{-t} (1-e^{-t})
\nn\\ &&+ e^{-2 t} \overline{\dot u(0)^2}^{\rm c} \ .
\eea 
It is in agreement with the results of \cite{DobrinevskiLeDoussalWiese2011b} for a quench in the driving velocity. 
Note that for any $t>0$ (\ref{decay}) behaves as $G(\lambda,t) \sim A(t) (-\lambda)^v$ with 
$A(t) = (1-e^{-t})^{-v} G_0(- \frac{1}{e^t-1})$, hence $P(\dot u,t) \sim A(t) \dot u^{v-1}/\Gamma(v)$
and the current at the origin vanishes. 

\subsubsection{Back to the Brownian force model}
Having recalled the properties of the ABBM model, which contains the information about the center of mass, 
we now reexamine the BFM which contains all spatial information. 

The Langevin equation (\ref{abbmlangevin}) for the ABBM model can be rewritten as
\be 
\eta \partial_t \dot u_t = \sqrt{ \dot u_{t}} \xi(t) + m^2 (\dot w_t - \dot u_{t})\ , 
\ee  
with $\overline{\xi(t) \xi(t')}=2 \sigma \delta(t-t')$  a Gaussian white noise. It describes the
original model (\ref{eq:IntroABBM}) only if $\dot w_t \geq 0$. 

Similarly, the BFM can be defined focusing on the evolution of the velocity, by
the following Langevin equation in the laboratory frame:
\be  \label{deffbm2}
\eta \partial_t \dot u_{xt} = \sqrt{\dot u_{xt}} \xi(x,t) + f_{xt} + (\nabla_x^2 - m^2) \dot u_{xt}
\ ,\ee  
with $\overline{\xi(x,t) \xi(x',t')}=2 \sigma \delta(t-t') \delta(x-x')$ uncorrelated Gaussian white noises,
with obvious generalization to an arbitrary elastic kernel $g_{xx'}^{-1}$. It does describe the motion in a stationary Brownian random-force landscape if (and only if) driving
is monotonous $\dot f_{xt} \geq 0$ for all times. However, from the discussion in Section (\ref{s:finitek}) and in \cite{DobrinevskiLeDoussalWiese2011b}, 
if one complements it with an initial condition
\be  \label{init1}
\dot u_{x,t=0}= 0\ , 
\ee  
it does also describes the motion in the  non-stationary Brownian random-force landscape
\be
\overline{F(x,u) F(x',u')}^{\rm c} = 2 \sigma \min(u,u') \delta(x-x')
\ ,
\label{194a}\ee
for $u,u' \geq 0$ with initial condition $u_{x t=0}=0$. This setting has advantages since 
the landscape is defined by uncorrelated BMs which all start as $F(0,x)=0$.
This avoids the construction of stationary BMs in a large box, as a limiting process. 
The catch is that in position theory it does not satisfy STS; this is seen on observables
such as $\int_0^t \rmd s\, \dot  u_{xs}$, whose averages are time dependent. If one adds a large box,
then these converge to the stationary BFM observables. 

If one focuses only on velocity observables, and forgets about the position
theory, the BFM  is uniquely defined by Eq.~(\ref{deffbm2}). If one drives for some time, the
memory of the initial joint distribution of velocities ${\cal P}[\{ u_{xt=0} \}]$ is lost.
In the case of a steady drive, $\dot f_{xt}=v$, the system evolves 
towards a time-translationally invariant steady state, e.g. for the 1-time
distribution,\be 
{\cal P}[\{ u_{xt} \},t] \to {\cal P}_{\rm steady}[\{ u_{x} \}]
\ ,\ee  
which generalizes (\ref{ste}) and is more complicated to calculate; (it requires solving the instanton equation with a 
space dependent source). This steady-state measure for the full velocity-field identifies with the one of
the elastic manifold in dimension $d \geq d_{\rm uc}$, and for small $v$ and $m$, as discussed in the previous sections.

In addition, the BFM is an interesting model to study by itself. It can
be solved in arbitrary space dimension $d$, and for arbitrary driving, from the general formula:
\be\label{magic3}
\overline{\rme^{\int_{xt}\lambda_{xt} \dot u_{xt}}} = \rme^{\int_{xt}
\tilde u_{xt}^{\lambda} \dot f_{xt}}\ .
\ee
which assumes (monotonous) driving from the far past, or formula (\ref{magic2}) for the initial condition (\ref{init1}). More details can be found in \cite{DobrinevskiLeDoussalWiese2011b},
including a formula for an arbitrary initial velocity distribution. 

\subsection{Spatial fluctuations} \label{sec:spatial} 

 We can now use the full power of the tree theory, i.e.\ the BFM, and
calculate space-dependent observables within mean-field theory.
The space-dependent instanton equation allows to go beyond the ABBM model,
which describes only the center of mass, and to compute spatial fluctuations. 

In addition, the results below are exact for the BFM {\it in any space dimension} $d$.
Most results  concern the BFM in the steady state, i.e.\ they are time-translational invariant,
as discussed in the previous section. Time-dependent non-stationary generalizations 
are left for the future.

\subsubsection{General considerations}

Let us write  for completeness the instanton equation for an arbitrary elastic kernel $g_{xx'}^{-1}$
\begin{equation} \label{mfnonlinearxtg}
  \int_{x'}( \eta \partial_t \delta_{xx'} - g^{-1}_{xx'}) \tilde u_{x't} + \sigma \tilde u_{xt}^2  + \lambda_{xt} = 0\ ,
\end{equation}
with $\sigma=-\Delta'(0^+)$. Below, we first perform our calculations using the local elasticity \begin{equation}
g^{-1}_{xx'}= (-\nabla^2+m^2)\delta_{xx'}\ .
\end{equation}
At the end we indicate how the formulae generalize to an arbitrary elastic kernel.

Time-independent, but space-dependent solutions of the instanton equation with a $\delta$-function source in
space were studied in Refs.\ \cite{LeDoussalWiese2008c,LeDoussalWiese2009a,LeDoussalPetkovicWiese2012}, and allowed to obtain the distribution of  local avalanche sizes. 
Finding solutions which are both time- and space-dependent is notably more difficult\footnote{Note the resemblance of the
instanton equation with the KPP-Fisher equation for front propagation.} 
and must be left for
future research. Here we analyze solutions which are ``almost" space independent,
i.e we  choose\be 
\lambda_{xt} = \lambda_t +  \mu_{xt} 
\ ,
\ee  
where the spatially dependent part $\mu_{xt}$ is small. The solution of (\ref{mfnonlinearxtg})
can then be obtained in an expansion in powers of $\mu$. We write here the two lowest orders:
\bea
&& \tilde u_{xt} = \tilde u_t^0 + \tilde u_{xt}^1 + \tilde u_{xt}^2 + O(\mu^3)  \label{219}\\
&& (\eta \partial_t - m^2 ) \tilde u_{t}^0 + \sigma( \tilde u_{t}^0)^2 = - \lambda_t \label{220a}\\
&& (\eta \partial_t + \nabla_x^2 - m^2 + 2 \sigma \tilde u_t^0 ) \tilde u_{xt}^1
= - \mu_{xt} \label{220}\\
&& (\eta \partial_t + \nabla_x^2 - m^2 + 2 \sigma \tilde u_t^0 ) \tilde u_{xt}^2 + \sigma (\tilde u_{xt}^1)^2 = 0 
\ .~
\label{221}~~~~~~~~~~~\eea 
The solutions of Eq.~(\ref{220a}) have been discussed in section \ref{s:jpdfcomv}. Since no general solution for all $\lambda_t$ exists, let us proceed with a solution of Eqs.\ (\ref{220}) and (\ref{221}), supposing we know \(\tilde u^0_t\):
\bea
 \tilde u_{xt}^1 &=&\int_{x',t'}\mu_{x't'}\, \mathbb{R}_{x't',xt}  \ , \label{222}\\
 \tilde u_{xt}^2 &=&\sigma \int_{x',t'}(\tilde u_{x't'}^1)^2 \mathbb{R}_{x't',xt} \,
\ .
\label{223}\eea 
We have introduced the dressed response kernel \(\mathbb{R}_{x't',xt} \), which will be
a fundamental object in the remainder of this article. It is solution of the equation\begin{equation}\label{3.21}
\left[- \eta \partial_t - \nabla_x^2 + m^2 - 2 \sigma \tilde u_{t}^0 \right]
\mathbb{R}_{x't',xt} = \delta^{d} (x-x') \delta (t-t')
\ .
\end{equation}
Note that since the instanton equation has the time-derivative reversed, 
we have reversed the order of the  arguments in $\mathbb{R}$, so that,
as defined, it has the usual causal structure of a response function.
Thus as noted in Eqs.\ (\ref{222}) and (\ref{223}) it ``acts from the right'', in contrast to the usual convention.

It is easy to express \(\mathbb{R} \) in Fourier space, i.e.\ $\mathbb{R}_{x't',xt}  = \int_k \mathbb{R}_{k,t',t} e^{i k(x'-x)}$ 
with
\be 
 \mathbb{R}_{k,t_2,t_1} = \frac{1}{\eta} e^{- \frac{1}{\eta} (k^2+m^2) (t_2-t_1) + 2 \frac{\sigma}{\eta} \int_{t_1}^{t_2} \rmd s \tilde u_s^0 } \,\theta(t_2-t_1) \ .
 \label{defmathR}
\ee
First, this allows us to obtain the avalanche statistics in the small-velocity stationary state, working,
as in Section (\ref{s:jpdfcomv}), to first order in $v$. Integrating Eq.~(\ref{219}) over space and time, we find 
\bea
  Z[\lambda_t+\mu_{xt}]  &=& {m^2} L^{-d} \int_{xt} \tilde u_{xt} =Z^0 + Z^1 + Z^2 ~~~~~~~~~~~\\
Z^0&=&m^2   \int_t  \tilde u_t^0\\
 Z^1&=& m^2  L^{-d} \int_{t,t'} \mathbb{R}_{q=0,t',t}  \int_{x'} \mu_{x't'} \\
 Z^2 &=& m^2 \sigma   L^{-d} \int_{t,t'} \mathbb{R}_{q=0,t',t}  \int_{x'} (\tilde u_{x't'}^1)^2
\ .
\eea 
To this order in $\mu$ we  thus obtain averages of the velocity field containing two
space-dependent velocities. Indeed, to 
first order in $v$, in the small-velocity limit, one can write\footnote{ Here and below we drop
the factor $v$ in the term $v + \dot u_t$ in the exponential since it is subdominant at small $v$.}
\bea
&& \overline{ \dot u_{x_1 t_1} \dot u_{x_2 t_2} \rme^{L^d \int \rmd t\, \lambda_t \dot u_t} } =
v \frac{ \delta^2  Z[\lambda_t+\mu_{xt}]}{\delta \mu_{x_1 t_1} \delta \mu_{x_2 t_2} }  L^{d} \nn\\
&& \qquad=  2 v m^2 \sigma \int_{x' t t'} \mathbb{R}_{x_1 t_1,x't'}  \mathbb{R}_{x_2 t_2,x't'} \mathbb{R}_{q=0,t',t}
\ .~~~~~~~~~~~~~
\eea
(The factor of 2 comes from the fact that $ \tilde u_{xt}^2= O(\mu ^2)$.) This is easier\footnote{Note that the first-order derivative does not yield any new information: It confirms $\overline{ \dot u_{x_1 t_1} + v}=v$
for $\lambda=0$.} expressed in Fourier space\footnote{\label{Fourier} We use, for an arbitrary function $A$  the
short-hand notation $\overline{u_{qt_1} u_{q't_2} A} = (2 \pi)^d \delta^{d}(q+q')
\overline{u_{qt_1} u_{-q t_2} A}$ for
translationally invariant correlations. Hence $\overline{\dot u_{x_1 t_1}
\dot u_{x_2 t_2} A} =
\int_{q} e^{i q(x_1-x_2)} \overline{u_{q t_1} u_{-q t_2} A}$ and, integrating
over $x_1,x_2$, one obtains the center of mass
$\overline{\dot u_{t_1} \dot u_{t_2} A} = L^{-d} \overline{u_{q t_1} u_{-q
t_2} A}|_{q=0}$, hence recovering Eq.~(\ref{second}).
Everywhere $\int_q=\int \frac{\rmd^dq}{(2\pi)^d}$.}, 
\bea \label{RRfourier}
&&\overline{ \dot u_{q, t_1} \dot u_{-q, t_2} e^{ L^d \int \rmd t \lambda_t \dot u_t} } \nn\\ 
&&\qquad = 2 v m^2 \sigma \int_{t t'} \mathbb{R}_{q,t_1,t'}  \mathbb{R}_{q,t_2,t'} \mathbb{R}_{q=0,t',t} 
\nn\\
&&\qquad=2 v m^2 \sigma \diagram{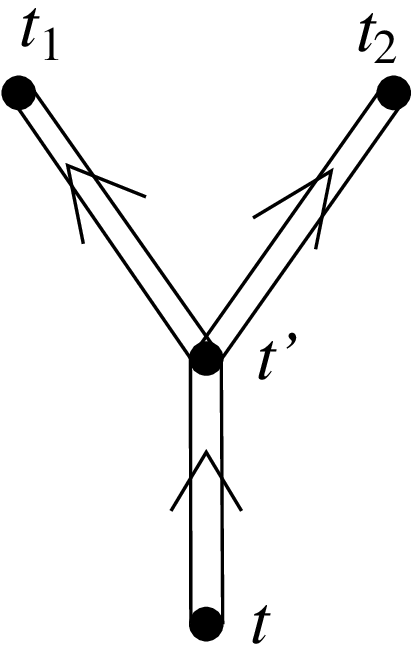}
\ .\eea
Note that we have  introduced a graphical notation that will be useful later, when calculating loop corrections. 
\\ A nice feature is that the source, which couples to the center of mass, is still quite general.  For $\lambda=0$, $\mathbb{R}$ reduces to the usual response function and we recover after integration over \(t\) and \(t'\)
\be 
 \overline{ \dot u_{q, t_1} \dot u_{-q, t_2} } = v \sigma \frac{1}{q^2 + m^2} \frac{1}{\eta} e^{-  \frac{1}{\eta} (m^2+q^2) |t_1-t_2|}
\ .\ee 
This is a finite-momentum generalization of Eq.\ (\ref{second}); a factor of \(2m^2\) has canceled.

Next, using the results of Section \ref{sec:interpret} we also obtain information about avalanches following
a small {\it local} step in the applied force at time $t_0$, i.e.\ $\delta f_{xt} = \delta f_x \theta(t-t_0)$,
\be  \label{super}
\overline{e^{\int_{xt} (\lambda_t + \mu_{xt}) \dot u_{xt}} } -1 = \int_{x} \tilde u^{\lambda+\mu}_{xt_0} \delta f_{x}  + O(\delta f^2)\ ,
\ee  
where 
$\delta f_{x}=\int_{x'}g^{-1}_{xx'} \delta w_{x'}$, where  we also work for $v\to 0^+$, but  due to the step the leading 
result is  non-vanishing, though of order zero in $v$. 

\subsubsection{Dressed response function, and space-dependent shape following a local force step} 
\begin{figure}[t]
\Fig{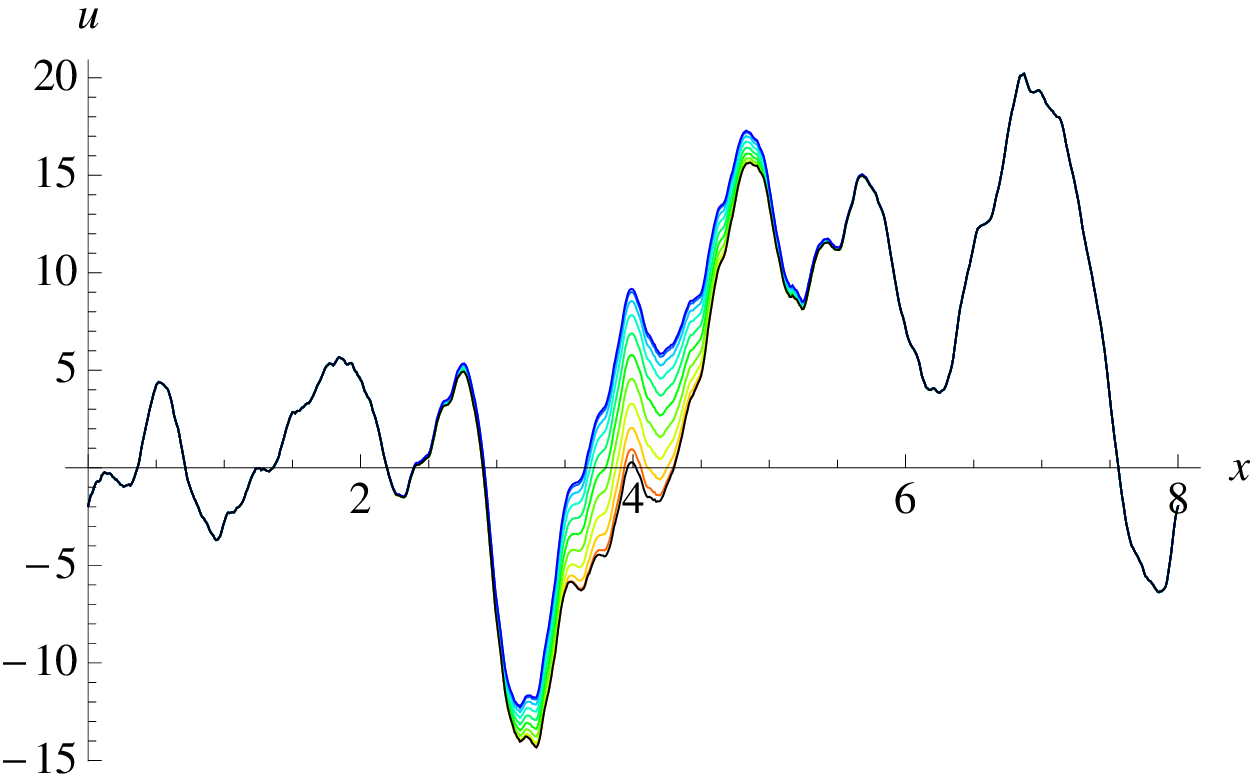}
\caption{Disorer-averaged unfolding of an avalanche following a local step in the force at $x_0=4$ and of duration $T=1$, 
according to formula (\ref{su1x}). Mean intermediate positions $u(x,t=0) + \int_{0}^{t} \langle \dot u_{xt} \rangle_{0T} $ are shown at $t$ multiples of  $T/10$.
The velocity $\dot u_{xt} \ge 0$, so the motion is towards the top of the plot. For the sake of illustration
we chose a random but fixed initial condition $u(x,t=0).$}\label{f:artist-1}
\end{figure}
Let us now apply our formulae to the case $\lambda_t = \lambda \delta(t-t_3),$ and pursue in dimensionless units.
We recall the instanton solution
\be 
\tilde u_t = \frac{\lambda}{\lambda + (1- \lambda) e^{t_3-t}} \theta(t_3-t)
\ .
\ee  
It leads to the dressed response function
for a single-time $\delta$-function source\bea\label{235}
 \mathbb{R}^{t_3}_{k,t_b,t_a} &=& e^{-  (k^2+1) (t_b-t_a) }  \theta(t_b-t_a) \nn \\
&& \times\left[ \frac{1 + \lambda (e^{t_b-t_3}-1) \theta(t_3-t_b)}{1 + \lambda (e^{t_a-t_3}-1)\theta(t_3-t_a)}\right]^2 
\ .~~~~~~~\eea 
Using Eq.~(\ref{super}) to first order in $\mu,$ and Eq.~(\ref{222}), we obtain
the (linear) response to a local step of the driving force at time $t_0$ \be 
\overline{\dot u_{x_1 t_1} e^{\lambda L^d} \dot u_{t_2}}  = \int_{x_0} \mathbb{R}^{t_2}_{x_1,t_1,x_0,t_0}  \delta f_{x_0} 
\ .\ee  
Taking $\lambda \to - \infty$ we obtain the average {\it local} avalanche shape
(i.e.\ the average velocity conditioned s.t.\ the avalanche starts at $t_0$ and ends at $t_2$) as
\bea \label{su1x}
 \left< \dot u_{x_1 t_1} \right>_{02} &=&  \frac{ \partial_{t_2} \lim_{\lambda \to - \infty} \int_{x_0} \mathbb{R}^{t_2}_{x_1t_1,x_0t_0}  \delta f_{x_0} }
{P _{\rm duration}(t_2-t_0) \int_x \delta f_x} \nn \\  
& =& L^d \frac{\int_{x_0}  e^{- \frac{(x_0-x_1)^2}{4 (t_1-t_0)}} \delta f_{x_0}}{[4 \pi (t_1-t_0)]^{d/2}\int_x \delta f_x} \left< \dot u_1 \right>_{02}\ , ~~~~~~~~
\\ 
  \label{su10} \left< \dot u_1 \right>_{02} &=& \frac{4 \sinh(\frac{t_2-t_1}{2}) \sinh(\frac{t_1-t_0}{2})}{\sinh(\frac{t_2-t_0}{2})}
\ .\eea  
\(\left< \dot u_1 \right>_{02}\) is the center-of-mass shape given in Eq.~(\ref{shapeshape}). Thus the avalanche
velocity spreads on average diffusively (for $d=d_{\rm uc}$) from the seed, i.e.\ the point where  the kick was applied. It looks even simpler
in Fourier space\footnote{The center-of-mass velocity $\du_t$ and the velocity of the zero mode (\(q=0)\) $\du_{0t}$ are in our conventions related via $L^d\du_t =\du_{0t} $.} 
\bea
 L^{-d} \left< \dot u_{q t_1} \right>_{02} &=& \frac{R_{q,t_1-t_0} \delta f_k}{R_{q=0,t_1-t_0} \delta f_{q=0}}  \left< \dot u_1 \right>_{02}\nn \\
& =& e^{-  q^2 (t_1-t_0) } \frac{\delta f_q}{\delta f_{q=0}}   \left< \dot u_1 \right>_{02} 
\eea
On figure \ref{f:artist-1} we have drawn the {\em mean } advance of an avalanche 
following a local step in the force. 

\subsubsection{3-time, 2-space point correlation}

Let us now compute the 3-time correlation,  in the steady state to lowest order in $v$, using the single-time source at $t_3$ and Eq.~(\ref{235}).
The $t$-integral in Eq.~(\ref{RRfourier}) is easily performed,
assuming that $t_3>t'$, \be 
\int_{t<t'} \mathbb{R}^{t_3}_{q=0,t',t} = \frac{1+\lambda (e^{-|t'-t_3|}-1)}{1-\lambda} 
\ .
\ee 
For $t_1<t_2<t_3$, the second integral over $t'$ leads to
\bea \label{resnew}
&& \overline{ \dot u_{q t_1} \dot u_{-q t_2} e^{\lambda  L^d \dot u_{t_3}} } \\
\nn&& =  v \frac{ \left[\lambda 
   (e^{t_1-t_3}-1)+1\right]^2 \left[\lambda 
   (e^{t_2-t_3}-1)+1\right]^2}{(\lambda -1)^4
   \left(q^2+1\right)} \nn \\
   && \times \, _2F_1\!\left(3,2 \left(q^2+1\right);2 q^2+3;\frac{\lambda 
   e^{t_1-t_3}}{\lambda -1}\right)e^{- \left(q^2+1\right) \left(t_2-t_1\right)}
\nn 
\eea
By analogy with the procedure used in Section \ref{s:lambda-to-inf}, page \pageref{s:lambda-to-inf}, taking now the limit $\lambda \to - \infty$ 
allows to select the contribution $v q'_{3,12} \delta(\dot u_3) {\calP}_{12,3}(\dot u_{1,q},\dot u_{2,-q})$ in the
3-times joint distribution $P(\dot u_{1,q},\dot u_{2,-q},\dot u_3)$. The normalization $v q'_{3,12}$ should be
the same as for zero momentum $q=0$, since if a piece of the manifold is moving, the center of mass is also moving.
It is equal to the
probability that the avalanche starts before $t_1$ and ends at $t_3$. As in Section  \ref{s:lambda-to-inf}, we
determine it as $q'_{3,12}=\int_{0}^\infty \rmd s_1 \int_{0}^\infty \rmd s_2 \lim_{\lambda_3 \to - \infty} \partial_{\lambda_1}\partial_{\lambda_2} \tilde Z_3|_{\lambda_1=-s_1,\lambda_2=-s_2}
= \ln(z_{31}/z_{21}),$ and we check that $\partial_{t_3} q'_{3,12} =  1/(e^{t_3-t_1}-1)= \int_{t_3-t_1}^\infty \rmd \tau P_{{\rm duration}}(\tau)
$ can be obtained also from the duration distribution. We can thus  take $\partial_{t_3} \lim_{\lambda \to - \infty}$
of Eq.\ (\ref{resnew}) to obtain the
conditional average
\bea
\lefteqn{ \langle \dot u_{q t_1} \dot u_{-q t_2} \rangle_{3}=\left(\partial_{t_3} q'_{3,12}\;\right)^{-1}\partial_{t_3} \lim_{\lambda \to - \infty}  \overline{ \dot u_{q t_1} \dot u_{-q t_2} e^{\lambda \dot u_{t_3}} }}\nn\\
&&=\frac{2 \left(e^{t_2}-e^{t_3}\right) e^{q^2 t_1-\left(q^2+1\right) t_2-4
   t_3}}{q^2+1} \nn\\&& \times \bigg\{ (q^2+1) e^{3 t_3} (e^{t_3}-e^{t_2})+ (e^{t_1}-e^{t_3})^2\nn \\
   && \times \Big[q^2 e^{t_1+t_3}+q^2
   e^{t_2+t_3}-(q^2-1) e^{t_1+t_2}-(q^2+1) e^{2
   t_3}\Big] \nn\\
&&\times\; _2F_1\left(3,2 (q^2+1);2 q^2+3;e^{t_1-t_3}\right) \bigg\}\ .
\label{256}
\eea
It is conditioned, s.t.\ the avalanche started before $t_1$ and ended at $t_3$. For $q=0$ it
reduces to $\langle \dot u_{0 t_1} \dot u_{0 t_2} \rangle_{3}=2 (1-e^{t_2-t_3})(1-e^{t_1-t_3})$. We can obtain the large-$q$ asymptotics using
the formula
\bea
&& {}_2F_1(a,b+x,c+x,z) \nn\\
&&~~~~= (1-z)^{-a} \left[1 + \frac{a(c-b)}{x} \frac{z}{z-1}  + O\left(x^{-2}\right)\right]
\ .~~~~~~~~~~~~
\eea
This yields 
\bea
 \langle \dot u_{q t_1} \dot u_{-q t_2} \rangle_{3} &\simeq_{q \to \infty}&
 \frac{\left(e^{t_3}-e^{t_2}\right) e^{q^2
   \left(t_1-t_2\right)-t_2-t_3}}{q^2 \left(e^{t_3}-e^{t_1}\right)}\nn\\
&& \times\left(2 e^{t_2+t_3}-e^{t_1+t_2}-e^{t_1+t_3}\right)~~~~~~~~~
\eea
Fixing $t_1$ and $t_3$, the function (\ref{256})  decays monotonically to zero for $t_2 \to t_3$. 
Depending on the value of $q$, it is either concave (small $q$) or convex (large $q$). 
   
\subsubsection{4-time,2-space point velocity correlations and asymmetry ratio}

To compute the average at a given $q$, conditioned to both a starting time \(t_0\) and a final time \(t_3\) for
the avalanche, we need the more general observable, for $t_0 <t_1 < t_2 < t_3$,
\be  \label{obs1}
 \overline{ e^{\lambda_0  L^d \dot u_0} \dot u_{q t_1} \dot u_{-q t_2} e^{\lambda_3  L^d \dot u_3} } 
\ .
\ee 
The calculation is more complicated and done in  appendix \ref{a:spatial-correlations} by considering
a source $\lambda_t = \lambda_0 \delta(t-t_0)+\lambda_3 \delta(t-t_3)$
and its associated dressed response function. The full result for (\ref{obs1})
is displayed in Eq.\ (\ref{e14}). An interesting observation is that at $q \neq 0$ {\it it is not invariant under
time reversal}, i.e.\ the simultaneous changes $t_0 \to - t_3$, $t_1\to - t_2$, $t_2 \to - t_1$, $t_3 \to - t_0$, and 
$\lambda_0 \leftrightarrow \lambda_3$. This invariance is recovered only at $q=0$. Hence at the
level of the tree theory there is no way to tell the arrow of time by watching the center-of-mass
velocity, but there is an arrow of time  for modes with non-zero $q$. This can already be seen on the 4-time
velocity correlation function  obtained from the expression (\ref{obs1}) by applying $L^{-2 d} \partial_{\lambda_0} \partial_{\lambda_3}|_{\lambda_0=\lambda_3=0}$. The general result (\ref{e18}) is bulky, so let us display it here  for $t_1=t_2$:
\bea 
\lefteqn{L^{2d}\overline{ \dot u_{-T/2} \dot u_{q, t_1} \dot u_{-q, t_1} \dot u_{T/2} } 
=  v  \frac{2
   (2 q^2+3) e^{-T} }{ (1 + q^2) (1 + 2 q^2)}} \nn\\&&+v
 \frac{2 q^2 e^{-2 (q^2+1) t_1-\left(q^2+2\right) T} \left[8
   (q^2+2) e^{t_1+\frac{T}{2}}-6 q^2-3\right] }{ (1 + q^2) (2 + q^2) (1 + 2 q^2) (3 + 2 q^2)}\nn\\
&&=   v  \Big[ 6 e^{-T}+q^2 \left(\frac{16}{3} e^{-t_1-\frac{3 T}{2}} -e^{-2 t_1-2 T}-14 
   e^{-T}\right)  \nn\\ &&~~~~~~~+O(q^4)\Big ] 
\ ,
\eea 
which is clearly not symmetric under $t_1 \to - t_1$, although it is for $q=0$.   Note that here we do not
know when the avalanche starts and ends, we only know that the duration is larger than $T$. 
We define the asymmetry ratio of the 4-time velocity correlation as
\begin{equation}
A(t_1):=\frac{\overline{ \dot u_{-T/2} \dot u_{q, t_1} \dot u_{-q, t_1} \dot u_{T/2}
}} {\overline{ \dot u_{-T/2} \dot u_{q, 0} \dot u_{-q, 0} \dot u_{T/2}
} }
\ .\label{asymmetry}\end{equation}
It is plotted on figure \ref{f:asymmetry}.\begin{figure}[t]
\Fig{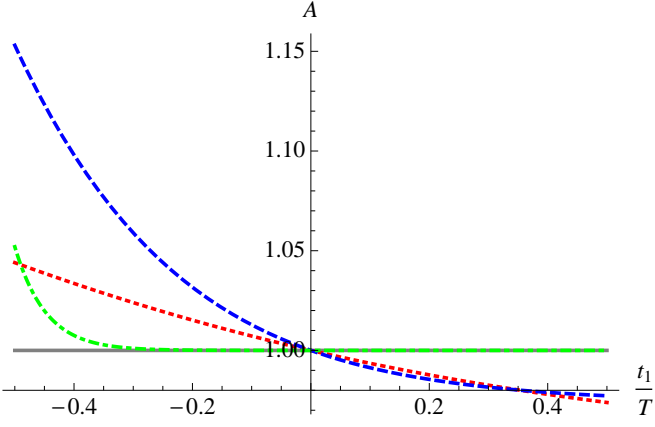}
\caption{Plot of the asymmetry ratio \(A\)  defined in equation
(\ref{asymmetry}). The different curves are for \(q^2=0\) (solid gray), \(q^2=0.2\)  (dotted red), \(q^2=1.703\) (dashed blue), and \(q^2=10\) (dot-dahed, green). The maximum of \(A\)  at $t_1=-T/2$ is attained for \(q^2=1.703\) (dashed blue) The plot is for \(T=1\).  }
\label{f:asymmetry}
\end{figure}\begin{figure}[b]
\Fig{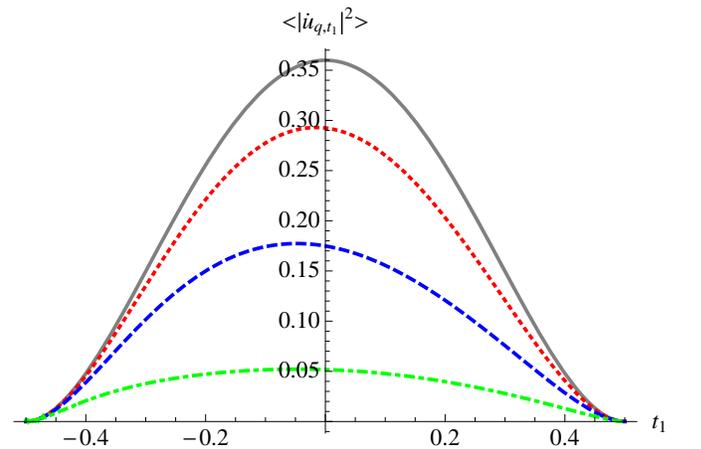}
\caption{Plot of  the conditional average \(\left< \dot u_{qt_1}\dot u_{-qt_1} \right>\)   given in Eq.~(\ref{Kayuu})  for an avalanche starting at time $-T/2$,
and ending at time $T/2$, in our dimensionless units. The different curves are for \(q^2=0\) (solid gray),
\(q^2=0.5\)
 (dotted red), \(q^2=2\) (dashed blue), and \(q^2=9\) (dot-dashed, green).
 The plot is for \(T=1\).}
\label{f:asymmetryvv}
\end{figure}\begin{figure}[t]
\Fig{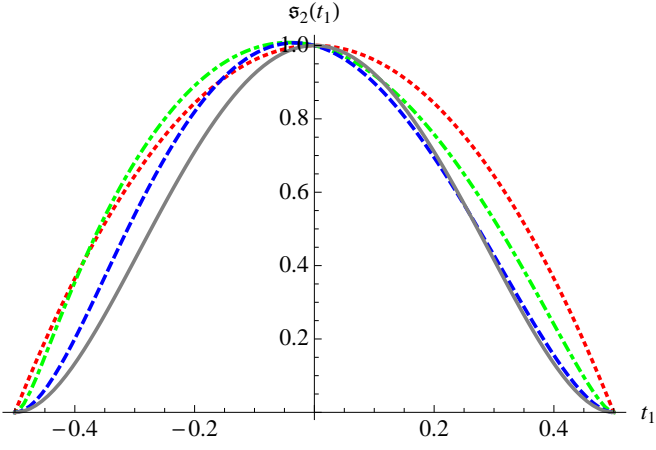}
\caption{Plot of the (normalized) second shape $\mathfrak{s}_{2}(t_1)$, i.e.\ the ratio of conditional second moments of the local velocity $\mathfrak{s}_{2}(t_1):= \left< \int_{q}\rme^{-q^{2} a^{2}} \dot u_{q,t_{1}} u_{-q,t_{1}} \right>\Big/\left< \int_{q}\rme^{-q^{2} a^{2}} \dot u_{q,0} u_{-q,0} \right> \)  for an avalanche starting at time $-T/2$,
and ending at $T/2$, normalized s.t.\ $\mathfrak{s}_{2}(0)=1$. The different curves are for \(a\to \infty \) (solid gray, equivalent to the same-colored curve on figure \ref{f:asymmetry}),
\(a=1 \)
 (blue dashed), \(a=0.1 \)
 (green dot-dashed), and \(a\to 0\) (dotted, red), which approaches a parabola. Both limiting curves for $a\to 0$ and $a\to \infty$ are  symmetric, while for generic values of $a$ they are not. The reason why for $a\to0$ the curve becomes symmetric is due to a diverging symmetric contribution to \(\left< \int_{q}\rme^{-q^{2} a^{2}} \dot u_{q,t_{1}} u_{-q,t_{1}} \right>\), {\em not} due to a vanishing of the asymmetric part.      }
\label{f:asymmetryvvx}
\end{figure}
Since the asymmetry ratio is larger at large \(q\) at the beginning of the avalanche, it implies that the local velocities in an avalanche are  higher in the beginning of an avalanche than at the end. Stated differently, the avalanches are more compact at the beginning and the parts which move move more quickly. This is consistent with our physical intuition that an avalanche starts at some seed $x_{\rm seed}$, grows quickly around that point, while at the end it is spatially extended, but stops more uniformly.

\subsubsection{ ``Second" shape of an avalanche at non-zero momentum}
We now obtain $\left< \dot u_{qt_1}  \dot u_{-qt_2} \right>_{0,3}$ i.e.\ the
shape fluctuation, or second shape, at non-zero wave vector for an avalanche which started at 
$t_0$ and ended at $t_3$. The times are chosen ordered as 
 \(t_0<t_1<t_2<t_3\). We compute it both for an avalanche (\textit{i})~generated
 by a uniform small force step at time $t=t_0$, i.e.\ $\delta f_{xt} = \delta f \theta(t-t_0)$ with $\delta f =m^2 \delta w$;  (\textit{ii})
 for an avalanche  in the stationary state to first order in $v$. The two protocols give the same
 result, as was  explained in Section \ref{sec:interpret}; it is based on the identity (\ref{iden}).
 We present the calculation of ({\it i}) in the main text;  (\textit{ii}) is more involved and is
 presented in  Appendix \ref{a:spatial-correlations}.

Let us consider the following velocity average following a uniform force step at time $t_0$,
\bea
&& \overline{ \dot u_{x_1 t_1} \dot u_{x_2 t_2} \rme^{L^d \lambda \dot u_{t_3}} } =
\int_{x_0}  \frac{ \delta^2 \tilde u_{x_0 t_0}}{\delta \mu_{x_1 t_1} \delta \mu_{x_2 t_2} } \delta f \nn\\ 
&& \qquad=  2 \delta w m^2 \sigma \int_{x' t t'} \mathbb{R}^{t_3}_{x_1 t_1,x't'}  \mathbb{R}^{t_3}_{x_2 t_2,x't'} \mathbb{R}^{t_3}_{q=0,t',t_0}\ . \nn
\eea
We have worked to linear order, i.e.\ up to terms or order $O(\delta f^2,v)$. In Fourier space, and dimensionless units, the latter reads
\begin{equation}\label{zz1}
\overline{ \dot u_{qt_1}  \dot u_{-qt_2} 
\rme^{L^d \lambda \dot{u}_{t_3} } } = 2  \delta w \int_{t_{0}}^{t_{1}}\rmd t'\,  \mathbb{R}^{t_3}_{q,t_1,t'}  \mathbb{R}^{t_3}_{q,t_2,t'} \mathbb{R}^{t_3}_{q=0,t',t_{0}} \ .
\end{equation}
The function $\mathbb{R}$ is given in Eq.~(\ref{235}), and as written,
we can drop the $\theta$-functions. Taking the limit $\lambda \to -\infty$ enforces the center-of-mass
velocity at the final time to be  $\dot{u}_{t_3}=0$, leading to \footnote{ Here we use the Kronecker symbol 
$\delta_{\dot u}=1$ or $0$ according
to whether $\dot u=0$ or not, i.e.\ the characteristic function for the event $\dot u=0$, which is dimensionless.} 
\begin{eqnarray}\label{zz2}
&& \overline{ \dot u_{qt_1}  \dot u_{-qt_2} \delta_{\dot u_3} } \label{246} \nn\\
&=&  \delta w \frac{2 (e^{t_1-t_3}-1)^2 (e^{t_2-t_3}-1)^2
   e^{(q^2+1) (t_0-t_1-t_2+t_3)}}{e^{q^2(t_0-t_3)}(e^{t_0-t_3}-1){}^2} \nn \\
&&\times \Bigg [\frac{e^{( t_0-t_3) ( 2q^2+1)} \,
   _2F_1\!\left(1,2;2-2q^2;\frac{1}{1-e^{t_0-t_3}}\right)}{
   (e^{t_0-t_3}-1)^2(1-2q^2)} \nn\\
&&~~~~~~- \frac{e^{(t_1-t_3) (2 q^2+1)} \,
   _2F_1\!\left(1,2;2-2q^2;\frac{1}{1-e^{t_1-t_3}}\right)}{
   (e^{ t_1-t_3}-1)^2(1-2q^2)}  \Bigg ]\ . \nn\\ 
\end{eqnarray}
This is a joint expectation value conditioned to the event that the avalanche ends before $t_3$.

As in Section \ref{s:jpdfcomv} we obtain the conditional average s.t. the avalanche ends exactly at $t_3$ by
taking a derivative $\partial_{t_3}$ of the above average (\ref{zz1}), and dividing by the
total probability $P_{\rm duration}(t_3-t_0) \delta f$ for the avalanche starting at $t_0$ and to end at
$t_3$, leading to \bea \label{2shapeq}
\lefteqn{\!\!\!\left< \dot u_{qt_1}  \dot u_{-qt_2} \right>_{0,3}=} \nn\\
&& 2\delta w (e^{t_1-t_3}-1)^2 (e^{t_2-t_3}-1)^2 e^{(q^2+1)
   (2t_3 -t_1-t_2 )} \nn\\
&&\Bigg\{ \Bigg[\frac{e^{(t_0-t_3)(2q^2+1)} \,
   _2F_1\left(1,2;2-2q^2;\frac{1}{1-e^{t_0-t_3}}\right)}{(2q^2-1)
   (e^{t_0-t_3}-1)^2}\nn\\
&&~~~~~-\frac{e^{(t_1-t_3)(2q^2+1)} \,
   _2F_1\left(1,2;2-2q^2;\frac{1}{1-e^{t_1-t_3}}\right)}{(2q^2-1)
   (e^{t_1-t_3}-1)^2}\Bigg]\times\nn\\
&& ~~\times\left[2q^2+
   \frac{2}{1-e^{t_2-t_3}}+\frac{2}{e^{t_0-t_3}-1}-\coth\! \left(\frac{t_1-t_3}{2}
   \right)\right]\nn\\ &&~~+\frac{e^{2q^2
   \left(t_0-t_3\right)+t_0+t_3}}{\left(e^{t_0}-e^{t_3}\right){}^2}-\frac{e^{2q^2
   \left(t_1-t_3\right)+t_1+t_3}}{\left(e^{t_1}-e^{t_3}\right){}^2}\Bigg\}
\label{Kayuu}\ .
\eea
The resulting function, for \(t_2=t_1\) and $T=1$ is plotted on figure \ref{f:asymmetryvv}. One sees again that higher wave-vectors $q$ are (slightly) skewed towards earlier times.

It is interesting to perform the same calculation in real space. One can either Fourier transform the above result (which is not
easy) or go back  to Eq.~(\ref{zz1}) and directly work in real space. Because of  divergences indicated below, 
we need to compute the more general observable, smoothed on a small region of space (i.e.\ for close-by 
points $x_1$, $x_2$): 
\be 
\int_{x_2}\left<\dot u_{x_1t_1}  \dot u_{x_2t_2} \right> \frac{\rme^{-(x_1-x_2)^2/(4a^2)}}{2a\sqrt{\pi}} = \int_q \left<\dot u_{q,t_1} \dot u_{-q,t_2}\right>e^{-(aq)^2 }
\ee Integrating over momentum directly in $d=4$ we obtain  \bea
&&  \int_q \left<\dot u_{q,t_1} \dot u_{-q,t_2}\right>e^{-(aq)^2 }=\frac{\delta w}{8\pi^2}\sinh^2\!\Big(\frac{t_0-t_3}{2}\Big) \nn\\
&&\times  \partial_{t_3} \bigg[\frac{\sinh^2(\frac{t_1-t_3}2)\sinh^2(\frac{t_2-t_3}2)}{\sinh^2(\frac{t_3-t_0}2)}\nn\\
&& \qquad ~\times   \int_{t_0}^{t_1} \rmd t' \frac{1
   }{
   \sinh^2(\frac{t_3-t'}2) (a^2-2 t'+t_1+t_2)^2} \bigg]~~~~~~~~~~~
   \label{265}
\eea 
For $t_1 < t_2$ we can set $a=0$ and obtain a finite result. However, for equal times $t_1=t_2$, there is
an ultraviolet divergence and the integral diverges like $1/a$ as $a\to 0$, hence we must keep $a>0$. This allows to define a (normalized) second
shape at time $t_1$ as the ratio \be 
\mathfrak{s}_2(t_1):=\frac{\int_q\left<\dot u_{q,t_1} \dot u_{-q,t_1}\right>e^{-(aq)^2
}} {\int_q\left<\dot u_{q,t_{\rm m}} \dot u_{-q,t_{\rm m}}\right>e^{-(aq)^2
}}\ , \quad t_{\rm m}:=\frac{t_0+t_3}2\ .
\ee 
This is  the second shape normalized to unity for $t_1=t_{{\rm m}}$ the mid-time of the avalanche. The result is
plotted on figure \ref{f:asymmetryvvx} where the integral over $t'$ in Eq.~(\ref{265}) was performed numerically. 
Note that upon normalization the limit $a \to 0$ exists (even if
both numerator and denominator diverge) and is a parabola. 
Another possibility to regularize the function is to chose $t_1<t_2$; the role of the parameter \(a^2\) is then replaced by 
the difference $t_2-t_1$.

 For the Brownian force model, 
the tree theory remains exact below $d_{\rm uc}=4,$ hence we can use the formula
in any $d$. Upon integration over momentum in $d<4,$ 
 the factor \( (a^2-2 t'+t_1+t_2)^{-2}\) is replaced by \( (a^2-2 t'+t_1+t_2)^{-d/2}\). 
  In dimensions $d<2$, the limit $a\to 0$ can be taken. 
  In smaller dimensions, the asymmetry is less pronounced. 
  This is expected, since for $d\to0$ we must recover the result for the particle, 
  where $\dot u^2\equiv \dot u_x^2$.  The same holds true for LR elasticity.   

\subsubsection{Arbitrary elastic kernel and non-local elasticity}

Finally we can now give the result for an {\it arbitrary} elastic kernel, $g_q^{-1}$. Since we often use
dimensionless units, we must first define $\tilde g_q^{-1} := g_{q=0} g_q^{-1} = g_q/m^2$. Thus one  has to substitute $q^2 \to \tilde g_q^{-1} -1 $ in all above equations 
containing $q$ explicitly, e.g. Eqs (\ref{resnew}), (\ref{256}), (\ref{246}), and (\ref{Kayuu}). 

The equations where $q$ has been integrated over,   such as (\ref{265}), have to be recalculated. 
There the changes to be made can be condensed to a change of the integration measure over momentum.
For the simplest form of a long-range elastic kernel this is explained in section \ref{s:LR}.

\section{Loop corrections}\label{s:loops}

Until now we found that the mean-field theory involves only the cusp parameter $\sigma = - \Delta'(0^+)$. 
As was the case for  static avalanches \cite{LeDoussalWiese2008c}, the small dimensionless parameter which controls the 
importance of the loop corrections (and thus the deviations from mean field) is
the second derivative of the (renormalized dimensionless) disorder correlator,  i.e.\ 
using the same notations as in \cite{LeDoussalWiese2008c} 
\bea
\label{A}
A&=& -m^{d-4}\Delta''(0^+)\ ,\\
  \label{alpha1}
\alpha &:=& - \epsilon \tilde I_2 m^{-\epsilon} \Delta''(0^+) = - \tilde \Delta''(0^+)\ , \\
\tilde I_2 &:=& \int \frac{\rmd^d q}{(2\pi)^2}\frac1{(1+q^2)^2}\ .
\eea  
The parameter $\alpha$ is of order   $O(\epsilon=d_{\rm uc}-d)$. 
Below we study first the 1-loop corrections using a simplified theory, which retains
only $\sigma$ and $\Delta''(0^+)$. This simplified theory streamlines  the calculations, and allows to derive, at
least heuristically, the result, which we then analyze. Finally we present a detailed
derivation from first principles. Note that the presentation here focusses on 
standard short-range elasticity, i.e.\ $d_{\rm uc}=4$. The generalization to LR elasticity
is straightforward, so we only detail  the main  features in section \ref{s:LR}, and give more explict formulas in appendix \ref{a:LR}.

\subsection{General framework}
\label{s:loops-general}
In order to compute the generating functions (\ref{genfunctionG}) and (\ref{genfunctionZ}) beyond
mean-field, let us start again with the dynamical action (\ref{msr}) of the velocity theory,
which we recall has the form 
$ {\cal S} =  {\cal S}_0 +  {\cal S}_{\mathrm{dis}}$, $ {\cal S}_0$ given in Eq.~(\ref{msr}) and 
\be 
 {\cal S}_{\mathrm{dis}} = - \frac{1}{2} \int_{xtt'} \tilde u_{xt} \tilde u_{xt'}  \partial_t \partial_{t'}  \Delta(v(t-t') + u_{xtt'}) 
\ .
\ee  
We now rewrite this term with no approximations as
\begin{align} \label{exactS} 
&{  {\cal S}_{\mathrm{dis}} = - \sigma \int_{xtt'} \tilde u_{xt} \tilde u_{xt}  (v + \dot u_{xt})} \\
&+ \frac{1}{2} \int_{xtt'} \tilde u_{xt} \tilde u_{xt'}  (v {+} \dot u_{xt}) (v {+} \dot u_{xt'}) \Delta''_{\rm reg}\big(v(t{-}t') {+} u_{xtt'}\big) \nn
\ .
\end{align} 
We have defined
\be 
\Delta(u)= - \sigma |u|  + \Delta_{\rm reg}(u)\ ,
\ee 
such that $\Delta''_{\rm reg}(u)$ is the second derivative of \(\Delta(u)\) without the $\delta$-function part; hence  $\Delta''_{\rm reg}(0)=\Delta''(0^+)$, and
\(\Delta_{\rm reg}(u)\) has a regular Taylor expansion in $|u|$ around zero starting at order $|u|^2$ . Below we
loosely denote $\Delta''(0)\equiv \Delta''(0^+)$ since the right and left second derivatives coincide.

\subsection{Simplified model}\label{sec:simplified}
The decomposition (\ref{exactS}) is exact. Now we make a simplification. We neglect the higher derivatives $\Delta^{(n)}(0^+)$ with $n \geq 3$.  
We will see below that this is sufficient to give the 1-loop result for the generating function {\it almost completely}, up to some
 subtleties that we discuss below. With this assumption, we  have $ {\cal S}_{\mathrm{dis}} =  {\cal S}_{\mathrm{dis}}^{\rm simp} + ...$, with\begin{eqnarray}
  {\cal S}_{\mathrm{dis}}^{\rm simp}  &=& -   \sigma  \int_{xt}  \tilde u_{xt}^2  (v+ \dot u_{xt})\nn  \\
&& +  \frac{1}{2}  \Delta''(0)  \int_{xtt'}  \tilde u_{xt}  \tilde u_{xt'}  (v+ \dot u_{xt})  (v+\dot u_{xt'})\ .~~~~~~~~~~~~~ 
\end{eqnarray}
We now work with this ``simplified" model, and discuss later on the
effects of the neglected terms.

The nice feature of this simplified model is that the new term can be written as an average over a fictitious (centered) 
Gaussian disorder $\eta_x$ with correlations
\be 
\langle \eta_x \eta_{x'} \rangle_\eta = m^{4-d} A \delta^d(x-x') \label{etadis}
\ ,
\ee  
where $A$ is dimensionless, and we will choose later $A=- m^{d-4} \Delta''(0)$. With these definitions one can write
\footnote{Note that the noise $\eta_x$ is unrelated to the friction $\eta$ despite the coincidence in notations.}
\be 
G[\lambda]=  \langle G_\eta[\lambda] \rangle_\eta
\ee  
with
\begin{align}
&\!\!\! G_\eta[\lambda]= \int D[\dot u] D[\tilde u] e^{-  {\cal S}_\eta +  \int_{xt} \lambda_{xt} (v+ \dot u_{xt}) }\\
&\!\!\!  {\cal S}_\eta =  {\cal S}_0 -   \sigma  \int_{xt}  \tilde u_{xt}^2   (v+\dot u_{xt})   - \int_{xt} \eta_x \tilde u_{xt} (v+\dot u_{xt})\ .
\end{align}  
For each realization of $\eta_x$, the theory has the same features as the mean-field theory (\ref{Stree}) of
Section \ref{simplified}. In particular, the total action (including the sources) is linear in the velocity field. Integrating over
the latter, as in Section (\ref{simplified}) one finds
\be  \label{Geta}
   G_\eta[\lambda] = e^{ v \int_{xt} \lambda_{xt} + \sigma (\tilde u^{\lambda\eta}_{xt})^2 + \eta_x \tilde u^{\lambda\eta}_{xt} } 
\ .
\ee 
The quantity $\tilde u^{\lambda\eta}_{xt}$ is now solution of the (modified)
instanton equation
\be 
 (\eta \partial_t + \nabla_x^2 - m^2) \tilde u_{xt} ^{\lambda\eta}+ \sigma (\tilde u^{\lambda\eta}_{xt})^2 = - \lambda_{xt}  - \eta_x \tilde u_{xt}^{\lambda\eta} 
\ ,
\ee  
which  has an additional ``random-mass'' term. Using this equation, Eq.\ (\ref{Geta}) can be written as
\bea
 G_\eta[\lambda] &=& e^{ v L^d Z_\eta[\lambda] }  \\
 Z_\eta[\lambda] &=& - L^{-d} \int_{xt} (\eta \partial_t + \nabla_x^2 - m^2) \tilde u_{xt}^{\lambda\eta} 
 \nn\\ &= &L^{-d} m^2 \int_{xt} \tilde u_{xt}^{\lambda\eta} 
\ .
\eea
To lowest order in $v$ we thus find
\be
\label{Zav} 
Z[\lambda] =  L^{-d} \partial_v \overline{ e^{\int_{xt} \lambda_{xt} \dot u_{xt}}}\Big|_{v=0^+} 
=\frac{ m^2} {L^{d}} \int_{xt} \langle \tilde u^{\lambda\eta}_{xt} \rangle_{\eta} 
\ .
\ee
As we discuss later, we will need to take $A<0$ at the fixed point, hence the sign of the random term (\ref{etadis}) is not
consistent with an additional real disorder. Since all we want to do here is perturbation theory in $\Delta''(0)$,
more precisely in the parameter $\alpha=0(\epsilon)$ defined in Eq.\ (\ref{alpha1}), this is immaterial. It should
be considered as a trick to simplify the perturbative calculations. 

\subsection{Perturbative solution}\label{q1}
\subsubsection{General equations and formal solution for arbitrary $\lambda_{xt}$ }\label{q2}

For simplicity we switch from now on to dimensionless units, which amounts to setting $\eta=m=\sigma=1$. 
We want to solve perturbatively in $\eta_x$ the equation \begin{equation}\label{q3}
\left[\partial_t + \nabla_x^2 - 1 \right] \tilde u^{\lambda\eta}_{xt} = - \lambda_{xt}  -   (\tilde  u_{xt}^{\lambda\eta})^2 - \eta_{x} \tilde u_{xt}^{\lambda\eta}
\ .
\end{equation}
We expand the solution in powers of $\eta_x$, denoting by   $\tilde  u^{n}_{xt}$ the term of order $O(\eta^n)$, 
\begin{equation}\label{q4}
\tilde u_{xt}^{\lambda\eta} = \tilde u_{xt}^0 +  \tilde u^{1}_{xt} + \tilde u^2_{xt} + ... 
\ .
\end{equation}
One must thus solve a hierarchy of equations, \begin{eqnarray}\label{3eq}
&&\!\!\!\! \left[\partial_t+ \nabla_x^2 - 1 \right] \tilde u^0_{xt} = - \lambda_{xt}  -  (\tilde u^0_{xt})^2 \ ,\qquad \\
&&\!\!\!\!\left[ \partial_t + \nabla_x^2 - 1 + 2 \tilde u^0_{xt}  \right] \tilde u^{1}_{xt} = - \eta_{x} \tilde u^{0}_{xt} \ , \\
&&\!\!\!\!\left[ \partial_t + \nabla_x^2 - 1 + 2 \tilde u^0_{xt} \right] \tilde u^2_{xt} = - (\tilde u^{1}_{xt})^2 - \eta_x \tilde u^1_{xt}\ . \qquad
\label{3.20}
\end{eqnarray}
The first line, for  order zero, is the usual (mean-field) instanton equation (\ref{mfnonlinearxt}). 
This perturbation problem is distinct, but similar, to the one studied in Section \ref{sec:spatial}. We introduce again the dressed response kernel (\ref{3.21}), now in dimensionless variables, 
\begin{equation}\label{3.21new}
\left[- \partial_t - \nabla_x^2 + 1 - 2 \tilde u^0_{xt}  \right]
\mathbb{R}_{x't',xt} =  \delta^{d} (x-x') \delta (t-t')
\ .
\end{equation}
It has the usual causal structure of a response function, and
obeys a backward evolution equation. It allows to rewrite the solution of the system of equations (\ref{3eq}) to (\ref{3.20}) as
\begin{eqnarray}\label{q5}
\tilde  u^{1}_{xt} &=& \int_{x'}\int_{t'>t} \eta _{x'}\,\tilde u^{0}_{x't'}  \mathbb{R}_{x't',xt} \label{u1} \ ,\\
\tilde u^{2}_{xt} &=&  \int_{x'}\int_{t'>t} \left[ (\tilde u^{1}_{x't'})^2 + \eta_{x'} \tilde u^1_{x't'} \right] \mathbb{R}_{x't',xt}~.~~~~~~~~
\label{u2}
\end{eqnarray}
Consider now the average (\ref{Zav}) over $\eta_x$ using (\ref{etadis}), i.e.\ in our (dimensionless) units $\langle \eta_{x}\eta_{y} \rangle_\eta = A \delta^{d} (x-y)$.
Since $ \langle\tilde  u^1_{xt} \rangle_\eta = 0$, the lowest-order correction is given by the average of $ \tilde u^{2}_{xt}$,
\be 
Z[\lambda] = Z_{\mathrm{tree}}[\lambda] +  L^{-d} \int_{xt} \langle\tilde  u^2_{xt} \rangle_\eta + O(A^2) 
\ .
\ee  
Inserting Eq.~(\ref{u1}) into Eq.~(\ref{u2}), and performing the average over $\eta$, one  finds
\begin{align}\label{q6}
\langle  \tilde u^{2}_{xt} \rangle_\eta &= A \int_{t<t_{1}<t_{2},t_{3}} \int_{x_{1}x'} \tilde  u^{0}_{x't_{2}} \tilde u^{0}_{x't_{3}} \nonumber \\
& \qquad\qquad \qquad \qquad  \times \mathbb{R}_{x' t_{2},x_{1}t_{1}}
 \mathbb{R}_{x't_{3},x_{1}t_{1}} \mathbb{R}_{x_{1}t_{1}, xt} \nonumber \\
&~~~ +  A \int_{t<t_{1}<t_{2}} \int_{x'}\tilde u^{0}_{x',t_{2}}
\mathbb{ R}_{x't_{2},x' t_{1}} \mathbb{R}_{x' t_{1}, xt} \ .
\end{align}
It admits the following graphical representation
\begin{align}\label{q7}
&\langle \tilde u^{2}_{xt} \rangle_\eta = \diagram{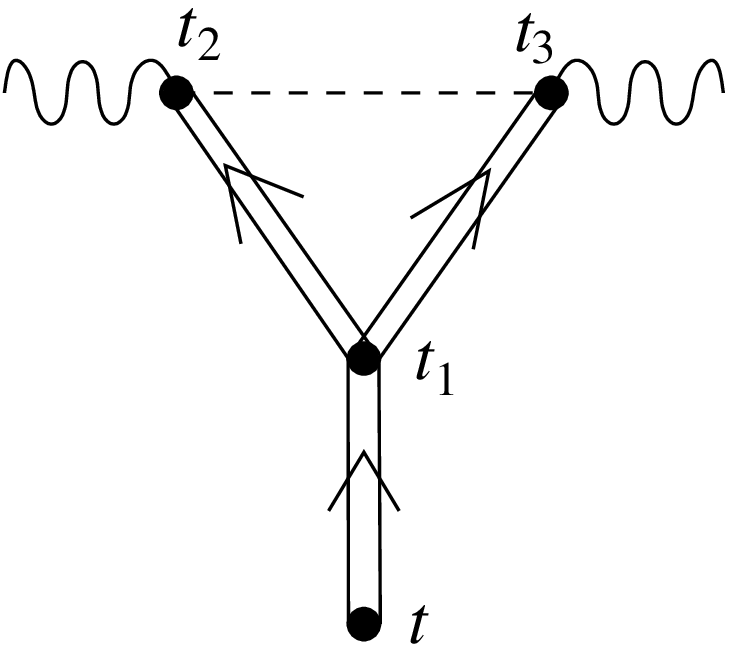}+\diagram{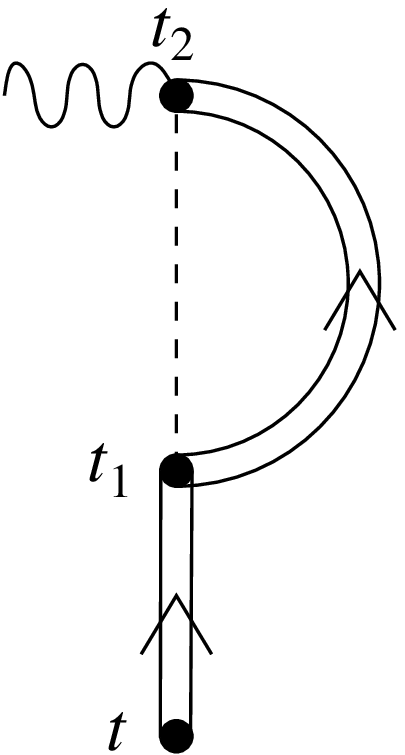}\ .\nonumber \\
\end{align}
The symbols are as follows: (\textit{i}) a wiggly line represents $\tilde u^{0}_{xt}$, the mean field-solution;
(\textit{ii}) a double solid line is a dressed response function $\mathbb{R}$,
advancing in time following the arrow (upwards), thus times are
ordered from bottom to top. Note that for the choice $\lambda_{t} =\lambda \delta(t)$, 
one has $\tilde u^0_{xt}\equiv \tilde u^0_t=0$ for $t>0$, hence the integrals only involve negative times. 

We now define the combination 
\begin{equation}\label{Phi}
\Phi (x',x,t):= \int_{t'>t} \tilde u^{0}_{x't'} \mathbb{R}_{x't',xt}\ ,
\end{equation}
in terms of which one can rewrite
\begin{equation}\label{tilde-u2}
\langle \tilde u^{(2)}_{xt} \rangle_\eta = \int_{t',x'} \left[\int_{y} \Phi
(y,x',t')^{2}+ \Phi (x',x',t') \right] \mathbb{R}_{x't',xt}
\ .
\end{equation}
In section \ref{s:corrections} we shall show that there is an additional term.

\subsubsection{Space-independent source, $\lambda_{xt} =\lambda_t$}\label{sec:spaceindep}

We now pursue the calculation in the case of a spatially uniform source $\lambda_{xt} =\lambda_t$, i.e.\ we study the center-of-mass
velocity.
Since then $\tilde u^0_{xt} = \tilde u^0_t$ is also uniform, we can express the solution of Eq.\ (\ref{3.21new}) -- as in 
Eq.\ (\ref{defmathR}) -- in momentum space\be \label{R-gen}
\mathbb{R}_{k,t_{2},t_{1}} =  e^{ - (k^2+1) (t_2-t_1) + 2 \int_{t_1}^{t_2} \rmd \tau \tilde u^{0}_\tau  } \theta (t_2-t_1)
\ .
\ee
The same is possible for Eqs.\ (\ref{Phi}) and (\ref{tilde-u2}), by defining $\Phi(x',x,t) = \int_k \Phi(k,t) e^{ i k (x'-x)}$ and 
\bea \label{Phik} 
 \Phi(k,t_1) &=& \int_{t_1<t_2}  \tilde u^0_{t_2} \;{\mathbb R}_{k,t_2,t_1}\ ,\\
\langle \tilde u^{2}_{xt} \rangle_\eta &=&  \langle \tilde u^{2}_{t} \rangle_\eta = A \int_k  {\cal J}_t (k) 
\ ,\\ \label{Jt}
 {\cal J}_t (k)   &=&  \int_{t_1>t}  \left[ \Phi(k,t_1)^2 + \Phi(k,t_1) \right]  {\mathbb R}_{k=0,t_{1},t} \ .\qquad\quad 
\eea
From Eq.\ (\ref{Zav}) we find that
$Z[\lambda]$ is then given by
\bea
 Z[\lambda] &=& Z^{\mathrm{tree}}[\lambda] + A \int \frac{\rmd^d k}{(2 \pi)^d} {\cal J} (k) \\
 {\cal J} (k) &=& \int_t  {\cal J}_t (k) \label{J}
\eea 
As discussed below,  some counter-terms are missing, and the correct formula
is obtained by ${\cal J} (k)  \to {\cal J} (k)  +  {\cal J}_{\rm ct} (k)$. 

We now consider the space dimension to be $ d\approx 4$, since we want to perform an $\epsilon=4-d$ expansion. Since \(A\sim \epsilon\), it is sufficient to calculate \(\int_k {\cal J}(k)\) in $d=4$. 
In that case, we note that for any isotropic integral one can write (recalling $A=-m^{d-4}\Delta''(0)$, and Eq.\ (\ref{alpha1}))
\be \label{Atoalpha}
A \int \frac{\rmd^d k}{(2 \pi)^d} = \frac{\alpha}{\epsilon \tilde I_2} S_d \int k^{d-1} \rmd k
= \frac{\alpha}{2} \int k^2 \rmd(k^2) +\dots\ .
\ee
We used that\be
\frac{\tilde I_{2}}{S_d}  = \frac{1}{2} \int_{0}^{\infty}\rmd (k^{2})\,
\frac{(k^{2})^{1-\epsilon /2}}{( k^{2}+1)^{2}} = \frac{1}{\epsilon}+\dots.
\ ,
\ee
where $\dots$ denotes higher-order terms in $\epsilon$ and $S_d$ the unit-sphere area divided by $(2\pi)^d$. 
\smallskip

\subsection{1-point velocity distribution}
\label{s:1-point}
\subsubsection{Generating function $Z(\lambda)$ and moments} 

We now specify to $\lambda (x,t) = \lambda \delta (t)$ to obtain the 1-point velocity distribution.

Let us recall the solution of the instanton equation \be
\tilde u^0_{xt} =\tilde u^0_t = \frac{e^t \kappa }{e^t \kappa -1 } \theta(-t) = 
\frac{1}{1-\kappa^{-1} \rme^{-t}} \theta(-t)\ . 
\ee
We found useful to define
\be
\label{8.195}
\kappa := \frac{-\lambda}{1-\lambda }\ , \qquad (1-\kappa)
(1-\lambda) =1 
\ ,
\ee
which we often use below as it simplifies the calculations. The  relevant interval
$\lambda \in\, ]{-}\infty,1[$ maps onto $\kappa \in\, ]{-}\infty,1[$ (with reversed boundaries).

From the previous section we have
\begin{eqnarray}\label{3.31}
Z (\lambda) &=&  Z_0(\lambda) + \frac{\alpha}2 \delta Z (\lambda ) \\
\delta Z (\lambda) &:=& \int_0^{\infty} k^2 \rmd(k^2)  \left[{\cal J} (k,\kappa)+{\cal J}_{\mathrm{ct}} (k,\kappa ) \right]
\ , \qquad 
\end{eqnarray}
where we denote  $Z^{\mathrm{tree}}=:Z_0$ and ${\cal J} (k)$ in (\ref{J}), (\ref{Jt}) by ${\cal J} (k,\kappa)$ to make the $\kappa$ dependence explicit. 
The calculation of $\delta Z (\lambda)$ then proceeds as follows. We need the dressed response only for $t_2<0$ (since 
$\tilde u^0$ vanishes at positive times). It reads
\be\label{b30}
{\mathbb R}_{k,t_{2},t_{1}} =  \frac{e^{- (k^2+1) (t_2-t_1)} (e^{t_2} \kappa -1)^2}{(e^{t_1} \kappa
   -1)^2} \theta(t_2-t_1) 
\ .
\ee
This yields for $t_1<0$\be \label{297}
\Phi (k,t_{1}) = \frac{e^{t_1} \kappa  \left[k^2 \kappa e^{t_1}  +e^{k^2 t_1} \left(k^2 (1-\kappa)-1\right)+1-k^2\right]}{k^2 (k^2-1) (e^{t_1} \kappa -1)^2} 
\ee
with $\Phi (k,t_{1}) =0$ for $t_1>0$. 
 \begin{widetext}
\begin{eqnarray}\label{298}
\int_{t<t_1<0} \Phi(k,t_1) {\mathbb R}_{k=0,t_{1},t}  &=& - \frac{\left[ k^2 (\kappa -1)+1\right] \kappa ^{-k^2} B_{\kappa }(k^2+1,0)+k^2 \kappa +\log
   (1-\kappa )}{k^2 (k^2-1)}\\
\int_{t<t_1<0} \Phi(k,t_1)^{2}\, {\mathbb R}_{k=0,t_{1},t} &=&\frac{1}{2 k^2
(k^2-1)^2} \bigg\{ 2 \left(2 k^2+1\right) \left[k^2 (\kappa -1)+1\right]^2 \kappa ^{-2 k^2} B_{\kappa }\left(2 k^2+1,0\right)\nonumber \\
&& -6   \label{299}
   \left(k^2+1\right) \left[k^2 (\kappa -1)+1\right] \kappa ^{-k^2} B_{\kappa }\left(k^2+1,0\right)\nonumber \\
&& +k^2
   \kappa  \left[2 k^2 (\kappa -1)+\kappa -4\right]-2 (k^2+2) \log (1-\kappa )\}
\ .
\end{eqnarray}\end{widetext} 
We have introduced the incomplete beta function $B_{\kappa} (a,b)$, defined as
\begin{equation}\label{b32}
B_{\kappa} (a,b) := \int_{0}^{\kappa} t^{a-1} (1-t)^{b-1}
\end{equation}
and related to the hypergeometric function $_2F_1$ via
\begin{equation}\label{b33}
B_{\kappa }(a,0) = \frac{\kappa ^a \, _2F_1(1,a;a+1;\kappa )}{a}
\ ,
\end{equation}
which can equivalently be used. Note that while $B_{\kappa }(a,0)$ has a branch cut for negative $\kappa$, it is a
spurious one since in our results only  the combination $\kappa^{-a} B_{\kappa }(a,0)$
appears, which is perfectly regular on the negative $\kappa$ axis. 

The final result for ${\cal J}(k,\kappa )$ is
\begin{eqnarray}\label{b31}
{\cal J}(k,\kappa )
&=& \frac{2 k^2+1}{2 k^2
(k^2-1)^2} \times  \\
&&\times \bigg\{-4 \left[k^2 (\kappa -1)+1\right] \kappa ^{-k^2} B_{\kappa }(k^2+1,0)\nonumber \\ 
&& \hphantom{\times \bigg[}+\left[ k^2    (\kappa -2) \kappa -2 \log (1-\kappa )\right]\nonumber \\
&&\hphantom{\times \bigg[} +2  \left[k^2 (\kappa -1)+1\right]^2  \kappa ^{-2 k^2} B_{\kappa }(2
   k^2+1,0) \bigg\}\,.\nn  \end{eqnarray}
The special cases we need are of the form
\begin{equation}\label{b35}
 \kappa ^{-x} B_{\kappa} (1+x,0) = \int_{0}^{\kappa} \rmd t\,
 \left(\frac{t}{\kappa} \right)^{x} \frac{1}{1-t}
\ .
\end{equation}
Taylor expanding the denominator \(1/(1-t)\), and then integrating leads to
a very useful series representation
\begin{equation}\label{b36}
 \kappa ^{-x} B_{\kappa} (1+x,0)  =\sum_{n=0}^{\infty} \frac{\kappa
 ^{n+1}}{n+x+1} = \kappa  \Phi (\kappa ,1,x+1)\ .
\end{equation}
$ \Phi$ is known by Mathematica as the HurwitzLerchPhi function. 
Using the above series expansion, one can easily obtain the small- and large-$k$ behaviour:\bea
&& {\cal J}(k,\kappa ) = - \kappa + \frac{1}{2} \kappa^2 + O(k^2)  \\
&&  {\cal J}(k,\kappa ) = - \frac{\kappa}{k^2} - \frac{\kappa + 2 \ln(1-\kappa)}{k^4} + O\left(\frac{1}{k^6}\right) 
\ .~~~~~~~~~
\eea 
Hence ${\cal J}(k,\kappa )$ is integrable with the measure $k^2 \rmd(k^2)$ at small $k$, but contains
a quadratic and a logarithmic divergence at large $k$. These will  have to be  cancelled 
by the counter-terms, leading to a finite result. We will show in section \ref{s:counter-terms} that the exact
expression for the counter-term is
\begin{equation}\label{b29}
{\cal J}_{\mathrm{ct}} (k,\kappa) = \frac{( 3+k^{2})\kappa +2 \ln (1-\kappa ) }{(1+k^{2})^{2}}
\ .\end{equation}
Using the series expansion (\ref{b36}), the integration over $k$ can
be performed, keeping a large-$k$ cutoff in the intermediate
expressions. The final result is after simplifications, and inclusion
of the counter-terms:
\begin{eqnarray}\label{delta-Z}
\delta Z(\lambda) &:=& \int_0^{\infty} k^2 \rmd(k^2)  \left[{\cal J} (k,\kappa)+{\cal J}_{\mathrm{ct}} (k,\kappa ) \right]  \nn \\
&=& \kappa^2 (1-\ln 4) + \sum_{n=3}^\infty a_n \kappa^n \\
 a_n &=& \frac{(n-3) (n-2)^2 \log (n-2)}{2 n^2}\nn \\
&& +\frac{6\log (2)-2 n (n+1) (\log
   (2)-1)}{ n^2 (n+1)} \nn \\
&& -\frac{(n-1) (n ((n-6) n+2)+6) \log (n-1)}{n^2
   (n+1)}\nn \\
&& +\frac{\left(n^2-8 n+3\right) \log (n)}{2 ( n+1)}\ . \label{an}
\end{eqnarray}
Note that $\lim_{n\to 2}a_{n} = 1-2 \ln 2$, i.e.\ the first term
$a_{2}$ follows the same relation, if the coefficients are properly
interpreted. For later convenience we set $a_{1}=0$.

It is also possible to calculate the cumulants of the velocity directly
in a perturbative expansion in the full disorder to 1-loop accuracy.  This  is performed in Appendix 
\ref{expansion-lambda} up to the third cumulant. We have checked that
this indeed agrees with our explicit series expansion up to order $\lambda^{3}$. 
As the reader will see, the calculation of the appendix increases formidably 
in difficulty with each new order, while the present  method allows to
sum these diagrams much more  efficiently.

It is interesting to give the lowest moments. In dimensionless units they read,  expanding
(\ref{delta-Z}) in powers of $\lambda$ using (\ref{an}) and (\ref{8.195}),\bea
 \overline{\dot u_t^2} &=& v\ , \\
 \overline{\dot u_t^2} &=& v \left[ 1 + \alpha (1 - \ln 4)\right] + O(v^2)\ , \\
 \overline{\dot u_t^3} &=& v \left[2 + \frac{\alpha}{2} (8 + 9 \ln 3 - 13 \ln 4)\right] + O(v^2)\ ,   \\
 \overline{\dot u_t^4} &= & v \left[6 + \frac{\alpha}{2}  \frac{9}{5} \left(20 - 132 \ln 2 + 69 \ln 3\right)\right] + O(v^2)
\ .~~~~~~~~~~~
\eea 
We recall the mean-field result $\overline{\dot u_t^p} = (p-1)!$ which follows from \(Z (\lambda)=-\ln(1-\lambda)\). 
The general formula for the moments $p \geq 2$ is easily obtained as 
\be
\overline{\dot u_t^p} = v(p-1)!  \left[1+  \frac{\alpha}{2}  p! \sum_{n=2}^p a_n \frac{(-1)^n}{(p-n)! (n-1)!} \right]\ee
Let us recall that  the small parameter $\alpha$ is  related to the second derivative at the 
fixed point and equals  (see (B12) of \cite{LeDoussalWiese2008c}):
\bea\label{alphavalue}
 \alpha &=& - \tilde \Delta'' (0^{+})  = - \frac{\epsilon -\zeta}{3} +O (\epsilon^{2}) \\
& =& - \frac{1 - \zeta_1}{3} \epsilon + O (\epsilon^{2})
\eea
with $\zeta_1=1/3$ for the RF class, and $\zeta_1=0$ for the periodic class, i.e.
\bea
&& \alpha = - \frac{2}{9} \epsilon \quad , \quad {\rm (RF=non periodic~ disorder) } ~~~~~\label{314}\\
&& \alpha = - \frac{1}{3} \epsilon \quad , \quad {\rm (periodic~ disorder ) }
\ .
\label{315}\eea 
Hence for  non-periodic depinning
\bea
&& \overline{\dot u_t^2} = v (1 + 0.0858432 \epsilon)  \\
&& \overline{\dot u_t^3} = v (2 + 0.014924 \epsilon ) \\
&& \overline{\dot u_t^4} = v (6  - 0.861764 \epsilon )
\eea 
More ambitiously, we will now determine the correction to the velocity
distribution in an avalanche.

\subsubsection{From $Z (\lambda)$ to $P (\dot u)$: Distribution of velocities
in an avalanche}\label{q8}
\begin{figure}[t]
\Fig{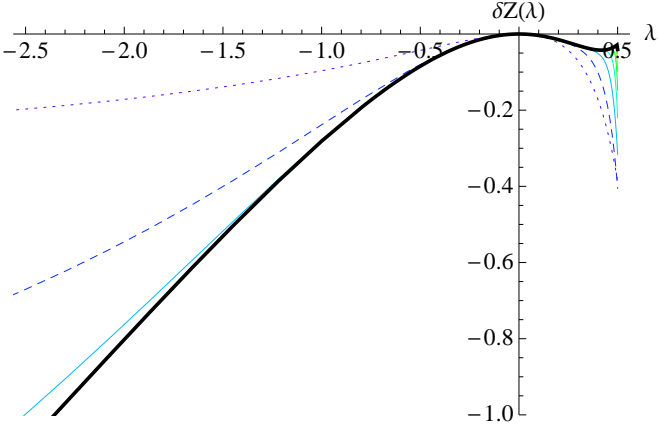}\caption{Plot of $\delta Z(\lambda)$ defined in equation
(\ref{delta-Z}). For real $\lambda$ the function  $\delta Z(\lambda)$ is defined for $\lambda \in ]-\infty,1[$, with a singularity at $\lambda=1$. The Taylor series in $\kappa$ has convergence
radius 1, which translates into convergence for $\lambda$ from
$-\infty$ to $1/2$. This is plotted with an upper bound on the series
$n_{\mathrm{max}}=2^i$, for $i=1,2,...,10$, starting with the dotted curve for $i=1$, dashed for $i=2$, the remaining ones solid, and finally \(i=10\) (fat). This
establishes convergence in that interval. The last 6 curves are indistinguishable, except at \(\lambda \to 1/2\). See figure \ref{q13} for the resulting function.
Note  that $\delta Z(\lambda)$ can be obtained numerically from the alternative formula (\ref{B6}) with excellent agreement.}
\label{f:delta-Z}
\end{figure}
The series for $\delta Z (\lambda)$, defined in Eq.~(\ref{delta-Z}),
as a series in $\kappa$, has convergence radius 1, since $a_n/a_{n+1} \to 1$ at large $n$, equivalent to
$\Re (\lambda)<\frac12$. This is demonstrated on Fig.\ \ref{f:delta-Z}. The physical singularity however
is outside of this interval, at $\lambda=1$.

We now obtain the avalanche-size distribution. As explained in section \ref{s:MF-1pt}, we have
\be 
P(\dot u) = (1- p' v) \delta(\dot u) + v p' {\calP}(\dot u) + O(v^2) 
\ee  
with
\be  \label{Ztrue} 
Z(\lambda) = p' \int_0^\infty \rmd \dot u (e^{\lambda \dot u} - 1) {\calP}(\dot u) 
\ .
\ee  
We have obtained the expansion of $Z(\lambda)$ to order $O(\alpha)$ in
the form (\ref{3.31}), hence
\bea
p' {\calP}(\dot u) &=&  \frac{1}{\dot u} e^{-\dot u} + \frac{\alpha}{2} \delta { {\cal P}}( \dot u)\ , \\
 \delta Z(\lambda) &=& \int_0^\infty \rmd\dot  u (e^{\lambda \dot u} - 1)\, \delta {{\cal P}}(\dot u) 
\ .
\label{322}\eea 
In the case where $\delta Z(\lambda)$ admits an inverse Laplace transform
we can also write
\be 
{\cal P}(\dot u) = {\cal P}_{\rm MF}(\dot u) + \frac{\alpha}{2} \delta { {\cal P}}(\dot u)\ . 
\ee 
For $\dot u>0$ the inversion reads\be  \label{LTinverse}
\delta {\cal P}(\dot u)  =  \int_{i \infty}^{-i \infty} \frac{\rmd \lambda}{2\pi  i} \delta Z(\lambda)
\rme^{-\lambda \dot u}
\ ,\ee 
the contour being closed to the right. Note that
\bea
 \delta Z(\lambda=0)=0 \quad &\Leftrightarrow& \quad \int \rmd\dot u\,  \delta {{\cal P}}(\dot u) = 0\ , \\
 \delta Z'(\lambda=0)=0 \quad &\Leftrightarrow& \quad \int \rmd \dot u\, \dot u\, \delta {{\cal P}}(\dot u) = 0
\ .\eea To construct the probability distribution we first note the inverse Laplace transform
\begin{eqnarray}\label{3.52}
\mbox{LT}^{-1}_{-\lambda \to \dot u} \kappa^{n}_{} &=& \delta(\dot u)  -n ~ _1F_1(1 + n, 2, - \dot u)\nn \\
&=& \delta(\dot u) +  \rme^{- \dot u} \partial_{\dot u} \mathrm{L}_{n} (\dot u)
\end{eqnarray}
in terms of the hypergeometric
function $_1F_1$, or equivalently the Laguerre-polynomial $\mathrm{L}_{n}$. 
For $\dot u>0$ it can be found by rewriting the contour integral (which with our 
conventions must be closed to the right):
\bea
&&  \int_{i \infty}^{-i \infty} \frac{\rmd \lambda}{2\pi  i} \left(\frac{-\lambda}{1-\lambda} \right)^{n} 
\rme^{-\lambda \dot u} \nonumber \\
&=& \left(\frac{\partial}{\partial {v}} \right)^{n} \int_{0}^{\infty}\rmd \alpha \int_{i \infty}^{-i
\infty} \frac{\rmd \lambda}{2\pi  i}  \rme^{-\lambda \dot u -\alpha (1-\lambda )} \frac{\alpha^{n-1}}{\Gamma (\alpha )} \nonumber \\
&=& \left(\frac{\partial}{\partial {\dot u}} \right)^{n} \frac{\dot u^{n-1}\rme^{-\dot u}}{\Gamma (n)} \nn 
\eea
leading to (\ref{3.52}). Thus we can now write the formal series
\begin{equation}\label{q9}
\delta {{\cal P}} (\dot u) = \sum_{n=2}^{\infty} a_{n} \rme^{-\dot u}  \partial_{\dot u} \mathrm{L}_{n} (\dot u)
\ .
\end{equation}
Unfortunately, this series is divergent.

This problem can be
cured as follows: We will subtract from the series (\ref{delta-Z})
terms which can be summed analytically, resulting in polylogarithmic
functions, and their derivatives, and inverting the latter via a
cut-integral. These terms are chosen to render the remaining sum
(quasi-)convergent. To this aim, we note 
\begin{equation}\label{q10}
\delta Z (\lambda) = \delta Z_{\mathrm{ser}} (\lambda) +\delta Z_{\mathrm{cut}} (\lambda) \ .
\end{equation}
We start by Taylor-expanding $a_{n}$ around $n=\infty$, 
\begin{eqnarray}\label{3.57}
a_{n}&=& \frac{
{1}-4 \log   (2n)}{2n}
+\frac{2}{n^2}
+\frac{6 \log
   (2)-\frac{37}{12}}{n^3}
+\frac{5-6 \log (2)}{n^4} \nn \\
&&+
\frac{6 \log (2)-\frac{101}{30}}{n^5}+O\Big(\frac{1}{n^{6}} \Big) \nn \\
&& =  -\frac{2\ln n}{n} +\sum_{j=1}^{\infty} \frac{b_{j}}{n^{j}}
\end{eqnarray}
with $b_{1}= \frac{1}{2}-2\ln 2$, $b_{2}=2$,  and an explicit formula
for  $b_j$ for $j \geq 3$ is given in Appendix \ref{a:bj}. We recall that
$a_2=1-2 \ln 2$ and that we set $a_1=0$ \footnote{Although this
may appear to impose an artificial constraint $\sum_{j=1}^\infty b_j=0$ 
it will be immaterial in what follows since we will use 
only a finite sum and add and subtract the same terms.}. 
\begin{figure}[t]
\Fig{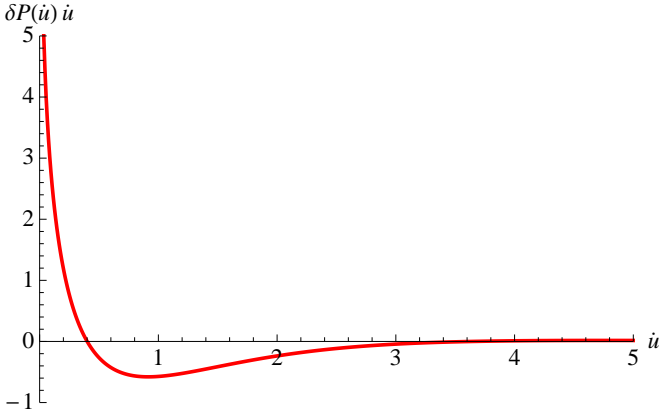}
\caption{\(\dot u \delta P(\dot u)\) as given by Eq.~(\ref{q11}),  or, equivalently by Eq.~(\ref{B7}).}
\label{fdeltaPofudot}
\end{figure}
Performing the
summation, we obtain (if the series converges) the alternative representation
\begin{equation}\label{3.56}
\delta Z (\lambda) = 2 \partial_{j} {\rm Li}_j(\kappa )\big|_{j=1} +
\sum_{j=1}^{\infty} b_{j} {\rm Li}_j(\kappa ) \ .
\end{equation}
${\rm Li}_j(\kappa )$ is the polylogarithm function, which is analytic on the 
complex plane with a cut on the real axis for $\kappa \in [1,\infty[$, which maps on the same interval for $\lambda \in [1,\infty[$ with reversed boundaries. It is  along this cut that we have to integrate. The  discontinuity there is
given by
\begin{equation}
\lim_{\epsilon \to0^+} {\rm Li}_j(\kappa +i\epsilon)-{\rm Li}_j(\kappa -i\epsilon)=
\begin{cases}
2\pi i \frac{(\ln \kappa )^{j-1}}{\Gamma(j)}\,,& \kappa >1 \,,\\
0\,, &\kappa <1\,.
\end{cases}
\label{disc}
\ .
\end{equation}
Note that this also holds true for the derivative w.r.t. $j$, and for
$j=1$, i.e.\ $\mbox{Li}_{1} (\kappa ) = -\ln (1-\kappa )$. 

Thus, the inverse Laplace transform (\ref{LTinverse}) becomes a
compact and simple cut-integral
\begin{align}\label{delta-P-v}
\delta {\cal P} (\dot u) &= -\int\limits_{1}^{\infty} \frac{\rmd \lambda}{\rme^{\lambda v}}
 \left[2\gamma_{\mathrm{E}} + 2\ln (\ln \kappa ) +
\sum_{j=1}^{\infty} \frac{b_{j}  (\ln \kappa )^{j-1}}{\Gamma (j)}
\right]
\end{align}
However, this series  also diverges. Therefore we choose
$j_{\mathrm{max}}$ as a cutoff, by defining
\begin{eqnarray}\label{q11}
\delta {\cal P} (\dot u) &=& \delta {\cal P}_{\mathrm{ser}} (\dot u) + \delta {\cal P}_{\mathrm{cut}} (\dot u)\\
\label{delta-P-magic}
\delta {\cal P}_{\mathrm{cut}} (\dot u) &=&   -\int\limits_{1}^{\infty} \frac{\rmd \lambda}{\rme^{\lambda v}}
 \left[2\gamma_{\mathrm{E}} + 2\ln (\ln \kappa ) +
\sum_{j=1}^{j_{\mathrm{max}}} \frac{b_{j}  (\ln \kappa )^{j-1}}{\Gamma (j)}
\right]\nn \\
\end{eqnarray}\begin{figure*}[t]
\Fig{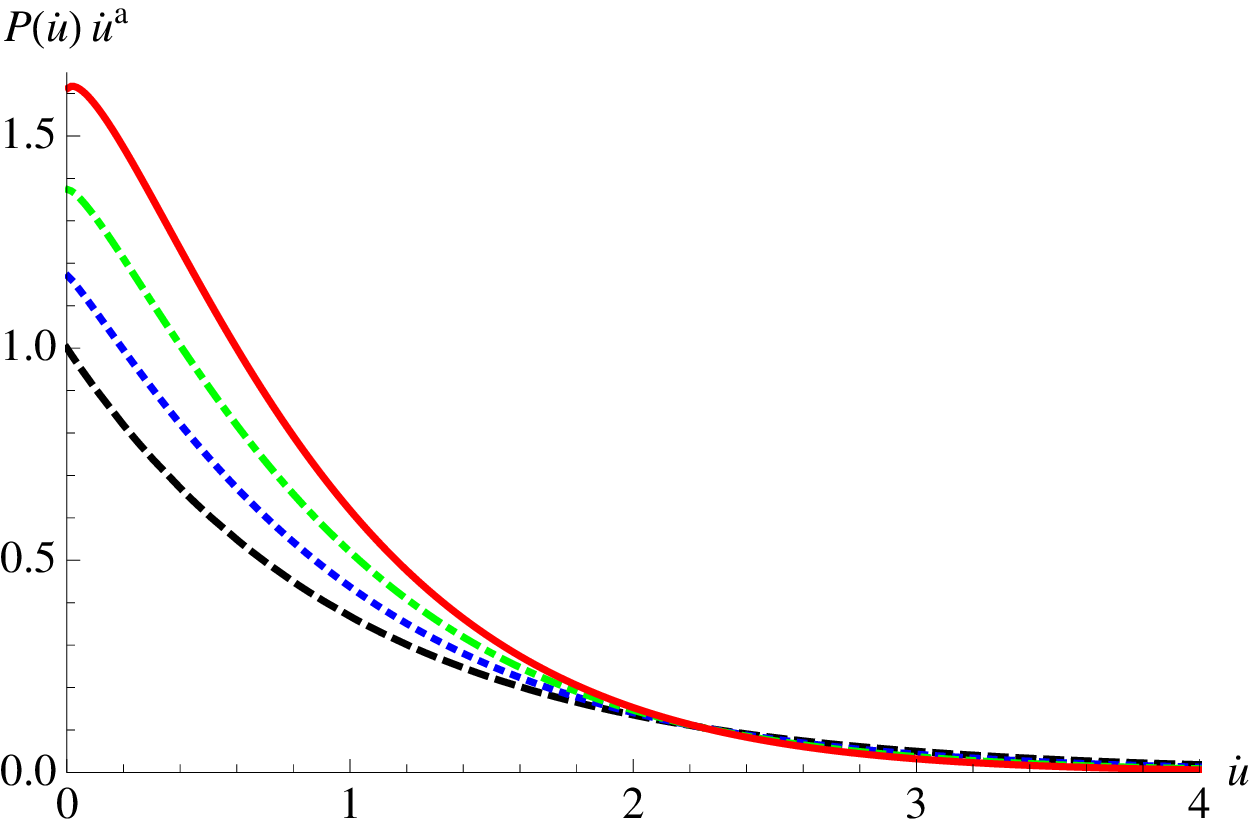}\Fig{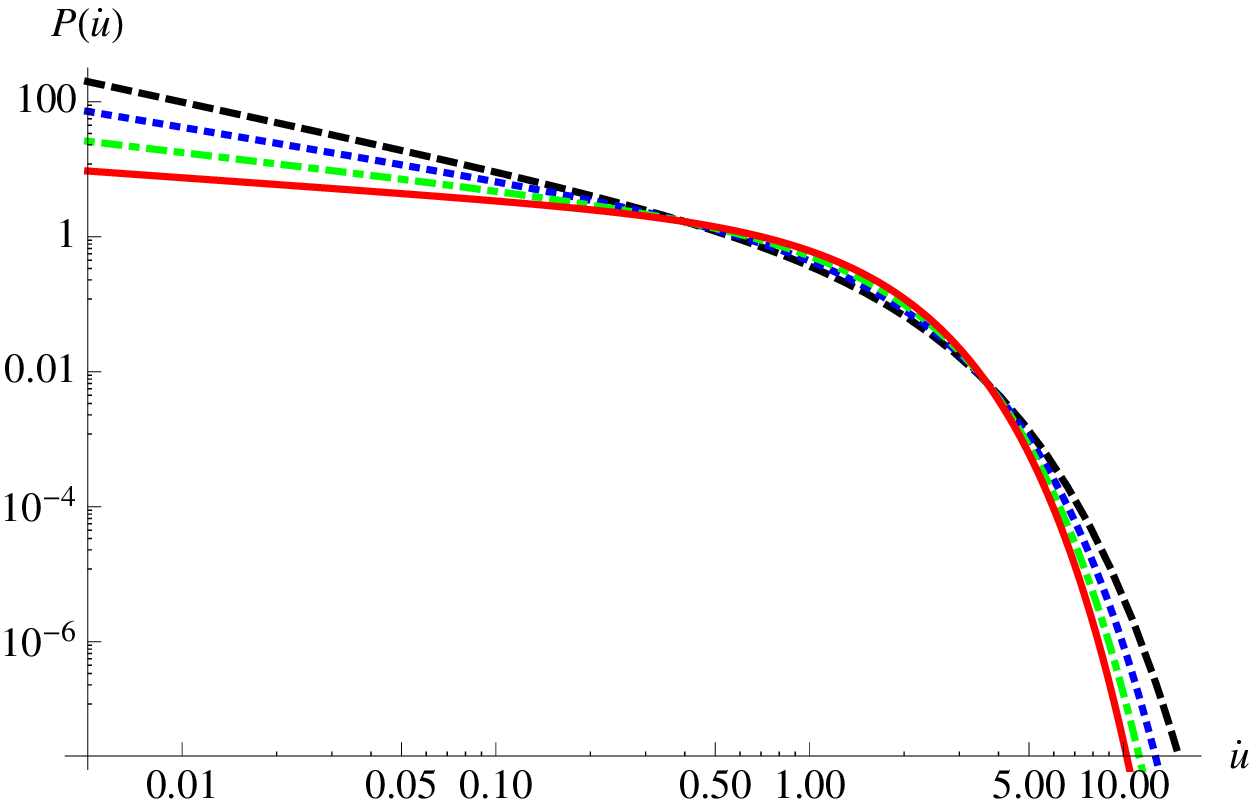}
\caption{Left: ${\cal P}(\dot u) \dot u^{\rm a}$ as a function of $\dot u^{\rm a}$ for RF-disorder; $\epsilon
=0$ i.e.\ MF (black dashed), \(\epsilon=1\) (blue, dotted), \(\epsilon=2\) (green, dash-dotted) and \(\epsilon=3\) (red, solid). Right: log-log plot of ${\cal P} (\dot u)$ as a function of $\dot u $, for the same values of $\epsilon$. For both plots resummation
formula (\ref{Pv-resum}) was used. We note that since only $\alpha$ appears as a parameter, at this order $\epsilon=3$ RF and $\epsilon=2$ RP are indistinguishable, see Eqs.~(\ref{314}), (\ref{315}).}
\label{fig:Pv}
\end{figure*}The coefficients $\tilde a_{n}$ are what remains of $a_{n}$ after
subtracting their asymptotic behavior, 
\begin{equation}\label{q12}
\tilde a_{n} := a_{n} +2 \frac{\ln n}{n} -  \sum_{j=1}^{j_{\mathrm{max}}} \frac{b_{j}}{n^{j}}\ .
\end{equation}
Especially note that $\tilde a_{1}$ becomes non-zero, even though $a_1=0$; in fact, this
coefficient grows rather quickly with $j_{\mathrm{max}}$, while the other
coefficients decay. 
\begin{equation}\label{deltaPser}
\delta {\cal P}_{\mathrm{ser}} (\dot u)  =  \sum_{n=1}^{\infty} \tilde a_{n} \rme^{-\dot u}  \partial_{\dot u} \mathrm{L}_{n} (\dot u)
\ .
\end{equation}
Both expressions, $\delta{\cal P}_{\mathrm{cut}} (\dot u)$ and $\delta
{\cal P}_{\mathrm{ser}} (\dot u)$ can be obtained numerically with good precision,
and seem to decay rapidly at large $\dot u$. One then checks that
the sum of the two, for any $\dot u$ in the bulk of the distribution, converges extremely
well versus the result at $j=j_{\mathrm{max}}$, e.g.\ for $\dot u=1$ excellent precision is already obtained
for $j_{\rm max=3}$. Of course, for a fixed $j_{\mathrm{max}}$ the sum over $n$ in (\ref{deltaPser}) should 
be stopped at $n$ not too large since it is an asymptotic series, which is ultimately divergent, but in practice the range of
convergence (with respect to $n_{\rm max}$) is rather  broad.

Practical values are $j_{\mathrm{max}} =15$, and
(\ref{deltaPser}) can also be stopped at $n=15$. With this choice, we find
that the precision is excellent and that all moments  
$\int_0^\infty \dot u^p \delta P(\dot u)\rmd \dot u$  between the fourth and 36th are at least given with
a relative precision of  $10^{-7}$, most even of $10^{-10}$.  
$j_{\mathrm{max}}$ should not be taken too large, since otherwise this shifts
too much 
weight into the moment $\tilde a_{1}$, leading to numerical problems (cancelation of large terms.)  
As an example, for $j_{\max}=15$, one has $\tilde a_{1}=-51.97$,
$\tilde a_{2}=0.002976$, $\tilde a_{3}= 1.359 \times 10^{-6}$, \dots, 
$\tilde a_{20} = 2.373 \times 10^{-15}$. There are no convergence problems at small or large \(\dot u\).

The final result for ${\cal P} (\dot u)$ is  
\begin{eqnarray}\label{Pv-final}
{\cal P}(\dot u) &=& {\cal P}_{0} (\dot u) + \frac{\alpha}{2} \delta {\cal P}(\dot u) + O
(\epsilon^{2}) \\ 
\label{Pv-resum}
&=&{\cal P}_{0} (\dot u) \exp \left(\frac{\alpha}{2}
\frac{\delta {\cal P} (\dot u)}{{\cal P}_{0} (\dot u)}  \right) + O (\epsilon^{2})\ ,\qquad 
\end{eqnarray}
where we remind the value of the small parameter $\alpha$ from
(\ref{alphavalue}). Note that the second formula (\ref{Pv-resum}), while being equivalent
to order $\epsilon$, has the  property to resum the
logarithmic behavior at small $v$ into the correct power-law
behavior. This is why we have chosen it in Fig. \ref{fig:Pv}.

\begin{figure}[b]
\Fig{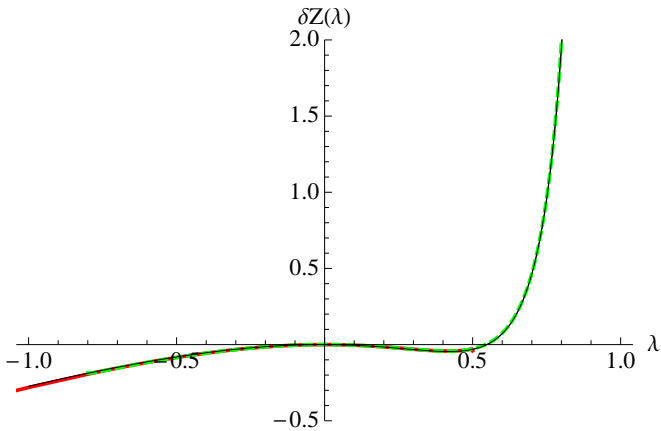}
\caption{(Color online) $\delta Z (\lambda)$, as obtained by (\ref{delta-Z}) (thick
red line), a reexpansion in $\lambda$ (dashed thick green line) and
numerical integration of Eq.\ (\ref{322}), using Eq.\ (\ref{q11}) (thin black line). All
functions agree in their respective area of convergence. }
\label{q13}
\end{figure}

\subsubsection{Small-velocity behaviour, the critical exponent ${\sf a}$}\label{q14}

Let us now obtain the small-velocity asymptotics of $P(\dot u)$ 
and extract the a-priori new critical exponent ${\sf a}$. It is controlled by
the asymptotics of $\delta Z (\lambda)$ at large negative $\lambda$,
i.e.\ $\lambda \to - \infty$. This corresponds to the behaviour for $\kappa \to 1^{-}$
of the series (\ref{delta-Z}). It is determined by the leading behaviour of $a_n$ at large $n$,
i.e.\ from the leading term
$a_{n}=-2\ln ( n)/n$ of (\ref{3.57}).  Resumming with this term alone, we obtain
\begin{eqnarray}
 \delta Z (\lambda)   &=&  \sum_{n=2}^\infty a_n\kappa^n \approx -2  \sum_{n=2}^\infty \frac{\ln (n)}{n} \kappa^n = 2 \partial_a \mbox{Li}_a(\kappa)\big|_{a=1} \nn\\&=& - \ln^{2}(1-\kappa)  +O \Big(\ln(1-\kappa)\Big) +\dots \nonumber \\
&=&  - \ln^{2}(1-\lambda ) + \dots 
\ .\label{Zasymp1}
\end{eqnarray} 
This yields  for $\lambda \to -\infty $
\begin{eqnarray}\label{b39}
Z(\lambda ) 
 &=&  Z_0(\lambda ) +\frac{\alpha}2\delta Z(\lambda)\nn\\&=&
 -\ln(1-\lambda)\left[1+   \frac{\alpha}{2}  \ln(1-\lambda) + \dots 
 \right]\ . ~~~~~~
\end{eqnarray}
It is easy to see that this is consistent with a modified critical behaviour at small velocities
\be  \label{finala} 
{\cal P}(\dot u) \sim_{\dot u \ll 1} \frac{1}{\dot u^{\sf a}}   \quad , \quad {\sf a} = 1 + \alpha + O (\epsilon^{2})
\ .
\ee  
To show this,  we start from the  trial  probability\be 
p' {\calP}_{\rm trial} (\dot u) = \frac{1}{\dot u^{1+x}} e^{-\dot u}
\label{343}\ee 
for which the associated $Z(\lambda)$ can be computed exactly via a Laplace transform, using (\ref{Ztrue}). Expanding the  result in small $x$ yields
\bea
 Z_{\rm trial}(\lambda)& =& \int_0^\infty \rmd\dot  u \frac{1}{\dot u^{1+x}} e^{-\dot u} (e^{\lambda \dot u}-1)  \nn\\&=& - \ln(1-\lambda) \nn \\
&&  + \left[ -  \gamma_{\rm E} +   \ln(1 - \lambda) - \frac12 \ln^2(1 - \lambda) \right] x \nn\\
&&+ O(x^2) 
\label{344}\eea 
the first term is $Z^{\mathrm{tree}}(\lambda)$ and the second one the correction. Comparing the behavior at large negative $\lambda$ of Eqs.\ (\ref{b39}) and (\ref{344}), we can thus identify $x = \alpha$, consistent with Eq.\ (\ref{finala}). Note that multiplying (\ref{343}) by a prefactor $C_x=1+O(x)$ 
or changing the exponential to $e^{-\dot u [1+ O(x)]}$ produces only $\sim x \ln(\lambda)$ terms,\ subdominant w.r.t. the $\ln^2(1-\lambda)$ at $\lambda \to - \infty$. 

Let us now discuss our results for the small-velocity exponent. Using (\ref{finala}), together with (\ref{314}) and (\ref{315}), we find
\bea
&& {\sf a} = 1 - \frac{2}{9} \epsilon + O (\epsilon^{2})   \quad  {\rm non periodic} \\
&& {\sf a} = 1 - \frac{1}{3} \epsilon + O (\epsilon^{2})   \quad  {\rm periodic}
\ .
\eea
Our predictions for the change of  ${\sf a}$ are thus  quite large, and tend to reduce the
exponent. A naive extrapolation to  $d=1$, $\epsilon=3$
(depinning of a line) would suggest ${\sf a} \approx 1/3$ significantly reduced from the mean-field value ${\sf a}_{\rm MF} = 1$. 
 Preliminary numerical results indicate that the exponent ${\sf a}$ may even be negative in $d=1$ \cite{KoltonPrivate}. A 2-loop  calculation (or higher) would  settle the question from an analytical point of view. 

We can compare the above formula to the one for the dynamical exponent to one loop
\be 
z = 2 + \alpha + O(\epsilon^{2}) 
\ .
\ee 
Hence we could also write
\be 
{\sf a} = z-1 + O(\epsilon^{2})
\ ,
\ee  
which holds for both periodic and non-periodic systems.  Again it would be interesting to obtain
the higher-loop corrections, since we did not find any general argument why they would be absent.

Finally the small-$\dot u$ behaviour can be studied more systematically. This is done in 
Appendix \ref{app:small} where we obtain the amplitude at small $\dot u>0$,\be \label{CC}
   P(\dot u) \approx \frac{C}{\dot u^{{\sf a}}}  \quad , \quad  C = 1 - \frac{\alpha}{2} (4 \gamma_{\rm E} + b_1) 
\ee
where $b_1= \frac{1}{2}-2\ln 2$ as defined above. This yields $C = 1- 0.711284 \alpha$ in good
agreement with our numerical Laplace-inversion. In principle this amplitude is universal and can be measured.

\subsubsection{The behavior of $\delta Z(\lambda)$ for $\lambda\to 1$, and tail of \({\cal P}(\dot u)\) at large $\dot u$}

The behavior of \(Z(\lambda)  \) in the limit of $\lambda \to 1$, which controls the
tail of ${\cal P}(\dot u)$ for $\dot{u}\to \infty$, is obtained in 
Appendix \ref{app:lambda1}. The final result is\bea
 Z(\lambda) &=& -\ln(1-\lambda) + \frac\alpha {2} \delta Z(\lambda)  \label{362bis}
\\
 \delta Z(\lambda) &=& \frac1 {8} \frac{1}{(1-\lambda)^2 [\ln(1-\lambda)]^2}
+ ... \label{362}
\eea
To obtain the
tail of ${\cal P}(\dot u)$,  one needs to inverse Laplace tranform $Z(\lambda)$.  Before doing
so, let us point out that this form is incompatible with the naive expectation of a stretched exponential 
at large velocity,\be 
P(\dot u) \sim_{\dot u \gg 1} \frac{C'}{\dot u^{{\sf a}'}} e^{- B \dot u^\delta} 
\ ,\ee  
with $C'=B={\sf a}'=\delta=1$ in mean field ($\epsilon=0$). While it would be hard to extract $B,C'$ and ${\sf a}'$, we could extract
$\delta$ as follows. Expanding near $\delta=1,$ we find
\be 
\frac{\alpha}{2} \delta {\cal P} (\dot u) = - (\delta-1)  \rme^{-\dot u}  \ln \dot u + O((\delta -1)^2) 
\ .
\ee  
This is equivalent to 
\be \label{q16}
\frac{\alpha}{2} \delta Z(\lambda) = (\delta-1) \frac{\ln (1-\lambda)+\gamma_{\rm E} \lambda }{ 1-\lambda} + O((\delta -1)^2)  
\ .
\ee 
Clearly, this is not of the form (\ref{362}).
Noting $s:=1-\lambda$,  we claim that Eq.~(\ref{362}) is equivalent to
\be 
 \delta {\cal P} (\dot u) \simeq \frac1 {8} e^{- \dot u} \dot u^2 f(\dot u)
\ee  
at large $\dot u$, where $f(\dot u)$ has a Laplace transform $\hat f(s):=\int_0^\infty \rmd \du f(\du) \rme^{-s \du} $ which behaves  at small $s$ as
\be 
\hat f(s) = \hat f(0) + \frac{1}{\ln s}\ .
\ee 
Indeed that would imply
\be 
{\rm  LT}_{\dot u \to s} e^{\dot u}  \delta {\cal P} (\dot u) = \frac1 {8} \partial_s^2 \hat f(s) \simeq \frac1 {8} \frac{1}{s^2 (\ln s)^2}
\ee 
for small $s$, which is exactly the result (\ref{362}). It is then easy to guess that 
\be 
f(\dot u) = \frac{1}{\dot u (\ln \dot u)^2}
=-\frac{\partial}{\partial \du} \frac{1}{\ln \du} \ee  
at large $\dot u$, for $\dot u > \dot u_0$. Indeed, the contribution 
for $\dot u > \dot u_0$ reads\bea
\hat f(0) - \hat f(s) &=& -\int_{\du_0}^{\infty} \rmd \du\, (1-e^{- s \dot
u}) \frac{\partial}{\partial \du} \frac{1}{\ln \du} \nn\\
& \simeq & \int_{ \dot u_0}^{\infty} \rmd \du\, s \,e^{- s \dot
u}  \frac{1}{\ln \du} \nn\\
& =& \int_{\dot u_0s}^{\infty} \rmd w\, e^{-w}  \frac{1}{\ln w-\ln s} \nn\\
& \simeq & -  \frac{1}{\ln s} 
\ .\eea
In the partial integration from the first to the second line we have dropped a term $(1-e^{- s \dot
u_0})/\ln \du_0$, which  is of order $s$. In the last step, we have used that for $s\to 0$, first  $\ln w-\ln
s\approx -\ln s$, and second $\dot u_0 s\to 0$.

For the velocity distribution
at large \(\dot u\), we thus finally obtain 
\begin{eqnarray}\label{zz12}
{\cal P}^{\rm{1-loop}}_{\dot{u}\to \infty } (\dot{u}) &=&
\frac{\rme^{-\dot{u} }}{\dot{u}} \left[1+ \frac{\alpha}{16}
\frac{\dot{u}^{2}}{\log ^2(\dot u)}\right] +O
(\alpha^{2})\label{}   
\nn\\ \label{372}
 &=& 
\frac{\rme^{-\dot{u} }}{\dot{u}} \exp \left( \frac{\alpha}{16}
\frac{ \dot{u}^{2}}{\log ^2(\dot u)} \right) +O
(\alpha^{2}) ~~~~~~~~~
\end{eqnarray}
We remind that $\alpha<0$, which has motivated us to write the result
in an exponentiated form. Other forms are  however possible, such
as corrections to the pre-exponential only. The form (\ref{372}) renders the tail stronger decaying; it is plotted on figure \ref{zz13}. 
In all cases, given the smallness of the correction, this tail will be hard to see in numerical
simulations. With the help of Eq.~(\ref{q11}), we have been able to evaluate $\delta {\cal P}(\du)$ up to $\du \approx 100$, while the alternative representation (\ref{B7}) works up to $\du \approx 10$. For these values of $\du$,  the tail-behavior (\ref{372}) is not yet reached.

\begin{figure}
\Fig{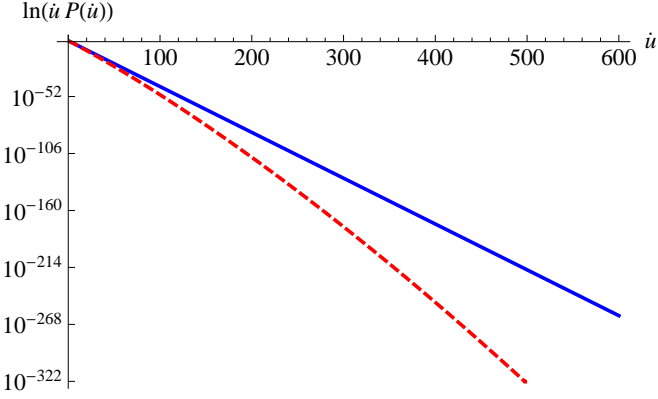}
\caption{The exponentiated  version (\ref{372}) of the function ${\cal
P}^{\mathrm{1-loop}}_{\dot{u}\to \infty } (\dot{u})$  for $\alpha
=-\frac{2}{3}$ (e.g.\ RF disorder in $d=1$) (red dashed line), compared to the mean-field result (i.e.\ $\epsilon=0$, blue, solid line). }
\label{zz13}
\end{figure}

\subsubsection{Alternative approach: Integrating over momentum first}
Our result for $\delta Z(\lambda)$ given in Eq.~(\ref{delta-Z}), was a compact series expansion from which one first had to extract the asymptotic behavior at large $\lambda$, before being able to perform the inverse Laplace transform. A complementary approach,  performed in detail in
appendix \ref{s:deltaZ-alt}, is to start from Eq.\ (\ref{Jt}), calculate $\Phi(k,t)$ as given in Eq.\ (\ref{297}),
and first integrate over $t$ and $k$, the final result for $\delta Z(\lambda)$,  given in (\ref{B6}), is now   an integral over $t_1$. (Recall that above in Eqs.\ (\ref{298}) and (\ref{299}) we integrated first over $t$, and $t_{1}$ leaving the $k$ integral for the end). We did not succeed in performing the final integral over $t_1$ analytically, although
it is easy to compute numerically. It confirms the above results for $\delta Z(\lambda)$. The advantage of this method is that the inverse-Laplace transform can be done explicitly, yielding a (relatively complicated) integral representation (as integral over $t_{1}$) of $\delta {\cal P}(\du)$  given in (\ref{B7}). 
It confirms all  statements made above, including the asymptotic behavior for small and large $\dot u$.

Note that  in Eqs.\ (\ref{298}) and (\ref{299}) one can interpret $t$ as the time of a kick (infinitesimal step in the force), 
or starting time of the avalanche, while time zero is the measurement time. The time
$t_1<0$ is an intermediate time, which must be integrated over the duration of the  avalanche. Hence, if we instead integrate over $k$ and then $t_1 \in [t,0]$ at fixed $t$ we obtain the joint probability that $\dot u(0)=\dot u$  and  the avalanche started at $t$. Although it is a straightforward generalization we will not give this
result here.

\subsection{Recovering the avalanche-size distribution to one-loop}

As discussed in section \ref{s:recover-stat}, to recover the avalanche-size distribution, one can use a 
source constant in time $\lambda_{xt}=\lambda$ during a large time window $T$. 
The avalanche-size generating function, noted here $Z_{S}(\lambda),$ is
obtained from the dynamic generating function studied here via $Z[\lambda] = T Z_{S}(\lambda)$.
In practice it amounts to suppressing the final time integral in the expression for $Z[\lambda]$.

For a source constant in time, the solution of the unperturbed instanton equation ($\eta_x=0$) is
\be
\tilde u^0 = Z_{S} ^{0}\equiv Z_{S}^0(\lambda) = \frac{1}{2} (1- \sqrt{1-4 \lambda}) 
\ .
\ee
The dressed response kernel then becomes
\be 
   \mathbb{R}_{k,t_{2},t_{1}} =  e^{ - (k^2+1- 2 Z_{S}^0) (t_2-t_1)   } \theta (t_2-t_1)
\ ,
\ee 
which is simply the bare response up to the replacement $m^2 \to m^2 - 2 Z_{S}^0(\lambda)$.
The formula (\ref{Phik}) then gives\be
\Phi(k,t_1) = Z_{S}  ^0\int_{t_1<t_2} {\mathbb R}_{k,t_2,t_1} = \frac{Z^0_{S}}{k^2+1-2 Z^0_{S}} 
\ .
\ee
Following the steps in Section \ref{sec:spaceindep}, this leads to
\bea
 Z_{S} &=& Z_{S} ^0+ Z_{S}^1+... \\
 Z^1_{S} &=& \langle \tilde u^{(2)}_{xt} \rangle_\eta = \frac{\alpha}{\epsilon \tilde I_2}  \int_k  {\cal J}_t (k) \\
 {\cal J}_t (k) &=&  \frac{1 }{1-2 Z^0_{S}}   {\cal J}^a_t (k)  
\ .
\label{382}\eea 
The coefficient $A=\frac{\alpha}{\epsilon \tilde I_2}$, and we have defined
\be \label{Jt2}
 {\cal J}^a_t (k) = \bigg(\frac{Z^0_{S}}{k^2+1-2 Z^0_{S}}\bigg)^{\!2}  + \frac{Z^0_{S}}{k^2+1-2Z^0_{S}} 
\ ,\ee
which is time independent. Graphically \Eq{382} can be written as in (\ref{q7}),
\be
Z_{S}^{1}= \frac{\alpha}{\epsilon \tilde I_2}  \left[ (Z^{0}_{S} )^{2} \diagram{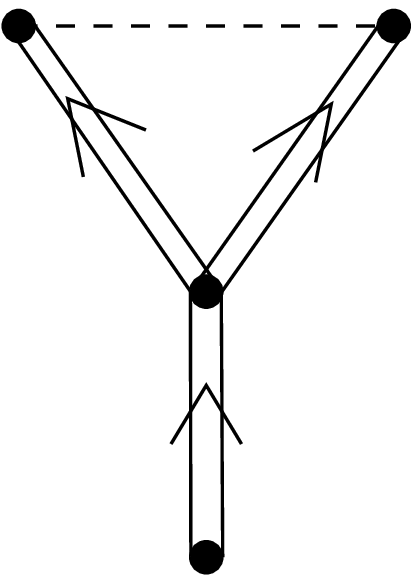} +
Z^{0}_{S} \diagram{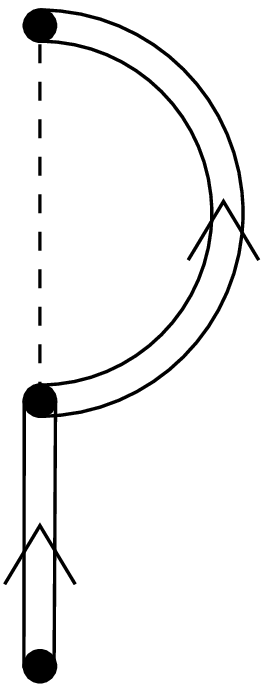}\right]
\ ,\ee
replacing the external wiggly lines of Eq.~(\ref{q7}) by the factors $Z^{0}_{S}$. 
Note that we have  recovered Eq.~(152) of \cite{LeDoussalWiese2011b} for the statics,
up to the two counter-terms discussed below. For pedagogical
purposes, we want to make further contact with the self-consistent equation obtained in \cite{LeDoussalWiese2008c}. 
To this aim we rewrite Eq.~(\ref{382}) as \bea
\frac{\alpha}{\epsilon
\tilde I_2}\int_k
{\cal J}_t^a (k) &=&
Z^1_{S} (1-2 Z_{S}^0) \nn\\
&=&  (Z_{S}-Z_{S}^0+...)(1-Z_{S}^0-Z_{S}+...) \nn\\
&=& Z_{S} -(Z_{S})^2
\ -\left[Z_{S}^0-(Z_{S}^0)^2\right]\nn\\&=&Z_{S}  -(Z_{S})^2
\ -\lambda\ .\label{384}\eea
Note that  by going from the first to the second line, we have added in each parenthesis a subdominant term. From the second to the third line, we have regrouped the terms, and finally from the third to the fourth line we used the exact relation \(Z_{S}^0-(Z_{S}^0)^2=\lambda\).  Eq.\ (\ref{384}) can thus be written as 
\bea
Z_{S} &=&\lambda+(Z_{S})^2
+\frac{\alpha}{\epsilon
\tilde I_2}\int_k{\cal J}_t^a (k)\ .\label{385}\\
&=&\lambda+(Z_{S})^2+ \frac{\alpha}{\epsilon \tilde I_2}  \left[ (Z^{0}_{S} )^{2} \diagram{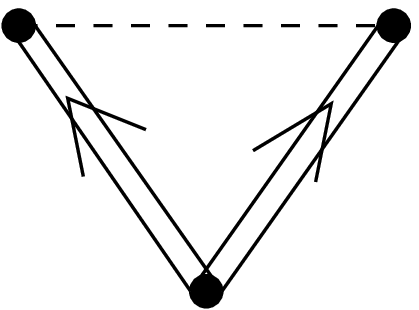} +
Z^{0}_{S} \diagram{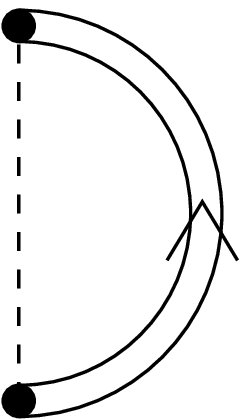}\right]\nn
\eea
where we recall the graphical interpretation of each term.
(The amputated lower response gave the factor of $1/(1-2 Z^{0}_{S})$.)
Comparison with formula (151) in  \cite{LeDoussalWiese2008c} shows that one recovers the result of the
static calculation, provided ({\em i}) one replaces in \(\mathcal{J}^a_t(k)\) the tree generating function \(Z_{S}^0\) by \(Z_{S}\), which does not make a difference at this order;  (\textit{ii}) one  adds to \Eq{385} two counter-terms,
discussed  below,
\bea && Z_{S} =\lambda+(Z_{S})^2
+\frac{\alpha}{\epsilon
\tilde I_2}  \int_k    \Big[{\cal J}^a_t (k)  + {\cal J}^a_{\rm ct} (k)  \Big]~~~~~~~~~\\
&& {\cal J}^a_{\rm ct} (k) = \frac{- 3(Z^0_{S})^2  }{(k^2+1)^2}  - \frac{Z^0_{S}}{k^2+1}
\ .
\eea 
In the statics these counter-terms appeared naturally by using everywhere the improved action. The first one comes from the renormalization of  $\Delta(u)$, thus all parameters which appear are renormalized ones. The second also appeared naturally in the statics from the definitions used there, while here it comes as a correction
from using the (over)simplified model, as is explained below.

\subsection{Counter-terms and corrections to the simplified theory}
\label{s:counter-terms}
\subsubsection{Counter-terms from renormalization}\label{s:counter-terms;ren}

In \cite{LeDoussalWiese2008c} the static avalanche-size distribution was computed using the 
improved action, i.e.\ in terms of the renormalized disorder $\Delta(u)$, which automatically
includes the counter-terms for the renormalization of the disorder. In the dynamics, there is an additional operator which is
marginal at $d=d_{\rm uc}$ and corresponds to the friction term in the dynamical action. Computing from
the start in terms of the renormalized friction $\eta$ is possible, but less convenient, hence  here we
perform the calculation first in terms of the {\em bare} disorder $\Delta_0(u)$ and the {\em bare} friction $\eta_0$, and then
reexpress at the end the result in terms of the {\em renormalized} disorder and friction. This yields 
an explicit derivation of the counter-terms.

We  start from the bare action given in Eqs.~(\ref{msr})~ff.  \begin{eqnarray}
  {\cal S} & =&    \int_{xt} \tilde u_{xt} (\eta_0 \partial_t -  \nabla_x^2 + m^2) \dot u_{xt} \nn \\
&& + \Delta_0'(0^+)  \int_{xt}  \tilde u_{xt}  \tilde u_{xt}  (v + \dot u_{xt})  \nn \\
&& +  \frac{1}{2}  \Delta_0''(0^+)  \int_{xtt'}  \tilde u_{xt}  \tilde u_{xt'}  (v + \dot u_{xt}) (v + \dot u_{xt'})\ . \nonumber 
\end{eqnarray}
Here the subscript zero denotes bare quantities.

The effective action to one loop, $\Gamma= {\cal S}+\delta  {\cal S}$, reads
\begin{align}\nonumber 
&{  \delta  {\cal S} =   \Delta_0''(0^+)  \int_{xtt'}  \tilde u_{xt} (v+ \dot u_{xt'}) \langle \tilde u_{xt'}  \dot u_{xt} \rangle} \nn \\
&-  2 \Delta_0'(0^+) \Delta_0''(0^+) \times \nn \\
&\times \nn  \Big[
\int_{x_1t_1x t t'}    \tilde u_{x_1t_1}  \tilde u_{xt} \dot u_{x t'} \langle \tilde u_{x_1t_1}   \tilde u_{xt'}   \dot u_{xt}  \dot u_{x_1 t_1} \rangle \\
&\hphantom{+\times } +  \int_{x_1t_1x tx't'}    \tilde u_{xt}  \tilde u_{xt'} \dot u_{x_1t_1} \langle \tilde u_{x_1t_1}  \tilde u_{x_1t_1}   \dot u_{xt'}   \dot u_{xt}  \rangle + \dots \Big]
\end{align}
where here averages $\langle...\rangle$  are w.r.t.\ $ {\cal S}_0$. Corrections to  $\Delta''(0^+)$ were omitted,  since they do not
matter to this order. From $\Gamma$ we can now identify the renormalized parameters. The second term
leads to $\Delta'(0^+)= \Delta_0'(0^+) +  \delta \Delta'(0^+),$ with
\begin{equation}\label{q17}
 \delta \Delta'(0^+) = - 3 \Delta_0'(0^+) \Delta_0''(0^+) \int_k \frac{1}{(k^2 + m^2)^2} 
\ .
\end{equation}
This is the correct FRG equation for $\Delta'(0^+)$ \cite{LeDoussalWiese2008c}. The first term gives
\begin{eqnarray}\label{first-term}
 \delta  {\cal S} &=&  v \Delta_0''(0^+) \int_k  \frac{1}{k^2 + m^2} \int_{xt}  \tilde u_{xt}
 \\
&& + \Delta_0''(0^+) \int_k  \frac{1}{ k^2 + m^2} \int_{xt}  \tilde u_{xt} \dot u_{xt} \nonumber \\
&& 
- \Delta_0''(0^+) \eta_0 \int_k  \frac{1}{( k^2 + m^2)^2} \int_{xt}  \tilde u_{xt} \partial_t \dot u_{xt} +\dots  \nonumber 
\end{eqnarray}
The last term gives, in agreement with \cite{LeDoussalWiese2008c}, the renormalized
$\eta = \eta_0 + \delta \eta$,
\begin{equation}\label{b23}
 \delta \eta = - \Delta_0''(0^+) \eta_0 \int_k  \frac{1}{(k^2 + m^2)^2} 
\ .\end{equation}
Reexpressing $Z(\lambda)$ instead in bare parameters as 
a function of  renormalized ones, defines the counterterms as
\be  \label{defct}
Z(\lambda;\eta_0,\Delta_{0})  = Z(\lambda;\eta,\Delta) + Z^{\rm ct}(\lambda;\eta,\Delta)
\ .
\ee  
Using that
\be 
Z(\lambda;\eta,\Delta) = Z_{\mathrm{tree}}(\lambda;\eta,\Delta) + \frac{\alpha}{2} \delta Z(\lambda)
\ ,
\ee  
where $\alpha \sim \Delta''(0)$, and given by \Eq{alpha1},
we only need to expand $Z_{\mathrm{tree}}$ to first order in the differences $\delta \Delta$ and
$\delta \eta$. Eq.~(\ref{Zt}) allows to restore units,
\be 
Z_{\mathrm{tree}}(\lambda;\eta,\Delta) = \frac{\eta m^2}{- \Delta' (0^{+})} \tilde Z_{\mathrm{tree}}\left(\frac{- \lambda \Delta' (0^{+})}{\eta m^2}\right)
\ .\ee  
Here $\tilde Z_{\mathrm{tree}}(\lambda)=-\ln(1-\lambda)$ and we remember that $\Delta' (0^{+})<0$. 
To compute the r.h.s of Eq.\ (\ref{defct}) we substitute $\eta \to \eta_0=\eta -\delta \eta$, $\Delta \to \Delta_0=\Delta - \delta \Delta$,  expand
to linear order in the differences, and in the final result we  replace, to this order, bare parameters by renormalized ones.
This gives
\bea
 Z^{\rm ct}(\lambda;\eta,\Delta) &=& \left(\frac{\delta \Delta' (0^{+})}{\Delta' (0^{+})} - \frac{\delta \eta}{\eta}\right)
 \frac{\eta m^2}{- \Delta' (0^{+})}\\
&& \times \left[\tilde Z_{\mathrm{tree}}(\mu) - \mu \tilde Z_{\mathrm{tree}}'(\mu)\right]\Big|_{\mu=\frac{- \lambda \Delta' (0^{+})}{\eta m^2}} \nn
\ .
\eea
We now switch back to  dimensionless units,
 setting $\eta \to 1$, $m \to 1$ and $- \Delta'(0^+) \to 1$. Using $(1-\lambda)
(1-\kappa) =1$, we then find
\be
Z^{\rm ct}(\lambda;\eta,\Delta) = -2 \Delta''(0^{+}) \int_{k}\frac{1}{(k^{2}+1)^{2}} 
\left[\ln(1-\kappa) + \kappa) \right] 
\ .
\ee
Comparing Eq.~(\ref{defct}) with Eqs.\ (\ref{3.31}), (\ref{alpha1}) and (\ref{Atoalpha}), we finally obtain\begin{equation}\label{q18}
{\cal J}_{\mathrm{ct}}^{\mathrm{RG}} (k,\kappa) = 2\Big[\kappa + \ln (1-\kappa)  \Big] 
\frac{1}{(1+k^{2})^{2}} 
\ .
\end{equation}
Note that to derive this counter-term we have used that $m=m_0$ i.e.\ that the
mass is not corrected, a property that we now discuss in  detail. 

\subsubsection{Corrections to the simplified theory}\label{s:corrections}

Let us examine  more closely the effective action derived
from the simplified theory, i.e.\ the  first two terms in Eq.\ (\ref{first-term}). 
We see that there appears an apparent correction to
$m^{2}$, obtained from the second line of Eq.\ (\ref{first-term}),
\begin{equation}\label{3.72}
\delta_{\rm simp} m^{2} = \Delta_{0}''(0^{+}) \int_{k}
\frac{1}{k^{2}+m^{2}}  \ .
\end{equation}
However we know from the STS symmetry that the mass {\em cannot} be corrected. 
The reason for this artifact is subtle. Let us go back to the exact theory (\ref{exactS}).
When computing the effective action, there is an additional term
\bea \label{new}
 \delta  {\cal S} &=& \int_{t'<t} \tilde u_{xt} \int_{t'}^t \rmd t_1 R_{t_1x, t'x} (\dot u_{xt} +v )( \dot u_{xt'}+ v)\nn\\ &&~~~~~~~~~~~~~~~~~~~~ \times \Delta'''_{\rm reg}(v(t-t')+u_{xt}-u_{xt'})\ , ~~~~~~~~
\eea
not present in the approximation $\Delta''_{\rm reg}(u)=\Delta''(0)$. Although it contains a
third derivative (which to this order is not supposed to matter), it gives a correction. To see this, we recognize that 
\bea
&&(\dot u_{xt} +v )( \dot u_{xt'}+ v)\Delta'''_{\rm reg}(v(t-t')+u_{xt}-u_{xt'}) \nn\\
&&~= -\partial_{t}
\partial_{t'}\Delta'_{\rm reg}(v(t-t')+u_{xt}-u_{xt'})\ . 
\eea
Inserting this relation into Eq.\ (\ref{new}), we obtain 
\bea
\delta  {\cal S} &=-& \int_{t'<t} \tilde u_{xt} \int_{0}^{t-t'} \rmd \tau R(\tau,x=0) \nn\\ &&\times \partial_{t}
\partial_{t'}\Delta'_{\rm reg}(v(t-t')+u_{xt}-u_{xt'})\ .~~~~~~~~~~
\eea
We now integrate by part w.r.t.\ $t'$:  there is no boundary term at $t'=t$ (since the $\tau$-integral then is zero); and there is no boundary term at $t'=-\infty$, since then \( \partial_{t}
\Delta'_{\rm reg}(v(t-t')+u_{xt}-u_{xt'})=0\). Thus only the upper bound of the $\tau$-integral contributes, and gives \be
\delta  {\cal S} =- \int\limits_{t'<t} \tilde u_{xt} R(t-t',x=0)
 \partial_{t}
\Delta'_{\rm reg}(v(t-t')+u_{xt}-u_{xt'})\ .
\ee
Thus we arrive at 
\bea
\delta  {\cal S} &=&- \int\limits_t \tilde u_{xt}(v+\dot u_{xt}) \nn\\
&& \times\int\limits_k  \int\limits_0^\infty \rmd\tau \rme^{-\tau (k^2+1)} \Delta''(v\tau +u_{xt}-u_{x,t-\tau})~~~~~~~~~~\eea
In the limit of small $v$, the term $\Delta''_{\rm reg}(v(t-t')+u_{xt}-u_{xt'})$ can be approximated by $\Delta''(0^+)$, thus  
\be
\delta  {\cal S} =-\Delta''(0^+) \int\limits_t \tilde u_{xt}(v+\dot u_{xt})
\int_k\frac{1}{k^2+1}
\ .\ee
Thus there is an additional term 
\begin{equation}\label{3.72b}
\delta_{\rm add} m^{2} =  - \Delta_{0}''(0^{+}) \int_{k}
\frac{1}{k^{2}+m^{2}}  \ ,
\end{equation}
This term  cancels the spurious mass correction. Note that in another derivation, given in  
appendix  \ref{s:first-principle}, both  terms (\ref{3.72}) and (\ref{3.72b}) appear. 

Two observations are in order: First of all, one can rewrite the two terms graphically as 
\be
\int \tilde u \dot u \left[\delta_{\rm simp} m^{2}+\delta_{\rm add} m^{2} \right] =\diagram{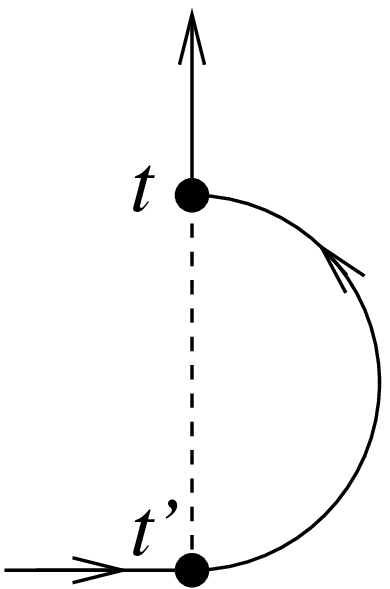} -\diagram{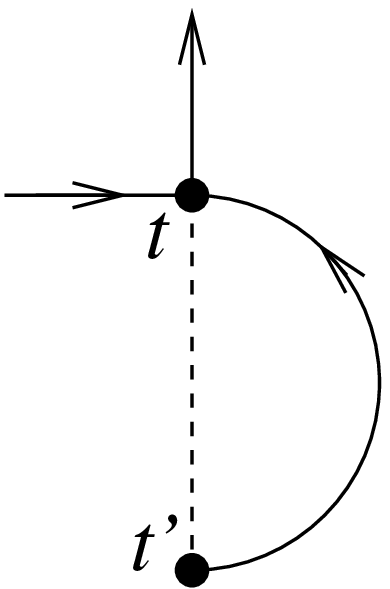}
\label{408}\ee
While the first one naturally arose in the velocity theory, it is the second one which we derived above. Their crucial difference is where the field $\dot u$ is sitting in time, as $\dot u_t$ at the same time $t$ as the response field $\tilde u_t$, or as $\dot u_{t'}$ at the earlier time $t'$. Thus there is no correction to \(m^2\) due to this cancellation, also known as the mounting property (and frequently used, see e.g.\ \cite{LeDoussalWieseChauve2003,FedorenkoLeDoussalWiese2006} ). 

Second, we have used that $\Delta''(u)$ decays to 0 for $u\to \infty$, i.e.\ short range disorder.

Finally, the additional loop correction (\ref{3.72b}) must be {\it added} to our calculation based until
now only on the simplified theory. It can be interpreted as an additional ``counter-term"
 to subtract (\ref{3.72}). To calculate it let us consider how this additional 
 term (\ref{3.72b}) contributes to $Z(\lambda)$. Indeed, it changes Eq.\
(\ref{3.21}) to 
\begin{align}\label{q19}
&\left[ \partial_t + \nabla_x^2 - 1 + 2 \tilde u^0_{xt} \right] \tilde u^2_{xt} \nn \\
&\qquad = - (\tilde u^1_{xt})^2 - \eta_x \tilde u^1_{xt} + \delta _{\rm add} m^{2} \tilde u^0_{xt} \ .
\end{align}
This is equivalent to an addition to $Z(\lambda),$ equal to\begin{eqnarray}\label{q20}
\int_t \delta \tilde u^{(2)}_{xt} &=& -\delta_{\rm add} m^2 \int_{t<t_1<0}  \tilde
u^0_{t_1} \,{\mathbb R}_{k=0,t_{1},t} \nn \\
&=&  - \Delta'' (0^{+}) \kappa  \int_{k} \frac{1}{k^{2}+m^{2}}
\ .
\end{eqnarray}
In terms of ${\cal J} (k,\kappa)$ it reads 
\begin{equation}\label{q21}
{\cal J}^{\delta m^{2}}_{\mathrm{ct}}(k,\kappa) = \kappa  \frac{1}{k^{2}+m^{2}}
\ .
\end{equation}
Both counter-terms together give, as already used in Eq.~(\ref{b29}), 
\begin{eqnarray}\label{q22}
{\cal J}_{\mathrm{ct}}(k,\kappa)&=&{\cal J}^{\delta
m^{2}}_{\mathrm{ct}}(k,\kappa)+{\cal
J}_{\mathrm{ct}}^\mathrm{RG}(k,\kappa) \nn \\
&=&  \frac{( 3+k^{2})\kappa +2 \ln (1-\kappa ) }{(1+k^{2})^{2}}\ .
\end{eqnarray}

\subsubsection{First-principle calculation in $u$-theory}

We note that the two terms in Eq.~(\ref{408}) naturally appear together in calculations based on the position field $u(x,t)$, instead of the velocity $\dot u(x,t)$, see e.g.\ Eq.~(3.22) and Fig.~9 on page 13 of \cite{LeDoussalWieseChauve2002}. The question thus arises whether one could construct the field theory directly for the position field instead of the velocity field, and whether this would give directly the combination (\ref{408}). As we show in appendix \ref{s:first-principle}, both answers are {\em ``yes''}.

\subsection{Distribution of velocities for long-ranged elasticity}\label{s:LR}
Although some systems with long-range elasticity are studied at their
upper critical dimension (usually interfaces with $d_{\rm uc}=2$), some 
require an $\epsilon$ expansion around $d_{\rm uc}$. This is the case
for instance for the contact-line or fracture fronts ($d=1$, $\mu=1$, \(d_{\rm uc}=2\)), i.e.\ 
$\epsilon=1$. We now indicate how the one-loop calculations of the previous sections
can be extended to these cases. 

It turns out that the details of the velocity distribution depend on the precise
form of the elasticity kernel at large scales. This was already the case for the
statics, and in \cite{LeDoussalWiese2008c} we established a general formula for the avalanche size to one loop
as a function of the elastic kernel. This formula was applied in \cite{LeDoussalWiese2009a}
in the case of the contact line. 

Although we sketch below the calculation for an arbitrary kernel, for simplicity we will concentrate on a kernel of  
the form
\be \label{defkernelLR} 
\epsilon(q) = g_q^{-1} = c (q^2 + \mu^2)^{\gamma/2}  \quad , \quad m^2 = c \mu^{\gamma}\ ;
\ee  
we set $c=1$ by a choice of units. The upper critical dimension $d_{\rm uc}=2 \gamma$ is identified by the large-$q$
divergence of
\be 
I_2 = \int_q g_q^2 = C_{d,\gamma} \mu^{-\epsilon} \frac{1}{\epsilon}
\ .\ee  
Here $\epsilon = d_{\rm uc}-d>0,$ and $C_{d,\gamma}=\epsilon \tilde I_2$ where $\tilde I_2=\int_q (q^2+1)^{\gamma/2}$. 
The rescaled disorder parameter is defined by
\be 
\alpha := - \tilde \Delta''(0) = \epsilon \int_q g_q^2 \Delta''(0) 
\ .\ee  
At the fixed point, it reaches, in the limit of small $m$ (small $\mu$), the same value as before, independent of $\gamma$,
\be 
\alpha = - \tilde \Delta^{*\prime \prime}(0) = - \frac{1}{3} (\epsilon-\zeta) + O(\epsilon^2)\ . 
\ee  
Note that the avalanche size becomes
\be
 S_m = m^{-4} \Delta'(0^+) = \mu^{-2 \gamma} \Delta'(0^+) = (\epsilon \tilde I_2)^{-1} \tilde \Delta'(0^+) \mu^{-d+\zeta} 
\ee 
and we refer to \cite{LeDoussalWiese2008c} for more details. We now use dimensionless units
meaning that we express $x$ in units of $1/\mu$, time in units of $\tau_m = \eta_m/m^2$, 
and all velocities in units of $v_m = m^d S_m/\tau_m$ (or $\tilde v_m = L^{-d} S_m/\tau_m$). 
In these dimensionless units the result for the center-of-mass velocity does not change at the tree level, i.e.\ for
mean-field. We will
write the 1-loop result for $Z(\lambda)$, or ${\cal P}(\dot u),$ in the form \bea
&& Z(\lambda) = Z_{\rm MF}(\lambda) +\alpha \frac2{d} \delta Z(\lambda)\ , \label{ZLR}\\
&& {\cal P}(\dot u) = \frac{1}{\dot u} e^{-\dot u} +\alpha \frac2{d} \delta {\cal P}(\dot u)\ , \label{PLR}
\eea
inserting the factor of $2/d$ for later convenience. For SR elasticity $d=d_{\rm uc}=4 $, and one recovers
the previous definition. 

The calculation of Section \ref{s:loops} is easily extended to an arbitrary kernel $g_k$. 
All we have to do is to replace $(k^2+1)$ by $g_k^{-1}$. Let us define,
from formulas (\ref{b31}) and (\ref{b29})\be 
f(y) := \big[ {\cal J}(k,\kappa) + {\cal J}^{\rm ct}(k,\kappa) \big]_{k^2 \to y-1} 
\ .\ee 
Then the result for $\delta Z(\lambda)$  is
\be 
\delta Z(\lambda)  = \frac{1}{\epsilon \tilde I_2} \frac{dS_d}{4} \int_0^\infty \rmd k^2 \, k^{d-2} f\Big(1/g_k\Big)
\ee For the choice $g_k^{-1}=(k^2+1)^{\gamma/2}$ on which we focus from now on, the calculation
can be brought in a form very similar to the case $\gamma=1$ as follows:
\bea
\delta Z(\lambda) &=&\frac{1}{\epsilon \tilde I_2} \frac{dS_d}{4} \int_0^\infty \rmd k^2 \, k^{d-2} f\Big((k^2+1)^{\gamma/2}\Big)
\\ 
&=&\frac{1}{\epsilon \tilde I_2}\frac{dS_d}{2\gamma} \int_1^\infty \rmd y\,y^{2/\gamma-1}  \, (y^{2/\gamma}-1)^{d/2-1} f(y)~~
\nn \eea
Taking the integral to the critical dimension $d=d_{\rm uc}$, and using that $d_{\rm uc}=2\gamma$ and that $\lim_{\epsilon \to 0} \epsilon \tilde I_2=S_{d_{\rm uc}}$ for any $\gamma$,
we arrive at 
\be
\delta Z(\lambda)\Big|_{d=d_{\rm uc}} = \int_1^\infty \rmd y  \,y^{4/d-1}  \, (y^{4/d}-1)^{d/2-1} f(y)
\label{415}\ee
The two cases of most interest are short-ranged elasticity ($\gamma=2$,  $d_c=4$), and long-ranged elasticity of the contact-line or fracture front ($\gamma=1$, \(d_c=2\)). For these cases, Eq.~(\ref{415}) reduces (after a shift from $y$ to $x+1$) to  
\bea
\delta Z(\lambda)\Big|^{\rm SR}_{d=4} &=& \int_0^\infty \rmd x  
 \,  x \,f(x+1) +O(\epsilon) \label{417}\\
\delta Z(\lambda)\Big|_{d=2}^{\rm LR} &=& \int_0^\infty \rmd x \, (x+1) f(x+1)+O(\epsilon)
\ \ \ \ \label{418}\eea
Hence the two calculations are very similar. 
For short-ranged elasticity, the results where given above. For long-ranged elasticity 
($\gamma=1$, \(d_c=2\)), we have plotted the resulting functions for $\delta Z(\lambda)$ and $\delta {\cal P}(\dot u)$ on figures \ref{f:deltaZLR} and  \ref{f:deltaPLR}. More details about the calculation 
and the results are presented in appendix \ref{a:LR}. In particular we find that the exponent of
the small-velocity behavior  changes to\be   
{\sf a} = 1 + 2\alpha + O (\epsilon^{2}) = 1 - \frac{2}{3} (\epsilon-\zeta) + O (\epsilon^{2})\ .  
\ee

\begin{figure}[t]
\Fig{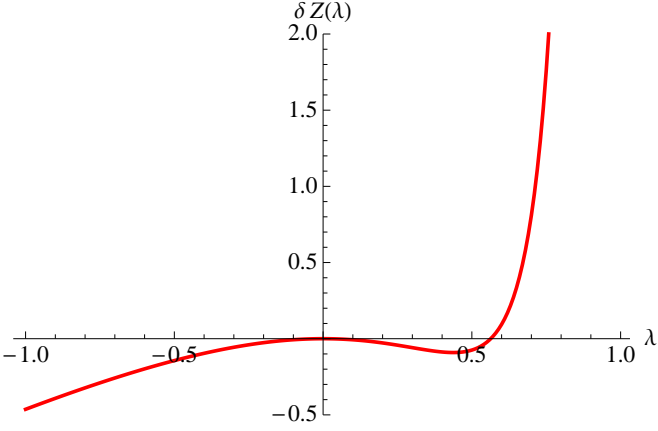}
\caption{\(\delta { Z}(\lambda)\) for LR elasticity ($\gamma=1$, $d_c=2$).
}\label{f:deltaZLR}
\end{figure}\begin{figure}[t]
\Fig{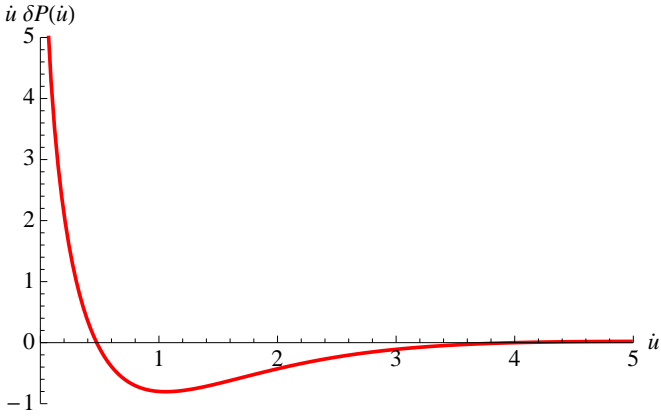}
\caption{\(\dot u\delta {\cal P}(\dot u)\) for LR elasticity  ($\gamma=1/2$,
$d_c=2$). The form of the curve is slightly different from the SR case, e.g.
it crosses zero for $\dot u$ slightly larger.}\label{f:deltaPLR}
\end{figure}

\section{First-principle calculation of generating functions to one loop in the position theory}
\label{s:first-principle}
\subsection{General framework}

Let us now go back to the more conventional formulation of pinned elastic systems formulated in the "position
theory", i.e.\ $u_{xt}$ rather than in the velocity variable $\dot u_{xt}$. 

Let us go back to the original equation of motion in the laboratory frame
\bea
  \int_{x',t'}R^{-1}_{xt,x't'} u_{x't'} &=& \int_{x'} g^{-1}_{xx'} w_{x' t}  + F(u_{xt},x) ~~~~~~~~\\
R^{-1}_{xt,x't'} &=& \delta_{tt'} ( \delta_{xx'} \eta \partial_{t'} - g^{-1}_{xx'} )
\ .
\eea
We want to compute an arbitrary generating function in the position theory \be  \label{G1}
G[\mu,w] = \overline{e^{\int_{xt} \mu_{xt} u_{xt}} } 
\ .
\ee  
It can be written as an expectation value with respect to the dynamical action $S[u,\hat u]$,
\bea 
 G[\mu,w] &=& e^{W[\mu,w]} =   \langle e^{\int_{xt} \mu_{xt} u_{xt} + \int_{xx't} \hat u_{xt} g^{-1}_{xx'} w_{x' t} }  \rangle_{ {\cal S}}  \nn \\
& =& \int {\cal D}[u] {\cal D}[\hat u] e^{-  {\cal S}[u,\hat u] + \int_{xt} \hat u_{xt} g^{-1}_{xx'} w_{x' t}  + \int_{xt} \mu_{xt} u_{xt}}\ . \nn  \\
&&  \label{G2} 
\eea 
This dynamical path integral is normalized to unity, $\int {\cal D}[u] {\cal D}[\hat u] e^{-  {\cal S}[u,\hat u]}=1$. The dynamical
action, now for the displacement \(u\), and a different response field \(\hat u\) instead of $\tilde u $ is\bea
  {\cal S} &=&  {\cal S}_0 +  {\cal S}_{\mathrm{dis}} \\
  {\cal S}_0 &=& \int_{xx'tt'} \hat u_{xt} R^{-1}_{xt,x't'} u_{x't'}  \\
  {\cal S}_{\mathrm{dis}} &=& - \frac{1}{2} \int_{xtt'} \hat u_{xt}  \hat u_{xt'} \Delta(u_{xt} - u_{xt'}) 
\ .
\eea
Note that here we have chosen to consider $w$ as a source and not included it in $ {\cal S}$, although this is
a matter of choice. (Disorder independent) initial conditions are easily specified considering the path integral
with fixed endpoints and convolving with the normalized initial distribution $P[\{u_{x,t=t_0}\}]$.
Non-zero temperature leads to an additional term $- \eta T \int_{xt} \hat u_{xt}^2$ in $ {\cal S}_0$.

To obtain an exact formula for the  observable $G[\mu,w],$ we need to
consider the effective action $\Gamma[u,\hat u],$ associated to $ {\cal S}[u,\hat u]$,
defined in the usual way as a Legendre transform,
\be
\Gamma[u,\hat u] + W[\mu,w] = \int_{xx't} \hat u_{xt} g^{-1}_{xx'} w_{x't} + \int_{xt} \mu_{xt} u_{x t} 
\ .~
\ee
Knowledge of $\Gamma$ allows to obtain our observable as
\be
G[\mu,w] = e^{\int_{xt} \mu_{xt} u_{xt}^{\mu,w} + \int_{xt} \hat u^{\mu,w}_{xt} g^{-1}_{xx'} w_{x' t}  - \Gamma[u^{\mu,w},\hat u^{\mu,w}]}  
\ ,\ee
in terms of the solutions $u^{\mu,w}_{xt}$ and $\hat u^{\mu,w}_{xt}$ of the ``exact" saddle-point equations
\be \label{Gsaddle}
\frac{\delta \Gamma}{\delta u_{xt}}[u,\hat u]= \mu_{xt} \quad , \quad \frac{\delta \Gamma}{\delta \hat u_{xt}}[u,\hat u] =\int_{x'} g^{-1}_{xx'} w_{x't}
\ .
\ee
These solutions are such that\bea
&& \hat u^{\mu,w}_{xt} = \langle \hat u_{xt} \rangle_{\mu,w} = \int_{x'} g_{xx'} \frac{\delta W}{\delta w_{x't}} \\
&&    u^{\mu,w}_{xt} = \langle u_{xt} \rangle_{\mu,w} =  \frac{\delta W}{\delta \mu_{xt}},
\eea
and thus $\hat u^{\mu,w}_{xt}$ vanishes when $\mu=0$. There are other interesting properties. The
covariance of the action under the STS transformation $u_{xt} \to u_{xt} + \phi_x$, $w_{xt} \to w_{xt} + g_{xx'} \phi_{x'}$
implies that $G[\mu,w + g \phi]=e^{\mu \phi} G[\mu,w]$, hence taking a derivative w.r.t. $w$ one finds
the property
\be  \label{identity0}
 \int_{t} g^{-1}_{xx'} \hat u^{\mu,w}_{x't} =  \int_{t} \mu_{xt}
 \ .\ee 
Note that because of the saddle point equation, in the derivative
\be 
\partial_{w_{xt}} W[\mu,w]  = \partial_{w_{xt}} \ln G[\mu,w]  = \int_{x'}g^{-1}_{xx'} \hat u^{\mu,w}_{x't} 
\ee  
one can differentiate only the explicit dependence on $w$.

The effective action can be computed in a loop expansion as follows. Consider $U := (\hat u,u)$ a shorthand notation for the fields.
Then for an action of the form
\be 
 {\cal S}[U] =  {\cal S}_0[U]  +  {\cal S}_{\mathrm{dis}}[U] 
\ee 
the associated effective action can be computed as\be 
\Gamma[\phi] =  {\cal S}_0[U] - \ln \langle e^{-  {\cal S}_{\mathrm{dis}}[U + \delta U]} \rangle^{\rm 1PI}_{ {\cal S}_0}
\ .
\ee  
Here $\langle ... \rangle^{\rm 1PI}_{S_0}$ indicates that averages over $\delta U$ should be 
performed using the action $S_0$ and that one keeps only graphs which are 1-particle irreducible 
w.r.t. the vertex $S_{\mathrm{dis}}$. Hence these diagrams are
sums of 1-loop diagrams.

\subsection{Tree calculation}

It is easy to see that, if one allows only for tree diagrams, one has
\be
\Gamma^{\mathrm{tree}}[U] = \Gamma_0 +  {\cal S}_0[U]  +  {\cal S}_{\mathrm{dis}}[U] = \Gamma_0   +  {\cal S}[U] 
\ ,
\ee
since the only 1PI tree diagram is  the vertex itself. We have defined $\Gamma_0 = \frac{1}{2} \Tr \ln  {\cal S}_0''$
which is just a constant. 

This leads to the tree approximation of $G[\mu,w]$,\be
G^{\mathrm{tree}}[\mu,w] = e^{\int_{xt} \mu_{xt} u_{xt}^{\mu,w} + \int_{xt} \hat u^{\mu,w}_{xt} g^{-1}_{xx'} w_{x' t} -  {\cal S}[u^{\mu,w},\hat u^{\mu,w}]} 
\ , \label{A17}
\ee
where in this Section the $u^{\mu,w}_{xt}$, $\hat u^{\mu,w}_{xt}$ are solution of the saddle-point equation
(\ref{Gsaddle}) with the replacement $\Gamma \to  {\cal S}$, and will also be denoted by 
$u_{xt}^{\mu,w,\mathrm{tree}}$, $\hat u_{xt}^{\mu,w,\mathrm{tree}}$ in the following. 
As is well known, this is the sum of all tree diagrams in perturbation theory of the non-linear part i.e.\ $S_{\mathrm{dis}}$.
It leads to the mean-field theory, as discussed below.

Note that because of the saddle-point equation, in the derivative
\be 
\partial_{w_{xt}} \ln G^{\mathrm{tree}}[\mu,w]  = \int_{t,x'} g^{-1}_{xx'} \hat u^{\mu,w,\mathrm{tree}}_{x't} 
\ee  
one can differentiate only the explicit dependence on $w$. Choosing e.g.\ 
$w_{xt}=v t$ one obtains
\bea
 Z^{\mathrm{tree}}[\mu]  &=& L^{-d} \partial_v G^{\mathrm{tree}}[\mu,w=vt]\Big|_{v=0^+}\nn \\
 &=& \int_{x'} g^{-1}_{xx'} \int_t t \hat u^{\mu,0^+}_{x't} 
\eea 
Here we have set $G^{\mathrm{tree}}[\mu,w=0]=1$, which is not necessarily true,
except if the system is prepared in the Middleton state, which we now  assume. 

\subsubsection{Tree saddle-point equations}

Let us now specialize to $g_q^{-1} = q^2 + m^2$. To tree level we need to solve the following saddle-point equations:\begin{eqnarray} \label{saddles}
&&  \eta_0 \partial_t u_{xt} + (m^2 - \nabla_x^2) (u_{xt} - w_{xt})
\nn\\ &&\qquad -  \int_{t'}  \hat u_{xt'} \Delta(u_{xt} - u_{xt'})  = 0    \\
&& (- \eta_0 \partial_t - \nabla_x^2 + m^2 ) \hat u_{xt} 
-  \int_{t'} \hat u_{xt}  \hat u_{xt'} \Delta'(u_{xt} - u_{xt'}) \nn\\ &&\qquad = \mu_{xt} 
\ .\end{eqnarray} 
Its solution is called $u^{\mu,w},\hat u^{\mu,w}$ only when needed, otherwise $u,\hat u$.  
At non-zero temperature there would be an  additional term
$- 2 \eta_0 T \hat u_{xt}$ on the r.h.s. of Eq.~(\ref{saddles}). Note that $\hat u^{\mu,w}$ vanishes for $\mu=0$. 
We now consider sources $\mu_{xt}$ which vanish at $t=\pm \infty$, hence we also assume that $\hat u_{xt}$ 
vanishes at $t=\pm \infty$. Note also that $u_{xt} \to u_{xt} + \phi(x), w_{xt} \to w_{xt} + (- \nabla_x^2 + m^2 ) \phi(x)$ is a symmetry of the equations (STS). We further have the remarkable property\be  \label{identity}
 (m^2 - \nabla_x^2) \int_{t}  \hat u_{xt} =  \int_{t} \mu_{xt}
 \ ,\ee 
using that $\Delta'(u)$ is an odd function. In the absence of disorder the solution is $\hat u = R^T \mu$ and $u = C \cdot \mu + R \cdot g^{-1} \cdot w$ with $C=2 \eta_0 T R^T R$ and one checks that $G[\mu]^{\mathrm{tree}}= e^{\mu R \cdot g^{-1} \cdot w + \frac{1}{2} \mu \cdot  C \cdot  \mu}$, as expected. 
Taking a time derivative of the first equation, one notes the structure\bea
&& ( R^{-1} + \tilde  \Sigma) \cdot \dot u = g^{-1} \cdot \dot w \\
&& ( R^{-1,T} + \tilde \Sigma) \cdot \hat u =  \mu \\
&& \tilde \Sigma_{xt,x't'} =-  \delta_{xx'}\delta_{tt'} \int_{t''} \hat u_{xt''} \Delta'(u_{xt} - u_{xt''})
~~~~~~~~~~~~~\eea
The scalar product ``$\cdot$'' denotes integration over the common space and time arguments.
We can now compute (\ref{A17}) by substituting the solution of (\ref{saddles}); again using   (\ref{saddles})   it can be simplified into the two equivalent forms\begin{eqnarray} \label{aat} 
 G[\mu]^{\mathrm{tree}} &=& e^{ \int_{xt} \mu_{xt} u^\mu_{xt} - \frac{1}{2} \int_{xtt'} \hat u^\mu_{xt}  \hat u^\mu_{xt'} \Delta(u^\mu_{xt} - u^\mu_{xt'})   } ~~~~~~~~\nn \\
& =& e^{\mu \cdot u - \frac{1}{2} \hat u\cdot R^{-1} \cdot u - \frac{1}{2} \hat u\cdot g^{-1}\cdot w }  
\ .\end{eqnarray}

\subsubsection{Expansion at small driving $w=0^+$}

The solution of the above saddle-point equations can be expanded in powers of $w_{xt}$, assuming that  $f_{xt}=(m^2 - \nabla_x^2) w_{xt}$ is a 
monotonous function of time for
each $x$. We find
\be 
u_{xt}^{\mu,w} = u^0_x + u^1_{xt} + ...  \  , \quad \hat u_{xt}^{\mu,w} = \hat u^0_{xt} + \hat u^1_{xt} + ...
\ .
\ee  
From Middleton's theorem we know that we should look for a solution of the saddle-point equation 
such that $u_{xt}^{\mu,w} - u_{xt'}^{\mu,w} \geq 0$ for $t-t'>0$, hence $u^1$ should satisfy this property.

\paragraph{Lowest order:}
 At lowest order, i.e.\ $w_{xt}=0^+$, the first saddle-point equation leads, 
using (\ref{identity}), to the quasi-static solution\footnote{Note that we expect that there are other solutions corresponding to a non-steady state, e.g.\ solutions with other prescribed boundary conditions. }
\begin{equation} 
    u^0_x  =   \Delta(0) (m^2 - \nabla^2)^{-2}_{xx'} \int_{t'} \mu_{x't'} 
\ .
\end{equation} 
while the second saddle-point equation leads to the ``instanton equation'' for $\hat u$,
\be
 (- \eta_0 \partial_t - \nabla_x^2 + m^2 ) \hat u^0_{xt}  + \sigma \hat u^0_{xt}  \int_{t'}  \hat u^0_{xt'} {\rm sgn}(t-t') 
 = \mu_{xt}
\ .
\ee
where here and below we denote
\be 
\sigma := - \Delta'(0^+)\ , 
\ee  
and we use
\bea
 \Delta'(u_{xt} - u_{xt'}) &=& - \sigma {\rm sgn}(t-t') + \Delta''(0) (u_{xt} - u_{xt'}) \nn \\
&& + O\big((u_{xt} - u_{xt'})^2\big)
\ .
\eea(ii) 
\paragraph{Next order: } To first order in $w_{xt}$ one finds
\bea
 u^1_{xt} &=&  \int_{x',t'} (R^{-1} + \Sigma)^{-1}_{xt,x't'} f_{x't'}   \\
 \hat u^1_{xt} &=& \int_{x',t'} \Delta''(0) \Big[(R^T)^{-1} + \Sigma^T\Big]_{xt,x't'}^{-1}\nn\\ && \times\int_{t_1} \hat u^0_{x't'}  \hat u^0_{x't_1} 
(u^1_{x't'}-u^1_{x't_1})  
\ .
\eea
We have defined
\bea
\Sigma_{xt,x't'} &=&  \delta_{xx'} 
 \sigma \left[ \delta_{tt'} \int_{t_1} {\rm sgn}(t-t_1) \hat u^0_{xt_1} - {\rm sgn}(t-t') \hat u^0_{xt'}  \right]\nn\\ \label{sigmadef}\\
\Sigma^T_{xt,x't'} &=& \Sigma_{x't',xt}
\eea
We also used that $\int_t \hat u^1_{xt} = 0$. Note that
\be 
\int_{t'} \Sigma_{xt,x't'} = 0\ .
\ee  

\subsubsection{Case $\int_t \mu_{xt}=0$ and connection to the velocity theory}

In the velocity theory one is interested in observables (\ref{G1}) such that 
\be  \label{muder}
\int_{t} \mu_{xt}=0 \quad , \quad \mu_{xt}=- \partial_t \lambda_{xt}
\ee  where
$\lambda_{xt}$ vanishes at $t=\pm \infty$. Then Eq.\ (\ref{identity}) implies 
that
\be  \label{uhatder}
\int_t \hat u_{xt}=0 \quad , \quad \hat u_{xt} = - \partial_t \tilde u_{xt}
\ ,
\ee 
where $\tilde u_{xt}$ vanishes at $t=\pm \infty$. Note that at the level of the MSR action one can rewrite
\be 
\int_{xt} \hat u_{xt} R^{-1}_{xt,x't'} u_{x't'} = \int_{xt} \tilde u_{xt} R^{-1}_{xt,x't'} \dot u_{x't'}
\ .
\ee 
The saddle-point equations in the velocity theory then read, after some integrations by part:
\bea
&& ( R^{-1} + \tilde \Sigma) \cdot \dot u = g^{-1} \cdot \dot w = \dot f \\
&& ( R^{-1,T} + \tilde \Sigma) \partial \tilde u = \partial \lambda \\
&& \tilde \Sigma_{xt,x't'} = \delta_{xx'}\delta_{tt'}  \\&&~~~~~~~~~~~~~~~~\times 
\Big[ - 2 \sigma \tilde u_{xt} + \int_{t''} \tilde u_{xt''} \dot u_{xt''} \Delta_{reg}''(u_{xt} - u_{xt''}) \Big] \nn
\ .
\eea 
To lowest order in $w$, i.e.\ for $w=0^+$ we obtain
\begin{eqnarray} 
  \dot u^0_{xt}  &=& 0\ ,  \\
   (\eta_0 \partial_t + \nabla_x^2 - m^2 ) \tilde u^0_{xt}
+ \sigma  (\tilde u^0_{xt})^2   &=&  - \lambda_{xt}   
\ ,~~~~~~~~~~
\end{eqnarray} 
which is exactly the instanton equation (\ref{mfnonlinearxt}), recovered here from first principles. In section \ref{simplified} we have obtained it by neglecting higher derivatives  than the first of $\Delta(u)$; we see
here that the contribution of these derivatives indeed vanishes if one looks at tree diagrams for
$w \to 0$. They do not vanish however to higher orders in $w$, or at non-zero velocity.

We  now go beyond the tree calculation and consider one-loop corrections.

\subsection{1-loop calculation}
\label{s:fond-1loop}
Now we compute $\Gamma[U]$ by including all  tree and one-loop diagrams. It is then easy to see that
\bea
&& \Gamma^{\rm tree+1-loop}[\phi] =  {\cal S}[U] + \Gamma^{1}[U]\ , \\
&& \Gamma^{1}[U] = \frac{1}{2} \Tr \ln  {\cal S}''[U] - \frac{1}{2} \Tr \ln  {\cal S}_0''[U]\ ,~~~~~~
\eea 
and we assume $\Gamma^{1}[U]$ to be small. Let us denote $\Lambda:=(g^{-1} w,\mu).$ The saddle-point equations are thus
\bea
&&  {\cal S}'[U^{\mathrm{tree}}]= \Lambda \label{vartree} \ ,\\
&&   {\cal S}'[U] + (\Gamma^{1})'[U] = \Lambda
\ ,\eea 
hence $U=U^{\mathrm{tree}}+ O(\Gamma^1)$. To compute
\be 
G = e^{\Lambda U -  {\cal S}[U] - \Gamma^{1}[U] } 
\ee  
we can thus consider $\Gamma^1$ as an explicit perturbation and 
to the same accuracy, i.e.\ neglecting terms of order $(\Gamma^1)^2$,
\be 
G = G^{\mathrm{tree}} e^{- \Gamma^1[U^{\mathrm{tree}}]} 
\ .
\ee  
Going back to our explicit notations,  we thus need to compute
\be 
G[\mu,w] = G^{\mathrm{tree}}[\mu,w] e^{- \Gamma^1[\hat u^{\mu,w,\mathrm{tree}}, u^{\mu,w,\mathrm{tree}}]}
\ .\ee  

\subsection{Explicit calculation}

From now on, we  focus on velocity observables, i.e.\ the case
\be  \label{mu0} 
\int_t \mu_{xt} = 0   \quad , \quad \mu_{xt} = - \partial_t \lambda_{xt}
\ ,\ee  
for which  (\ref{muder}) and (\ref{uhatder}) hold, and will be used extensively below. One thus has 
that $\Gamma_1=0$ for $w=0^+$. In this section $U=(\hat u,u)$ denotes $U^{\mathrm{tree}}=(\hat u_{xt}^{\mu,w,\mathrm{tree}},u_{xt}^{\mu,w,\mathrm{tree}}), $
and all derivatives are taken at the tree saddle point. 

To compute $Z(\lambda)$ we need to expand to first order in $w$. The small-$w$ dependence of
$U^{\mathrm{tree}}$, denoted $U$ here, can be obtained from (\ref{vartree}):
\bea
 U &=&U^0+U^1 \cdot w + O(w^2) \\
 U^1 &=&  ( {\cal S}'')^{-1}  (g^{-1} \cdot w,0) 
\ .~~~~~~~
\eea
We need to compute
\begin{eqnarray}
\Gamma_1 &=& \frac{1}{2} \Tr\big ( \ln  {\cal S}''[U]\big) -\Tr\big ( \ln R^{-1} \big)\nn \\
&=&  \frac{1}{2}  ( {\cal S}'')^{-1}_{\alpha \beta}  {\cal S}'''_{\alpha \beta \gamma} U^1_\gamma + \dots \nn \\
&=&  \frac{1}{2}  ( {\cal S}'')^{-1}_{\alpha \beta}  {\cal S}'''_{\alpha \beta \gamma} ( {\cal S}'')^{-1}_{\gamma \hat u} \cdot g^{-1} \cdot w + O(w^2)\ .\label{G1sum} 
~~~~~~~~\end{eqnarray}
For now, we ignore  the quadratic substraction. Greek indices denote either $\hat u$ or $u$ and all contractions
are implicit. 

At this stage this is still general enough to treat a non-uniform $\lambda_{xt}$. However for
simplicity we will now focus  on the case of a uniform $\lambda_{xt} = \lambda_t$, i.e.\ on center-of-mass observables. The saddle-point solution is then uniform and we
denote $\hat u_{xt} = \hat u^0_{t} = - \partial_t \tilde u^0_{t}$.  It is independent of $w$. 

We need first $ {\cal S}''$, the matrix of second derivatives. It is computed in Appendix \ref{app:second}
for general $U$, then specified for $U^{\mathrm{tree}}$ for general $\mu$. Here we need it only in the case
(\ref{mu0}), and for a uniform $\lambda$, hence we can use $\int_t \hat u^0_{t}=0$ and it  simplifies further into
\begin{eqnarray}
  {\cal S}''_{\hat u \hat u} &=& 0 \\
 ( {\cal S}''_{uu})_{xt,x't'} &=& \delta_{xx'} \Delta''(0) \hat u_{t}^0 \hat u_{t'}^0  \\
  {\cal S}''_{\hat u u} &=& R^{-1} + \Sigma \\
   {\cal S}''_{u \hat u} &=& (R^T)^{-1} + \Sigma^T 
\ .~~~
\end{eqnarray}
The ``self-energy'' $\Sigma$ is 
 defined in (\ref{sigmadef}), and reads
\bea
 \Sigma_{xt,x't'}&=& \delta_{xx'} \Sigma_{tt'} \\
 \Sigma_{tt'} &=& \sigma [ \delta_{tt'} \int_{t_1} {\rm sgn}(t-t_1) \hat u^0_{t_1} - {\rm sgn}(t-t') \hat u^0_{t'}  ] \nn \\
&=& - \sigma [ 2 \delta_{tt'} \tilde u^0_t - {\rm sgn}(t-t') \partial_{t'} \tilde u^0_{t'} ]
\ .\eea 
Note that
\be  \label{propsigma} 
\int_{t'} \Sigma_{t,t'} = 0
\ .
\ee  
The first component is actually $\Delta(0),$ but can never appear for velocity observables; hence we dropped it.
To pursue, we define the dressed response
\be 
   {\cal R} = (R^{-1} + \Sigma)^{-1}
\ .\ee 
In Fourier
\be 
({\cal R}_{k})_{ tt'}\equiv {\cal R}_{k tt'} := \big(R^{-1}_{k} + \Sigma\big)_{tt'}^{-1} 
\ ,
\ee 
with $(R^{-1}_{k})_{tt'}=R^{-1}_{ktt'}$. This dressed response
is related to the one defined in (\ref{defmathR}) and  (\ref{R-gen}), 
\begin{equation}\label{b43}
 \mathbb{R}_{ktt'}: =   \theta (t-t')\,\rme^{-(k^2+1) (t-t') + 2 \int^{t}_{t'} \rmd s  \tilde u^0_s  }
\ .\end{equation}
Namely one has
\be 
 {\cal R}_{k tt'}  \approx (\partial_t)^{-1} \mathbb{R}_{ktt'} \partial_{t'}
\ ,
\ee  
where the $\approx$ means that it is true up to a zero mode. The correct identity,
proven in Appendix \ref{app:dressed}, reads\begin{eqnarray}\label{b44}
\int_{t'}{\cal R}_{k tt'}\phi_{t'} &=& \int_{t'}  (\partial_t)^{-1}
\mathbb{R}_{ktt'}\partial_{t'} \left(\phi_{t'}-\phi_{-\infty} \right)
+\frac{1}{k^{2}+1} \phi_{-\infty } \nn \\
&& 
\eea
upon acting on a test function $\phi_t$. This implies the following property
\be  \label{propertymagic}
\partial_{t} {\cal R}_{k tt'}  =  {\mathbb R}_{k tt'}\partial_{t'}
\end{equation}
used extensively below. The above relations  arise because we
are working in the position theory in a case where we compute velocity
observables. 

We now need the inverse second-derivative matrix. One can first invert the $2 \times 2$
block structure
\begin{eqnarray}
( {\cal S}'')^{-1}_{\hat u \hat u} &=& - ( {\cal S}''_{u \hat u})^{-1}  {\cal S}''_{uu} ( {\cal S}''_{\hat u u})^{-1} = - {\cal R}^T  {\cal S}''_{uu} {\cal R}
\nn \\
 ( {\cal S}'')^{-1}_{\hat u u} &=& ( {\cal S}''_{u \hat u})^{-1}  = {\cal R}^T \nn \\
 ( {\cal S}'')^{-1}_{u \hat u} &=& ( {\cal S}''_{\hat u u})^{-1} = {\cal R} \nn
 \\
 ( {\cal S}'')^{-1}_{uu} &=& 0\ ,
 \end{eqnarray}
where the inversions on the r.h.s.\ refers only to the space and time dependence. Given that in addition one has $ {\cal S}'''_{\hat u \hat u \hat u}=0,$ there are only
three distinct terms in the sum (\ref{G1sum}) of order $O(w), $ and which we denote
\begin{eqnarray}
\delta \Gamma_1 &=&   \frac{1}{2}   ( {\cal S}'')^{-1}_{\hat u \hat u}  {\cal S}'''_{\hat u \hat u u} ( {\cal S}'')^{-1}_{u \hat u} \cdot g^{-1} \cdot w 
\nn \\
&&  +  ( {\cal S}'')^{-1}_{\hat u u}  {\cal S}'''_{\hat u u u} ( {\cal S}'')^{-1}_{u \hat u} \cdot g^{-1} \cdot  w\nn \\
&&
  +  ( {\cal S}'')^{-1}_{\hat u u}  {\cal S}'''_{\hat u u \hat u} ( {\cal S}'')^{-1}_{\hat u \hat u} \cdot g^{-1} \cdot w  \ .\end{eqnarray}
We now specify   to a uniform $w_{xt} = w_t$. The third derivative tensor is
computed in  Appendix \ref{app:third}. It is important to note that $ {\cal S}_{uu}''$ and all components of
$ {\cal S}'''$ are local in space, i.e.\ $ {\cal S}'''_{xt,x't',x_1,t_1} = \delta_{xx'x_1}  {\cal S}'''_{t,t',t_1}$. We can then make the
momentum structure more explicit, using
the above second-derivative matrix, and write
\bea \label{3Gamma}
&& \delta \Gamma_1 = \delta \Gamma_1^{(1)} + \delta \Gamma_1^{(2)}  + \delta \Gamma_1^{(3)} \\
&& \delta \Gamma_1^{(1)}  =  - \frac{1}{2} m^2 \int_k \big[  {\cal R}^T_k \cdot   {\cal S}''_{uu} \cdot {\cal R}_k \big]_{tt'} 
\big[  {\cal S}'''_{\hat u \hat u u} \cdot  {\cal R}_0 \cdot  w\big]_{tt'}  \nn  \\
&&  \delta \Gamma_1^{(2)}  = m^2 \int_k  \big[ {\cal R}^T_k \big]_{tt'}  \big[  {\cal S}'''_{\hat u u u} \cdot  {\cal R}_0 \cdot  w\big ]_{tt'}  \nn \\
&&  \delta \Gamma_1^{(3)}  = - m^2 \int_k \big[ {\cal R}^T_k \big]_{tt'}  \big[  {\cal S}'''_{\hat u u \hat u} \cdot {\cal R}^T_0 \cdot  {\cal S}''_{uu}  \cdot  {\cal R}_0 \cdot  w \big]_{tt'}  \nn
\ .
\eea
All three terms  are matrices in the time indices only, i.e.\ 
\bea \label{3der2}
&& \big[ {\cal S}''_{uu}\big]_{tt'} = \Delta''(0) \hat u_{t}^0 \hat u_{t'}^0 \ ,\\
&& \big[ {\cal S}'''_{\hat u \hat u u}\big]_{tt't_1} = \sigma (\delta_{t t_1} - \delta_{t' t_1}) {\rm sgn}(t-t')\ , \nn \\
&& \big[ {\cal S}'''_{\hat u u u}\big]_{tt't_1} = - \Delta''(0) \big[(\delta_{t't_1}  - \delta_{tt_1}) \hat u^0_{t'}  - \delta_{tt'} \hat u^0_{t_1}  \big]\ , \nn
\eea 
and $[ {\cal S}'''_{\hat u u \hat u}]_{tt't_1}=[ {\cal S}'''_{\hat u \hat u u}]_{tt_1t'}$. Note that we can define
\be 
{\cal R}_t := \int_{t'} {\cal R}_{0tt'} w_{t'}
\ee  
and replace it above since it appears on the right in all three
terms (\ref{3Gamma}).

We now specify to the choice of most
interest for us here, namely the driving at small but finite constant velocity $w_{t'}=v t'$. In that case
${\cal R}_t$ is not a well-behaved expression,
since it may contain an additive term in the position of the parabola. Fortunately,
in the calculation below, using  (\ref{propertymagic}) only the following combination
will appear:\begin{equation}\label{8.352}
\partial_t {\cal R}_t = v \int_{t'} \mathbb R_{0tt'} = v \int_{t'<t} e^{-m^2 (t-t') + 2 \int^{t}_{t'} \rmd s \tilde u^0_s  }
.\end{equation}
In particular,\begin{equation}\label{b45}
\lim_{t\to -\infty} \partial_t {\cal R}_t  =v\int_{t_{2}<t}R_{0,tt_{2}} = \frac{v}{m^{2}} 
\ ,
\end{equation}
since $\tilde u^{0}_{s} \to 0$ for $s\to -\infty$. 

It is shown in  Appendix \ref{a:ThirdDiagram} that the third term vanishes,
\begin{equation}
 \delta \Gamma^{(3)}_{1} = 0
\ .
\end{equation}
Hence we only need to compute two contributions. 

Substituting (\ref{3der2}) into (\ref{3Gamma}) we compute the first term,
\begin{eqnarray}
 \delta \Gamma_1^{(1)} &=& - \frac{1}{2} \Delta''(0)  \sigma
m^2 \int_{k,t,t',t_1,t_2} {\cal R}_{k t_1t}  \partial_{t_1} \tilde u^0_{t_1} \partial_{t_2}  \tilde u^0_{t_2} \nn\\
&&~~~~~~~~~~~~~~~~~~~~~~~~~~~~\times {\cal R}_{k t_2 t'}
 ({\cal R}_{t}-{\cal R}_{t'}) {\rm sgn}(t-t') \nn\\
&  =& - \frac{1}{2} \Delta''(0)  \sigma
m^2 \int_{k,t,t',t_1,t_2} 
  {\mathbb{R}}_{kt_1t} {\mathbb R}_{kt_2t'}  
  \tilde u^0_{t_1}  \tilde u^0_{t_2} 
    \nn\\
&& ~~~~~~~~~~~~~~~~~~~~~~~~~~~~\times  \partial_{t}  \partial_{t'}({\cal R}_{t}-{\cal R}_{t'}) {\rm sgn}(t-t') \nn \\
  & =&   \Delta''(0)  \sigma
m^2 \int_{k,t,t_1,t_2} 
 \mathbb{R}_{kt_1t}\mathbb{ R}_{kt_2t} 
  \tilde u^0_{t_1}  \tilde u^0_{t_2} 
   \partial_{t} {\cal R}_{t} 
  \nn\\ &=& v \Delta''(0)  \sigma
m^2 \int_{k,t'<t<0} {\mathbb R}_{0tt'} \Phi(k,t)^{2}
\ .\end{eqnarray}
To obtain the second line we have integrated by part over $t_1$ and $t_2$ and used (\ref{propertymagic}).
No boundary terms are generated since $\tilde u_t$ vanishes at $t=\pm \infty$. We used that $\partial_{t}  \partial_{t'}  ({\cal R}_{t}-{\cal R}_{t'})
{\rm sgn}(t-t') = - \partial_t \partial_{t'} {\cal R}_{t'} {\rm sgn}(t-t') = - 2 \partial_{t'} {\cal R}_{t'} \delta(t-t')$, i.e.\ there is 
a factor of $2,$  not  $4$. This is the first term obtained in Eq.~(\ref{tilde-u2}).

Graphically, this can be written as 
\begin{eqnarray}
\delta \Gamma_1^{(1)}&=& 2\ \parbox{3.2cm}{\begin{tikzpicture}
\draw[double distance=1.7pt] (0,1) -- (0,0);
\draw[-to,shorten >=10pt,double distance=2pt] (0,1) -- (0,0);
\draw[double distance=2pt] (1,1) -- (1,0);
\draw[-to,shorten >=10pt,double distance=2pt] (1,1) -- (1,0);
\draw[double distance=2pt] (2,0) -- (1,1);
\draw[-to,shorten >=14pt,double distance=2pt] (2,0) -- (1,1);
\draw [dashed] (0,0) --  (1,0);
\draw [dashed] (0,1) --  (1,1);
\node (t1) at (-.3,0) {$t_{1}$};
\node (t) at (-.3,1) {$t$};
\node (t2) at (1.3,0) {$t_{2}$};
\node (tp) at (1.4,1) {$t'$};
\node (w) at (2.4,0) {$\displaystyle w_{t_{5}}$};
\fill (0,0) circle (2pt);
\fill (0,1) circle (2pt);
\fill (1,0) circle (2pt);
\fill (1,1) circle (2pt);
\fill (2,0) circle (2pt);
\end{tikzpicture}} +
\ \parbox{3.2cm}{\begin{tikzpicture}
\draw[double distance=2pt] (0,1) -- (0,0);
\draw[-to,shorten >=10pt,double distance=2pt] (0,1) -- (0,0);
\draw[double distance=2pt] (1,1) -- (0,0);
\draw[-to,shorten >=14pt,double distance=2pt] (1,1) -- (0,0);
\draw[double distance=2pt] (02,0) -- (1,1);
\draw[-to,shorten >=14pt,double distance=2pt] (2,0) -- (1,1);
\draw [dashed] (0,0) --  (1,0);
\draw [dashed] (0,1) --  (1,1);
\node (t1) at (-.3,0) {$t_{1}$};
\node (t) at (-.3,1) {$t$};
\node (t2) at (1.3,0) {$t_{2}$};
\node (tp) at (1.4,1) {$t'$};
\node (w) at (2.4,0) {$\displaystyle w_{t_{5}}$};
\fill (0,0) circle (2pt);
\fill (0,1) circle (2pt);
\fill (1,0) circle (2pt);
\fill (1,1) circle (2pt);
\fill (2,0) circle (2pt);
\end{tikzpicture}}\nn\\
&& +
\ \parbox{3.2cm}{\begin{tikzpicture}
\draw[double distance=2pt] (0,1) -- (1,0);
\draw[-to,shorten >=14pt,double distance=2pt] (0,1) -- (1,0);
\draw[double distance=2pt] (1,1) -- (1,0);
\draw[-to,shorten >=10pt,double distance=2pt] (1,1) -- (1,0);
\draw[double distance=2pt] (2,0) -- (1,1);
\draw[-to,shorten >=14pt,double distance=2pt] (2,0) -- (1,1);
\draw [dashed] (0,0) --  (1,0);
\draw [dashed] (0,1) --  (1,1);
\node (t1) at (-.3,0) {$t_{1}$};
\node (t) at (-.3,1) {$t$};
\node (t2) at (1.3,0) {$t_{2}$};
\node (tp) at (1.4,1) {$t'$};
\node (w) at (2.4,0) {$\displaystyle w_{t_{5}}$};
\fill (0,0) circle (2pt);
\fill (0,1) circle (2pt);
\fill (1,0) circle (2pt);
\fill (1,1) circle (2pt);
\fill (2,0) circle (2pt);
\end{tikzpicture}}
\ .
\end{eqnarray}
Only the first term is non-zero. 

For $ \delta \Gamma_1^{(2)}$, we find
\begin{eqnarray}
  \delta \Gamma_1^{(2)} &=& m^2 \int_{k,t,t'} {\cal R}_{ktt'} [ {\cal S}'''_{\hat u u u}]_{t'tt_1} {\cal R}_{t_1} \\
& =&  - m^2 \Delta''(0) \int_{k,t,t'} {\cal R}_{ktt'} \partial_{t} \tilde u^0_{t} ( {\cal R}_{t'}
- {\cal R}_{t}) \nn\\
& =&  m^2 \Delta''(0) \int_{k,t,t'}    \tilde u^0_{t}  \partial_{t}[{\cal R}_{ktt'} ( {\cal R}_{t'}
- {\cal R}_{t}) ]\nn\\
& =&  m^2 \Delta''(0) \int_{k,t,t'} \left[   \tilde u^0_{t}
{\mathbb R}_{ktt'}  \partial_{t'} {\cal R}_{t'}
 -  \tilde u^0_{t}  {\cal R}_{ktt'} \partial_{t}
{\cal R}_{t} \right]\ . \nn
\end{eqnarray}
We have used that  ${\cal R}_{tt}=0$, absence of boundary terms $[{\cal R}_{ktt'} \tilde u^0_{t} ( {\cal R}_{t'}
- {\cal R}_{t})]_{t=-\infty}^{t=+\infty}=0$ and $\partial_t {\cal
R}_{t'}=\partial_{t'}{\cal R}_{t}=0$. We have also employed (\ref{propertymagic}). 

Now we can use (\ref{b44}) with $\phi_t=1$ which gives $\int_{t'} {\cal R}_{ktt'} =  \int_{t'} R_{ktt'}  = (1+k^2)^{-1}$
and obtain, using (\ref{8.352}) and (\ref{b45})
\begin{eqnarray}\label{c3}
  \delta \Gamma_1^{(2)} &=& m^2 v\Delta''(0)\bigg[ \int_k \int_{t_{2}<t'<0}
 {\mathbb R}_{0t't_{2}} \Phi(k,t')~~~~~~~~~~~~~\nn \\
 &&~~~~~~~~~~~~~~~~~~ -\int_{k} \frac{1}{k^{2}+1}
 \int_{t_{2}<t}  \tilde u^0_{t}\, {\mathbb R}_{0tt_{2}}   \bigg]
\ .
\end{eqnarray} 
The first term is exactly the term proportional to the single
$\Phi(k,t)$ in our previous calculation (\ref{tilde-u2}).
The last term can be calculated, recalling the definition $\kappa :=-\frac{\lambda}{1-\lambda }$
\begin{eqnarray}\label{c6}
 \delta \Gamma_1^{(2b)} &=& - m^2 v\Delta''(0)\int_{k} \frac{1}{k^{2}+1} \int_{-\infty}^0 dt_2 \int_{t_2}^0 dt 
   \tilde u^0_{t}\, {\mathbb R}_{0tt_{2}}  \nn \\
& =& m^2 v\Delta''(0) \int_{k}   \frac{\kappa}{k^{2}+1}
 \end{eqnarray}
Graphically, this can be written as
\begin{equation}
 \delta \Gamma_1^{(2)} = \parbox{3.2cm}{\begin{tikzpicture}
\draw[double distance=2pt] (1,1) arc (0:-180:0.5);
\draw[->,double distance=2pt] (1,1) arc (0:-90:0.5);
\draw[double distance=2pt] (2,0) -- (1,1);
\draw[-to,shorten >=14pt,double distance=2pt] (2,0) -- (1,1);
\draw [dashed] (0,1) --  (1,1);
\node (t) at (-.3,1) {$t$};
\node (tp) at (1.4,1) {$t'$};
\node (w) at (2.4,0) {$\displaystyle w_{t_{5}}$};
\fill (0,1) circle (2pt);
\fill (1,1) circle (2pt);
\fill (2,0) circle (2pt);
\end{tikzpicture}}+ \parbox{3.2cm}{\begin{tikzpicture}
\draw[double distance=2pt] (1,1) arc (0:-180:0.5);
\draw[->,double distance=2pt] (0,1) arc (-180:-90:0.5);
\draw[double distance=2pt] (2,0) -- (1,1);
\draw[-to,shorten >=14pt,double distance=2pt] (2,0) -- (1,1);
\draw [dashed] (0,1) --  (1,1);
\node (t) at (-.3,1) {$t'$};
\node (tp) at (1.4,1) {$t$};
\node (w) at (2.4,0) {$\displaystyle w_{t_{5}}$};
\fill (0,1) circle (2pt);
\fill (1,1) circle (2pt);
\fill (2,0) circle (2pt);
\end{tikzpicture}}
\end{equation}
We can now put all together and obtain
\be 
Z(\lambda) = Z^{\mathrm{tree}}(\lambda) - \lim_{v \to 0} \frac{\Gamma^1}{v} 
\ ,
\ee  
which coincides with the result (\ref{J}), (\ref{Jt}) apart from the
additional contribution $A \kappa \int_k (1+k^2)^{-1}$. This contribution, equivalent to (\ref{q20}) and (\ref{q21}),
exactly cancels the $O(\kappa)$ in $Z(\lambda)$ to one loop, as it
should and automatically removes the quadratic divergence. It
is thus exactly the quadratic counter-term. While in Section \ref{s:corrections}, it came via some manipulations on the seemingly vanishing term $\Delta'''\big(v(t-t')+u_{xt}-u_{xt'}\big)$, in the  present calculation
it appears automatically, and is related to the zero mode of the velocity theory.

\section{Conclusion}

In this  article  we  presented in  detail the novel tools and methods which allow to
calculate the statistics of  velocities in an avalanche for the prototypical model of an elastic interface
driven in a random environment. It is the extension to the dynamics of our work on  static avalanches, and the
quasi-statics reveals to be closely connected, albeit different, from the statics. The dynamical observables are much richer
as we aim to calculate  many-time correlations. The problem of how to define an avalanche, and the steady state
measure for avalanche statistics, is addressed and allows to make progress. At the same time connections
to avalanches following a kick, or non-stationary avalanches are discussed. The Middleton theorem,
which allows to order all realized  configurations in time, plays a crucial role at all stages of the derivation. 

Our construction starts by identifying the correct mean-field theory, valid in space dimensions $d \geq d_{\rm uc}$.
We discover that it is given, up to renormalization of a few parameters, by a simple tree theory, itself equivalent to a non-linear instanton
 equation. This tree theory is interesting in itself. For the center of mass of the interface it {\em exactly} reproduces  the ABBM model;  it
settles an important question concerning the validity of the ABBM\  model, introduced before as a toy model. The full space-time statistics of the
velocity field is found to be given by the Brownian force model (BFM). This model is  exactly solvable,  reducing the problem
to solving a space-time-dependent instanton equation. Our methods allow to obtain a host of new results for the
probability distributions at several times and a number of results at non-zero wave vector $q$, which go beyond the ABBM model.
A salient result is the time asymmetry of the avalanche shape, which, within mean field, manifests itself at the local level (or at non-zero $q$) but
not for the center of mass. The universality of our results is discussed and quantified.

Continuing to 1-loop order, we obtain the distribution of instantaneous velocities in an avalanche for an elastic manifold, as e.g.\ a magnetic domain wall, 
driven through disorder. These results are new, and have never been addressed before. They are the basis for further work on the avalanche duration and shape, beyond mean-field theory.

Many of the results of the present article can be confronted to experiments, and for this purpose we have extended
them to long-range elastic kernels which are ubiquitous in nature. There are numerous experimental systems at their upper
critical dimensions (e.g.\ magnets) and non-zero $q$ observables have not been measured and discussed previously. For 
other classes of systems below their upper critical dimension, the techniques introduced here provide a novel and at present
the only way to attack them.

Let us list a few important prospects for the future. Since we now know how to describe the space-time structure of avalanches within the mean field
theory, using the Brownian force model, it would be interesting to develop analytical and numerical techniques  to solve its evolution, and solve 
the space-time dependent 
instanton equation beyond what has been done here. This should yield a detailed description of the space-time processes
involved in an avalanche, and shed light on their physics.  Avalanches have similarities as well as differences with branching processes, and  the 
spatial shape of an avalanche is  an important observable for experiments. We have voluntarily focused on the small
driving-velocity limit, since at present the FRG is better controlled in that limit, but an important challenge is to understand
the finite-$v$ behavior, and in particular whether the $v$-dependent avalanche exponents present in the 
ABBM model survive beyond mean-field theory. Other more far-reaching issues are to treat non-monotonous driving, hysteresis
and to extend the theory for systems which do not obey in an obvious way  Middleton's theorem.

\acknowledgements
We are grateful to Alexander Dobrinevski for numerous useful remarks. We thank Andrei Fedorenko, Alejandro Kolton and Alberto Rosso for stimulating discussions. 
This work was supported by  ANR Grant No.\ 09-BLAN-0097-01/2.
We thank the KITP for hospitality and partial support
through NSF Grant No.~PHY11-25915.
\appendix

\section{Laplace inversion for a time window} \label{app:laplinv}

We give here the inverse Laplace transform (\ref{laplaceU}) in a series representation.
By inspection we find that for any finite $T$ the LT 
has simple poles on the negative real axis at $s=s_n < -1/4$, $n=1,...$ the closest one to zero
crosses over from $s_1(T) = - 1/T$ at small $T$ to $s_1 = -1/4$ at large $T$. 
Since all $s_n<-1/4$ we can write $s=- \frac{1+x^2}{4}$. Noting $x=\tan \psi$ the poles
are solutions of $- \psi_n = \frac{T}{4} \tan \psi_n - n \pi/2$. The function $s_n(T)$
is better represented as a function of $s_n$, 
\be 
T = \frac{4 \big[ \frac{n \pi}{2} - {\rm arctan}(\sqrt{- 1-4 s_n} )\big]}{\sqrt{- 1-4 s_n}} \leftrightarrow s=s_n(T) 
\label{B2a}\ee  
represented in Figure \ref{f:TofSn}.
\begin{figure}\Fig{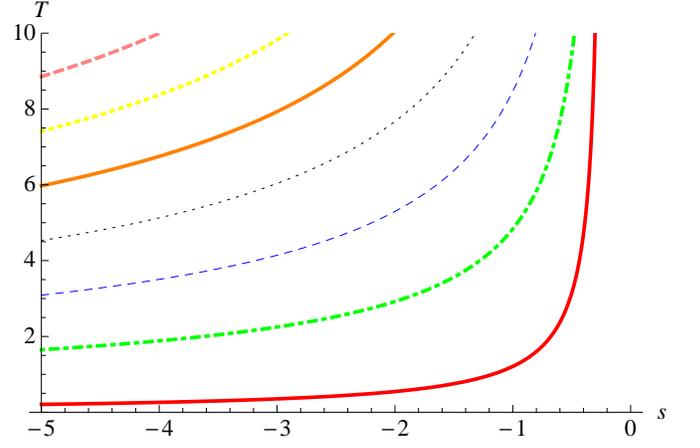}
\caption{The function \(T(s)\) defined in (\ref{B2a}) for \(n=1\) (red, thick, lower right curve) up to \(n=7\) (upper left curve).}
\label{f:TofSn}
\end{figure}
Now we can compute the residues and using the equation satisfied by the poles
we find, amazingly, that they are 
all simply all equal to $1/T$. Hence
\be 
{\calP}(U) = \frac{1}{T U} \sum_{n=1}^\infty e^{- |s_n(T)| U} 
\ .
\ee
The small-$T$ behavior of the poles is 
\bea
 |s_1(T)| &=& \frac{1}{T}+ \frac{1}{6} + O(T)\ , \\
  |s_n(T)| &=& \frac{(n-1)^2 \pi^2}{T^2}+ \frac{2}{T} + \left(\frac{1}{4} - \frac{1}{\pi^2 (n+1)^2}\right) \nn\\
&&+ O(T)\ . 
\eea 
Hence at small $T$ we get
\be 
{\calP}(U) \approx \frac{1}{T U} e^{- U/T} 
\ ,\ee 
consistent with the velocity distribution, as discussed in the text. For large $T$ the poles behave as
\be 
|s_n(T)| =
\frac{1}{4}+\frac{\pi ^2 n^2}{T^2}-\frac{8 \left(\pi ^2
   n^2\right)}{T^3}+\frac{48 \pi ^2 n^2}{T^4}+O\left(\frac{1}{T^5}\right)\ . \nn
\ee 
To leading order at large $T$ one can keep only the first two  terms, and
approximate the sum by an integral, which reproduces the
correct asymptotic result
\be 
{\calP}(U) \approx \frac{1}{T U} \int_0^\infty \rmd n\,  e^{\frac{U}{4} -\frac{\pi ^2 n^2}{T^2} U} = \frac{1}{2 \sqrt{\pi} U^{3/2}} e^{-U/4} 
\ ,\ee  
equal to the avalanche-size distribution as discussed in the main text.

\section{Irrelevant operators and response function}
\label{app:frg} 
The effective action of the position theory in the laboratory frame can be written in an expansion in powers of the response field $\hat u$ as
\be
\Gamma[\hat u, u] = \sum_{p=1}^\infty \frac{1}{p!} \int_{x_i,t_i} \hat u_{x_1 t_1}...\hat u_{x_p t_p} \Gamma_{\hat u_{x_1 t_1}..\hat u_{x_p t_p}}[u]\ .
\ee
  The  term $p=1$ expanded to linear order, $\Gamma_{\hat u_{x t}}[u]={\cal R}^{-1}_{x-x',t-t'} u_{x't'} + O(u^2)$ 
defines the exact inverse response function. Expanding the latter in time derivatives
defines the renormalized dynamical parameters, more conveniently expressed in the frequency domain,
\be 
   {\cal R}_{q=0,\omega}^{-1} := m^2 + \eta i \omega + \sum_{n=2}^\infty D_n (i \omega)^n
\ .\ee 
Similarly, in the limit $v=0^+$ the  local time-persistent   part of the term $p=2$ 
defines the renormalized second cumulant of the disorder 
$\Delta(u)$, \be
 \lim_{t \gg t' } \int_{xx'} \Gamma_{\hat u_{x t},\hat u_{x' t'}}[\{u_{xt}=u_t \}]= L^d \Delta(u_{t}-u_{t'})\ .
\ee
Similar definitions hold for the $p$-th disorder cumulant $\hat C^{(p)}$ from the term or order $p$ in $\Gamma$. 
All renormalized quantities depend implicitly on $m$. 

The main point is that near $d=d_{\rm uc}$ and in the limit $m \to 0,$ the only relevant terms, i.e.\ operators in  $\Gamma$  are 
$\eta$ and $\Delta(u),$ irrespective of the details of the bare model. For $d=d_{\rm uc}-\epsilon$, $\epsilon>0$, all other pieces of $\Gamma$ are irrelevant, i.e.\ higher orders in
$\epsilon$. For $d=d_{\rm uc}$ they are higher powers in $1/\ln(\Lambda/m)$.

In Refs.\ \cite{LeDoussal2008,LeDoussalWiese2008c,LeDoussalWiese2011b}, this property was discussed in detail for the disorder-part of $\Gamma$, for instance that 
the dimensionless (i.e.\ rescaled by
the appropriate power of $m$) higher cumulants $\hat C^{p} = O(\epsilon^p)$ for $p \geq 3$, and similarly, that the 
non-local part of the second disorder cumulant is $O(\epsilon^2)$. Since the local second cumulant $\Delta = O(\epsilon)$,
it implies that the renormalized disorder $\hat V = O(\sqrt{\epsilon})$ is local and gaussian, 
and that all other disorder operators are irrelevant. 

Let us thus discuss here the dynamical part of $\Gamma$, and consider the
dynamical coefficients $D_n$, as examples of irrelevant operators. For concreteness we restrict to SR elasticity with $d_{\rm uc}=4$.
The perturbative correction to the inverse response function reads, to lowest order in the disorder (see e.g. \cite{BalentsLeDoussal2003})
\be 
\delta R^{-1}_{q=0,\omega} = - \int_q \int_0^\infty \frac{\rmd t}{\eta} e^{- (q^2 + m^2) \frac{t}{\eta}} (1- e^{- i \omega t}) \Delta''(0^+)\ , 
\ee  
which leads in the limit of $v \to 0^+$ to 
\bea \label{PT0}\delta \eta &=& - \eta I_2 \Delta''(0^+)\ , \\
 \delta D_n &=& (-1)^n \eta^n I_{n+1} \Delta''(0)\  \\
I_n&=&\int_k^{\Lambda} (k^2 + m^2)^{-n}\ .
\label{PT}
\eea 
$\Lambda$ is an UV cutoff. For $d<6$, to which we restrict, 
we have $\lim_{\Lambda\to \infty} I_n = m^{d-2 n} \tilde I_n$ with $\tilde I_n = \int_k (k^2 + 1)^{-n}$; it is  
well-defined for $n \geq 3$. We define
$\tilde I_2 = (4 \pi)^{d/2} \Gamma(2-\frac{d}{2})$ as the analytical continuation
to any $d$, with $I_2=m^{d-4} \tilde I_2$ for $d<4;$  the integral
$I_2$ becomes UV divergent for $d \geq 4$. 

One now defines the
dimensionless inverse response function at $q=0$, with times scaled using the characteristic time 
$\tau_m = \eta/m^2$,
\bea \label{defr}
 R^{-1}(\omega) &=& m^2 f(i \omega \tau_m) \\
f(y) &=& 1+ y + \sum_{n=2}^\infty \tilde D_n y^n\\ 
\tilde D_n &=& D_n m^{2n-2} \eta^{-n}\ .
\eea 
The $\tilde D_n$ are dimensionless. Let us now discuss the two relevant cases:
\medskip

\leftline{(i) $d \leq 4$:}

\medskip
Using $- m \partial_m I_{n+1} = (2 n+2-d) \tilde I_{n+1} m^{d-2n-2}$, 
Eqs.~(\ref{PT0})--(\ref{PT}) lead to the RG equation, up to $O(\epsilon^2)$,
\bea \label{frgeta}
 - m \partial_m \eta &=& - \eta \tilde \Delta''(0^+) \\
 - m \partial_m \tilde D_n &=& - 2( n-1) \tilde D_n \nn \\
&& + (-1)^n (2 n +2 -d) \frac{\tilde I_{n+1}}{\epsilon \tilde I_2} \tilde \Delta''(0^+)\ , ~~~~~~~~~~~~~~~~
\eea
using the rescaled disorder (\ref{rescaledDelta}). Since for $d \leq 4$ the behavior of $\tilde \Delta(u)$ is universal for small $m$, so are 
the behavior of $\eta$ and of the coefficients $\tilde D_n$. The first equation gives $\eta \sim m^{2-z}$, i.e.\ $\tau_m \sim m^{-z}$ with 
\be  \label{zagain}
z=2 - \tilde \Delta^{* \prime \prime}(0^+) = 2 -  \frac{1-\zeta_1}{3}  \epsilon + O(\epsilon^2)\ .
\ee 
The exponent $z$ is the dynamical exponent, with $z<2$. In the second equation we can use \cite{LeDoussalWiese2008c}
\be
\frac{\tilde I_{n+1}}{\epsilon \tilde I_2}  = \frac{\Gamma(n+1-d/2)}{2 \Gamma(n+1) \Gamma(3-d/2)} \stackrel{d \to 4}{-\!\!\!\longrightarrow} \frac{1}{2n(n-1)} \ .
\ee
Hence for $d<d_{\rm uc}$ the scaled dynamical coefficients converge as $m \to 0$ to universal fixed point values,
given to lowest order in $\epsilon$ by
\be \label{limD}
\tilde D_n \to \frac{(-1)^n}{2 n (n-1)}  \tilde \Delta^{* \prime \prime}(0^+)  = \frac{(-1)^n}{2 n (n-1)} \frac{1-\zeta_1}{3}  \epsilon + O(\epsilon^2)\ . 
\ee
For $d=d_{\rm uc}=4$ one finds analogously
\be
\tilde D_n \simeq \frac{(-1)^n}{2 n (n-1)}  \frac{\hat \Delta^{* \prime \prime}(0^+)}{\ln(\Lambda/m)} = \frac{(-1)^n 4 \pi^2}{n (n-1)}  \frac{1-\zeta_1}{3 \ln(\Lambda/m)} 
\ .\ee
Hence at the upper critical dimension the dimensionless coefficients $\tilde D_n$ decay to zero at small $m$, thus
the model is faithfully described by the BFM and ABBM mean-field equations of motion, with (only two) parameters, $\tau_m$ and $\sigma_m$.\ The behavior is universal, and largely independent of  details of the bare model. For $d< d_{\rm uc}$ the model
is not described by mean-field theory, but by a new universal fixed point which is studied in Section \ref{s:loops}. 
We can obtain the inverse response function for $d=d_{\rm uc}-\epsilon$ by  
inserting (\ref{limD}) into (\ref{defr}) and summing over $n \geq 2$,
\be 
   f(y) = 1+y +  \frac{1}{2} ((y+1) \log (y+1)-y)  \tilde \Delta^{* \prime \prime}(0^+)\ .  
\ee  
Thus the final result for the inverse response function  to one loop, i.e.\ $O(\epsilon)$ accuracy is 
\be
{\cal R}^{-1}_{q,\omega} = q^2 + m^2 \left( 1 + i \omega \tau_m\frac{z}2\right)^{\frac{2}{z}}  + O(\epsilon^2)\ .
\ee
We used the result (\ref{zagain}) for the dynamical exponent $z$. The behavior for large $\omega \tau_m \gg 1$, i.e.\ in the limit of
small mass $m$, is ${\cal R}^{-1}_{q,\omega} \sim (i \omega)^{2/z}$ as expected from scaling. This provides
a derivation of the dynamical exponent at finite frequency. 

\medskip

(ii) $d > 4$ 

\medskip

The FRG flow of the disorder for this case was discussed  in \cite{BalentsLeDoussal2004} (Appendix H) and \cite{LeDoussalWiese2008c} (Appendix B). There are two phases: (a) if the (smooth) bare disorder is small, it remains smooth under coarse graining, i.e.\ there is no metastability, no cusp, and no avalanches.
(b) if the bare disorder is larger than a threshold, $\tilde \Delta(u)$ acquires a cusp, but flows back to zero
as $\tilde \Delta(u) \sim (\frac{m}{\Lambda})^{d-4} A(u),$ where $A(u)$ is non-universal, and equivalently, 
$\Delta(u) \sim \Lambda^{4-d} A(u)$ is non-universal. Alternatively, if one considers a non-smooth and weak bare disorder (i.e.\ with a cusp 
in $\Delta_0(u)$), then perturbation theory converges, schematically $\Delta - \Delta_0 \sim I_2 O(\Delta_0^2)$ where
$I_2 \sim \Lambda^{d-4} - m^{d-4}$ since $I_2$ is now UV convergent and dominated by the UV cutoff
(see \cite{LeDoussalWiese2008c} for details). 

Since the rescaled disorder $\tilde \Delta$ flows to zero as $m \to 0$, the FRG equations (\ref{frgeta})  shows
that $\eta$  converges to a non-zero value $\eta_{\rm R}$ as $m \to 0$, hence $z=2$. 
The value of $\eta_{\rm R}= \eta_0 \exp( - \int_0^\Lambda \frac{dm}{m} \tilde \Delta_m''(0^+))$ obtained from (\ref{frgeta}) 
is non-universal, since the flow of the disorder is itself non-universal. The coefficients $\tilde D_n$, on the 
other hand, using (\ref{frgeta}) converge to zero as
\be 
\tilde D_n \sim \frac{(-1)^n}{2 n (n-1)}  \left(\frac{m}{\Lambda}\right)^{d-4} A''(0^+)\ ;
\ee  
hence for $m \to 0$ the model is well described by the ABBM model with constant but non-universal parameters $\eta$ and $\sigma$.

\section{A differential equation for $Z (\lambda)$}\label{q25}

We give a very general argument of how to calculate $Z (\lambda)$,
without calculating the instanton. This method works for all
first-order instanton equations.

The instanton equation away from
the source reads 
\begin{equation}\label{2.1}
\partial_{t} \tilde u (t) = \tilde u (t)-\tilde u (t)^{2} =: f (\tilde u (t))\ ,
\end{equation}
where we have allowed for a possible generalization to an arbitrary
function $f(\tilde u)$. 
To obtain $Z (\lambda)$, one has to integrate its solution
\begin{eqnarray}\label{2.2}
Z (\lambda) &=& \int_{-\infty}^{t (\lambda) }\rmd t\, \tilde u (t) \\
\tilde u (t (\lambda)) &=&\lambda \label{2.3}
\ .
\end{eqnarray}
Note that the translational zero-mode in time of $\tilde u (t)$ is not
fixed in (\ref{2.1}), but  by the condition (\ref{2.3}). Compared
to the standard solution, there is an arbitrary change in the
time of measurement. Taking a derivative w.r.t.\ $\lambda$ of the last
two equations 
yields 
\begin{eqnarray}\label{q26}
&&\frac{\rmd Z (\lambda)}{\rmd \lambda}  = \frac{\rmd t (\lambda)}{\rmd
\lambda} \tilde u \big(t (\lambda)\big) \\
&&\partial_{t }{\tilde u} (t) \Big|_{t=t (\lambda)}    \frac{\rmd t (\lambda)}{\rmd
\lambda} =1
\ .
\end{eqnarray}
Combining these two equations yields 
\begin{equation}\label{q27}
\frac{\rmd Z (\lambda)}{\rmd \lambda} = \frac{ \tilde u (t )}{
\partial_{t} \tilde u (t )}\bigg |_{t=t (\lambda )} =  \frac{ \tilde u (t )}{
f ( \tilde u (t ) )}\bigg |_{t=t (\lambda )}\ ,
\end{equation}
where in the last step we used the instanton equation (\ref{2.1}). 
Using (\ref{2.3}) we find the simple result 
\begin{equation}\label{q28}
\frac{\rmd Z (\lambda)}{\rmd \lambda} = \frac{\lambda}{f (\lambda)}\ .
\end{equation}
If $f (\tilde u) = \tilde u-\tilde u^{2}$, the case usually
considered, we arrive at 
\begin{equation}\label{q29}
\frac{\rmd Z (\lambda)}{\rmd \lambda} = \frac{1}{1-\lambda}
\ .
\end{equation} 
The solution is 
\begin{equation}\label{q30}
Z (\lambda) = -\log (1-\lambda )\ ,
\end{equation}
where the integration constant is fixed by demanding that $Z (0)=0$.

\section{More details on the ABBM model} 
\label{a:AABBM}
In this appendix we use dimensionless units. Let us rewrite Eq.~(\ref{fp}) as
\begin{eqnarray} \label{abbmeq}
&& \partial_t Q = - \partial_{\sf v} J \\
&& J({\sf v},t) = - (\partial_{\sf v} ({\sf v} Q) - ({\sf v}-v) Q) 
\ ,
\end{eqnarray}
where $J({\sf v},t) $ is the current of probability. 
The equation for the eigen-modes is
\begin{equation}
   -s  Q = \partial_{\sf v} (\partial_{\sf v} ({\sf v} Q) + ({\sf v}-v) Q) 
\ .
\end{equation}
Let us first discuss the case $v>0$.
The general solution is
\be 
Q({\sf v}) = {\sf v}^{v-1} e^{-{\sf v}} \left[C_1 L_s^{-1+v}({\sf v}) + C_2 U(-s,v,{\sf v}) \right]\ , 
\ee 
given in terms of the Laguerre polynomials and confluent hypergeometric functions.
The Laguerre polynomials can only have $s=n=0,1,2,..$ since for different values they
do not decay fast enough at infinity 
For these integer values of $s$ the two solutions become linearly dependent. These 
Laguerre solutions for all $s=n$ have the peculiarity that {\it the current vanishes at the origin}, i.e
$J({\sf v}=0,t)=0$, more precisely $J({\sf v}=0,t) \simeq {\sf v}^v$ at small ${\sf v}$ for all $n \geq 1$. 
In addition the current vanishes everywhere for $n=0$. For the
hypergeometric solution the current is $J({\sf v}=0,t)= \frac{\Gamma(v)}{\Gamma(-s)}$.

In their work \cite{AlessandroBeatriceBertottiMontorsi1990,AlessandroBeatriceBertottiMontorsi1990b} ABBM retained the solution with zero current at the origin, hence the
solution which vanishes for ${\sf v} \to \infty$,
\be
Q_n({\sf v}) = {\sf v}^{v-1} e^{-{\sf v}} L^{a=v-1}_{n}({\sf v}) \quad , \quad s=n=0,1,2,...\ .  
\ee
They thus obtained the normalized propagator \cite{AlessandroBeatriceBertottiMontorsi1990,AlessandroBeatriceBertottiMontorsi1990b},
\bea \label{prop}
&& Q_v({\sf v},t|{\sf v}_1,t) = {\sf v}^{v-1} e^{-{\sf v}}\nn \\
&& ~~~~~~~~~~~~~~~~~~~\times  \sum_{n=0}^\infty \frac{n!}{\Gamma(v+n)} L^{v-1}_{n}({\sf v}) L^{v-1}_{n}({\sf v}_1) e^{- n t}\ ,~~~~~~~~~~~
\eea
a formula  valid for $v>0$. The term $n=0$ is 
$Q_0({\sf v})={\sf v}^{v-1} e^{-{\sf v}}/\Gamma(v)$ and integrates over ${\sf v}>0$ to unity, the others to zero. 
Hence $\int_0^\infty \rmd{\sf v\,} Q({\sf v},{\sf v}_1,t)=1$. Since the current vanishes at the origin for all times (i.e
the total probability for ${\sf v}>0$ remains unity), for large times the probability reaches the
stationary state which has zero current everywhere $Q({\sf v},{\sf v}_1,t) \to Q_0({\sf v})$. 

Let us now consider  $v=0^+$. One then finds that (i) the Laguerre polynomials must again
be of integer order  to behave well at infinity (one has $L^{-1}_0({\sf v})=1$, $L^{-1}_1({\sf v})=-{\sf v}$, and so on).
(ii) The Laguerre solution corresponding to $n=0$ behaves as $e^{-v}/v,$ hence is not normalizable. (iii) 
The Laguerre solutions for $n=1,2,..$  have a {\it non-zero current at the origin.} (iv) The hypergeometric
solution does not behave well at the origin $\sim 1/v$ unless $s$ is positive and integer, in which case it
again becomes identical to the Laguerre solutions. The only possible solution for the propagator thus 
seems to be
\be  \label{prop2}
Q_{v=0}({\sf v},{\sf v}_1,t) = {\sf v}^{-1} e^{-{\sf v}} \sum_{n=1}^\infty n L^{-1}_{n}({\sf v}) L^{-1}_{n}({\sf v}_1) e^{- n t}
\ ,
\ee 
which is the limit of (\ref{prop}) for $v=0^+$, where the term $n=0$ has dropped because its prefactor $1/\Gamma(v)$
vanishes.

On the other hand, inspired by our result from the text, we  found that there is another
expression for the propagator at $v=0^+$, namely
\begin{eqnarray}
&& Q({\sf v}_2,{\sf v}_1,t)= \tilde Q({\sf v}_2,{\sf v}_1,t) +  \delta({\sf v}_2) e^{-\frac{(1-z) {\sf v}_1}{z}} \\
&& \tilde Q({\sf v}_2,{\sf v}_1,t)  = {\sf v}_1 e^{{\sf v}_1} \frac{\sqrt{1-z}}{z} e^{ - \frac{{\sf v}_1 + {\sf v}_2}{z} } \frac{I_1(2 \frac{\sqrt{1-z}}{z}  \sqrt{{\sf v}_1 {\sf v}_2})}{ \sqrt{{\sf v}_1 {\sf v}_2}}\ . \nn
\end{eqnarray}
We recall $z=1-e^{-t}$ and that $Q$ satisfies  (\ref{abbmeq})  
with as ${\sf v} \to {\sf v}_1$,
\begin{equation}
Q({\sf v},{\sf v}_1,t) \approx {\sf v}_1 \frac{e^{ - \frac{(\sqrt{{\sf v}_1} - \sqrt{{\sf v}_2})^2}{t} }}{\sqrt{4 \pi t} ({\sf v}_1 {\sf v}_2)^{3/4}}  \approx
\delta({\sf v}_2-{\sf v}_1)\ . 
\end{equation}
We have checked with Mathematica that $\int_{0^-}^\infty \rmd{\sf v\,} Q({\sf v},{\sf v}_1,t)=1$, and that the $\delta$-function piece
in (\ref{solu2abbm}) is crucial for this probability conservation. 

It turns out that the two expressions (\ref{prop2}) and (\ref{solu2abbm})  coincide for ${\sf v}>0$, i.e.\ $Q({\sf v},{\sf v}_1,t)=Q_{v=0}({\sf v},{\sf v}_1,t)$
as we have checked numerically with excellent accuracy (the convergence of the sum over $n$ is very good). However the $\delta$ function in (\ref{solu2abbm}) is not reproduced. Hence the terms $n \geq 1$  now have a finite integral over ${\sf v}$. This integral  does not add up to 1. Somehow the $n=0$ term is replaced, for $v=0$ by a delta function, multiplied by the factor $e^{-\frac{{\sf v}_1}{e^t-1}}$. This factor takes into consideration the absorption at zero, which is now present.

Other boundary conditions at $v>0,$ such as absorbing ones, can be studied, which we leave for the
future.

\section{Checks of the 3-time formula for MF (ABBM)}
\label{a:3-time-formula}
We now want to check the 3-times correlation. We use the formula 
\be 
\int_0^\infty \rmd v e^{-v} I_1(2 a \sqrt{v}) I_1(2 b \sqrt{v})  = I_1(2 a b) e^{a^2 + b^2} 
\ ,
\ee 
which yields:
\bea
&& \int \rmd v_2 e^{\lambda_2 v_2} \tilde Q(v_3,v_2,z') \tilde Q(v_2,v_1,z) 
\nn\\
&& ~~= \sqrt{\frac{v_1}{v_3}} e^{v_1} \frac{\sqrt{1-z''}}{z''} \frac{\tilde \gamma}{\gamma} 
e^{\frac{v_1}{z} (\frac{1-z}{z \gamma} -1)} e^{\frac{v_3}{z'} (\frac{1-z'}{z' \gamma} -1)} \nn
\\
&& ~~~~~~~\times I_1\left(2  \frac{\sqrt{1-z''}}{z''} \sqrt{v_3 v_1}\right) 
\ .\eea
Herer $\tilde \gamma=\frac{1}{z}+\frac{1}{z'} -1$, $\gamma=\tilde \gamma - \lambda_2$
and $1-z''=(1-z)(1-z')$. For $\lambda_2=0$ we find 
\be 
\int\rmd v_2\,  \tilde Q(v_3,v_2,z') \tilde Q(v_2,v_1,z) 
=  \tilde Q(v_3,v_1,z'') 
\ ,
\ee 
as expected for a propagator. Other useful identities are
\bea
&&\!\!\!\int_0^\infty \rmd v_3 e^{\lambda_3 v_3} \tilde Q(v_3,v_2,z') = e^{v_2 ( 1- \frac{1}{z'})} \left(e^{ v_2 \frac{1-z'}{z'(1-z'\lambda_3)} } -1\right)\nn\\ \\
&&\!\!\! \int_0^\infty \rmd v_1 e^{\lambda_1 v_1} \tilde Q(v_2,v_1,z) \frac{e^{-v_1}}{v_1}  = \frac{1}{v_2} e^{- \frac{v_2}{z}} \left( e^{v_2 \frac{1-z}{z(1-z\lambda_1)} } -1\right)\nn\\
\eea
This allows to obtain
\bea\label{QQ}
&&\!\!\!\! \int_{v_1,v_2,v_3>0} e^{\lambda_1 v_1 + \lambda_2 v_2 + \lambda_3 v_3} \times\tilde Q(v_3,v_2,z') \tilde Q(v_2,v_1,z) \frac{e^{-v_1}}{v_1} \nn\\ &&= \ln( 1- \lambda_1 z'' - \lambda_2 z' + \lambda_1 \lambda_2 z z')  \nn
\\
&& ~~~+ \ln( 1- \lambda_2 z - \lambda_3 z'' + \lambda_2 \lambda_3 z z')
 - \ln(z''-\lambda_2 z z') 
\nn\\ &&~~~- \ln\Big(1 - \lambda_1 -\lambda_2 -\lambda_3 + \lambda_1 \lambda_2 z + \lambda_1 \lambda_3 z'' + \lambda_2 \lambda_3 z'
\nn\\ && ~~~~~~~~~~~~~- \lambda_1 \lambda_2 \lambda_3 z z'\Big)
\ .\eea
We recognize that the last logarithm is $\tilde Z_3$.
Taking the three derivatives $\partial_{\lambda_1} \partial_{\lambda_2} \partial_{\lambda_3}$ 
gets rid of the other terms, and shows that 
\be
 v_1 v_2 v_3 \tilde Q(v_3,v_2,z') \tilde Q(v_2,v_1,z={\mathrm{LT}}^{-1}_{-\lambda_i \to v_i}
\partial_{\lambda_1} \partial_{\lambda_2} \partial_{\lambda_3} \tilde Z_3
\ .~~~~
\ee
Since  the latter expression is also the inverse LT of 
$q'_{123}  v_1 v_2 v_3 {\calP}(v_1,v_2,v_3)$, and  since neither function contains
 a $\delta$ function, we obtain\be
q'_{123}  {\calP}(v_1,v_2,v_3) = \tilde Q(v_3,v_2,z') \tilde Q(v_2,v_1,z) \frac{e^{-v_1}}{v_1}
\ .
\ee
This shows that the 3-time velocity probability can be written as a product
of 2-time propagators, i.e.\ the 3-time velocity probability at tree level (i.e.\ in the ABBM model) is Markovian.

\section{Spatial correlations in the tree theory} 
\label{a:spatial-correlations}
Here we give further calculational details concerning Section \ref{sec:spatial}, in particular we work in the
steady state to lowest order in $v$ and in dimensionless units. The results are exact for the tree theory, i.e.\ the BFM in any $d$, or for
SR disorder in the mean-field theory.  For the 3-point function to first order in $v$, computed in the text, let us indicate the
following integral formula, useful to generate a series expansion in $q$ (with $t_1<t_2<0$):
\begin{widetext}
\be 
 \overline{ \dot u_{q t_1} \dot u_{-q t_2} e^{\lambda L^d \dot u_0} }  =  v \frac{2}{1-\lambda} e^{-(1+q^2)(t_1+t_2)} \Big[1+\lambda(e^{t_1}-1)\Big]^2 \Big[1+\lambda(e^{t_2}-1)\Big]^2
\int_{-\infty}^{t_1} \rmd t' \frac{e^{2 (1+q^2) t'}}{\big[1+\lambda (e^{t'}-1)\big]^3}\ . 
\ee 
Let us now detail the calculation of the 4-time correlation function, from which we will also extract the
avalanche shape in the stationary state. Consider the source $\lambda_t = \lambda_0 \delta(t-t_0)+\lambda_3 \delta(t-t_3)$ 
for  $t_0 <t_1 < t_2 < t_3$. 
In dimensionless units, this gives
\be
\tilde u_t = \frac{1}{1 + \frac{1-\lambda_3}{\lambda_3} e^{t_3-t}} \theta(t_0<t<t_3)  + \frac{1}{1 - \frac{\lambda_3 \lambda_0 e^{t_0} - (1-\lambda_3)(1-\lambda_0) e^{t_3}}{
(1+\lambda_0)\lambda_3 e^{t_0} + \lambda_0 (1-\lambda_3) e^{t_3}} e^{t_0-t} } \theta(t<t_0) 
\ .
\ee
The dressed response function has six sectors. We indicate only those  needed:
\bea
&& \mathbb{R}^{t_0,t_3}_{k,t_b,t_a} = e^{-  (k^2+m^2) (t_b-t_a) } \theta(t_b-t_a) \Phi^2 \\
&& \Phi = \frac{ (1-\lambda_0)(1-\lambda_3) + \lambda_0 (1-\lambda_3) e^{t_b-t_0} - \lambda_0 \lambda_3 e^{t_0-t_3} + (1+\lambda_0) \lambda_3 e^{t_b-t_3} }{ (1-\lambda_0)(1-\lambda_3) + \lambda_0 (1-\lambda_3) e^{t_a-t_0} - \lambda_0 \lambda_3 e^{t_0-t_3} + (1+\lambda_0) \lambda_3 e^{t_a-t_3} } \quad , \quad t_a<t_b<t_0<t_3 
 \nn \\
&& \Phi = \frac{ 1+ \lambda_3 (e^{t_b-t_3} -1) }{ (1-\lambda_0)(1-\lambda_3) + \lambda_0 (1-\lambda_3) e^{t_a-t_0} - \lambda_0 \lambda_3 e^{t_0-t_3} + (1+\lambda_0) \lambda_3 e^{t_a-t_3} } \quad , \quad t_a<t_0<t_b<t_3 \\
&& \Phi = \frac{ 1+ \lambda_3 (e^{t_b-t_3} -1) }{1+ \lambda_3 (e^{t_a-t_3} -1)  } \quad , \quad t_0<t_a<t_b<t_3
\ .
\eea 
We now use
\be
 \int_{t<t'} \mathbb{R}^{t_0,t_3}_{q=0,t',t} = \left\{
 \begin{array}{l}
 \displaystyle
  \frac{(1+ \lambda_3 (e^{t'-t_3}-1))(1+\lambda_0 (e^{t_0-t'}-1))}{(1-\lambda_0)(1-\lambda_3) - \lambda_0 \lambda_3 e^{t_0-t_3}} \quad~~~~~~~~~~ \mbox{ for }
t'>t_0 \\
  \displaystyle 1 + \frac{\lambda_0 (1-\lambda_3) (e^{t'-t_0}-1) + e^{t'-t_3} \lambda_3 (1+\lambda_0)}{(1-\lambda_0)(1-\lambda_3) - \lambda_0 \lambda_3 e^{t_0-t_3}} \quad \mbox{ for } t'< t_0 \end{array} \right.
\ .
\ee
We must split the integral into two parts,\be 
\overline{ e^{\lambda_0 L^d \dot u_0} \dot u_{q t_1} \dot u_{-q t_2} e^{\lambda_3 L^d \dot u_3} } =
2 v \Big( \int_{t'<t_0< t_1} +  \int_{t_0 < t'<t_1} \Big) \mathbb{R}^{t_0,t_3}_{q,t_1,t'}  \mathbb{R}^{t_0,t_3}_{q,t_2,t'} \Big[\int_t \mathbb{R}^{t_0,t_3}_{q=0,t',t} \Big]
\ .
\ee 
The result is
\bea
  \overline{ e^{\lambda_0 L^d \dot u_0} \dot u_{q t_1} \dot u_{-q t_2} e^{\lambda_3 L^d \dot u_3} } &=& v \frac{\left(\left(\lambda _3-1\right) e^{t_3}-\lambda _3
   e^{t_1}\right){}^2 \left(\left(\lambda _3-1\right) e^{t_3}-\lambda _3
   e^{t_2}\right){}^2}{\left(\left(\lambda _0-1\right) \left(\lambda_3-1\right) 
   e^{t_3}-\lambda _0 \lambda _3 e^{t_0}\right){}^4} \nn \\
   && \times { \bigg( } \frac{e^{\left(-q^2-1\right) \left(-2 t_0+t_1+t_2\right)} \,
   _2F_1\left(3,2 \left(q^2+1\right);2 q^2+3;\frac{e^{t_0} \left(\lambda
   _0+1\right) \lambda _3-e^{t_3} \lambda _0 \left(\lambda
   _3-1\right)}{e^{t_0} \lambda _0 \lambda _3-e^{t_3} \left(\lambda
   _0-1\right) \left(\lambda _3-1\right)}\right)}{q^2+1} \nn \\
  \nn && + \frac{e^{\left(-q^2-1\right) \left(t_1+t_2\right)-3 t_3}
   \left(\left(\lambda _0-1\right) \left(\lambda _3-1\right)
   e^{t_3}-\lambda _0 \lambda _3 e^{t_0}\right){}^3}{\left(\lambda
   _3-1\right){}^3 \left(2 q^4+3 q^2+1\right)} \nn\\
   && \times \bigg\{    \left(\lambda _0-1\right) \left(2 q^2+1\right) [ \left(-e^{2
   \left(q^2+1\right) t_0}\right) \, _2F_1\left(3,2 \left(q^2+1\right);2
   q^2+3;\frac{e^{t_0-t_3} \lambda _3}{\lambda _3-1}\right)\nn \\
   && + e^{2 \left(q^2+1\right) t_1} \,
   _2F_1\left(3,2 \left(q^2+1\right);2 q^2+3;\frac{e^{t_1-t_3} \lambda
   _3}{\lambda _3-1}\right) ] \nn\\
   && +2 \lambda _0 \left(q^2+1\right) \bigg[e^{2
   \left(q^2+1\right) t_0} \, _2F_1\left(3,2 q^2+1;2
   \left(q^2+1\right);\frac{e^{t_0-t_3} \lambda _3}{\lambda
   _3-1}\right)\nn \\
   && -e^{2 q^2 t_1+t_0+t_1} \, _2F_1\left(3,2 q^2+1;2
   \left(q^2+1\right);\frac{e^{t_1-t_3} \lambda _3}{\lambda
   _3-1}\right)\bigg] \bigg\} \bigg) \label{e14}
    \eea 
We checked that for $q=0$ this expression yields $\partial_{\lambda_2} \partial_{\lambda_1} \tilde Z_4(\lambda_0,\lambda_1,\lambda_2,\lambda_3)|_{\lambda_2=\lambda_1=0}$ and 
that for $\lambda_0=0$ it yields (\ref{resnew}).
This expression is not invariant by time reversal
i.e.\ by simultaneous changes $t_0 \to - t_3$,$t_1\to - t_2$, $t_2 \to - t_1$, $t_3 \to - t_0$ 
$\lambda_0 \leftrightarrow \lambda_3$. It is invariant however, at $q=0$.
The non-invariance by time reversal can already be seen on the 
4-point function, taking $\partial_{\lambda_0} \partial_{\lambda_3}$:
\bea
&& \overline{ \dot u_{-T/2} \dot u_{q, t_1} \dot u_{-q, t_2} \dot u_{T/2} } 
= v L^{-2d} \frac{2 e^{-(q^2+2) T- (q^2+1) (2 t_1+3 t_2)}}{(1 + q^2) (2 + q^2) (1 + 2 q^2) (3 + 2 q^2)} 
\nn \\
 && ~\times \bigg[ 2 \left(2 q^6+9 q^4+13 q^2+6\right) e^{\left(\left(3 q^2+2\right)
   t_1+\left(2 q^2+3\right) t_2+\left(q^2+1\right) T\right)}  +4
   \left(q^2+2\right) q^2 e^{\left(q^2+2\right) t_1+2 \left(q^2+1\right)
   t_2+\frac{T}{2}} \nn\\
   && ~~~~~~+4 \left(q^2+2\right) q^2 e^{\left(q^2+1\right) t_1+\left(2
   q^2+3\right) t_2+\frac{T}{2}}-3 \left(2 q^2+1\right) q^2
   e^{\left(q^2+1\right) \left(t_1+2 t_2\right)}+\left(2 q^4+7 q^2+6\right)
   e^{\left(q^2+1\right) \left(3 t_1+2 t_2+T\right)}   \bigg] \label{e18}
\ .~~~~~~~~~~~
\eea
This function is not symmetric by $t_1 \to - t_2$ and $t_2 \to - t_1$. 

If we take the limit $\lambda_0,\lambda_3 \to - \infty$ we obtain $\overline{ \delta_{\dot u_0} \dot u_{q t_1} \dot u_{-q t_2} \delta_{\dot u_3} }$
which we do not reproduce here. One can check that the first
hypergeometric term yields zero, although the limit is quite delicate.
Taking $- \partial_{t_0} \partial_{t_3}$ and dividing by the duration distribution we find our final result:
\bea \nn
&& \langle \dot u_{q t_1} \dot u_{-q t_2} \rangle_{03} =
\frac{e^{-t_3}}{(e^{t_3}-1)^2}  \bigg\{ \\
&& 2 \left(e^{t_2}-e^{t_3}\right){}^2 e^{-\left(q^2+1\right)
   \left(t_1+t_2\right)} \left[2 e^{t_1+t_3} \left(e^{2 q^2
   t_1}-1\right)-e^{\left(2 q^2+1\right) t_1}-e^{2 t_3} \left(e^{\left(2
   q^2+1\right) t_1}-1\right)+e^{2 t_1}\right]\nn \\
 &&  + \frac{\left(e^{t_3}-1\right) \left(e^{t_3}-e^{t_1}\right)
   \left(e^{t_3}-e^{t_2}\right) e^{\left(-q^2-1\right)
   t_1-\left(q^2+1\right) t_2-3 t_3}}{2 q^4+3 q^2+1} \nn \\
   && \times \bigg[ \left(1-2 q^2\right) e^{t_1+t_2}+\left(1-2 q^2\right) e^{t_1+2
   t_3}+\left(1-2 q^2\right) e^{t_2+2 t_3}+\left(2 q^2-3\right)
   e^{t_1+t_2+t_3}+\left(2 q^2+1\right) e^{3 t_3} \nn\\
   && ~~~~~+\left(2 q^2+1\right)
   e^{t_1+t_3}+\left(2 q^2+1\right) e^{t_2+t_3}-\left(2 q^2+3\right) e^{2
   t_3} \bigg]\nn \\
   && \times 
  \bigg[ \left(2 q^2+1\right) \left(-e^{2 \left(q^2+1\right) t_1}\right) \,
   _2F_1\!\left(3,2 \left(q^2+1\right);2 q^2+3;e^{t_1-t_3}\right)+\left(2
   q^2+1\right) \, _2F_1\left(3,2 \left(q^2+1\right);2
   q^2+3;e^{-t_3}\right)\nn \\
   && ~~~~~+2 \left(q^2+1\right) e^{t_3} \left(e^{\left(2
   q^2+1\right) t_1} \, _2F_1\!\left(3,2 q^2+1;2
   \left(q^2+1\right);e^{t_1-t_3}\right)-\, _2F_1\left(3,2 q^2+1;2
   \left(q^2+1\right);e^{-t_3}\right)\right) \bigg] \bigg\}
\eea 
where we have set $t_0=0$ for simplicity; the general case is obtained setting $t_i \to t_i-t_0$, $i=1,2,3$. \medskip

\end{widetext}
{~~~~~}
\newpage 
{~~~~~}
\newpage

\section{Behaviour of the 1-loop correction $\delta Z(\lambda)$ near $\lambda=1$} 
\label{app:lambda1}
\begin{figure}[b]
\Fig{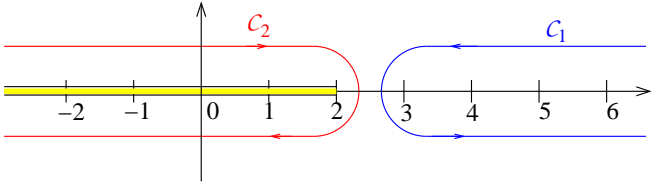}
\caption{The complex-$n$ plane with the contours ${\cal C}_{1}$ of
Eq.\ (\ref{zz4}) and ${\cal C}_{2}$ of Eq.\ (\ref{zz5}). The branch cut starting at $n=2$ is indicated.}
\label{f:C12}
\end{figure}

Here we indicate how we extract the behavior of $\delta Z(\lambda)$ near $\lambda=1$.
We recall our result \begin{equation}
\label{delta-Z-r1}
\delta Z(\lambda) = \sum_{n=2}^\infty a_n \kappa^n
\end{equation}
with \(a_n\) given in Eq.\ (\ref{an}), and repeated here 
\begin{eqnarray}
 a_n &=& \frac{(n-3) (n-2)^2 \log (n-2)}{2 n^2}\nn \\
&& +\frac{6\log (2)-2 n (n+1) (\log
   (2)-1)}{ n^2 (n+1)} \nn \\
&& -\frac{(n-1) (n ((n-6) n+2)+6) \log (n-1)}{n^2
   (n+1)}\nn \\
&& +\frac{\left(n^2-8 n+3\right) \log (n)}{2 ( n+1)}\ , \label{an-rep}\\
a_{2} &=&\lim_{n\to2} a_{n} =  1-\ln 4\ . 
\end{eqnarray}
From the relation $(1-\lambda)(1-\kappa)=1$, in order to get
\(Z(\lambda)  \) in the limit of $\lambda \to 1$, which controls the
tail of ${\cal P}(\dot u)$ for $\dot{u}\to \infty$, we need this
expression for $\kappa \to -\infty$.   However, the series expansion
has a convergence radius in $\kappa$ of  only $1$, equivalent to
 $\lambda <1/2$. A first thing one can do, is to re-express this series in $\lambda$:
\be \label{356}
\delta Z(\lambda) = \sum_{n=2}^\infty a_{n} \kappa^{n} = \sum_{p=2}^\infty c_{p} \lambda^{p}
\ .
\ee 
The formula for the coefficients \(c_p\) is 
\be 
c_p = (p-1)! \sum_{n=2}^p a_n \frac{(-1)^n}{(p-n)! (n-1)!} 
\ .
\ee
The convergence radius of $\delta Z(\lambda)$, as a series of
$\lambda$, is 1. While this is useful for intermediate values of
$\lambda$, it does not allow to study the singularity for $\lambda \to
1$. In order to analyze the latter, we now derive an expansion of
$\delta Z (\lambda)$ in powers of $-1/\kappa $.   
We start with 
\begin{eqnarray}\label{zz4}
\delta Z (\lambda ) &=& \sum_{n=2}^{ \infty} a_{n} (-1)^{n}
(-\kappa)^n \nn \\
&=&a_{2}\kappa^{2} +\oint_{{\cal C}_1} \frac{\rmd n}{2\pi  i} \frac{\pi}{\sin(\pi n)}
a_{n} (-\kappa)^{n}
\end{eqnarray}
The contour starts at $\infty +i\delta $, goes to $3+ i \delta $
passes left of $3$ and then goes to  $\infty -i\delta $, for any $0< \delta <1
$, see figure \ref{f:C12}. The formula uses the residue theorem, and
that the residue of $ \frac{\pi}{\sin(\pi n)}  $ at  integer $n$ is
$(-1)^{n}$. Two remarks are in order:  $a_{n}$ has three different
branch-cut singularities, starting at $n=2$, $n=1$, and
$n=0$, and going to $n=-\infty$. Singling out the term $a_{2}$ avoids crossing the branch-cut
starting at $n=2$,
which would not be allowed. Second, one could try to move the explicit
factor of $(-1)^{n}$ into $(-\kappa )^{n}$. This does not work, for
two reasons: First of all, $\pi /\sin (n\pi)$, when prolonged to the
complex plane, converges exponentially fast. This would not be the
case for $\pi\cot (n\pi)$, to be used to produce the non-alternating
sign. Worse, $\kappa^{n}$, for negative $\kappa$, when prolonged to
the complex plane, actually diverges in the lower half-plane. This is
why we use the formula as is.

Having an integral representation for $\delta Z (\lambda)$, we can now
prolong analytically for $\kappa \to -\infty$, by deforming the
contour of integration to ${\cal C}_{2}$, which starts at $-\infty
+i\delta$, goes to $2+i \delta$, then passes at the right of $2$, and
finally goes from $2-i\delta$ to $-\infty -i\delta$; see again figure
\ref{f:C12}. This gives 
\begin{equation}\label{zz5}
\delta Z (\lambda ) =
 \oint_{{\cal C}_2} \frac{\rmd n}{2\pi  i} \left[\frac{\pi}{\sin(\pi n)}
a_{n} (-\kappa)^{n} - \frac{a_{2}}{n-2} (- \kappa)^{2}  \right]
\end{equation}
Note that while the integral representation (\ref{zz4}) is convergent for
$-1<\kappa<0$, the representation (\ref{zz5}) is valid for $-\infty
<\kappa<-1$;  the smaller $\kappa$, the better the convergence.  
We have checked the integral representation (\ref{zz5}) for $\kappa
=-8$, i.e.\ $\lambda =8/9$ numerically. Then both (\ref{zz5}) and the $\lambda$-series (\ref{356}) 
 give $\delta Z (8/9)= 8.17538$, with a relative error of
$10^{-7}$. Therefore trusting our integral representation, we can now
analyze it for large negative $\kappa$. Then it will be dominated by
the contribution at the beginning of the cut singularity of $a_{n}$,
which starts at $n=2$, see the first term of (\ref{an-rep}), and the
corresponding plot \ref{f:C12}.
Therefore for large negative $\kappa $, the integral (\ref{zz5}) is
given by 
\begin{eqnarray}\label{zz6}
\delta Z (\lambda ) &\simeq&
 \oint_{{\cal C}_2} \frac{\rmd n}{2\pi  i}  \frac{- (n-2)\log (n-2)}{8}
(-\kappa)^{n} \nn\\
&=& \frac{\kappa^2}{8 [\ln (-\kappa)]^2} +O\!\left(\frac{\kappa^2}{[\ln(- \kappa)]^3} \right)
\label{360}\end{eqnarray}
One can obtain more subleading terms by expanding $a_n$ to higher powers in $(n-2)$. Doing this, we find 
\begin{eqnarray}\delta Z(\lambda) =\kappa^2 &\!\Bigg[&\! \frac{1}{8[\log (-\kappa
   )]^2}+ \frac{1}{2[\log
   (-\kappa )]^3} + \frac{21+2\pi^2}{16[\log
   (-\kappa )]^4} \nn\\ 
&&+ \frac{15+4\pi^2}{4[\log
   (-\kappa )]^5}  + \frac{585+210\pi^2 +14 \pi^4}{48[\log
   (-\kappa )]^6}\nn\\ 
&& +\bigg(45{+}\frac{75 \pi ^2}{4}{+}\frac{7 \pi ^4}{2}\bigg)  \frac{1}{[\log
   (-\kappa )]^7}+...\Bigg] ~~~~~~
\end{eqnarray} 
We can test this series against the integral (\ref{360}): We find for $\kappa=-10^{10}$ that $\delta Z=2.887\times  10^{16}$ with a relative error of \(10^{-4}\). For $\kappa =-10^{100}$, we find \(\delta Z = 2.400 \times 10^{194}\),
with a relative error of \(10^{-9}\). For $\kappa =-10^{1000}$, we find \(\delta Z = 2.361 \times 10^{1992}\), with a relative error of \(10^{-6}\) (probably from the numerical integration). 

Expressed in terms of $\lambda$, our final result is given in Eq.~(\ref{362}) of the main text.

\section{An alternative approach to express the 1-loop contributions  $\delta Z(\lambda)$ and $\delta P(u)$.}
\label{s:deltaZ-alt}
Here we calculate the 1-loop correction $\delta Z(\lambda)$ by first integrating over momentum. More precisely we 
start from Eq.\ (\ref{Jt}), calculate $\Phi(k,t)$ as given in Eq.\ (\ref{297}), but instead of Eq.\ (\ref{298}) and (\ref{299}) we first integrate over $t$, and then $k$, leaving the $t_1$-integral for the end. In order to be able to perform the $k$-integration, we have to introduce counter-terms right away. The term involving $\Phi(k,t)$, with the necessary counter-term \({\cal J}_{\rm ct}^{(1)}(k,\kappa,t_{1})\) becomes \begin{eqnarray}
&& {\cal J}^{(1)}(\kappa,t_{1})\nn\\
&& =\int_{0}^{\infty}(k^{2}) \rmd k^{2} \left[{\cal J}_{\rm ct}^{(1)}(k,\kappa,t_{1})+\int_{-\infty}^{t_{1}}\rmd t\,\Phi(k,t)\mathbb{R}(k,t_{1}) \right] \nn\\
&&=\frac{\kappa ^2 e^{2 t_1} \text{Ei}\left(-t_1\right)}{1-\kappa 
   e^{t_1}}-\kappa  \text{Ei}\left(t_1\right)\nn \\
&&~~~   -\frac{\kappa  e^{t_1}
   \left(\kappa +\kappa  e^{t_1} \left(2 t_1-1\right)-t_1\right)}{t_1
   \left(\kappa  e^{t_1}-1\right)}\label{B1}\\
&&{\cal J}_{\rm ct}^{(1)}(k,\kappa,t_{1})\nn\\
&&=\frac{\kappa  e^{t_1} \left(2 \kappa 
   e^{t_1}-1\right)}{\left(k^2+1\right)^2 \left(\kappa 
   e^{t_1}-1\right)}+\frac{\kappa  e^{t_1}}{k^2+1}-\frac{\kappa  e^{k^2
   t_1+t_1}}{k^2+1}\ .   ~~~~~~~~~~~~~~~
\label{B2}\end{eqnarray}
The second term involving $\Phi(k,t)^2$, with the necessary
counter-term \({\cal J}_{\rm ct}^{(2)}(k,\kappa,t_{1})\) becomes  \begin{eqnarray}
&& {\cal J}^{(2)}(\kappa,t_{1}) \nn\\ 
&&= \int_{0}^{\infty}(k^{2}) \rmd k^{2} \left[{\cal J}_{\rm ct}^{(2)}(k,\kappa,t_{1})+\int_{-\infty}^{t_{1}}\rmd
t\,\Phi(k,t)^2\,\mathbb{R}(k,t_{1}) \right] \nn\eea
   \bea
&&=\frac{\kappa ^2 e^{2 t_1}}{ (\kappa  e^{t_1}-1)^3} \bigg[ \kappa  e^{2 t_1} \text{Ei}\left(-2 t_1\right) \left(\kappa +2 \kappa 
   t_1-2\right)\nn\\&&~~~-2 \kappa  e^{t_1} \text{Ei}\left(-t_1\right)
   \left(\kappa  e^{t_1}+\kappa  e^{t_1} t_1-e^{t_1}-1\right)\nn\\&&~~~+ \kappa
   ^2-2 \kappa ^2 e^{t_1}+2 \kappa  e^{t_1}- \log \left(t_1\right)-
   \gamma_{\rm E} -1+ \log (2) \bigg]~~~~~~~\label{B3}\\
&&{\cal J}_{\rm ct}^{(2)}(k,\kappa,t_{1})=\frac{\kappa ^2 e^{2 t_1}}{(k^2+1)^2 (\kappa 
   e^{t_1}-1)}
\ .\end{eqnarray}
Several checks are in order: First, the two counter-terms, when integrated over $t_1$ reproduce the one given earlier in Eq.~(\ref{b29}),\bea
&&\int_{t_1<0} {\cal J}_{\rm ct}^{(1)}(k,\kappa,t_{1})+
 {\cal J}_{\rm ct}^{(2)}(k,\kappa,t_{1}) \nn\\
 &&~~~~~~~~~= \frac{\kappa
 (3+k^2)+2 \log (1-\kappa )}{(k^2+1)^2}\ .
\eea
Second, both \({\cal J}^{(1)}(\kappa,t_{1})\) and \({\cal J}^{(2)}(\kappa,t_{1})\) have a finite limit for $t_1\to 0$. This is why the last term in Eq.\ (\ref{B2}) was added, even though the $k$-integral would have been convergent without the term at fixed $t_1$.

We thus have found an integral-representation for $\delta Z(\lambda)$ as defined  in Eq.\ (\ref{delta-Z}), with the same counter-terms,
\be
\delta Z(\lambda) = \int_{t_1<0} {\cal J}^{(1)}(\kappa,t_{1})+{\cal J}^{(2)}(\kappa,t_{1})
\label{B6}\ .\ee
The two contributions were given in Eqs.~(\ref{B1}) and (\ref{B3}).

We now note that all terms in Eq.~(\ref{B6}) are algebraic functions of $\kappa$, and thus of $\lambda$. Hence the inverse-Laplace transform is possible. Replacing $t_{1}$ by $t$ to alleviate the notations, this becomes
\begin{widetext}
\bea\label{B7}
\delta {\cal P}(\dot u) &=&\displaystyle\int_{t<0} e^{-\dot u}f_1(t)+{\rme^{-\textstyle\frac{\dot u}{1-\rme^t}}} \left[\frac{ f_2(t)}{(\rme^t-1)^4} + \frac{ f_3(t)\dot u}{(\rme^t-1)^5}+ \frac{ f_4(t)\dot u^2}{(\rme^t-1)^6} \right ]\\ 
f_1(t)&=& e^t (2 t+3)
   \text{Ei}(-t)-e^t (2 t+1) \text{Ei}(-2
t)+\text{Ei}(t)+e^t
   \left(2-\frac{1}{t}\right)-e^{-t}+\frac{1}{t}+2 \\
f_2(t)   &=& \Big[\left(2 e^t-8 e^{2 t}+12 e^{3 t}\right) t+e^t-4 e^{2
   t}+6 e^{3 t}-6 e^{4 t}\Big] \text{Ei}(-2
   t)\nn\\&&+\left[\Big(-2 e^t+8 e^{2 t}-12 e^{3 t}\right) t-3
   e^t+10 e^{2 t}-7 e^{3 t}+6 e^{4 t}\Big]
   \text{Ei}(-t)\nn\\&&-\Big[\ln (t/2)+\gamma_{\rm E}\Big]  \left(2 e^{2 t}+e^{3
   t}\right)+e^{-t}+13 e^t-12 e^{2 t}+4 e^{3 t}+\frac{3 e^t-3
   e^{2 t}+e^{3 t}-1}{t}-6 \\
f_3(t)&=&\Big[\left(8 e^{3 t}-2 e^{2 t}\right) t-e^{2 t}+4 e^{3 t}-6
   e^{4 t}\Big] \text{Ei}(-2 t)+\Big[\left(2 e^{2 t}-8 e^{3
   t}\right) t+2 e^{2 t}-2 e^{3 t}+6 e^{4 t}\Big]
   \text{Ei}(-t)\nn\\&&-\Big[\ln (t/2)+\gamma_{\rm E}\Big]  \left(e^{2
t}+2 e^{3
   t}\right)+6
   e^t-9 e^{2 t}+4 e^{3 t}-1\\
f_4(t) &=&   \Big[e^{3 t} t+\frac{e^{3 t}}{2}-e^{4 t}\Big] \text{Ei}(-2
   t)+\Big[e^{4 t}-e^{3 t} t\Big]
   \text{Ei}(-t)+\frac{e^t}{2}-e^{2 t}+\frac{e^{3
   t}}{2}-\frac{1}{2}   e^{3 t}\Big[\log
   (t/2)+\gamma_{\rm E}\Big]
\eea\end{widetext}
This is a closed expression for   $\delta {\cal P}(\dot u)$. We can now check all our statements made in the main text. First of all, we reproduce the plot on figure \ref{fdeltaPofudot}.  

For the small-$\dot u$ behavior, we remark that the integral (\ref{B7}) is dominated by the terms proportional to $\rme^{-\dot u/(1-\rme^{-t})}$, in the limit of small $t$. The leading contribution comes from expanding $f_2(t)$ for small  $t$, and reads 
\be
\delta {\mathcal{P}}_{f_2}(\dot u)\simeq -2\int_{t<0}\rme^{-\textstyle\frac{\dot
u}{t}} \frac{\ln t}{t^2} = -2\,\frac{\log (\dot u)+\gamma_{\rm E} }{\dot u}
\ .\ee 
Note that $f_3(t)$ and $f_4(t)$ could also contribute at the same order, but they have no term proportional to $\ln t$, thus they only correct the subleading term $\sim 1/\dot u$
leading to the final result
\be
\delta {\mathcal{P}}(\dot u) = -2\,\frac{\log (\dot u)+2 \gamma_{\rm E}+ \frac{1}{4}-\ln 2 }{\dot u} + O(\ln u) 
\ .\ee 
To obtain a systematic  expansion  one rescales $t \to \dot u t$ and integrates term by term in $t$
 the series expansion at small $\dot u$. This confirms the predictions
given in Eq.~(\ref{finala}) for the exponent ${\sf a}$, and for the constant $C$ in Eq.~(\ref{CC}).

\section{Long-ranged elasticity $\gamma=1$}\label{a:LR}
In this appendix, we calculate all relevant quantities for LR-elasticity \(\gamma=1\), \(d_{\rm c} =2\),
with the kernel defined in the main text.
We found in Eqs.~(\ref{ZLR}) and (\ref{418})  that 
\bea\label{A1}
Z^{\rm LR}(\lambda) &=& Z_0(\lambda) + \alpha\, \delta Z^{\rm LR}(\lambda)
\\ 
\delta Z^{\rm LR}(\lambda)&=& \int_0^\infty \rmd x \, (x+1) f(x+1)+O(\epsilon)\ ,
~~~~~\label{A2}
\eea
where $f(x)$ is defined in the text, 
in other words, the calculation is identical to the short-range case, except that when integrating over $k$, we have to  replace $\int \rmd(k^2) k^2$ by $\int \rmd(k^2) (1+k^2)$.
This replacement can be performed before or after the time integral.

\subsubsection{First method} 

In this method, we first integrate over $t$ leading to
formulas (\ref{b31}) and (\ref{b29});  then we integrate over $k$ with the modified measure.
The series expansion is then given by  
\be
\delta Z^{\rm LR}(\lambda) = \sum_{n=2}^\infty a_n^{\rm LR} \kappa^n
\ee
with 
\bea
a^{\rm LR}_2&=&-\ln(2)\\
a^{\rm LR}_{n>2}&=& -\frac{2 \left(n^2+n-6\right) \log (2)}{n^2 (n+1)}\nn\\&&+\frac{(n-4) (n-3)
   (n-2) \log (n-2)}{2 n^2}\nn\\
&& 
   +\frac{(2-n) (6 + 2 n - 7 n^2 + n^3) \log
   (n-1)}{n^2 (n+1)}\nn\\
&&+\frac{(n-1) (n^2-9 n+2) \log (n)}{2 n (n+1)}
\ .\eea
For $n\to \infty$, the leading behavior is
\be
a_n^{\rm LR}=\frac{-2 \log (n)-\frac{3}{2}-2 \log (2)}{n}+O(n^{-2})
\ee
Comparing with Eqs.~(\ref{3.57}) and (\ref{Zasymp1}) shows that 
\be
\delta Z^{\rm LR}(\lambda) = -\ln^2 (1-\lambda)+ ... ~~\mbox { for }\lambda \to -\infty\ .
\ee
Thus \begin{eqnarray}\label{b39-2}
Z^{\rm LR}(\lambda ) 
 &=&  Z_0(\lambda ) +{\alpha} \delta Z^{\rm LR}(\lambda)\nn\\&=&
 -\ln(1-\lambda)\left[1+  {\alpha} \ln(1-\lambda) + \dots 
 \right]\ . ~~~~~~
\end{eqnarray}
This is consistent with a modified critical behavior at small velocities,  
\be  \label{finalaLR} 
{\cal P}_{\mu=1}^{\rm LR}(\dot u) \sim_{\dot u \ll 1} \frac{1}{\dot u^{\sf a}}   \quad , \quad {\sf a} = 1 + 2\alpha + O (\epsilon^{2})
\ .
\ee 
The behavior for $\kappa \to -\infty$ (i.e.\ \(\lambda\to 1\)) now reads
\begin{eqnarray}\label{zz6LR}
\delta Z_{\mu=1}^{\rm LR} (\lambda ) &\simeq&
 \oint_{{\cal C}_2} \frac{\rmd n}{2\pi  i}  \frac{\log (n-2)}{4}
(-\kappa)^{n} \nn\\
&=& \frac{\kappa^2}{4 \ln (-\kappa)} +O\!\left(\frac{\kappa^2}{[\ln(- \kappa)]^2} \right)
\label{360-bis}\ .\end{eqnarray}
This implies a different tail than in the SR case.

\subsubsection{Second method} 

We find, analogously to Eqs.\ (\ref{B1}) and (\ref{B3}), the integral representation\be
 \delta Z^{\rm LR}(\lambda) = \int_{t_1<0} {\cal J}^{(1)}(\kappa,t_1)+{\cal J}^{(2)}(\kappa,t_1) \label{AA3}
 \ .\ee
The contributing terms are 
\bea
&&{\cal J}^{(1)}(\kappa,t_1) =-\frac{\kappa  e^{t_1} }{2 t_1 (\kappa 
   e^{t_1}-1 )}  \\
   && 
~~\times\left[2 \kappa  e^{t_1}
\left(2 t_1
   \text{Ei}(-t_1\right)-1)+2 \kappa -2 \gamma_{\rm E} 
   t_1-2 t_1 \log \left(-t_1\right)\right]\nn\\  
&&\!\!\!{\cal J}^{(2)}(\kappa,t_1) = \frac{\kappa ^2 e^{2 t_1}}{ (\kappa 
   e^{t_1}-1)^3} \nn\\
&&\times \Big[\text{Ei}\left(-t_1\right) \left(-2 \kappa ^2 e^{2 t_1}-4
   \kappa ^2 e^{2 t_1} t_1+4 \kappa  e^{t_1}+4 \kappa  e^{2
   t_1}\right)\nn\\ &&~~~~+\text{Ei}\left(-2 t_1\right) \left(\kappa ^2
   e^{2 t_1}+4 \kappa ^2 e^{2 t_1} t_1-4 \kappa  e^{2
   t_1}\right)+2 \kappa ^2\nn\\ &&~~~~-4 \kappa ^2 e^{t_1}+2 \kappa ^2 e^{2
   t_1}-(\gamma_{\rm E}  +\log \left(-t_1\right))\left(2 \kappa  e^{t_1}+1\right)\nn\\ &&~~~~+\log (2)
   \left(2 \kappa +2 t_1+1\right)\Big]\ .
\eea 
\begin{widetext}
Inverse-Laplace transforming Eq.~(\ref{AA3}) yields an integral representation for ${\cal P}^{\rm LR}(\dot u) $, 
\be
{\cal P}^{\rm LR}(\dot u) = {\cal P}_{0 }(\dot u) + \alpha\, \delta  {\cal P}^{\rm LR}(\dot u)
\ee
\bea
\delta {\cal P}^{\rm LR}(\dot u) &=&\displaystyle\int_{t<0} e^{-\dot u}f^{\rm LR}_1(t)+{\rme^{-\textstyle\frac{\dot
u}{1-\rme^t}}} \left[\frac{ f^{\rm LR}_2(t)}{(\rme^t-1)^4} + \frac{ f^{\rm LR}_3(t)\dot u}{(\rme^t-1)^5}+
\frac{ f^{\rm LR}_4(t)\dot u^2}{(\rme^t-1)^6} \right ]\\ 
f^{\rm LR}_1(t)&=& -e^t (4 t+1) \text{Ei}(-2 t)+4 e^t (t+1)
   \text{Ei}(-t)+\frac{(  e^{-t}-1) (2 e^t
   t-2 t+e^t)}{t} \\
f^{\rm LR}_2(t) &=&\Big[(4 e^t-16 e^{2 t}+24 e^{3 t}) t+e^t-4 e^{2
   t}+6 e^{3 t}-12 e^{4 t}\Big] \text{Ei}(-2t) \nn \\
&&   +\Big[\left(-4 e^t+16 e^{2 t}-24 e^{3 t}\right) t-4
   e^t+12 e^{2 t}-2 e^{3 t}+12 e^{4 t}\Big]
   \text{Ei}(-t)\nn\\&&+2 e^{-t}+30 e^t-32 e^{2 t}+12 e^{3 t}+\frac{3
   e^t-3 e^{2 t}+e^{3 t}-1}{t}+\left(e^t-4 e^{2 t}-6 e^{3
   t}\right)\big[ \log (-t)+\gamma_{\rm E}\big] \nn\\&&+\left(8 e^{2 t}+e^{3 t}\right) \log
   (2)+\left(4 e^{2 t}+2 e^{3 t}\right) t \log (2)-12 \\
f^{\rm LR}_3(t) &=& \Big[(16 e^{3 t}-4 e^{2 t}) t-e^{2 t}+4 e^{3 t}-12
   e^{4 t}\Big] \text{Ei}(-2 t)+\Big[(4 e^{2 t}-16
   e^{3 t}) t+2 e^{2 t}+4 e^{3 t}+12 e^{4 t}\Big]
   \text{Ei}(-t)\nn\\&&+12
   e^t-18 e^{2 t}+8 e^{3 t}-\left(e^{2 t}+8 e^{3 t}\right)
   \Big[\log (-t)+\gamma_{\rm E}\Big]+\left(7 e^{2 t}+2 e^{3 t}\right) \log (2)+\left(2
   e^{2 t}+4 e^{3 t}\right) t \log (2)-2~~~~~~~~~~~~~~~\\   
f^{\rm LR}_4(t)& =& \left(2 e^{3 t} t+\frac{e^{3 t}}{2}-2 e^{4 t}\right)
   \text{Ei}(-2 t)+\left(-2 e^{3 t} t+e^{3 t}+2 e^{4 t}\right)
   \text{Ei}(-t)+e^t-2 e^{2 t}+e^{3 t}\nn\\&&-\frac{3}{2} \gamma_{\rm E}e^{3 t}+\left(e^{2 t}+\frac{e^{3 t}}{2}\right) \log
   (2)-\frac{3}{2} e^{3 t} \log (-t)+e^{3 t} t \log (2)   
\eea 
\end{widetext}
The analysis of the small $\dot u$ behavior gives  $\delta {\cal P}^{\rm LR}(\dot u) \simeq -2  \frac{\ln \dot u}{\dot u} $,
hence is consistent with the above result (\ref{finalaLR}).

\section{Second-order derivatives $S''$ and third order derivatives $S'''$}
\label{app:second}

\subsection{Second-derivative matrix}

We give here the matrix of second derivatives of the action:
\begin{eqnarray}
  {\cal S}''_{u_{xt} u_{x't'}} &=&\delta_{xx'} \Big[ - \hat u_{xt}
\delta_{tt'} \int_{t_1} \hat u_{xt_1} \Delta''(u_{xt} - u_{xt_1})\nn \\
&&\qquad  ~ +  \hat u_{xt}  \hat u_{xt'} \Delta''(u_{xt} - u_{xt'})  \Big]
  \\
  {\cal S}''_{\hat u_{xt} u_{x't'}}& =& \delta_{tt'} (\eta_0 \partial_{t'}
- \nabla_x^2 + m^2 ) \nn \\
&& -
 \delta_{xx'} \Big[ \delta_{tt'} \int_{t_1}  \hat u_{xt_1} \Delta'(u_{xt}-u_{xt_1})
\nn \\
&&\qquad \quad - \Delta'(u_{xt}-u_{xt'}) \hat u_{xt'}  \Big] \\
  {\cal S}''_{u_{xt} \hat u_{x't'}} &=& \delta_{tt'} (- \eta_0 \partial_{t'}
- \nabla_x^2 + m^2 ) \nn \\
&& - 
 \delta_{xx'} \Big[ \delta_{tt'} \int_{t_1}  \hat u_{xt_1} \Delta'(u_{xt}-u_{xt_1})
\nn \\
&&\qquad \quad + \Delta'(u_{xt}-u_{xt'}) \hat u_{xt}  \Big] \\
  {\cal S}''_{\hat u_{xt} \hat u_{x't'}} &=& - \delta_{xx'} \Delta(u_{xt}-u_{xt'})
  \label{secondder}
\end{eqnarray}
We will need it at the tree saddle point and to lowest order in $w$, i.e.\
for $w=0^+$, where according to
the previous section $u=u^0=0$, and $\hat u=\hat u^0$. Hence
\begin{align}\label{S2w0}
&  {\cal S}''_{u_{xt} u_{x't'}} =\delta_{xx'} \Delta''(0) \Big[ - \hat u^0_{xt}
\delta_{tt'} \int_{t_1} \hat u^0_{xt_1} + \hat u^0_{xt}  \hat u^0_{xt'} 
\Big]\  \nn  \\
&  {\cal S}''_{\hat u_{xt} u_{x't'}} = ( R^{-1} + \Sigma)_{xt,x't'} \nn \\
& {\cal S}''_{u_{xt} \hat u_{x't'}} =  \Big( (R^{T})^{-1} + \Sigma^T\Big)_{xt,x't'}
\nn  \\
&  {\cal S}''_{\hat u_{xt} \hat u_{x't'}} = - \delta_{xx'} \Delta(0)\ .
\end{align}

\subsection{Third-derivative tensor} \label{app:third}

In the text we need the third derivative tensor only at the tree saddle point
with $w=0^+$. It can 
be obtained from (\ref{secondder})

\begin{eqnarray}
&&\!\!\! \int_{t_{1}}-  {\cal S}'''_{\hat u_{x t} \hat u_{x' t'} u_{x_1t_1}} u^1_{x_1t_1}
\\
&& = \delta_{x x'} \Delta'(0^+) (u^1_{x t}-u^1_{x t'}) {\rm sgn}(t-t')
\nn \\
&&\!\!\! \int_{t_{1}}-  {\cal S}'''_{\hat u_{x t} u_{x' t'} u_{x_1t_1}} u^1_{x_1t_1}
\\
&&  = \delta_{x x'} \Delta''(0) 
\Big[\delta_{t t'} \int_{t_2} \hat u^0_{xt_2} (u^1_{x t} - u^1_{x t_2}) -
\hat u^0_{xt'} 
(u^1_{x t} - u^1_{x t'})\Big]\nn  \\
&&\!\!\! \int_{t_{1}}-  {\cal S}'''_{\hat u_{x t} u_{x' t'} \hat u_{x_1t_1}} \hat
u^1_{x_1t_1} 
 \\
&& =\delta_{xx'} \Delta'(0^+)   \Big[ \delta_{tt'} \int_{t_2}  \hat
u^1_{xt_2} {\rm sgn}(t-t_2) -  {\rm sgn}(t-t') \hat u^1_{xt'}  \Big] \nn
\end{eqnarray}

Consider now the uniform case $\mu_{xt}=\mu_t$ and $\hat u^0_{xt}=\hat u^0_t$.
Then
$ {\cal S}'''_{xt,x't',x_1,t_1} = \delta_{xx'x_1}  {\cal S}'''_{t,t',t_1}$ with:
\bea
&& [ {\cal S}'''_{\hat u \hat u u}]_{tt't_1} = \sigma (\delta_{t t_1} - \delta_{t'
t_1}) {\rm sgn}(t-t') \\
&& [ {\cal S}'''_{\hat u u u}]_{tt't_1} = - \Delta''(0) (\delta_{tt't_1} \int_{t_2}
\hat u^0_{t_2} \nn \\
&& - \delta_{tt'} \hat u^0_{t_1}  - \delta_{tt_1} \hat u^0_{t'} 
+ \delta_{t't_1} \hat u^0_{t'} ) \nn \\
&& [ {\cal S}'''_{\hat u u \hat u}]_{tt't_1}=[ {\cal S}'''_{\hat u \hat u u}]_{tt_1t'} \nn
\eea

\section{Dressed response functions for velocity observables in the position theory} 
\label{app:dressed} 

We note that with notation  $\phi_{kt}\to \phi_{t}$:
\begin{align}\label{b40}
& \int_{t'} 
(R^{-1} + \Sigma)_{tt'} \phi_{t'}  \\
&\qquad = \partial_{t} \phi_t + k^2 \phi_t + \phi_t  - \int_{t'} {\rm sgn}(t-t') \hat u^0_{t'} (\phi_{t'}-\phi_{t}) \nn 
\end{align}
Hence for a smooth function $\phi_t$:
\begin{eqnarray} \label{b41}
\lefteqn{\!\!\!\!\!\int_{t'} \partial_t (R^{-1} + \Sigma)_{tt'} \phi_{t'}} \nn \\
&=& \partial_{t} \dot \phi_t + k^2 \dot \phi_t + \dot \phi_t  + \int_{t'} {\rm sgn}(t-t') \hat u^0_{t'} \dot \phi_{t} \nn \\
&=& (\partial_t + k^2 + 1 - 2 \tilde u^0_t ) \partial_t \phi_t\nn  \\
& =& e^{2 \int^t \rmd t ' \tilde u^0_{t'}} (\partial_t + k^2 + 1) e^{- 2 \int^t dt' \tilde u^0_{t'}} \partial_t \phi_t\qquad 
\ .
\end{eqnarray}
Hence  apart from a zero-mode in time,
\begin{equation}\label{q33}
 (R^{-1} + \Sigma)_{tt'} = (\partial_t)^{-1} e^{2 \int^t dt' \tilde u^0_{t'}} (\partial_t + k^2 + 1) e^{- 2 \int^t dt' \tilde u^0_{t'}} \partial_{t'} \ .
\end{equation}
The zero-mode can be treated as follows. Consider the constant vector
$\phi_{t}=\phi_{-\infty}=\mbox{const}$. Then because of (\ref{propsigma}) one has
\begin{equation}\label{b42}
\int_{t'} (R^{-1} + \Sigma)_{tt'}  \phi_{t'} = (k^{2}+1) \phi_{-\infty}
\ .
\end{equation}
This implies that the vector $\phi_t = \phi_{-\infty}$ is an eigenvector of $R^{-1} + \Sigma$, with
eigenvalue $k^2+1$. Hence one also has\be 
\int_{t'} (R^{-1} + \Sigma)^{-1}_{tt'}  \phi_{t'} = \frac{1}{k^{2}+1} \phi_{-\infty}
\ .
\ee  
This yields
\begin{eqnarray}\label{Rktt'}
\!\lefteqn{\int_{t'}{\cal R}_{k tt'}\phi_{t'} = \int_{t'} (R^{-1} +
\Sigma)^{-1}_{tt'}\phi_{t'}} \nonumber \\
\!&=& (\partial_t)^{-1} e^{2 \int \limits^t \rmd t'
\tilde u^0_{t'}} (\partial_t + k^2 + 1)^{-1} e^{- 2 \int\limits^t \rmd t' \tilde
u^0_{t'}} \partial_t [\phi_{t}-\phi_{-\infty}] \nn \\
&& +\frac{1}{k^{2}+1}\phi_{-\infty }
\ .
\end{eqnarray}
Using the definition of $\mathbb{R}_{ktt'}$ given in Eqs.\ (\ref{R-gen}) and (\ref{b43}), we can rewrite
Eq.~(\ref{Rktt'}) to get the fundamental equations
\begin{eqnarray}\label{b44a}
\int_{t'}{\cal R}_{k tt'}\phi_{t'} &=& \int_{t'}  (\partial_t)^{-1}
\mathbb{R}_{ktt'}\partial_{t'} \left(\phi_{t'}-\phi_{-\infty} \right)
\nn\\&&+\frac{1}{k^{2}+1} \phi_{-\infty }  \\ \label{b44b}
\partial_{t} {\cal R}_{k tt'} &=& {\mathbb R}_{k tt'}\partial_{t'}
\ .
\end{eqnarray}
The subtraction of $\phi_{-\infty}$ from $\phi_{t'}$ is noted for clarity reasons only.
A first corollary is 
\begin{equation}
\tr \Big( \ln (R^{-1} + \Sigma)_{tt'}  \Big) = \tr \left( \ln R^{-1}  \right)\ .
\end{equation}
Similarly one finds that 
\begin{equation}\label{c1}
 \Big((R^T)^{-1} + \Sigma^T\Big)_{\!tt'}^{\!-1} = \Big( (R^{-1} + \Sigma)_{tt'}
 \Big)^{\!T} =  \Big(R^{-1} + \Sigma\Big)^{\!-1}_{\!t't}\ .
\end{equation}

\section{Third diagram $\delta \Gamma_{1}^{(3)}$}\label{a:ThirdDiagram}
We now turn to the third contribution $\delta \Gamma_{1}^{(3)}$:
\bea
 \delta \Gamma^{(3)}_{1} &=& - m^{2}\int {\cal R}_{k t't}  {\cal S}'''_{\hat u_t
u_{t'} \hat u_{t_1}} {\cal R}_{0t_2t_1}  {\cal S}''_{u_{t_2} u_{t_4}} {\cal R}_{t_4}
\nn \\
 & =& m^{2}\sigma \Delta''(0) \int_{tt't_{1}t_2t_4} {\cal R}_{k t' t} [\delta_{tt'}-
 \delta_{t't_{1}} ] \sgn(t-t_{1}) \nn \\
 && ~~~~~~~~~~~~~~~~~~~~~~~~~~~~~\times {\cal R}_{0 t_2 t_{1}} \hat u^0_{t_2}  \hat u^0_{t_4} {\cal R}_{t_4}
\eea
Using that ${\cal R}_{k tt}=0$, and exchanging $t$ and $t'$ we get
\begin{eqnarray}
 \delta \Gamma^{(3)}_{1} &=&   m^{2}\sigma \Delta''(0) \int_{tt't_2t_4} {\cal
R}_{k t t'} {\rm sgn}(t-t')  {\cal R}_{0 t_2 t} \nn\\
&& ~~~~~~~~~~~~~~~~~~~~~~~~~~~~~~~\times \partial_{t_2} \tilde u^0_{t_2} \partial_{t_4} \tilde u^0_{t_4}
{\cal R}_{t_4} \nn\\
& =& v m^{2}\sigma \Delta''(0) \int_{tt't_2t_4 t_{5}} \partial_{t}[ {\cal
R}_{k t t'} {\rm sgn}(t-t') ] \nn\\ && ~~~~~~~~~~~~~~~~~~~~~~~~~~~~~~~~~\times {\mathbb R}_{0 t_2 t}  \tilde u^0_{t_2} \tilde
u^0_{t_4}  {\mathbb R}_{0t_4t_{5}}
\ ,~~~~~~~~~~
\end{eqnarray}
where we have used (\ref{propertymagic}) and (\ref{8.352}). Now we use that
 \begin{eqnarray}
&&\!\!\!\! \!\!\!\!  \partial_{t}[ {\cal R}_{k t t'} {\rm sgn}(t-t')] \nn\\
&& =  {\mathbb R}_{k t t'} \partial_{t'}{\rm sgn}(t-t') + {\cal R}_{k t t'}
\partial_{t} \sgn(t-t')\nn\\
&&= -2 \delta(t-t') \left[{ {\mathbb R}_{k t t'} - {\cal R}_{k t t'}  }\right]
\ .
\end{eqnarray}
Since ${\cal R}_{k tt}={\mathbb R}_{k t t}=0$ we find
\begin{equation}
 \delta \Gamma^{(3)}_{1} = 0\ .
\end{equation}
Graphically, this can be written as 
\begin{eqnarray}
 \delta \Gamma^{(3)}_{1} &=& 
\parbox{5.3cm}{\begin{tikzpicture}
\draw[double distance=2pt] (0.5,0.5) arc (0:360:0.5);
\draw[->,double distance=2pt] (0,1) arc (-270:-90:0.5);
\draw[double distance=2pt] (1,1) --  (2,1);
\draw[-to,shorten >=10pt,double distance=2pt] (1,1) -- (2,1);
\draw[double distance=2pt] (4,1) --  (3,1);
\draw[-to,shorten >=10pt,double distance=2pt] (4,1) -- (3,1);
\draw [dashed] (0,1) --  (1,1);
\draw [dashed] (2,1) --  (3,1);
\node (t) at (0,.75) {$t$};
\node (t1) at (1.15,.75) {$t_{1}$};
\node (t2) at (2.1,.73) {$t_{2}$};
\node (t4) at (3.1,.73) {$t_{4}$};
\node (w) at (4.4,1) {$\displaystyle w_{t_{5}}$};
\fill (0,1) circle (2pt);
\fill (1,1) circle (2pt);
\fill (2,1) circle (2pt);
\fill (3,1) circle (2pt);
\fill (4,1) circle (2pt);
\end{tikzpicture}}
\nn\\&& +
\parbox{5.cm}{\begin{tikzpicture}
\draw[double distance=2pt] (1,1) arc (0:-180:0.5);
\draw[->,double distance=2pt] (0,1) arc (-180:-90:0.5);
\draw[double distance=2pt] (1,1) --  (2,1);
\draw[-to,shorten >=10pt,double distance=2pt] (1,1) -- (2,1);
\draw[double distance=2pt] (4,1) --  (3,1);
\draw[-to,shorten >=10pt,double distance=2pt] (4,1) -- (3,1);
\draw [dashed] (0,1) --  (1,1);
\draw [dashed] (2,1) --  (3,1);
\node (t) at (-.2,1) {$t$};
\node (tp) at (1.15,.75) {$t'$};
\node (tp) at (2.1,.73) {$t_{2}$};
\node (tp) at (3.1,.73) {$t_{4}$};
\node (w) at (4.4,1) {$\displaystyle w_{t_{5}}$};
\fill (0,1) circle (2pt);
\fill (1,1) circle (2pt);
\fill (2,1) circle (2pt);
\fill (3,1) circle (2pt);
\fill (4,1) circle (2pt);
\end{tikzpicture}}
\end{eqnarray}
These terms are zero: The first term is the response at
equal times. The second term, when viewed in standard diagrammatics
can be mounted, moving one arrow-head from $t_{2}$ to $t_{4}$, or vice
versa. So it is expected to be zero anyway.

\section{1-loop expansion for the lowest cumulants}\label{q32}

\subsection{Expansion in $\lambda$ of  \(Z(\lambda)\)}\label{expansion-lambda}

Let us first recall the result for the one loop contribution to $Z(\lambda)$ to all orders in $\kappa$
derived via perturbation of the instanton equation and displayed in
Eq.~(\ref{b31}). Here we reexpress it as a function of $\lambda$ and display it up to
  to order 4 in $\lambda$,
  \begin{eqnarray}
 Z(\lambda) &=& Z_0(\lambda) + A \int_k  {\cal J}(k,\lambda) + {\cal J}^{\rm ct}(k,\lambda)  \\
 {\cal J}(k,\lambda) &=& \frac{1}{1+k^2} \lambda +  \frac{2(3 + k^2)}{(1 + k^2) (2 + k^2)} \frac{\lambda^2}{2} \nonumber \\
&& + 
\frac{2 (108 + 128 k^2 + 47 k^4 + 6 k^6)}{(1 + k^2) (2 + k^2) (3 + 
   k^2) (3 + 2 k^2)} \frac{\lambda^3}{3!}\nn  \\
   && +\frac{ 6 (16 + 13 k^2 + 2 k^4) (45 + 22 k^2 + 4 k^4)} {(1 + k^2) (2 + 
   k^2) (3 + k^2) (4 + k^2) (3 + 2 k^2)} \frac{\lambda^4}{4!} \nonumber \\
&&+ O(\lambda^5) \label{seriesla}
\ .
\end{eqnarray}
The counter-term has the expression\bea
  {\cal J}^{\rm ct}(k,\lambda) &=& -\frac{\lambda}{k^2+1}-\frac{2 \left(k^2+2\right)}{\left(k^2+1\right)^2} \frac{ \lambda^2}{2!}
 \\
 && -\frac{2 \left(3 k^2+7\right) }{ \left(k^2+1\right)^2} \frac{\lambda^3}{3!}
 -\frac{12 \left(2 k^2+5\right)}{
   \left(k^2+1\right)^2} \frac{\lambda^4}{4!} +O(\lambda^5). \nn
  \end{eqnarray}
As requested for a counter-term, in the sum \({\cal J}(k,\lambda) + {\cal J}^{\rm
ct}(k,\lambda)\), the terms proportional to  $1/k^2$ and $1/k^4$  at large $k$ cancel, and one is left with
\bea
&& {\cal J}(k,\lambda) + {\cal J}^{c.t.}(k,\lambda) \nn \\
&& = \left[ -\lambda ^2+\frac{\lambda
   ^3}{2}+\frac{5 \lambda
   ^4}{4}+O\left(\lambda ^5\right) \right] \frac{1}{k^6} + O\left(\frac{1}{k^8}\right) \eea

\subsection{Diagrammatic calculation of the lowest-order cumulants}
\label{yyyyyyy}
We recall from Section \ref{simplified}  that
\be 
\sum_{n=1}^\infty \overline{\dot u_t^n}^{\rm c} \frac{\lambda^n}{n!} = v Z(\lambda) + O(v^2) 
\ .
\ee  
The cumulants, or equivalently the moments, were computed at tree level 
up to $n=5$ (and arbitrary times), in Section \ref{s:tree} i.e.\ using only the local
cubic vertex. Here we compute the
1-loop correction to this result, at equal times, and 
show how the  result (\ref{b31}), after re-expansion in $\lambda$,  is recovered. The diagrammatic rules are
those of the simplified theory, which has (i) a cubic, local-in-time  vertex
proportional to $\sigma=-\Delta'(0^+);$ (ii) a non-local-in-time quartic vertex
proportional to $\Delta''(0)$, which comes from\ the (simplified) interaction
\bea \label{Ssimpl}
  {\cal S}_{\mathrm{dis}}^{\rm simp} &=&  -\sigma \int_{xt} \tilde u_{xt} \tilde u_{xt} (v + \dot u_{xt}) \nn\\
&& + \frac{1}{2} \Delta''(0) \int_{xt} \tilde u_{xt} \tilde u_{xt'} (v+\dot u_{xt}) (v+\dot u_{xt'})
\ .
~~~~~~~\eea 
Due to the quartic vertex, 1-loop diagrams are now possible in contrast
to the cubic theory, which has only tree diagrams. Since we use dimensionless
units below, we set $\sigma \to 1$ and $\Delta''(0) \to -A$. Note that we have written the action
(\ref{Ssimpl}) in the co-moving frame to make apparent the $v$ terms, but the calculation
can also be made in the laboratory frame; then one must remember that
$\dot u$ has an average $v$.

Let us first discuss the two lowest orders and their diagrammatic representation.

To order $\lambda$ there is a single diagram
\begin{equation}
\overline{\dot u_t} = v \int_k \frac{1}{1+k^2}  = \diagram{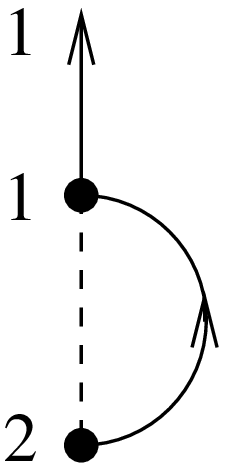}
\ .
\end{equation}
This term involves the vertex $\Delta''(0)$ represented by the dashed lines. It is also
the usual representation of the disorder vertex $\Delta(u)$ and identifies to it
whenever there are 2 entering legs. Since all our contribution are $O(v)$ the $v$ has been
chosen in the lowest $v+\dot u$ field, which will be the case in all diagrams written in this section. 
Propagators with arrows are bare
response functions, $1/(k^2+1) e^{-k^2(t-t')}$ in Fourier. External arrows are in the same number as $n$ in $\dot u^n$ to match the external
$\dot u$ fields. External legs are at zero momentum (since we compute
center-of-mass velocity moments) but internal ones carry momentum, to be
integrated over  (1-loop diagrams). 

To order $\lambda^{2}$ (two outgoing lines), there are 4 contributions: 
\begin{align}
\overline{\dot u_t^2} &= v \int_k
   \frac{2(3 + k^2)}{(1 + k^2) (2 + k^2)} & && \nn\\
  &  = v(  2 I_1 + 2 I_2 + 4 I_3 + 4 I_4 ) &&&\\
D_1 &=\diagram{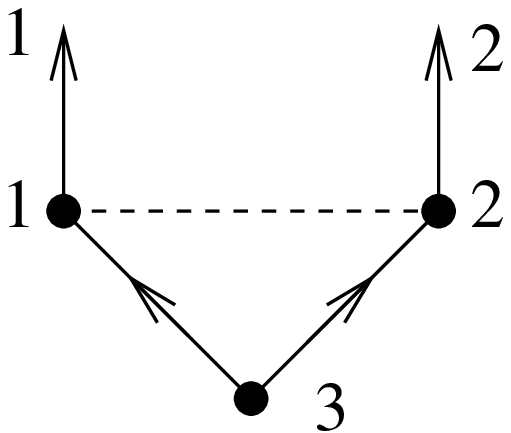}\ ,\qquad  &I_1 &= \frac{1}{2} \frac{1}{(1+k^2)(2+k^2)} \nn\\
D_2 &=\diagram{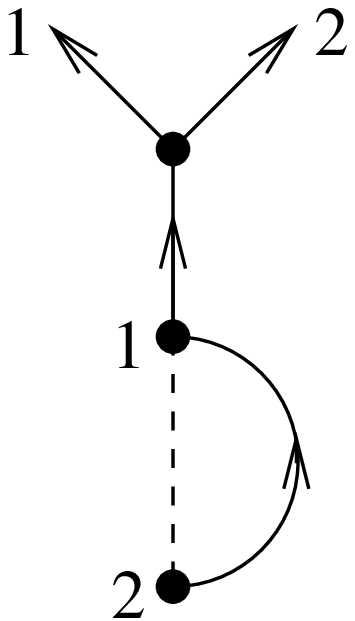}\ ,
\qquad &I_2& = \frac{1}{2} \frac{1}{1+k^2} \\
D_3 &=\diagram{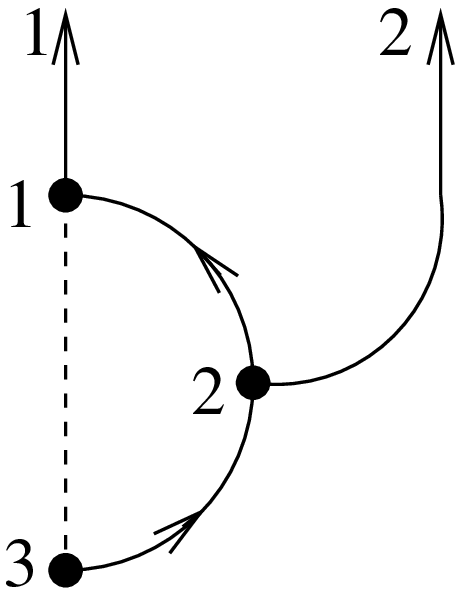}\ , \qquad &I_3&=\frac{1}{2} \frac{1}{(1+k^2)(2+k^2)} \\
D_4 &= \diagram{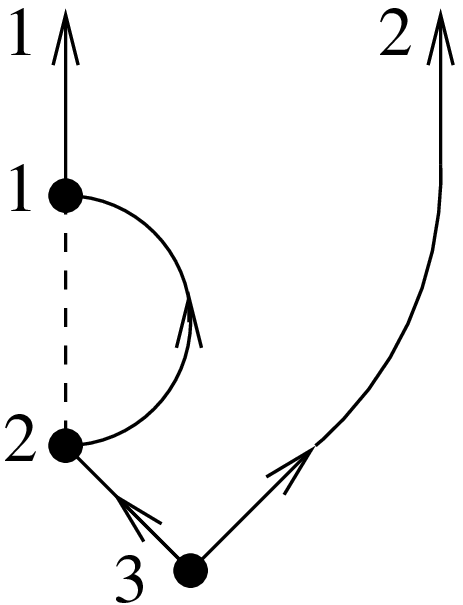}\ ,\qquad &I_4& = \frac{1}{4} \frac{1}{2+k^2}\ .
\end{align}
We see that both the cubic and the quartic vertices  appear. One can check
that the sum of  these terms with their indicated weights reproduces
(\ref{seriesla}).

At third order, one has 
 \begin{eqnarray}
  \overline{\dot u_t^3} &=& v \int_k
\frac{2 (108 + 128 k^2 + 47 k^4 + 6 k^6)}{(1 + k^2) (2 + k^2) (3 + 
   k^2) (3 + 2 k^2)}  \nn\\&=& v \int_k
   -\frac{16}{k^2+2}+\frac{5}{k^2+3}-\frac{8}{2 k^2+3}+\frac{21}{k^2+1}\nn \\
&  =& v  \sum_{i=1}^{11} d_i I_i  \ .
\end{eqnarray}
This comes from 11 diagrams: 
\begin{align}
T_1&=\diagram{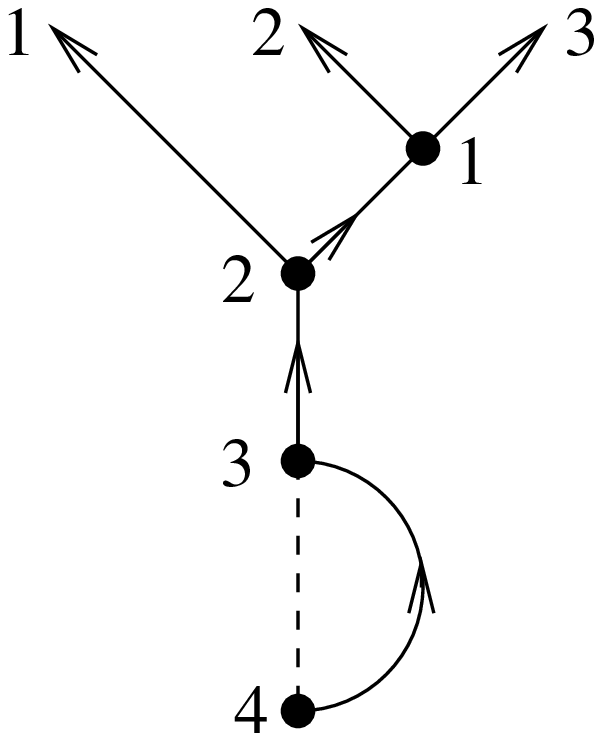}\quad & d_1 = 12 \nn \\
  I_1 &= \frac{1}{6(1+k^2)}    \\
T_2&=\diagram{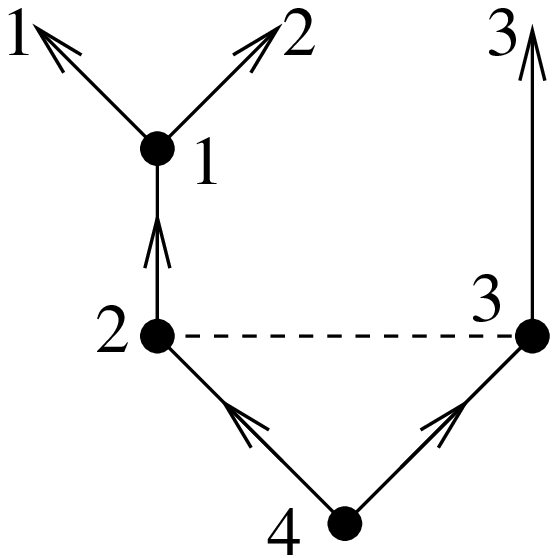}   & d_2  = 12 \nn \\
I_2 &= \frac{4+k^2}{6(1+k^2)(2+k^2)(3+k^2)} \\
T_3&=\diagram{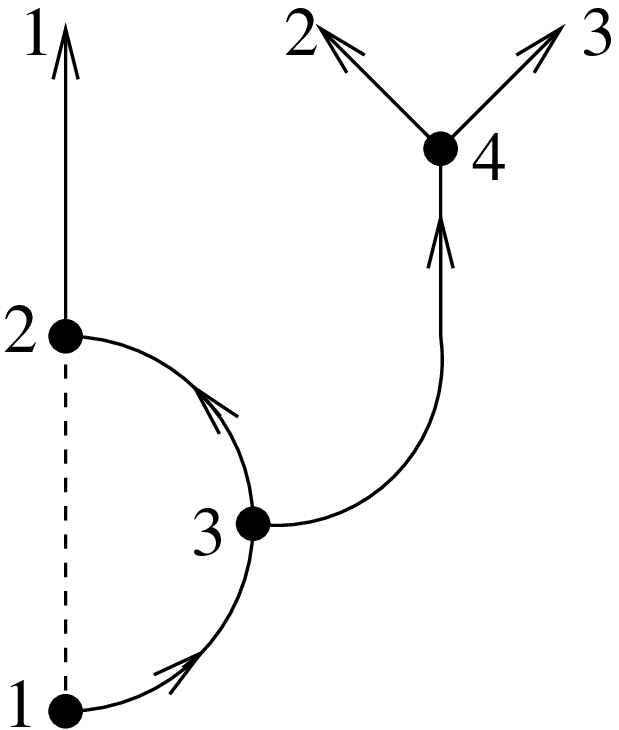} & d_3  = 12\nn \\
 I_3 &= \frac{5+k^2}{6(1+k^2)(2+k^2)(3+k^2)}   \\
T_4&=\diagram{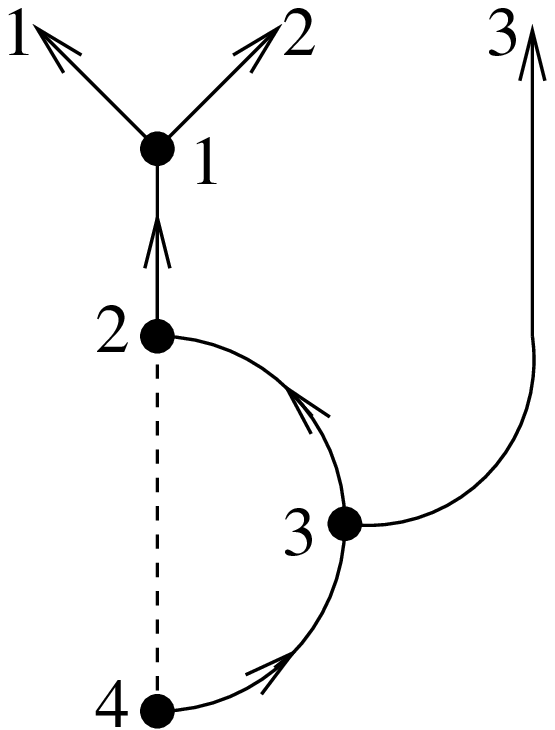} \quad  & d_4  = 12 \nn \\
I_4 & = \frac{1}{6(1+k^2)(2+k^2)}   \\
T_5&=\diagram{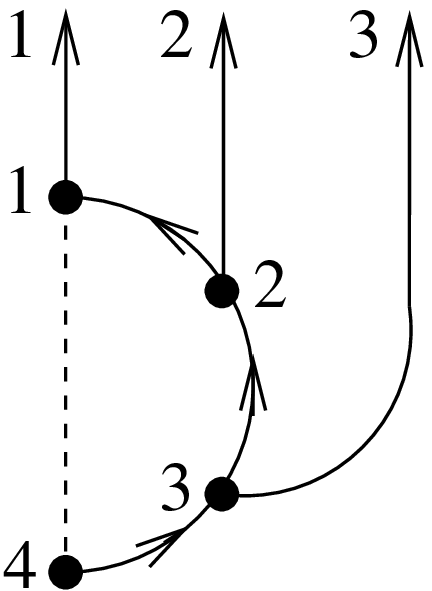} \quad &d_5  = 24\nn \\
I_5 &= \frac{1}{3(1+k^2)(2+k^2)(3+k^2)}  
\end{align}
\begin{align}
T_6&=\diagram{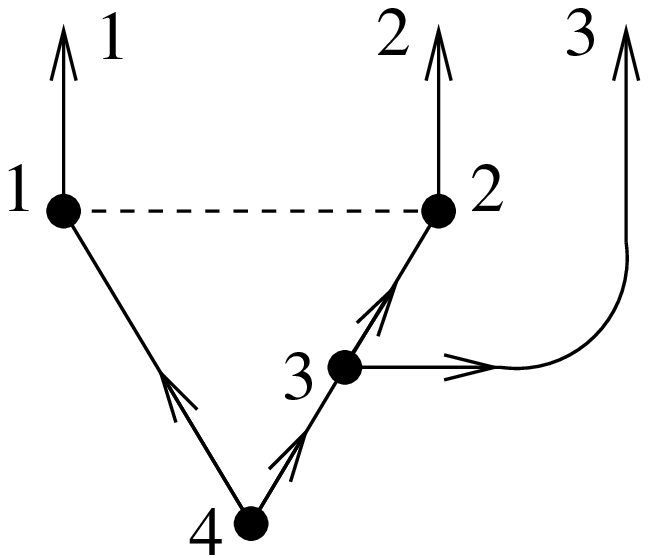} \quad &d_6  = 24\nn \\
I_6 &= \frac{7+4 k^2}{6(1+k^2)(2+k^2)(3+k^2)(3+2 k^2)}   \\
T_7&=\diagram{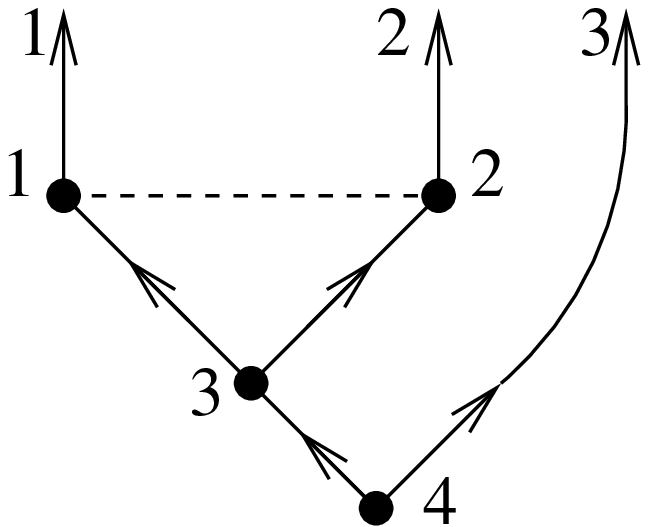} \quad &d_7 = 12\nn \\
I_7 &= \frac{1}{3(3+k^2)(3+2 k^2)}   \\
T_8&=\diagram{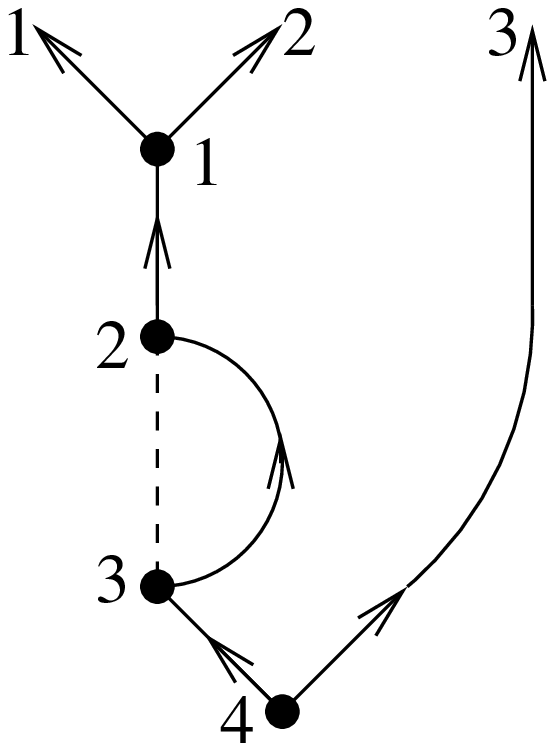} \qquad &d_8 =  12 \nn \\
&I_8 = \frac{1}{12(2+k^2)}     \\
T_9&=\diagram{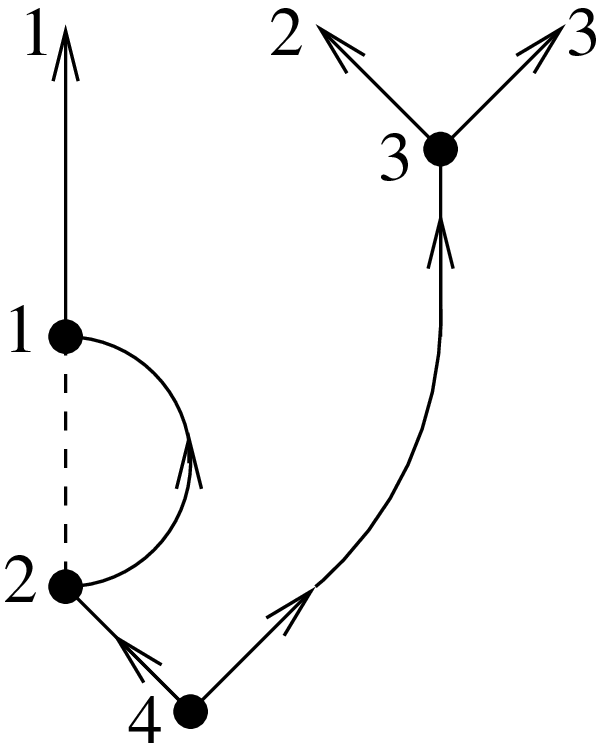} &d_9  = 12\nn \\
 I_9 &= \frac{19+5 k^2}{36(2+k^2)(3+k^2)}   \\
T_{10}&=\diagram{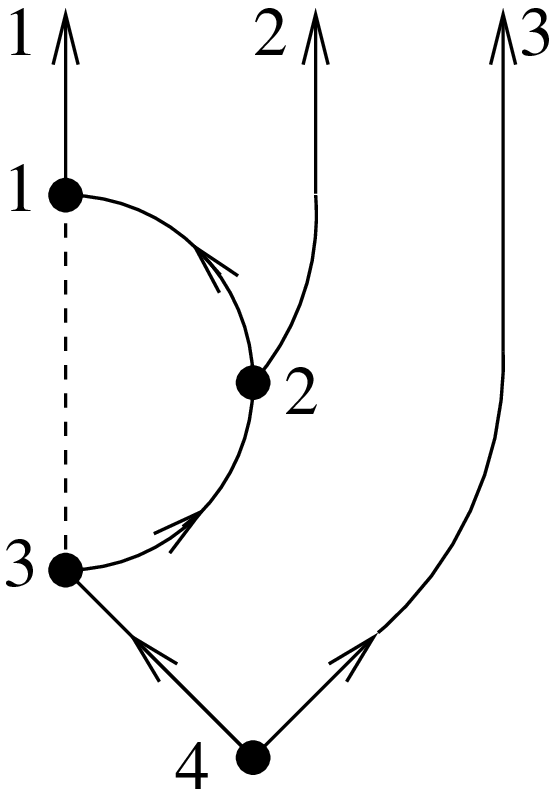} \qquad &d_{10}  = 24 \nn \\
 I_{10} &= \frac{1}{6(2+k^2)(3+k^2)} \\
T_{11}&=\diagram{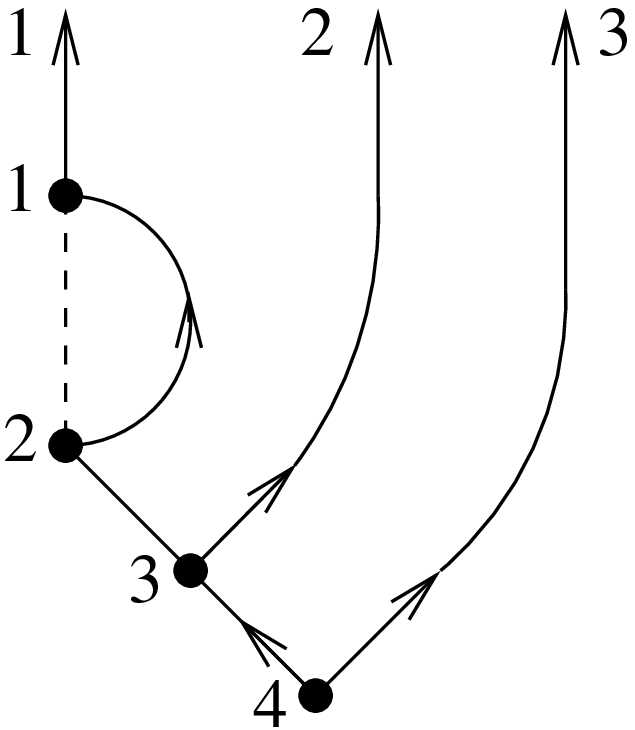} \qquad &d_{11} = 24 \nn \\
I_{11} &=\frac{1}{18(3+k^2)}  
 \end{align}
This calculation illustrates how the complexity increases formidably with the order,
and how powerful the algebraic method developed in section \ref{s:loops} is in summing these contributions.

\section{Series expansion of the $a_n$}
\label{a:bj}

The $b_j$ defined in the text can be obtained, for $j \geq 3$ as
\bea
 b_k &=& \frac{-16 + 2^k \times 7 - k [10 - 2^k + 3 k(k+1)] }{k (k+1) (k-1)(k-2)}\nn \\
&& + 6\, \Phi_{{\rm L}}(-1,1,k-2) 
\ ,
\eea
where $\Phi_{{\rm L}}(a,b,c)$ is the Lerch-$\Phi$ function. 

\section{Small-velocity behaviour} \label{app:small} 
Let us discuss  in more detail the expansion of $\delta {\cal P}(\dot u)$ at small
$\dot u$, looking also at subdominant terms. Denoting $s:=-\lambda$ and thus
$\kappa=s/(1+s)$, we can expand at large $s$,
\be 
{\rm Li}_j(\kappa) = -  \ln(1-\kappa) \frac{(\ln \kappa)^{j-1}}{\Gamma(j)} + \phi_j(\kappa)\ ,
\ee  
for $j=1,2,...$, where $\phi_j(\kappa)$ is analytic around $\kappa=1$ and $\phi_j(1)=\zeta(j)$.
Hence 
\bea
 {\rm Li}_j\left(\frac{s}{1+s}\right) &=&  \ln(1+s) \frac{[- \ln(1+ \frac{1}{s})]^{j-1}}{\Gamma(j)} + \zeta(j) \nn\\
&& + \sum_{p=1}^\infty \frac{d_{jp}}{s^p} \nn \\
& =& \frac{\ln s}{\Gamma(j) s^{j-1}} \left(1 + O\left(\frac{1}{s}\right)\right) + \zeta(j) + O\left(\frac{1}{s}\right)
\nn\eea 
We also have
\bea
 -2 \sum_{n=1}^\infty \frac{\ln n}{n} \left(\frac{s}{1+s}\right)^{\!\!n} &=& - (\ln s)^2 + \ln s\left (2 \gamma_{\rm E}  - \frac{1}{s} + ...\right ) \nn \\
&& + K + O\left(\frac{1}{s}\right) 
\ .
\eea 
Hence we find for large $s$
\bea
 \delta Z(\lambda=-s) &=& - (\ln s)^2 + (2 \gamma_{\rm E} + b_1) \ln s + \frac{\ln s}{s} (b_2 - 1) \nn \\
&& + O\left(\frac{\ln s}{s^2}\right)
+ {\rm analytic}
\ .
\eea 
We have the following Laplace transforms:\bea
&&{\mathrm{LT}}_{\dot u \to s} \frac{\ln \dot u}{\dot u} = \frac{1}{2} (\ln s)^2 + \gamma_{\rm E} \ln s  + {\rm analytic} ~~~~~~~~~~~~\\
&&{\mathrm{LT}}_{\dot u \to s} \frac{1}{\dot u} = - \ln s  \\
&&{\mathrm{LT}}_{\dot u \to s} \dot u^n \ln \dot u = - n! \frac{\ln s}{s^{n+1}} + \frac{B_n}{s^{n+1}} 
\eea 
for $n =0,1,...$, where in the first two lines the Laplace Transform is defined via the correctly subtracted formula.
We can surmise that
\bea
 \delta P(\dot u) &=& - \frac{4 \gamma_{\rm E} + b_1}{\dot u} -  2 \frac{\ln \dot u}{\dot u} \left[1   + \frac{1}{2}  \dot u \Big( b_2-1 + O(\dot u)\Big) \right] \nn \\
&& + K + O(\dot u)\ . 
\eea

\section{Adiabatically switching on of the disorder}\label{c7}
In this appendix, we recuperate the missing terms of the velocity theory, as discussed in section \ref{s:corrections}. It is suggestive from the discussion in that section, that these terms could be boundary terms, lost in a partial integration in time. Since the theory is causal, the time in question is  $t\to -\infty$; physically it is  related to the preparation of the system: Remind, that we crucially use that we are in the Middleton state. 

 In order  to be on the safe side, we could switch on the disorder adiabatically slowly, which will suppress any boundary terms at time $t=-\infty$, since there is no disorder at that time. 

Let us start from
the equation of motion for the velocity 
(for short-ranged elasticity, and a source $w_{xt}$ constant in space)\begin{equation}\label{c8}
 (\partial_t - \nabla_x^2 + 1) \dot u_{xt} = \partial_t \left[ F(v t + u_{xt},x) g_t  \right] + m^2 \delta \dot w_t 
\end{equation}
We have added   an {\em  adiabatic} factor
$g_{t} $ which can e.g.\ be chosen as 
\begin{equation}\label{gt-example}
g_{t}=\rme^{-\delta t}\ ,\qquad \mbox{with} \ \delta \to 0 .
\end{equation}
Note that the exact form is not crucial, but this particular choice will
simplify some of the ensuing calculations, since $g_{t} =g_{t-t'}  g_{t'} $. 
This gives
\begin{eqnarray}
 -  {\cal S} &=& - {\cal S}_0 -  {\cal S}_{\mathrm{dis}}  \\
-  {\cal S}_0 &=& \int_{xt} \tilde u_{xt} (\partial_t - \nabla_x^2 + 1) \dot u_{xt} \\
-  {\cal S}_{\mathrm{dis}} &=& 
\frac{1}{2} \int_{xtt'}  \tilde u_{xt}  \tilde u_{xt'}  \partial_t \partial_{t'} [ \Delta(v (t-t') + u_{xt}-u_{xt'}) g_t g_{t'} ] \nn \\
&=&-  {\cal S}_{\mathrm{dis}}^{(0)}-  {\cal S}_{\mathrm{dis}}^{(1)}-  {\cal S}_{\mathrm{dis}}^{(2)}\\
-  {\cal S}_{\mathrm{dis}}^{(0)} &=& 
\frac{1}{2} \int_{xtt'}  \tilde u_{xt}  \tilde u_{xt'}  g_t g_{t'}  \partial_t \partial_{t'}  \Delta(v (t-t') + u_{xt}-u_{xt'})\nn \\\\
-  {\cal S}_{\mathrm{dis}}^{(1)} &=& 
 \int_{xtt'}  \tilde u_{xt}  \tilde u_{xt'}\dot  g_{t} g_{t'} \partial_{t'}  \Delta(v (t-t') + u_{xt}-u_{xt'})  \\
-  {\cal S}_{\mathrm{dis}}^{(2)} &=& 
\frac{1}{2} \int_{xtt'}  \tilde u_{xt}  \tilde u_{xt'} \dot g_t \dot g_{t'}   \Delta(v (t-t') + u_{xt}-u_{xt'}) 
\end{eqnarray}
We now study corrections to $ {\cal S}_{\mathrm{dis}}$, which may intervene in our generating function $\rme^{\lambda \dot u(0)}$. Noting that all diagrams contain response-functions which decay in
time at least exponentially fast, or more precisely faster as
\begin{equation}\label{c9}
|R_{ktt'}| \le \rme^{-|t-t'|m^{2}}
\ ,\end{equation}
we have two types of  diagrams for our
new perturbation expansion (for the case of interest $\delta \to 0$): 
\begin{enumerate}
\item [(i)] {\em Connected Diagrams:} The disorder  vertex at time $t$ is attached to $t=0$ via a
string of response functions; then  we can make the replacement
$g_{t}\to 1$, and $\dot{g_{t}} \to 0 $. Especially this reproduces all
 diagrams of the velocity theory. Only the vertex \(S_{\rm dis}^{(0)} \) contributes. E.g.\ all diagrams given in appendix \ref{yyyyyyy} are of this form. 
\item [(ii)] {\em Disconnected Diagrams:} If the disorder at time $t$ is {\em not} attached to $t=0$ via a
string of response functions, then the integral over $\dot{g}_{t}
$ may produce a factor of $\int_{t}\dot{g_{t}} =1 $, even though $\dot g_t\sim \delta$. As a 
consequence, $-t$ is of order $1/\delta$, and all  response functions connected via a string of response functions to $t$ may have both
time-arguments at very large negative times, and thus are to be evaluated in
the {\em  flat} background $\tilde u_{t}^{0}=0$ (since  $\left< \tilde
u_{t} \right>\to 0$ for $t\to -\infty$.) (For an example see below). 
\end{enumerate}
We now discuss the leading-order correction. It comes from
a term with  {\em one} $\partial_t g_{t}$, i.e.\ from $-
 {\cal S}_{\mathrm{dis}}^{(1)}$: 
\begin{eqnarray}\label{c10}
-  {\cal S}_{\mathrm{dis}}^{(1)} &=&  \int_{xtt'}  \tilde u_{xt}  \tilde
u_{xt'}\dot  g_{t'} g_{t} \partial_{t}  \Delta(v (t-t') +
u_{xt}-u_{xt'})   \nonumber \\
&=& \int_{xtt'}  \tilde u_{xt}  \tilde
u_{xt'}\dot  g_{t'} g_{t}  (v+\dot u_{xt}) \Delta'\left(\int_{t'}^{t}
\rmd \tau [v  +
\dot u_{x\tau }] \right)    \nn\\
\end{eqnarray}
In order to conform to the rules discussed above, $\tilde u_{t}$
must somehow be connected to $t=0$, whereas $\tilde u_{t'}$ may
not. This gives the only  possible diagram
\begin{equation}\label{c11}
-  {\cal S}_{\mathrm{dis}}^{(1)}  \to  
\parbox{5.cm}{\begin{tikzpicture}
\draw[double distance=2pt] (1,1) arc (0:-180:0.5);
\draw[->,double distance=2pt] (0,1) arc (-180:-90:0.5);
\draw[dotted] (1,1) --  (2,1);
\draw[-to,shorten >=10pt,double distance=2pt] (2,1) -- (3,1);
\draw (2-.2,1+.2) -- (2,1);
\draw [dashed] (0,1) --  (1,1);
\node (t) at (-.2,1) {$t'$};
\node (tp) at (1.15,.75) {$\tau $};
\node (tp) at (2.1,.73) {$t$};
\fill (0,1) circle (2pt);
\fill (1,1) circle (2pt);
\fill (2,1) circle (2pt);
\end{tikzpicture}} 
\end{equation}
The times are $t'<\tau \ll t <0$, where only $t-\tau $ will become
very large, $\sim \frac{1}{\delta }$. Therefore we can set $\dot g_{t'} = g_{t'-\tau} 
\dot g_{\tau -t}  g_{t} \approx  \dot g_{t-\tau }$, and the
ensuing integral $\int_{\tau <t}\dot g_{\tau -t} = 1$. The dotted line
indicates this factor of $\int_{\tau <t}\dot g_{\tau
-t}$. Furthermore, since both times $\tau$ and $t'$ are very negative, 
the response function ${\mathbb{R}_{k\tau t'}} \to {{R}_{k\tau t'}}$. This gives 
\begin{eqnarray}\label{c12}
-  {\cal S}_{\mathrm{dis}}^{(1)} &\to& 
\parbox{5.cm}{\begin{tikzpicture}
\draw (1,1) arc (0:-180:0.5);
\draw[->] (0,1) arc (-180:-90:0.5);
\draw[dotted] (1,1) --  (2,1);
\draw (2-.2,1+.2) -- (2,1);
\draw[-to,shorten >=10pt,double distance=2pt] (2,1) -- (3,1);
\draw [dashed] (0,1) --  (1,1);
\node (t) at (-.2,1) {$t'$};
\node (tp) at (1.15,.75) {$\tau $};
\node (tp) at (2.1,.73) {$t$};
\fill (0,1) circle (2pt);
\fill (1,1) circle (2pt);
\fill (2,1) circle (2pt);
\end{tikzpicture}} \nn \\
&= &  \int_{k} \int_{t'<\tau<t}  (v+\dot u_{xt})  R_{k\tau t'} \dot g_{\tau
-t}   \Delta'' (0^{+}) \tilde u_{xt}+{O} (\delta) \nonumber \\
&=&   \Delta'' (0^{+}) \int_{k}\frac{1}{k^{2}+1} \int_{t}
(v+\dot{u}_{xt} )\tilde u_{xt} +{O} (\delta) 
\end{eqnarray}
At leading order we now have to replace the remaining fields by their expectations; we
also 
drop the term of ${O} (\delta) $:
\begin{eqnarray}\label{c13}
-  {\cal S}_{\mathrm{dis}}^{(1)} &\to& 
\parbox{5.cm}{\begin{tikzpicture}
\draw (1,1) arc (0:-180:0.5);
\draw[->] (0,1) arc (-180:-90:0.5);
\draw[dotted] (1,1) --  (2,1);
\draw[double distance=2pt] (1,2) --  (2,1);
\draw[-to,shorten >=10pt,double distance=2pt] (1,2) -- (2,1);
\draw[-to,shorten >=10pt,double distance=2pt] (2,1) -- (3,1);
\draw [dashed] (0,1) --  (1,1);
\node (v) at (.7,1.97) {$\dot w_{t_{2}}$};
\node (t) at (-.2,1) {$t'$};
\node (tp) at (1.15,.75) {$\tau $};
\node (tp) at (2.1,.73) {$t$};
\fill (0,1) circle (2pt);
\fill (1,1) circle (2pt);
\fill (2,1) circle (2pt);
\end{tikzpicture}} \nn \\
&=  &  v \Delta'' (0^{+}) \int_{k}\frac{1}{k^{2}+1}
\int_{t'<t<0} \mathbb{R}_{0tt_{2}}\,\tilde u^{0}_{t} \nonumber \\
&=&  v \Delta'' (0^{+}) \kappa  \int_{k}\frac{1}{k^{2}+1}
\end{eqnarray}
This is exactly the additional term found in Eq.~(\ref{3.72b}), or in the more rigorous derivation in Eq.~(\ref{c6}).

We also note that $ {\cal S}_{\mathrm{dis}}^{(2)}$ can not contribute (at
least at leading order), since we need to gain 2 {\em free} time
integrals. That implies that both response-fields must be contracted
inside the interaction, which is impossible due to causality.
However there will be a contribution at 2-loop order. 

Further we note that, in spirit, the above derivation is similar to the one given in section \ref{s:corrections}: In both cases, it was important that the second derivative of the disorder $\Delta''(v(t-t')+u_t-u_{t'})$, decays, as a function of the time-distance $t-t'$, to 0, which allows for a partial integration (eating up the time derivative $\dot g_t$).  

\hfill

\tableofcontents

\end{document}